\documentstyle[amsfonts,amssymb,epsfig]{elsart}

\newcommand{\ds}{{\vphantom{\dagger}}}

\newcommand{\etalia}{{\em et al.}}         
\newcommand{\ia}{\emph{i.a.}}              
\newcommand{\ie}{i.e.}                     
\newcommand{\eg}{e.g.}                     
\newcommand{\gc}{g.c.}                     
\newcommand{\Eq}[1]{Eq.~(\ref{#1})}        
\newcommand{\Eqs}[1]{Eqs.~(\ref{#1})}      
\newcommand{\Sec}[1]{Sec.~\ref{#1}}        
\newcommand{\Secs}[1]{Secs.~\ref{#1}}      
\newcommand{\Fig}[1]{Fig.~\ref{#1}}        
\newcommand{\Figs}[1]{Figs.~\ref{#1}}      
\newcommand{\Ref}[1]{Ref.~\cite{#1}}      
\newcommand{\Refs}[1]{Refs.~\cite{#1}}    
\newcommand{\App}[1]{App.~\ref{#1}}        
\newcommand{\ii}{{\rm i}}    
\newcommand{\re}{{\mathop{\mathrm{Re}}}}   
\newcommand{\ssg}{{\mathop{\mathrm{g}}}}   
\newcommand{\ssL}{{\mathop{\mathrm{L}}}}   
\newcommand{\ssR}{{\mathop{\mathrm{R}}}}   
\newcommand{\ssD}{{\mathop{\mathrm{D}}}}   
\newcommand{\tun}{{\mathop{\mathrm{tun}}}} 
\newcommand{\orth}{{\mathop{\mathrm{orth}}}} 
\newcommand{\sound}{{\mathop{\mathrm{sound}}}} 
\newcommand{\typ}{{\mathop{\mathrm{typ}}}} 
\newcommand{\orb}{{\mathop{\mathrm{orb}}}} 
\newcommand{\eff}{{\mathop{\mathrm{eff}}}} 
\newcommand{\tot}{{\mathop{\mathrm{tot}}}} 
\newcommand{\pure}{{\mathop{\mathrm{pure}}}} 
\newcommand{\ave}{{\mathop{\mathrm{ave}}}} 
\newcommand{\inel}{{\mathop{\mathrm{inel}}}} 
\newcommand{\bulk}{{\mathop{\mathrm{bulk}}}} 
\newcommand{\grain}{{\mathop{\mathrm{grain}}}} 
\newcommand{\so}{{\mathop{\mathrm{so}}}} 
\newcommand{\pert}{{\mathop{\mathrm{pert}}}} 
\newcommand{\tildez}{{\mathop{\mathrm{p}}}} 
\newcommand{\llangle}{{\mathop{\langle\!\langle}}} 
\newcommand{\rrangle}{{\mathop{\rangle\!\rangle}}} 
\newcommand{\tildezp}{{\mathop{\tilde {\mathrm{p}}}}} 
\newcommand{\can}{{\mathop{\mathrm{can}}}} 
\newcommand{\qp}{{\mathop{\mathrm{qp}}}} 
\newcommand{\gdc}{{\mathop{g_{\mathrm{dim}}}}}
\newcommand{\kF}{{\mathop{k_{\mathrm{F}}}}}
\newcommand{\vF}{{\mathop{v_{\mathrm{F}}}}}
\newcommand{\eF}{{\mathop{\varepsilon_{\mathrm{F}}}}} 
\newcommand{\kB}{{\mathop{k_{\mathrm{B}}}}}  
\newcommand{\Vol}{{\mathop{\mathrm{Vol}}}}   
\newcommand{\Ec}{{\mathop{E_{\mathrm{C}}}}}  
\newcommand{\KDelta}{d}  
\newcommand{\Tk}{{\mathop{T_{\mathrm{K}}}}}  
\newcommand{\ETh}{{\mathop{E_{\mathrm{Thouless}}}}}
\newcommand{\Ddiff}{{\mathop{D_{\mathrm{diff}}}}}
\newcommand{\ltr}{{\mathop{l_{\mathrm{tr}}}}}
\newcommand{\Epot}{{\mathop{E_{\mathrm{pot}}}}}
\newcommand{\Vg}{{\mathop{V_{\mathrm{g}}}}}  
\newcommand{\Tc}{{\mathop{T_{\mathrm{c}}}}}  
\newcommand{\muB}{{\mathop{\mu_{\mathrm{B}}}}} 
\newcommand{\VD}{{\mathop{V_{\mathrm{D}}}}}  
\newcommand{\Qex}{{\mathop{Q_{\mathrm{ex}}}}}
\newcommand{\Qexsq}{{\mathop{Q^2_{\mathrm{ex}}}}}
\newcommand{\Nex}{{\mathop{N_{\mathrm{ex}}}}}
\newcommand{\oneup}{{\mathop{ {1 \! \uparrow}}}}
\newcommand{\onedown}{{\mathop{ {1 \! \downarrow}}}}
\newcommand{\twoup}{{\mathop{ {2 \! \uparrow}}}}
\newcommand{\twodown}{{\mathop{ {2 \! \downarrow}}}}
\newcommand{\Gine}{{\mathop{\Gamma^{\mathrm{inel}}_\varepsilon}}}%
\newcommand{\Gbulke}{{\mathop{\Gamma^{\mathrm{bulk}}_{\varepsilon,
{\mathrm{GR}}}}}}       
\newcommand{\Gfinitee}{{\mathop{\Gamma^{\mathrm{finite}}_{\varepsilon,
{\mathrm{GR}}}}}}       
\newcommand{\hNex}{{\mathop{\hat N_{\mathrm{ex}}}}} 
\newcommand{\Nexsq}{{\mathop{N^2_{\mathrm{ex}}}}}
\newcommand{\E}{{\mathcal{E}}}               
\newcommand{\op}[1]{{\hat{#1}}}              
\newcommand{\bbalpha}{{\B}}         

\newcommand{\cond}{{\mathrm{cond}}}             
\newcommand{\PBCS}{{\mathrm{PBCS}}}          
\newcommand{\PP}{{\mathrm{P}}}               
\newcommand{\pb}{{\mathrm{pb}}}              
\newcommand{\GC}{{\mathrm{BCS}}}             
\newcommand{\BCS}{{\mathrm{BCS}}}            
\newcommand{\Pauli}{{\mathrm{Pauli}}}        
\newcommand{\MF}{{\mathrm{gc}}}            
\newcommand{\G}{{\mathrm{G}}}              
\newcommand{\F}{{\mathrm{F}}}              
\newcommand{\red}{{\mathrm{red}}}          
\newcommand{\ren}{{\mathrm{ren}}}          
\newcommand{\normal}{{\mathrm{normal}}}    
\newcommand{\elel}{{\rm{el-el}}}      
\newcommand{\el}{{\rm{el}}}           
\newcommand{\ML}{{\mathrm{ML}}}            
\newcommand{\Vac}{{\mathrm{Vac}}}          
\newcommand{\ex}{{\mathrm{exact}}}         
\newcommand{\omegaD}{{\mathop{\omega_{\mathrm{D}}}}} 
\newcommand{\baromegaD}{{\mathop{\bar \omega_{\mathrm{D}}}}} 
\newcommand{\CC}{{\mathrm{CC}}}      
\newcommand{\gol}{\stackrel{\scriptscriptstyle >}{\scriptscriptstyle
<}}                                        
\newcommand{\llogg}{\stackrel{\scriptscriptstyle <}{\scriptscriptstyle
>}}                                        
\newcommand{\N}{{\mathcal N}}          
\newcommand{\I}{I}          
\newcommand{\B}{B}          
\newcommand{\U}{U}          
\newcommand{\Hsw}{{\mathop{H_{\mathrm{sw}}}}}  
\newcommand{\Hspin}{{\mathop{\hat H_{\mathrm{spin}}}}}  
\newcommand{\anis}{{\mathop{{\mathrm{anis}}}}}  
\newcommand{\dbcsm}{discrete BCS model} 
\newcommand{\som}{spin-orbit model} 
\newcommand{\magrev}{magnetization reversal} 

\begin{document}

\begin{frontmatter}

\title{Spectroscopy of discrete energy levels in ultrasmall metallic grains}


\author[Karlsruhe]{Jan von Delft\thanksref{jansupport}}
\and \author[Cornell]{D. C. Ralph\thanksref{dansupport}}
\address[Karlsruhe]{Institut f\"ur Theoretische Festk\"orperphysik, \\
    Universit\"at   Karlsruhe, D-76128 Karlsruhe,   Germany}
\address[Cornell]{Laboratory of Atomic and Solid States 
Physics, Cornell
    University, \\Ithaca, New York 14853  
 \vspace{3mm}\\  {\rm March 30, 2000 
}}
\thanks[jansupport]{\emph{E-mail address:}
vondelft@th.physik.uni-bonn.de}
\thanks[dansupport]{\emph{E-mail address:} ralph@msc.cornell.edu}

\begin{abstract}
  We review recent experimental and theoretical work on ultrasmall metallic
  grains, i.e.\ grains sufficiently small that the conduction electron energy
  spectrum becomes discrete.  The discrete excitation spectrum of an
  individual grain can be measured by the technique of single-electron
  tunneling spectroscopy: the spectrum is extracted from the current-voltage
  characteristics of a single-electron transistor containing the grain as
  central island.  We review experiments studying the influence on the
  discrete spectrum of superconductivity, nonequilibrium excitations,
  spin-orbit scattering and ferromagnetism.  We also review the theoretical
  descriptions of these phenomena in ultrasmall grains, which require
  modifications or extensions of the standard bulk theories to include the
  effects of level discreteness.
\end{abstract}

\end{frontmatter}

\newpage

\tableofcontents

\section{Introduction}
\label{sec:introduction}

One of the most fundamental features of quantum mechanics is the fact
that the energy spectrum of a system of particles confined to a small
region is {\em discrete\/} or quantized, with the typical spacing
between energy levels increasing with decreasing system size.  In
atomic and nuclear physics, spectroscopic techniques for measuring and
analyzing such discrete spectra have for decades been a major source
of detailed information on the forces between the particles and the
correlations which they experience.  In condensed matter physics,
however, it has been much more difficult to
spectroscopically study the discrete spectrum of an individual
sample, since system sizes were typically so large that discrete
eigenenergies could not be resolved on the energy scale set by the
temperature.

This changed in the course of the last 15 years due to advances in
microfabrication techniques, which made it possible to study
individual systems of mesoscopic or nanoscopic dimensions, whose
characteristic length scales range from a few $\mu$m down to a few nm.
In the early 1990's, semi-conductor devices were used to fabricate the
first ``quantum dots'', i.e.  droplets of charge confined to a
2-dimensional region so small (radius of order 50~nm) that discrete
levels in the conduction electron spectrum \cite{johnson92,goldman}
could be resolved at dilution refridgerator temperatures in the
10-100~mK range. The technique by which this was done is called
\emph{single-electron-tunneling spectroscopy:} the dot is connected to
two leads via electrostatically-defined tunnel barriers to form a
so-called single-electron transistor (SET), the current-voltage
characteristics of which are measured and analyzed.  Under certain
conditions the conductance shows well-defined resonances which can be
associated with tunneling through discrete eigenstates of the dot.
Since quantum dots exhibited various features familiar from atomic
physics, such as energy shells that feature magic numbers and are
filled according to Hund's rules, etc., they are often aptly referred
to as ``artificial atoms'' (see e.g.\ the collection of reviews in
\cite{curacao}, in particular that by Kouwenhoven \etalia\ 
\cite{kouwenhoven-curacao}).

In the mid 1990's, a similar advance was achieved with \emph{metals}
\cite{rbt95,rbt96a,rbt96b,black-thesis,ralph-curacao,rbt97}, when
Ralph, Black and Tinkham (RBT) succeeded to perform
single-electron-tun\-ne\-ling spectroscopy on \emph{individual
  ultrasmall metallic grains} (of radii $r \lesssim 5$~nm and mean
level spacings $d \gtrsim 0.1$meV): by attaching such a grain via
oxide tunnel barriers to two leads they constructed a single-electron
transistor having the grain as central island, and showed that a
well-resolved, discrete excitation spectrum could indeed be extracted
from the conductance.  This opened up a new frontier in the study of
electron correlations in metals, since the ability to resolve discrete
energy levels allows the nature of electron correlations to be studied
in unprecedented detail.  During the last several years,
single-electron-tunneling spectroscopy of ultrasmall metallic grains
has been used to probe superconducting pairing correlations in Al
grains \cite{rbt96a,rbt97}, nonequilibrium excitations
\cite{rbt97,agam97a,agam97b,agam98} and spin-orbit interactions
\cite{rbt95,rbt96b,salinas99,davidovich99,davidovich00} in normal
grains, and ferromagnetic correlations in Co grains
\cite{desmicht98,gueron99}.  A brief overview of some of the early
experiments is given in \Ref{ralph-curacao} by Ralph \etalia, a more
thorough one in the thesis of Black \cite{black-thesis}.

Although tunnel-spectroscopic studies of metallic grains are similar
in spirit to those of semiconductor quantum dots, there are a
number of important differences:
\begin{enumerate}
\item Metals have much \emph{higher densities of states} than
  semiconductors (because the latter have smaller electron densities
  and effective masses), hence metals require much smaller sample
  sizes ($\lesssim 10$~nm) before discrete levels become resolvable.
\item Consequently, metallic grains have much \emph{larger charging
    energies} than quantum dots, which is an advantage when
  fluctuations in electron number are to be minimized.  On the other
  hand, this also implies that for quantum dots the electron number
  can be varied over a much larger range than for metallic grains, 
 which is useful for analyses
  of statistical properties.
\item For metallic grains, the ability to study a range of materials
  throughout the periodic table, including samples doped with
  impurities (which tend to trap or deplete electrons in quantum dots) and
  alloys, allows some control over the strength and type of electron
  interactions to be studied.  In particular, one may study
    \emph{superconductivity} and itinerant \emph{ferromagnetism}.
\item For metallic grains, the tunnel barriers to the leads are
  insulating oxide layers and hence insensitive to applied voltages,
  whereas for quantum dots they are electrostatically defined and hence
  \emph{tunable}.  Tunability is usually an advantage, but not
  always: for example, \emph{nonequilibrium effects} are easier to
  study quantitatively for a metallic grain than a quantum dot,
  because for the latter a large source-drain voltage lowers the
  tunnel barrier in poorly-controlled ways.
\item \emph{Spin effects} are easily probed in metallic grains by
  applying a magnetic field and studying the Zeeman-splitting of
  time-reversed pairs of states, which for nm-scale metallic grains is
  a much stronger effect than that of the applied field on orbital
  properties.  In contrast, in semiconductors the latter are dominant
  over spin effects.
\item For the same reason, \emph{spin-orbit} effects are more easily
  studied in metallic grains than in quantum dots.
\end{enumerate}

A particularly interesting feature of experiments on ultrasmall
metallic grains is that they probe the ways in which finite-size
effects modify a system's characteristic correlations relative to
their bulk properties, both due to mesoscopic fluctuations and via
level discreteness.  Such modifications arise when the new energy
scale characterizing the spectrum's discreteness, namely the
single-particle mean level spacing $d = 1 / \N (\varepsilon_F) \sim 1
/ \mbox{Vol}$ [where $\N (\varepsilon)$ is the density of states per
spin species], becomes comparable to the energy scale characterizing
the correlations in bulk systems (such as the energy gap in
superconductors or the Kondo temperature in magnetic alloys).  Such
quantum finite size effects in metals had attracted considerable
attention in the past, but could hitherto only be studied in
ensemble-averaged quantities (for reviews, see \cite{wyder,halperin}).
Spectroscopic studies of discrete spectra of {\em individual\/} grains
yield significant new information.  Let us briefly mention the most
important examples, organized according to the section in which they
will be discussed in detail later.

RBT were able to determine the \emph{number parity} (even or odd) of a
given grain [\Sec{sec:normal-grain-magfield}], by studying the
evolution of the discrete spectrum in an applied magnetic field: For
an odd grain, the ground state energy was observed to Zeeman-split in
a magnetic field $H$, as expected for a spin-1/2 Kramers doublet;
moreover, for pure Al grains, the Land\'e $g$ factor extracted from
the size of the splitting was close to the expected value of $g^\pure
= 2$. In contrast, for an even grain the ground state is a
nondegenerate spin singlet, and accordingly no ground state splitting
was observed.

Parity effects were also observed in RBT's experiments on largish ($r
\gtrsim 5$ nm) Al grains \cite{rbt96a,rbt97,vondelft96}: 
an even grain had a distinct spectroscopic
gap $(\gg d)$ but an odd grain did not, which is clear evidence for the
presence of \emph{superconducting pairing correlations} in these
grains [\Sec{sec:superconductivity}].  The spectroscopic gap for even
grains was driven to zero by an applied magnetic field, hence the
paramagnetic breakdown of pairing correlations could be studied in
detail. The corresponding theory was worked out by Braun, von Delft,
Ralph and Tinkham \cite{braun97,braun99}.

In RBT's smallest grains ($r \lesssim 3$ nm), however, no such
distinct spectroscopic gap could be discerned.  This observation
revived an old but fundamental question [\Sec{sec:crossover}]:
\emph{What is the lower size limit for the existence of
  superconductivity in small grains?}\/, and thereby stimulated a
large number of theoretical investigations
\cite{vondelft96,braun97,braun99,smith96,balian-short,balian-long,%
bonsager98,matveev97,Rossignoli-98,Rossignoli-99a,Rossignoli-99b,%
Rossignoli-00,mastellone98,berger98,braun98,dukelsky99a,dukelsky99b,%
braun-vieweg,sierra99,vondelft-ankara99,dukelsky99c,tian99,%
tanaka99,dilorenzo99}.  Anderson \cite{anderson59} had addressed this
question already in 1959, arguing that ``superconductivity would no
longer be possible'' when the mean level spacing $d$ becomes larger
than the bulk gap, to be denoted by $\tilde \Delta$, because then $
\tilde \Delta$ looses its special significance as gap in an otherwise
continuous spectrum.  RBT's new experiments stimulated a number of
theoretical attempts to quantitatively describe the \emph{crossover}
from the bulk limit $d \ll \tilde\Delta$, where superconductivity is
well-developed, to the fluctuation-dominated regime of $d \gg \tilde
\Delta$, where pairing correlations survive only in the form of weak
fluctuations.  Describing this crossover constituted a conceptual
challenge, since the standard grand-canonical mean-field BCS treatment
of pairing correlations
\cite{vondelft96,braun97,braun99,smith96,balian-short,balian-long,%
  bonsager98,matveev97} breaks down for $d \gtrsim \tilde \Delta$.
This challenge elicited a series of increasingly sophisticated
canonical treatments of pairing correlations
\cite{mastellone98,berger98,braun98,%
dukelsky99a,dukelsky99b,braun-vieweg,sierra99,%
vondelft-ankara99,dukelsky99c}, based on a simple reduced
BCS-Hamiltonian for discrete energy levels, which showed that the
crossover is completely smooth, but, interestingly, depends on the
parity of the number of electrons on the grain, as pointed out by von
Delft \etalia\ \cite{vondelft96}.  Very recently, the main conclusions
of these works were confirmed \cite{sierra99} using an exact solution
of the reduced BCS model, discovered by Richardson in the context of
nuclear
physics in the 1960s \cite{richardson63a,richardson63b,richardson64,%
  richardson65a,richardson65b,richardson66,richardson66-b,richardson67,%
  richardson77}.  (The existence of this solution came as a surprise
-- in the form of a polite letter from its inventor -- to those
involved with ultrasmall grains, since hitherto it had apparently
completely escaped the attention of the condensed-matter community.)

An interesting finite-size effect was also revealed in
\emph{nonequilibrium} grains [\Sec{sec:nonequilibrium}], for which BRT
observed the excitation spectrum to consist of \emph{clusters of
  resonances}, with the spacing between clusters comparable to
free-electron-estimates of the single-particle mean level spacing $d$,
but the spacings between subresonances of the same cluster much
smaller than $d$. A theory for this effect was developed by Agam
\etalia\ \cite{agam97a,agam98}, who showed that clusters of resonances
can be caused by nonequilibrium excitations on the grain, provided
that mesoscopic fluctuations of the matrix elements of the
electron-electron interaction are sufficiently strong.  Such
fluctuations are neglected in the so-called ``orthodox model'' that is
commonly used to describe single-electron transistors, but are
expected, according to the general theory of disordered interacting
electron systems (\eg\ \cite{agam98}), to become significant for
sufficiently small grains, and can be described using random matrix
theory.  The experimental observation of such clusters thus
constitutes a beautiful and direct illustration of the importance of
mesoscopic fluctuations in ultrasmall grains.

A further example of such mesoscopic fluctuations was published by
Salinas \etalia\ \cite{salinas99}, who experimentally studied the
effect of the \emph{spin-orbit interaction} [\Sec{sec:spin-orbit}] in
Al grains doped with Au, and by Davidovi\'c and Tinkham
\cite{davidovich99,davidovich00}, who studied Au grains. These authors
observed effective $g^\eff$ factors significantly smaller than
$g^\pure = 2$, as expected, since the spin-orbit interaction mixes
states with opposite spin. Interestingly, the measured $g^\eff$
factors were also found to vary from one Kramers doublet to the next,
which is a clear signature for mesoscopic fluctuations.  Very
recently, a theory for the statistics of the fluctuations of $g^\eff$
was worked out by Matveev, Glazman and Larkin \cite{matveev00} and by
Brouwer, Waintal and Halperin \cite{brouwer00}.

During the last year, Gu\'eron \etalia\ \cite{gueron99} published the
first set of detailed experimental results on \emph{ferromagnetic Co
  grains} [\Sec{sec:ferromagnetism}].  Among the novel features found
for these are a strong asymmetry between the tunneling probabilities
for spin-up or spin-down electrons, a larger-than-expected density of
low-lying excitations, and hysteretic behavior of the excitation
spectrum as function of an applied magnetic field. The properties of
the hysteresis loop reflect the changes in the direction of the
grain's magnetic moment as the applied field is ramped. The latter
features hold great promise for being a potential tool with which to
study the dynamics of magnetization reversal in individual nm-scale
ferromagnets.  No detailed theory for these experiments exists at
present.

An ultrasmall grain containing a single magnetic impurity can in
principle be used to study the effect of level discreteness on the
\emph{Kondo effect} [\Sec{sec:kondo-box}].  A theory for such a
``Kondo box'' was worked out by Thimm, Kroha and von Delft
\cite{Thimm}.  They found that when the mean level spacing in the
grain becomes larger than the Kondo temperature, the Kondo resonance
is strongly affected in a way that depends on the parity of the number
of electrons on the grain, and that should be detectable by
tunnel-spectroscopic measurements.

The present review summarizes the above developments.  The guiding
principle in the choice of topics was to focus on the experimental
studies of discrete spectra in ultrasmall metallic grains that have been
carried out since 1995 using single-electron-tunneling spectroscopy,
and on those theoretical developments that were directly inspired by
them. The notation to be used throughout is introduced in
\Sec{sec:setspectroscopy}, which describes how tunneling spectroscopy
works in practice and reviews the well-known orthodox theory for
single-electron transistors (also reviewed in, \eg,
\Refs{averin-likharev,grabert92,schoen98}), emphasizing those features
that have special relevance for ultrasmall grains. The contents of
\Secs{sec:normal-grain-magfield} to \ref{sec:kondo-box},
which for the most part can be read independently
of each other, were already outlined above. The main conclusions
of each section are summarized concisely in \Sec{sec:summary}, which
also gives an outlook towards directions for further work.

For each subject, an attempt was made to give a detailed account of
the arguments and calculations that have a direct bearing on
understanding experimental data, and to qualitatively explain the
theoretical ideas required for their interpretation.  Theoretical
developments beyond those of direct relevance to experiment are
usually either summarized or only briefly mentioned, but seldom
reproduced in detail.  The fact that more than a third of the review
is devoted to superconductivity in ultrasmall grains is a reflection
of the number of papers that have appeared on this subject since 1995.

\newpage \section{Single-electron-tunneling spectroscopy}
\label{sec:setspectroscopy}

The experiments in which Ralph, Black and Tinkham (RBT) succeeded to
resolve discrete energy levels of an ultrasmall
metallic grain used the technique of single-electron-tunneling
spectroscopy.  In \Sec{subsec:ultrasmallset} we explain the
idea behind this technique and illustrate its capabilities by showing
some representative data sets (without delving into the interesting
physics contained in this data, which will be done in later sections).
In \Sec{subsec:theoretical background} we estimate theoretically under
which conditions discrete states should be resolvable, in
\Sec{sec:theoryofultrasmallSET} develop a theoretical description for
ultrasmall single-electron transistors (SETs), 
and in \Sec{subsec:experimentaldetails} discuss
experimental details concerning fabrication and measurement
techniques.

\subsection{Ultrasmall single-electron transistor}
\label{subsec:ultrasmallset}

In the first generation of experiments of 1995
\cite{rbt95,rbt96a,rbt96b}, a grain made from Al (a superconducting
material) was connected to two metal leads via high-resistance tunnel
junctions, with capacitances $C_\ssL$ and $C_\ssR$, say.  In the next
generation of 1997 \cite{rbt97}, the grain was also coupled
capacatively to a gate, with capacitance $C_\ssg$.  The resulting
device, schematically depicted in \Fig{fig:sample}(a), has the
structure of a SET, with the grain as central island.  The circuit
diagram for an SET is shown in \Fig{fig:circuit-diagram}(b).  Applying
a bias voltage $V$ between the two leads causes a tunnel current $I$
to flow between the leads through the grain, via incoherent sequential
tunneling through the tunnel junctions.  The current can be influenced
by changing the gate voltage $\Vg$ (hence the name ``transistor''),
which tunes the electrostatic potential on the grain and thereby also
its average number of electrons $N$.  (For devices without a gate
these two quantities cannot be tuned and instead have some
sample-dependent, fixed value. For such devices, set $C_\ssg = 0$ in
all formulas below.)

The physics of SETs had been clarified in the early 1990s
\cite{curacao} through extensive studies of lithographically defined
SETs of {\em mesoscopic\/} size, \ie\ with {\em micron}-scale central
islands. The fundamentally new aspect of RBT's work was that their
SETs, made by a novel fabrication technique (described in
\Sec{subsec:experimentaldetails}), were {\em nanoscopic\/} in size:
they had ultrasmall grains with radii between 15nm and 2nm as central
islands, which were thus several orders of magnitude smaller in volume
than in previous experiments.  This had two important consequences:
\begin{figure}[t]
\centerline{\epsfig{figure=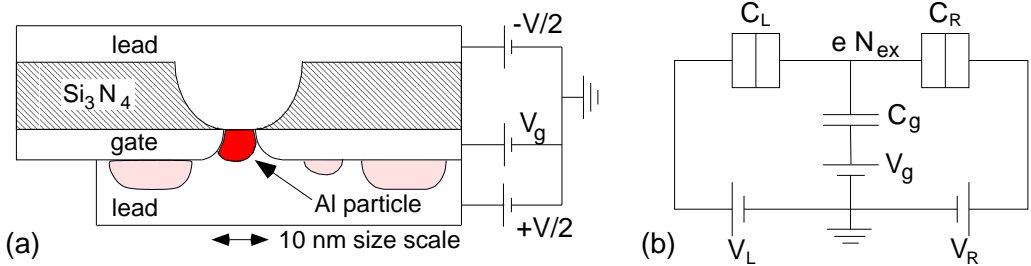,%
width=0.98\linewidth}}
  \caption[Schematic depiction of an ultrasmall SET]{
   (a) Schematic cross section of the ultrasmall SETs 
       studied by RBT in \protect\cite{rbt97}, and (b)  
   the corresponding circuit diagram.
 }
   \label{fig:sample} \label{fig:circuit-diagram}
\end{figure}
\begin{enumerate} 
\item {\em The grain's charging energy\/} $\Ec \equiv e^2/2C$ {\em was
    much larger than for mesoscopic SETs\/}, ranging roughly between 5
  and 50~meV (where $C \equiv C_\ssL + C_\ssR + C_\ssg$). $\Ec$ is the
  scale that determines the energy cost for changing $N$ by one.
  Since for ultrasmall grains it far exceeds all other typical energy
  scales of the SET, such as those set by the bias voltage ($V
  \lesssim 1$~mV), the temperature ($T \lesssim 4.2$~K) and the bulk
  superconducting gap for Al ($\Delta_\bulk = 0.18$~meV), fluctuations
  in electron number are strongly suppressed.
\item {\em Discrete eigenstates of the conduction electron energy
    spectrum became resolvable\/} -- their mean level spacing $d$
  ranged from 0.02 to 0.3~meV, which is in order-of-magnitude
  agreement with the free-electron estimate of $d = 2 \pi^2 \hbar^2/(m
  \kF \Vol)$ for the single-particle level spacing. 
 Such $d$-values are much larger than $\kB T$ for the
  lowest temperatures attained (around $T \simeq 30 \mbox{mK}$), but
  on the order of $\Delta_\bulk$.  However, the number of conduction
  electrons for grains of this size is still rather large (between
  $10^4$ and $10^5$).
\end{enumerate}
Since the two scales $\Ec$ and $d$ differ by at least an order of
magnitude, they manifest themselves in two distinct and easily
separable ways in the low-temperature $I$-$V$ curves of the devices,
shown in \Fig{fig:generic-IV} for a series of different $\Vg$
values:
\begin{enumerate}
\item When $V$ is varied on a large scale of tens of mV
  [\Fig{fig:generic-IV}], the $I$-$V$ curves have a typical
  ``Coulomb-staircase form'' characteristic of SETs: zero current at
  low $|V|$ (the ``Coulomb blockade'' regime), sloping steps equally
  spaced in $V$, and step thresholds sensitive to $\Vg$.  This proves
  that the tunnel current flows only through {\em one\/} grain. The
  maximal width of the flat step of zero current around $|eV|=0$ is
  governed, in order of magnitude, by $\Ec$
  and typically varies between 5 and 50~mV.  As $V_g$ is varied, the
  $I$-$V$ curves periodically repeat, with a period   $e/C_\ssg$.
\item When $V$ is varied on the much smaller scale of a few mV around
  the threshold of the Coulomb blockade regime and the temperature is
  sufficiently low ($T \ll d$), the $I$-$V$ curves have a step-like
  substructure, shown in \Fig{fig:set-transport}(b) [see also
  \Fig{fig:drago-set}(b)].  As first pointed
  out by Averin and Korotkov \cite{averin90}, such small steps in the
  $I$-$V$ curve are expected to occur whenever the voltage drop across
  one of the tunnel junctions equals the threshold energy at which the
  rate for tunneling across that junction into or out of one of the
  grain's {\em discrete energy eigenstates\/} becomes nonzero, since
  this opens up another channel for carrying current across that
  junction.  Correspondingly, the differential conductance $({\rm d}I
  / {\rm d} V)$ curves contain a series of fine peaks, see
  \Fig{fig:set-transport}(b).  Under certain conditions described in
  \Sec{sec:extractingspectrum}(a), the distances between these peaks
  directly reflect the energy differences between eigenenergies of
  same-$N$ eigenstates of the grain.  {\em Such conductance curves
    thus directly yield the grain's fixed-$N$ excitation spectrum}, or
  more precisely, the set of energy differences
  \begin{equation}
    \label{eq:fixed-N-spectrum}
    \delta \E_{\alpha \alpha'}^N = \E^N_{\alpha} - \E^N_{\alpha'}
  \end{equation}
  for those eigenstates $| \alpha \rangle_N$ of the $N$-electron grain that
  are accessible final states for a tunneling process that removes or
  adds an electron to the grain if its initial ground state has $(N+
  1)$ or $(N-1)$ electrons, respectively.
\end{enumerate}
The magnetic-field dependence of the fixed-$N$ excitation spectrum can
be obtained by simply tracing the motion of the conductance peak
positions as a magnetic field is turned on (at fixed $\Vg$). Examples
are given in \Fig{fig:kramers} in \Sec{sec:kramers}, 
\Fig{fig:sc-magneticfield} in \Sec{sec:sc-measuredgap},
\Fig{fig:spin-orbit} in \Sec{subsec:spin-orbit-exp}
and \Fig{fig:gueron} in \Sec{sec:Co-experiments}. If a gate is
present, two further very interesting options exist: Firstly, by
tuning $\Vg$ by an amount large enough ($\simeq \Ec/e$) to change $N$
by one unit, the {\em influence on the spectrum of the parity\/} of
the number of electrons on the grain can be studied.  Secondly, by
tuning $\Vg$ such that the Coulomb blockade regime is large or small,
so that the $V$-threshold at which current begins to flow is large or
small, {\em nonequilibrium effects can be maximized or minimized},
respectively, depending on whether one chooses to study them or not.
\begin{figure}[t]
\centerline{\epsfig{figure=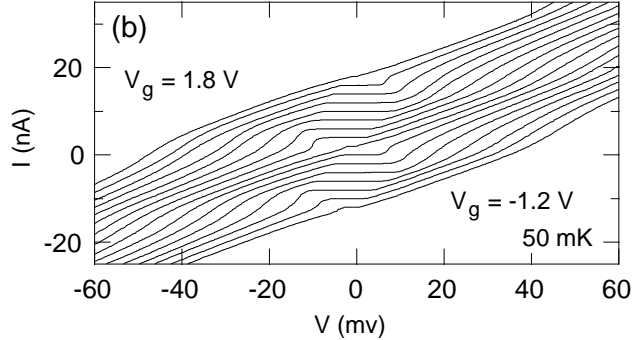,%
width=0.6\linewidth}}
  \caption[Current-voltage characteristics of ultrasmall Al 
  SETs]{Current-voltage curves for
    an ultrasmall SET \protect\cite{rbt97} at 50~mK, artificially
    offset on the vertical axis for a set of equally-spaced values of
    $\Vg$ between $-1.2$ and 1.8~V. The $I$-$V$ curves display
    Coulomb-staircase structure on a bias voltage scale of tens of mV.
    By fitting these to standard SET theory \cite{hanna91},
    the SET's
    basic parameters can be determined: $R_\ssL = 3.5~{\rm M}\Omega$,
    $R_\ssR = 0.2~{\rm M}\Omega$, $C_\ssL = 3.5$~aF, $C_\ssR =
    9.4$~aF, $C_\ssg = 0.09$~aF, $\Ec = 46$~meV. The grain 
    radius and mean level spacing are estimated
    as  $r \simeq 4.5$~nm and $d \simeq 0.45$~meV, using
    assumptions stated in \Sec{subsubsection:devicecharacterization}.}
\label{fig:generic-IV}
\end{figure}

\begin{figure}[t]
  \begin{center}
\epsfig{figure=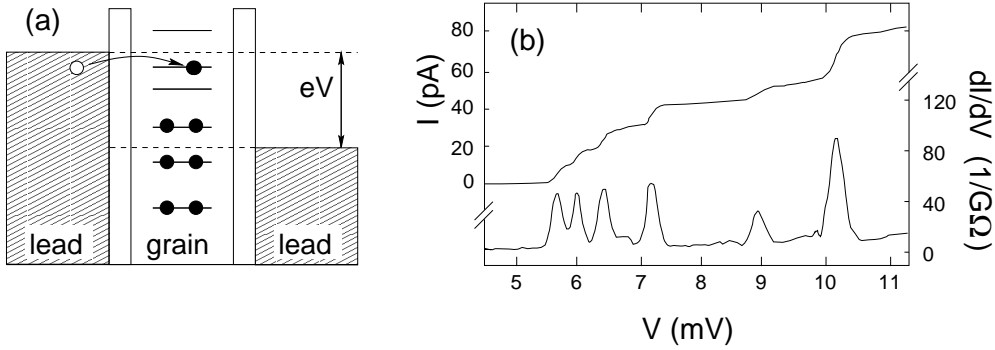,width=0.95\linewidth}
    \caption[${\d} I/{\d} V$-measurement]{(a) Cartoon 
      of an ${\d} I/ {\d} V$-measurement: The number of available
      transport channels through discrete states (here three) is
      determined by the bias voltage $V$. (b) The current and
      differential conductance as functions of bias voltage $V$ for
      one of RBT's ultrasmall grains; beyond the Coulomb-blockade
      threshold at around 5.5~mV, the current displays fine steps and
      the conductance fine peaks, on a voltage scale of a few mV,
      reflecting the grain's discrete eigenspectrum.}
    \label{fig:set-transport}
  \end{center}
\end{figure}

\subsection{Conditions under which discrete
states are resolvable}
\label{sec:theoreticalestimates}
\label{subsec:theoretical background}

Having shown in \Fig{fig:set-transport} an example of the data by
which RBT demonstrated that discrete energy levels of an ultrasmall
metallic grain can be resolved, let us take a step back and recall the
theoretical arguments, mainly due to Averin and Korotkov
\cite{averin90}, for why this should be possible. This requires
comparing various characteristic energy scales of a 3D grain and their
dependence on its size, such as the mean level spacing $d$, the
charging energy $\Ec$, the Thouless energy $\ETh$, and the amount of
level broadening due to inelastic relaxation processes (with rate
$ \Gamma^\inel$) and tunneling (with rate $\Gamma_\tun$).

The mean level spacing $d = 1/ \N(\eF)$
can be estimated  using the free-electron expression for 
the density of states at the Fermi surface of a 3D grain:
\begin{eqnarray}
  \label{eq:d-estimate}
d = {2 \pi^2 \hbar^2
\over m \kF \Vol} = {1.50 \, 
 \mbox{eV} . \mbox{nm}^2 \over  \kF \Vol} \; ,
\end{eqnarray}
where, for example,
$\kF = 17.5$ $ {\rm nm}^{-1}$ for Al and 
$\kF = 12.1$ $ {\rm nm}^{-1}$ for Au.
 For the sake of
order-of-magnitude estimates below,
we shall use $\kF = 10$ $ {\rm nm}^{-1}$ 
and crudely assume the grain to have
linear dimension $r$, volume $r^3$ and contact area $r^2$ (via tunnel
junctions) with each lead.

Next we consider the charging energy $\Ec = e^2/2C$. 
The capacitances $C_\ssL$ and $
C_\ssR$ to the left and right leads
are each of order $C_\Box r^2$, where $C_\Box$ is the
capacitance per unit area of the tunnel junctions, typically of order
$C_\Box \simeq 0.05 {\rm aF/nm}^2$ \cite{Lu98}; 
if a gate is present, $C_\ssg$
is typically an order of magnitude smaller. We therefore
have 
\begin{eqnarray}
  \label{eq:estimate-Ec}
\Ec \simeq {e^2 \over 4  C_\Box r^2}
 \simeq{ 0.8 \, \mbox{eV} . \mbox{nm}^2 \over r^2 } \; .
  \end{eqnarray}

Although the mean level spacing increases faster ($r^{-3}$) with
decreasing size than the charging energy ($r^{-2}$), they become
comparable only for grains of almost atomic size ($r \simeq 2$~\AA).
For nm-scale grains (say $r = $ 15 -- 2 nm), $d$ is still
significantly smaller (0.04 -- 20 meV) than $\Ec$ (4 -- 200 meV).
Moreover, the total number of conduction electrons $N \propto \eF/d$
is still very large ($\simeq$ 300\,000 -- 500, for $\eF \simeq
10$~eV).  As long as $r$ is substantially larger than the Thomas-Fermi
screening length $\lambda_{\rm TF} \simeq 1/\kF$, nm-scale grains can
still be regarded as metallic.

Since $d \ll \Ec$ for nm-scale SETs, the \emph{large-scale} features
of their current-voltage characteristics [\eg\ \Fig{fig:generic-IV}]
can be correctly described \cite{averin90} using continuous spectra
for both island and leads and adopting the standard ``orthodox''
theory \cite{averin-likharev} for Coulomb-blockade phenomena in
metallic SET's.  For additional \emph{fine-structure} due to a
discrete spectrum to be resolvable, three more
conditions must be met:\\
(i) To avoid thermal smearing, $T \ll d$ is required.
\\
(ii) $\Gamma_\alpha^r$, the tunneling rate out of a given discrete
state $|\alpha \rangle$ on the grain into lead $r$~(=L,R), must be
small enough that the tunneling-induced level widths $\hbar
\Gamma_\alpha^r$ don't cause neighboring levels to overlap, \ie\ 
$\hbar \Gamma_\alpha^r \ll d$; but this condition is equivalent to one
which must hold anyway for Coulomb blockade effects to occur
\cite{grabert92,schoen98}, namely $G_r \ll 2 e^2 / h$, where $G_r =
e^2 \N(\eF) \Gamma_\tun^r$ is, roughly, the total conductance across
junction $r$, and $ \Gamma_\tun^r$ the average of the
$\Gamma_\alpha^r$'s near $\eF$.
\\
(iii) For an excited state on the grain with energy $\varepsilon$
above the ground state to be resolvable, its inelastic relaxation rate
$\Gamma^\inel_\varepsilon$ must be small enough that the corresponding
line-width is less than the level spacing, $\hbar
\Gamma^\inel_\varepsilon \lesssim d$.  The question when this is
satisfied depends (neglecting phonons) on the grain's so-called
``dimensionless conductance'' $\gdc$, defined for finite systems by
\cite{AGKL97}
\begin{eqnarray}
  \label{eq:g-dimensionlessconductance}
  \gdc & \equiv &  \ETh / d ,  
\\
  \label{eq:define-Thouless}
\ETh & \simeq & \left\{ \begin{array}{ll}
{\hbar \Ddiff / r^2} =
{(0.25 \, \kF  \ltr /  r^2)} \, 
  \mbox{meV}   . \mbox{nm}^2  \quad
& \mbox{(diffusive)} \; , 
\\
 {\hbar \vF / (a 2 r)} = 
{(38 \, \kF  / a \, r)} \, 
  \mbox{meV}   . \mbox{nm}^2  \quad
& \mbox{(ballistic)} \; .
\end{array} \right. 
\end{eqnarray}
Here the Thouless energy $ \ETh$ is the inverse time for an electron
near $\eF$ to travel once across the system, $\Ddiff = \vF
\ltr/3 $ is the diffusion constant, 
$\ltr$ the transport mean free
path, and $a$ a geometrical constant.
\Eqs{eq:define-Thouless} are order-of-magnitude estimates\footnote{The
  \label{f:Ethouless}
  expressions for $\gdc$ and $\ETh$ given by various authors differ by
  factors of order unity.  (For $\gdc$ we followed \cite{AGKL97}, and
  for $\ETh$ \cite{matveev00} for the diffusive and
  \cite{Falko94,davidovich99} for the ballistic case.)  Note that the
  ballistic definition for $\ETh$ is similar to the diffusive one,
  except that the grain size determines the effective mean free path
  ($\ltr \to 3r/2a$).  The proper choice of $a$ depends on the precise
  geometry and the amount of disorder present \cite{agam98}; in
  practice, it is somewhat of a ``fudge factor''.  For a hemispherical
  grain, Davidovi\'c and Tinkham \cite{davidovich99} used $a=3$ (the
  number of dimensions); \label{f:drago} they argue that if the
  electronic motion in the grain is ballistic and surface scattering
  diffusive, an electron's mean-free path is approximately $2r$, but
  to diffuse through the entire particle volume, it should scatter
  from the surface about three times \cite{davidovich99}. For a
  pancake-shaped grain with thickness $z$, radius $r$ and diffusive
  dynamics, Agam \etalia\ \cite{agam97a} argued that $a \propto
  (r/z)^2$.
  } for diffusive or ballistic 3D grains, for which $\gdc \propto r$ or
$r^2$, respectively.  It was shown in \Refs{Sivan94,AGKL97} that for
grains so small\footnote{ 
\label{f:estimate-gdim}
For a ballistic, hemispherical Al grain, for example,
\Eqs{eq:d-estimate} to (\ref{eq:define-Thouless}) yield $\gdc = (16/a)
(r/\mbox{nm})^2$, which, for $a=3$, gives $\gdc \le 1$ if $r \le
0.4$~nm.  An experimental estimate \cite{agam97a} of $\gdc$ for two
ballistic Al grains with $\Vol \approx 40$ and $100 \,\mbox{nm}^3$
yielded values of $\gdc \approx 5$ in both cases, as discussed in
\Sec{sec:beyound-orthodox-model}.  If these grains are assumed to be
hemispherical, this would imply very large $a$ factors of $\approx 24$
and 42; this might imply that an assumption of pancake-shaped grains
would be more appropriate, for which $a$ does contain a large factor
$(r/z)^2$, cf.\ footnote~\ref{f:Ethouless} and, in
\Sec{sec:beyound-orthodox-model}, footnote~\ref{f:pancake-estimate}.}
that $\gdc \lesssim 1 $, the spectrum of low-lying excitations
consists of $\delta$-function like peaks with $\hbar
\Gamma^\inel_\varepsilon \ll d$ that can definitely be resolved
\cite{AGKL97}, whereas for larger grains with $g \gg 1$, the condition
$\hbar \Gamma^\inel_\varepsilon \lesssim d$ can be met as long as
$\varepsilon \lesssim \ETh$.  For grains with $d \ll \ETh \ll \Ec$,
the fine-structure due to level discreteness will thus be smeared out
for $\varepsilon \approx \Ec$, but should be clearly resolvable for
$\varepsilon \lesssim$ several units of $d$.  (For a more detailed
discussion of $\Gamma^\inel_\varepsilon$, including the effect of
phonons, see \Sec{sec:estimating-relaxation}.)

For future reference, we remark here that the parameter $\gdc$ also
controls the importance or not of corrections to
the orthodox model, which can be shown
\cite{Agam95,Blanter96,BMM-97,Blanter-Mirlin-97,Aleiner-Glazman-97,%
agam98} 
to be small as $1/\gdc $ [see \Sec{sec:beyound-orthodox-model}].

Finally, note that the amount of nonequilibrium effects on the grain
is controlled by the ratio of $ \Gamma_\tun^r$ to
$\Gamma^\inel_\varepsilon$.  If $ \Gamma_\tun^r \ll
\Gamma^\inel_\varepsilon$, nonequilibrium effects are negligible,
since an electron tunneling into an excited state of the grain has
plenty of time to relax before the next tunneling event.  Conversely,
nonequilibrium effects should become very important for $
\Gamma_\tun^r \gg \Gamma^\inel_\varepsilon$.  Indeed, they have been
observed directly \cite{rbt97,agam97a,agam97b,agam98} and are
discussed in \Sec{sec:nonequilibrium}.

\subsection{Theoretical description of an ultrasmall SET}
\label{sec:theoryofultrasmallSET}

In this section we set up a general formalism for describing transport
through ultrasmall metallic grains, in order to show explicitly how
level discreteness causes steps in the tunneling current and to
clarify precisely what information can be extracted from the latter.
Since ultrasmall grains may, due to their discrete states, be regarded
as ``metallic quantum dots'', we can carry over much of the formalism
developed in the literature for quantum dots; indeed, in spirit and
notation we shall very closely follow a review by Schoeller entitled
``Transport Theory of Interacting Quantum Dots''
\cite{schoeller-curacao}. Although this material may be familiar to
many readers, it is needed here to establish the notation to be used
throughout this review.

\subsubsection{Charging energy and Coulomb blockade}
\label{sec:chargingenergy}

We start by deriving the charging energy for the grain, using the
so-called ``orthodox'' or ``Coulomb-blockade'' model
\cite{averin-likharev}, which assumes the electrostatic potential to
be {\em homogeneous} on the grain.  This assumption was argued to be
reasonable as long as the Thomas-Fermi screening length $\lambda_{\rm
  TF} \simeq 1/\kF$ is very much shorter than the grain's linear
dimensions; a more rigorous condition is $\gdc \gg 1$
\cite{Agam95,Blanter96,BMM-97,Blanter-Mirlin-97,Aleiner-Glazman-97,%
  agam98}.
Nevertheless, deviations from the
predictions of the orthodox model can be expected in principle, and
indeed were observed in some of RBT's smallest grains with $\gdc \simeq
5$ (see \Sec{sec:nonequilibrium}).

Consider the SET shown in \Fig{fig:circuit-diagram}(b), and let $\Epot
(\Nex)$ denote the electrostatic work required to add $\Nex$ excess
electrons with a total charge of $\Qex = e \Nex$ (with $e <0$) to a
grain with initial random off-set charge $Q_0$, while the
time-independent voltages $V_\ssL$, $V_\ssR$ and $V_\ssg$ of the left
and right leads and the gate electrode, respectively, are held fixed.
Within the orthodox model, $\Epot (\Nex) = \int_{Q_0}^{\Qex+Q_0} {\d} Q
\, V(Q)$, where $V(Q)$ is the electrostatic potential of the grain for
given grain charge $Q$. It is determined by $C_r[V_r - V(Q)] = Q_r$,
where $Q_r$ is the screening charge on capacitor $r$ (= L,R,g) and
$Q_\ssL + Q_\ssR + Q_\ssg = - Q$.  (In the absence of a gate one has
$Q_\ssg=0$ and hence should set $C_\ssg = 0$.) Using the definitions
$C \equiv C_\ssL + C_\ssR + C_\ssg$ and $q_\ssD \equiv \sum_{r = {\rm
    L,R,g}} C_r V_r$, we obtain $V(Q) = ( q_\ssD + Q)/C$ and thus
\begin{eqnarray}
  \label{eq:Epot}
  \Epot(\Nex) = \Qex (q_\ssD + Q_0) /C + \Qexsq / 2C  = 
 e \VD \Nex \, + \,\Ec \Nexsq  \; .
\end{eqnarray}
Since the first term on the r.h.s. is linear in the number of excess
electrons, $\VD \equiv (q_\ssD + Q_0)/C $ can be viewed as the
electrostatic potential on the grain (the subscript D, for ``dot'', is
used to conform to Schoeller's notation \cite{schoeller-curacao}).
For gated devices, $Q_0$ may be absorbed into $q_\ssD$ by shifting
$\Vg$ by an amount $-Q_0 / C_\ssg $, but for gateless ones, it must be
treated as a fit parameter.  The second term on the r.h.s. of
\Eq{eq:Epot} represents the Coulomb interaction energy of the $\Nex$
excess electrons due to their mutual repulsion, and its scale is set
by the charging energy $ \Ec \equiv e^2/2C$.

$\Epot$ is often written as $\Ec (\Nex - n_\ssD)^2 - \Ec n_\ssD^2$,
where $n_\ssD \equiv - e \VD/2 \Ec$, since this makes it clear that
for a given value of $n_\ssD$, the system will adjust $\Nex$ to be the
integer closest to $n_\ssD$, in order to minimize $\Epot$. The
particle number on the grain can thus be controlled in discrete units
by varying $n_\ssD$ via the gate voltage $\Vg$.
Since the electrostatic energy difference 
\begin{eqnarray}
  \label{eq:U+-}
  \delta E_{\rm pot}^\pm (\Nex)
 \equiv \Epot (\Nex \pm 1 ) - \Epot (\Nex)  
\end{eqnarray}
between grains with $\Nex \pm 1$ or $\Nex$ excess electrons vanishes
when $n_\ssD$ is tuned to lie half-way between $\Nex \pm 1$ and
$\Nex$, half-integer values of $n_\ssD$ are called ``degeneracy
points''.  Transport is possible through a grain whose ground state
has $\Nex$ electrons only if, roughly speaking, $\mbox{min}[\delta
E_{\rm pot}^\pm (\Nex)] \lesssim \mbox{max}[\kB T, |eV|]$, \ie\ only
if the ``Coulomb barrier'' presented by \Eq{eq:U+-} can be overcome by
the temperature or bias voltage.  If both of these are small ($\kB
T,|eV| \ll \Ec$), a complete suppression of transport through the
grain, the so-called ``Coulomb blockade'', occurs far away from the
degeneracy points; in particular, the grain's low-temperature,
linear-response conductance shows ``Coulomb oscillations'' as function
of $\Vg$, \ie\ a series of peaks, with a uniform $n_\ssD$-spacing of
1, \ie\ $\Vg$-spacing of $e/ C_\ssg$.  In between degeneracy points,
low-temperature transport is possible only with a large bias voltage,
which will, in general, lead to nonequilibrium effects.
%
%


\subsubsection{General Hamiltonian}
\label{sec:generalhamiltonian}

To describe transport through the grain, we shall adopt a 
Hamiltonian $\hat H = \hat H_\ssL + \hat H_\ssR + \hat H_\ssD + 
\hat H_\tun$ 
of the following rather general form:
\begin{eqnarray}
  \label{eq:generalhamiltonian-leads}
  \hat H_r &=& \sum_{k \sigma} (\varepsilon_{kr} + e V_r)
  c^\dagger_{k \sigma r} c_{k \sigma r} \qquad (r = \ssL, \ssR) \; , 
\\
  \label{eq:generalhamiltonian-dot}
  \hat H_\ssD &=& \sum_\alpha 
( e V_\ssD  N_{{\rm ex},\alpha} + \E_\alpha  ) 
|\alpha \rangle \langle \alpha | \; , 
\\
  \label{eq:generalhamiltonian-tunnel-l}
  \hat H_\tun &=& \sum_{r = {\rm L,R}} \sum_{k l \sigma} 
T^r_{kl \sigma}  \,  c^\dagger_{k \sigma r} c_{l \sigma \ssD}
 \; + \, \mbox{(h.c.)}  \;  ,
\\
  \label{eq:generalhamiltonian-tunnel-ss}
  &=& \sum_{r = {\rm L,R}} \sum_{k \sigma, \alpha \alpha'} 
T^r_{k\sigma,\alpha \alpha'}  \,  
c^\dagger_{k \sigma r} |\alpha \rangle \langle \alpha' |
 \; + \, \mbox{(h.c.)}  \; , 
\\
\label{eq:Tchange-basis}
T^r_{k\sigma,\alpha \alpha'}& = & \sum_l
T^r_{kl\sigma } \, \langle \alpha | c_{l \sigma \ssD} 
| \alpha' \rangle \; .
\end{eqnarray}
Here $c_{k \sigma r}^\dagger$ creates an electron in lead $r$, in a
single-particle state with spin $\sigma$, kinetic energy
$\varepsilon_{kr}$ (measured relative to the Fermi energy
$\varepsilon_{k_{\rm F} r}$ of that lead) and electrostatic energy $e
V_r$.  $|\alpha \rangle$ is a many-body eigenstate of the {\em
  isolated\/} grain in the absence of tunneling, with definite total
electron number $N_\alpha$.  The sum on $\alpha$ in
\Eq{eq:generalhamiltonian-dot} is not restricted to a fixed-$N_\alpha$
Hilbert space, but is over {\em all\/} eigenstates in the isolated
grain's Fock space.  The eigenvalue $ eV_\ssD N_{{\rm ex},\alpha} +
\E_\alpha$ of $|\alpha \rangle$ has been split into two parts: the
first contains, via $V_\ssD$, all dependences on the external voltages
arising from $\Epot$ of \Eq{eq:Epot}, while $\E_\alpha$ is the
voltage-independent remainder and includes the interaction
contribution ($\Ec N^2_{{\rm ex},\alpha}$) to $\Epot$.  In the absence
of many-body correlations, \eg\ for a grain made from a normal metal,
one could take, as Averin and Korotkov did \cite{averin90},
\begin{eqnarray}
  \label{eq:H-dot-normal}
 && \hat H^\normal_\ssD  =  
  \Epot (\hNex) + \sum_{l\sigma}
\varepsilon_{l\sigma \ssD} c^\dagger_{l \sigma \ssD} c_{l \sigma \ssD}
 ,   \\
\label{eq:H-dot-normal-eigenstates}
&& |\alpha \rangle  = | \{ n^{(\alpha)}_{l \sigma \ssD} \} \rangle
\, , \qquad \quad {\cal E}_\alpha =  \Ec  N^2_{{\rm ex},\alpha} +
\sum_{l \sigma} \varepsilon_{l \sigma \ssD} \, n^{(\alpha)}_{l \sigma \ssD} 
 \; .
\end{eqnarray}
Here $c^\dagger_{l \sigma \ssD}$ is a creation operator for one of the
grain's single-particle states with energy $\varepsilon_{l\sigma \ssD}$,
and each many-body eigenstate $|\alpha \rangle$, with eigenenergy $ e
V_\ssD N_{{\rm ex},\alpha} + {\cal E}_\alpha $, is specified by a set
$\{ n^{(\alpha)}_{l \sigma \ssD} \}$ of single-particle occupation
numbers.  However, since such a single-particle description fails,
\eg, for superconducting or ferromagnetic grains, we shall 
treat the $|\alpha \rangle$ as general many-body eigenstates
below.

The tunneling term $\hat H_\tun$ has been written in two forms:
\Eq{eq:generalhamiltonian-tunnel-l} describes the tunneling of
electrons between {\em single-particle\/} states in the leads and
grain, with matrix elements $T_{kl\sigma }^r$ for tunneling events
involving lead $r$; and in \Eq{eq:generalhamiltonian-tunnel-ss},
$\hat H_\tun$ is represented in the $|\alpha \rangle$ eigenbasis, with
corresponding matrix elements $ T^r_{k\sigma,\alpha \alpha'}$
given by \Eq{eq:Tchange-basis}. 
When doing perturbation theory in $\hat H_\tun$, it turns out
\cite{schoeller-curacao} that the $T^r_{k l \sigma }$ matrix elements
always occur in the combination
\begin{eqnarray}
  \label{eq:spectralGamma}
{2 \pi \over \hbar} \sum_{k}
   T^{r \ast}_{kl \sigma} T^r_{kl' \sigma} \, 
\delta( \omega - \varepsilon_{kr} ) \, \simeq \, \delta_{ll'} \,
\Gamma_{l \sigma}^r \; ,
\end{eqnarray}
which represents the total tunneling rate from lead $r$ across barrier
$r$ into the single-particle state $|l \sigma \rangle$ on the grain.
On the right-hand side we have neglected all off-diagonal terms, since
for $l \neq l'$ the sum on $k$ would involve matrix elements with
randomly varying phases that average to zero, and also the $\omega$
dependence, which typically is of order $\omega/\! \eF$ (since $\eF$
sets the energy scale for changes in the leads' density of states).

\subsubsection{Tunneling current and master equation}
\label{sec:tunnelingcurrent}

The operator for the current carried by electrons tunneling between
lead $r$ and the grain, $ \hat I_r= - e \, {\partial \hat N_r /
  \partial t}$ (defined to be positive if $N_r$ increases), is
\begin{eqnarray}
  \label{eq:tunnelingcurrentoperator}
  \hat I_r  
= \ii e \mbox{[} \hat N_r , \hat H \mbox{]} 
= \ii e \sum_{k \sigma, \alpha \alpha'}  
T^r_{k\sigma,\alpha \alpha'}  \, 
c^\dagger_{k \sigma r} |\alpha \rangle \langle \alpha' |  
 \; + \, \mbox{(h.c.)}  \; .
\end{eqnarray}
Its steady-state expectation value $I_r = \mbox{Tr}_{\ssL, \ssR,
  \alpha} (\hat \rho \hat I_r)$ in general requires knowledge of the
system's full density matrix $\hat \rho$. However, since the
Hamiltonian is quadratic in the lead degrees of freedom, the latter
 can be 
integrated out, so that the remaining degrees of freedom are described
by the reduced density matrix $\hat P = \mbox{Tr}_{\ssL,\ssR} (\hat
\rho) $ for the grain. Its diagonal elements $P_\alpha = \langle
\alpha | \hat P | \alpha \rangle$ give the probability to find the
grain in state $|\alpha \rangle$.  They satisfy, and may be found by
numerically solving, a normalization condition and master equation of
the form
\begin{equation}
\label{generalmasterequation}
  \sum_{\alpha'} P_{\alpha'} = 1 \; , \qquad 
  0 = \sum_{\alpha' \neq \alpha} 
  \left(\Sigma_{\alpha \alpha'} P_{\alpha'} - \Sigma_{\alpha' \alpha}
 P_\alpha  \right) \; \quad
\mbox{for \, each \,} \alpha,
\end{equation}
where $\Sigma_{\alpha \alpha'}$ (for $\alpha \neq \alpha'$) is the
total transition rate from initial state 
$|\alpha' \rangle$ to final state $|\alpha \rangle$.
Furthermore, the current $I_r$ can be written as
\begin{eqnarray}
  \label{eq:currentexpectation}
  I_r &=&  e \sum_{\alpha \alpha'} \sum_{p=1}^\infty
 p \left( \Sigma_{\alpha \alpha'}^{r +p} - 
\Sigma_{\alpha \alpha'}^{r-p} \right) P_{\alpha'} \; ,
\end{eqnarray}
where $ \Sigma_{\alpha \alpha'}^{r + p }$ ($ \Sigma_{\alpha
  \alpha'}^{r -p }$) is that part of the total rate $\Sigma_{\alpha
  \alpha'}$ that involves the coherent transfer of a total of $p$
electrons onto (from) the grain from (onto) lead $r$.  Of course,
charge conservation ensures that $I_\ssL = I_\ssR$.

\Eqs{generalmasterequation} and (\ref{eq:currentexpectation}) have
intuitively plausible forms, but the calculation of the rates $
\Sigma_{\alpha \alpha'}$ and $ \Sigma_{\alpha \alpha'}^{r \pm p }$ is
in general a highly non-trivial task. The standard strategy is to
perform an expansion in powers of $\hat H_\tun$, the most general and
systematic formulation of which is the so-called real-time
diagrammatic approach developed by Schoeller, Sch\"on and K\"onig
\cite{schoeller-curacao,real-time-1,real-time-2,real-time-3,%
  real-time-4,schoeller-habil,koenig-thesis}.  Fortunately, for our present
purpose of analysing tunnel-spectroscopic measurements on ultrasmall grains,
we may restrict ourselves to the simplest possible situation: the tunnel
barriers are so large and the tunneling current so small that it suffices to
consider only {\em sequential\/} tunneling\footnote{An exception will be
  encountered in \Sec{sec:sub-gap-structures}, where cotunneling needs to be
  considered.}  of electrons, described by lowest-order perturbation theory in
$H_\tun$.  For this case one has \cite{schoeller-habil,koenig-thesis}
\begin{eqnarray}
  \label{eq:totalrates}
  \Sigma_{\alpha \alpha'} = \sum_{r = \ssL, \ssR} \sum_{p = \pm }
  \Sigma_{\alpha \alpha'}^{rp} \; ,
\end{eqnarray}
where $\Sigma_{\alpha \alpha'}^{r\pm } \equiv \Sigma_{\alpha
  \alpha'}^{r\pm 1}$ can be calculated using the golden rule,
\begin{eqnarray}
  \label{eq:sigmar+}
  \Sigma_{\alpha \alpha'}^{r+} &=& {2 \pi \over \hbar} \sum_{k\sigma}
  f(\varepsilon_{kr}) |T_{k \sigma, \alpha' \alpha}^{r \ast} |^2 
  \delta(\E_\alpha - \E_{\alpha'} + e \VD - \varepsilon_{kr} - e V_r)
\\
\label{eq:sigmar+short}
 &= & f (\E_\alpha - \E_{\alpha'} - e \bar V_r) \, 
\Gamma_{\alpha \alpha'}^{r+} 
\rule[-4mm]{0mm}{1mm}
\; , 
\\
  \label{eq:sigmar-}
  \Sigma_{\alpha \alpha'}^{r-} &=& {2 \pi \over \hbar} \sum_{k\sigma}
  \left[1 - f(\varepsilon_{kr})\right] |T_{k \sigma, \alpha \alpha'}^{r} |^2
  \delta(\E_\alpha - \E_{\alpha'} - e \VD + \varepsilon_{kr} + e V_r) 
\\
\label{eq:sigmar-short}
 &=&  f(\E_\alpha - \E_{\alpha'} + e \bar V_r)
 \, \Gamma_{\alpha \alpha'}^{r-} \; , \rule[-4mm]{0mm}{1mm} 
\\
\label{eq:Gamma-short}
\Gamma_{\alpha \alpha'}^{r+} 
&=& \sum_{l \sigma} \Gamma_{l\sigma} ^r 
\, |\langle \alpha |c_{l\sigma \ssD}^\dagger |\alpha' \rangle|^2 \; , 
\qquad 
\Gamma_{\alpha \alpha'}^{r-} = \sum_{l \sigma} \Gamma_{l\sigma}^r  \, 
|\langle \alpha |c_{l\sigma \ssD} |\alpha' \rangle|^2 \; .
\end{eqnarray}
Here $f(E) = 1/(\e^{E/\kB T} + 1)$ is the Fermi function and $\bar V_r
= V_r - V_\ssD$ is the voltage drop (electrostatic potential
difference) between lead $r$ and the grain. For the standard case of a
``symmetric'' circuit with $V_\ssL = - V_\ssR = V/2$, which we shall
henceforth adopt, it is given by
\begin{equation}
  \label{eq:barVr-explicitly}
  \bar V_{\ssL/\ssR} = \left[ \pm V(C_{\ssR/\ssL} + C_\ssg/2) 
   - V_\ssg C_\ssg  - Q_0 \right]/C \; ,
\end{equation}
which reflects the capacitative division of the bias voltage $V (
= V_\ssL - V_\ssR = \bar V_\ssL - \bar V_\ssR)$ across the two barriers.
\Eqs{eq:sigmar+short} and (\ref{eq:sigmar-short}) show that when $V$
is swept, at a fixed gate voltage $\Vg$ and at temperatures much lower
than the typical spacing $d$ between the eigenenergies $\E_\alpha$, a
rate $\Sigma_{\alpha \alpha'}^{r \pm }$ will be switched from ``off''
(exponentially small) to ``on'' (of order $\Gamma_{\alpha \alpha'}^{r
  \pm}$) each time $e \bar V_r$ passes through a threshold at which
one of the inequalities
\begin{equation}
  \label{eq:Vrthresholds}
  \pm \, e \bar V_r \ge \E_\alpha - \E_{\alpha'} \; 
\end{equation}
\begin{figure}[t]
\begin{center}
  \epsfig{figure=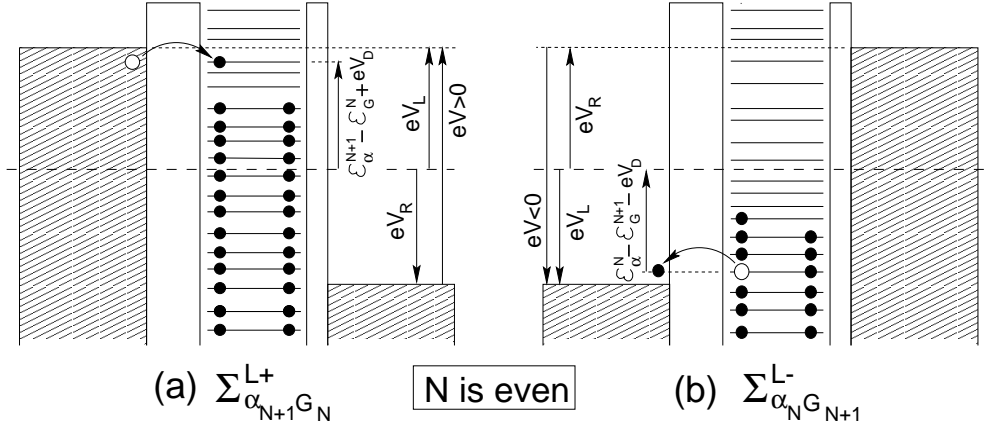,width=0.95\linewidth}
\end{center}
  \caption[Schematic energy diagram for tunneling rates]{
   \label{fig:energydiagram} 
   Schematic illustration of the threshold conditions [see
   \Eq{eq:Vrthresholds}] for bottleneck rates to become non-zero,
   (a) $e V_r \ge \E_\alpha^{N+1} - \E_G^N + e \VD$ for $\Sigma^{r
     +}_{\alpha_{N+1} G_N}$ and (b) $-e V_r \ge {\cal E}_\alpha^N -
   \E_G^{N+1} - e \VD$ for $\Sigma^{r -}_{\alpha_{N} G_{N+1}}$, for
   $r=L$ and the case that the left barrier is larger than the right
   ($\Gamma^L \gg \Gamma^R$).  The long-dashed line indicates the
   equilibrium, $V=0$ chemical potential of the L and R leads relative
   to which $eV_r$ and ${\cal E}_\alpha - \E_G \pm e\VD$ are drawn,
   using upward or downward arrows for positive and negative energies,
   respectively.  Note that excitation energies are always positive:
   (a) ${\cal E}^{N+1}_\alpha - \E_G^N + e \VD$ for particle-like
   excitations produced when an electron tunnels onto the grain; and
   (b) $\E^N_\alpha - \E^{N+1}_{G} - e \VD$ for hole-like excitations
   when it tunnels off. Filled circles depict the {\em final\/}
   electron configuration $|\alpha \rangle$ {\em after\/} such a
   tunneling process, \ie\ the electron number in the initial ground
   state before tunneling is drawn to be even $(N)$ in (a) and odd
   $(N+1)$ in (b).
}
\end{figure}
becomes true. Intuitively speaking, this occurs each time the energy gained by
an electron when leaving (entering) lead $r$, namely $\pm e V_r$, becomes
greater than the energy needed to enter (leave) the grain while inducing the
transition $|\alpha' \rangle \to | \alpha \rangle$, namely $\E_\alpha -
\E_{\alpha'} \pm e \VD$.  \Figs{fig:energydiagram} (a) and (b) illustrate this
condition for the rates $\Sigma^{L+}$ and $\Sigma^{L-}$, respectively, and the
case that the initial state is the ground state, $|\alpha'\rangle =
|G\rangle$.

The standard expressions for a \emph{meso}scopic (\ie\ large) SET with
a normal island, described by $\hat H_\ssD^\normal$ of
\Eq{eq:H-dot-normal}, can be recovered as follows from the above
formulas: firstly, one uses a factorized form for the island's density
matrix, with $P_\alpha = P_{N} P_{\alpha_N}$ (and $\sum_{\alpha_N}
P_{\alpha_N} = 1$).  Secondly, one makes the replacement
\begin{eqnarray}
  \label{eq:getbackstandard}
  \sum_{\alpha_{N\pm 1} \alpha'_{N} }
   \left( \Sigma^{r \pm}_{\alpha_{N\pm 1} \alpha'_{N}} \right) 
   P_{\alpha'_N} 
   & \rightarrow &
    \sum_{l\sigma} f(\pm \varepsilon_{l\sigma \ssD} + 
      \delta E^{\pm}_{\rm pot} 
     \mp e V_r)  f ( \mp \varepsilon_{l \sigma \ssD}) 
     \Gamma^r_{l \sigma} \\
     & = & 2 \Gamma^r \N(\eF) 
        \left[ { - (\delta E_{\rm pot}^\pm   \mp e V_r ) \over
    1 - e^{(\delta E_{\rm pot}^\pm 
      \mp e  V_r )/ \kB T} } \right] ,
\end{eqnarray}
where $\delta E_{\rm pot }^{\pm}$ [\Eq{eq:U+-}] is the change in
electrostatic energy for the transition $\Nex \to \Nex \pm 1$.  For
the second line we took the continuum limit for the energies
$\varepsilon_{l \sigma \ssD}$, and assumed that $\Gamma^r_{l \sigma} =
\Gamma^r$ is independent of $l$ and $\sigma$.

\subsubsection{Extracting the fixed-$N$ excitation spectrum}
\label{sec:extractingspectrum}
\label{subsubsection:excitationspectrum}

The very large charging energies ($\Ec \gtrsim 5$meV, $\gg T,eV$) of
ultrasmall grains ensure that for a fixed value of gate voltage $\Vg$,
at most two kinds of charge states are involved in electron transport
through the grain, \ie\ $P_\alpha = 0$ unless $N_\alpha$ equals $N$ or
$N+1$ (say).  To get a non-zero current, one needs non-zero rates for
tunneling both onto and off the grain, through different barriers, say
$\Sigma_{\alpha_{N+1}\alpha'_{N}}^{r+}$ and $\Sigma_{\alpha_{N}
  \alpha'_{N+1}}^{r'-}$.  To be specific, these will be of type
\begin{equation}
  \label{eq:eV><0}
\Sigma^{\ssL +}  , \, \Sigma^{\ssR -}  \; \mbox{for} \;
eV > 0  \quad \mbox{and} \quad
\Sigma^{\ssL -}  , \,  \Sigma^{\ssR +}  \; \mbox{for} \;
eV < 0 \quad \mbox{[with} \; e<0 \mbox{]},
\end{equation}
since the direction of electron tunneling is always from the higher
toward the lower chemical potential.  One can distinguish two
situations, depending on whether (a) only one or (b) both barriers are
bottlenecks that limit transport.

(a) {\em One bottleneck.}--- If all the tunneling-on rates
$\Sigma^{r+}$ are generally, say, much smaller than any of the
tunneling-off rates $\Sigma^{r'-}$, then barrier $r$ is the bottleneck
for transport. The probability to find $N$ or $N+1$ electrons on the
grain will then be close to unity or zero, respectively.  Moreover,
for $T \ll d$ (and neglecting nonequilibrium effects, to be discussed
in \Sec{sec:nonequilibrium}), the $N$-electron ground state $|G_N
\rangle$ will be the overwhelmingly most probable, \ie\ $P_{G_N}
\simeq 1$ and $P_\alpha \simeq 0$ for all other $\alpha$.  In this
case the current through barrier $r$ 
[\Eq{eq:currentexpectation}] is simply $I_r \simeq e
\sum_{\alpha_{N+1}} \left( \Sigma_{\alpha_{N+1} G_{N}}^{r+}\right)$.
If $V$ is swept at fixed $V_\ssg$, the current will thus show a
steplike increase and the differential conductance a peak each time a
rate $\Sigma_{\alpha_{N+1} G_{N}}^{r+}$ is switched on, \ie\ each time
$e \bar V_r$ increases past one of the threshold values
$\E^{N+1}_\alpha - \E^N_G$ of (\ref{eq:Vrthresholds}), see
\Fig{fig:energydiagram}(a).  Denote this threshold value by $\bar
V_r^\alpha$ and the corresponding threshold bias voltage [related to
it via (\ref{eq:barVr-explicitly})] by $V_\alpha$.  The voltage
differences $V_{\alpha \tilde \alpha} = V_\alpha - V_{\tilde \alpha}$
between any two conductance peaks can directly be translated into one
of the excitation energies of the {\em fixed-$(N+1)$}\/ excitation
spectrum, namely (for bottleneck barrier $r=$L/R)
\begin{equation}
  \label{eq:fixed-N-excitation-spectrum1}
\E^{N+1}_\alpha - \E^{N+1}_{\tilde \alpha} = \, e \, 
(\bar V_{\ssL/\ssR }^{\alpha} 
- \bar V_{\ssL/\ssR }^{\tilde \alpha }) \, 
  = \pm e V_{\alpha \tilde \alpha} (C_{\ssR/\ssL} + C_\ssg/2)/C \; .
\end{equation}
The case in which all rates $\Sigma^{r-}$ are much smaller than any
rate $\Sigma^{r'+}$ [\Fig{fig:energydiagram}(b)]
is completely analogous to the one just discussed.
Then the current is $I_r \simeq -e \sum_{\alpha_{N}} 
\left( \Sigma_{\alpha_{N}
    G_{N+1}}^{r-} \right) $, and the fixed-$N$ excitation energies are
given by (for $r=$L/R)
\begin{equation}
  \label{eq:fixed-N-excitation-spectrum2}
\E^N_\alpha - \E^N_{\tilde \alpha} = -e \, (\bar V_{\ssL/\ssR }^{\alpha} 
- \bar V_{\ssL/\ssR }^{\tilde \alpha })
  = \mp e V_{\alpha \tilde \alpha} (C_{\ssR/\ssL} + C_\ssg/2)/C \; .
\end{equation}
\Eqs{eq:fixed-N-excitation-spectrum1} and
(\ref{eq:fixed-N-excitation-spectrum2}) form the basis of RBT's
tunnel-spectroscopic measurements of discrete spectra: a conductance
curve showing a series of peaks as function of voltage can be
replotted as function of energy by rescaling the horizontal axis using
the voltage-to-energy conversion factors\footnote{If no gate is
  present ($C_\ssg = 0$) or if $C_\ssg \ll C_\ssL, C_\ssR$
(as in the device of \Fig{fig:generic-IV}), this factor reduces to $e
  C_{\ssR/\ssL} / (C_\ssL + C_\ssR)$.}  $ e(C_{\ssR/\ssL} +
C_\ssg/2)/C$ in \Eqs{eq:fixed-N-excitation-spectrum1} or
(\ref{eq:fixed-N-excitation-spectrum2}). The rescaled curves, which we
shall henceforth generically call ``fixed-$N$ excitation spectra'',
where $N$ is the number of electrons in the {\em final\/} state {\em
  after\/} the bottleneck tunneling process, then allow the
corresponding excitation energies to be read off directly from the
peak spacings.

(b) {\em Two bottlenecks.}--- The situation is more involved if the
tunneling-on and tunneling-off rates $\Sigma^{r+}$ and $\Sigma^{r-}$
are comparable in magnitude so that both barriers are bottlenecks:
Then some of the $P_{\alpha_N}$ and $P_{\alpha_{N+1}}$ will be
comparable too, implying the same for the total probabilities to find
$N$ or $N+1$ electrons on the grain.  The conductance will then show
peaks corresponding to the switching-on of both $\Sigma^{r+}$ and
$\Sigma^{r'-}$ rates, at both $e \bar V_r$ and $e \bar V_{r'}$ values
corresponding to both $\E^{N+1}_\alpha - \E^N_{\alpha'}$ and
$\E^N_\alpha - \E^{N+1}_{\alpha'}$ energy differences. It would thus
be rather difficult to extract a purely fixed-$N$ or fixed-$(N+1)$
excitation spectrum from the resulting mixed set of peak spacings
(though in principle not impossible: if a magnetic field is used to
switch the leads from being superconducting to normal, peaks due to
$\Sigma^{r+}$ or $\Sigma^{r'-}$ shift by different amounts, see
\Eq{eq:vshift-due-to-Delta} in \Sec{sec:sc-leads}; or, if a gate is
present, peaks due to $\Sigma^{r+}$ or $\Sigma^{r'-}$ move in opposite
$V$-directions if $\Vg$ is changed, see \Eq{eq:DeltaVgDeltaV} in
\Sec{sec:effects-of-gate}).

\subsection{Experimental details}
\label{subsec:experimentaldetails}

This section is devoted to experimental details such as device
fabrication and characterization, and what effects superconducting
leads or a gate voltage have on the tunneling spectra. In particular,
we explain how one can tell whether all conductance peaks correspond
to tunneling across the {\em same\/} grain, whether a given peak has
barrier L or R as bottleneck, and how the corresponding capacitance
ratios can be determined very accurately [this knowledge is needed in
\Eqs{eq:fixed-N-excitation-spectrum1} or
(\ref{eq:fixed-N-excitation-spectrum2})].

\subsubsection{Device fabrication}
\label{subsubsection:devicefabrication}
The devices of RBT shown schematically in \Fig{fig:sample}(a) were
fabricated as follows \cite{rbt97}: First electron-beam lithography
and reactive-ion etching were used to make a bowl-shaped hole in a
suspended silicon nitride membrane, with an orifice between 5 and
10~nm in diameter \cite{ralls89}.  The gate electrode was formed by
evaporating 12~nm of Al onto the flat [bottom in \Fig{fig:sample}(a)]
side of the membrane.  Plasma anodization and deposition of insulating
SiO were then used to provide electrical isolation for the gate.  Next
an Al electrode which fills the bowl-shaped side [top in
\Fig{fig:sample}(a)] of the nitride membrane was formed by evaporation
of 100~nm of Al, followed by oxidation in 50~mTorr ${\rm O}_2$ for
45~s to form a tunnel barrier near the lower opening of the
bowl-shaped hole. A layer of nm-scale grains was created by depositing
2.5~nm of Al onto the lower side of the device; due to surface tension
the metal beaded up into separate grains less than 10~nm in diameter
\cite{GiaeverZeller-68,ZellerGiaever-69}. In approximately 25\% of the
samples [determined as those exhibiting the typical SET
``Coulomb-staircase'' structure exemplified by \Fig{fig:generic-IV}],
a single grain formed under the nm-scale tunnel junction to contact
the top Al electrode.  Finally, after a second oxidation step to form
a tunnel junction on the exposed surface of the grain, a lower
electrode is formed by evaporating 100~nm of Al to cover the grain.

The resulting device is an SET whose central island is a nm-scale
grain fully coated by an insulating oxide layer, sitting on the lower
lead electrode, encircled by a gate electrode, with the bowl part of
the upper lead electrode at a fixed distance above it like ``an STM
tip cast in concrete''.

Recently, Davidovi\'c and Tinkham succeeded in contacting
leads to an ultrasmall grain with a radius
as small as $r \approx 1$~nm,   using a somewhat different
fabrication technique, described in detail in 
\cite{davidovich98,davidovich99}.

\subsubsection{Device characterization}
\label{subsubsection:devicecharacterization}

\begin{figure}
\centerline{\epsfig{figure=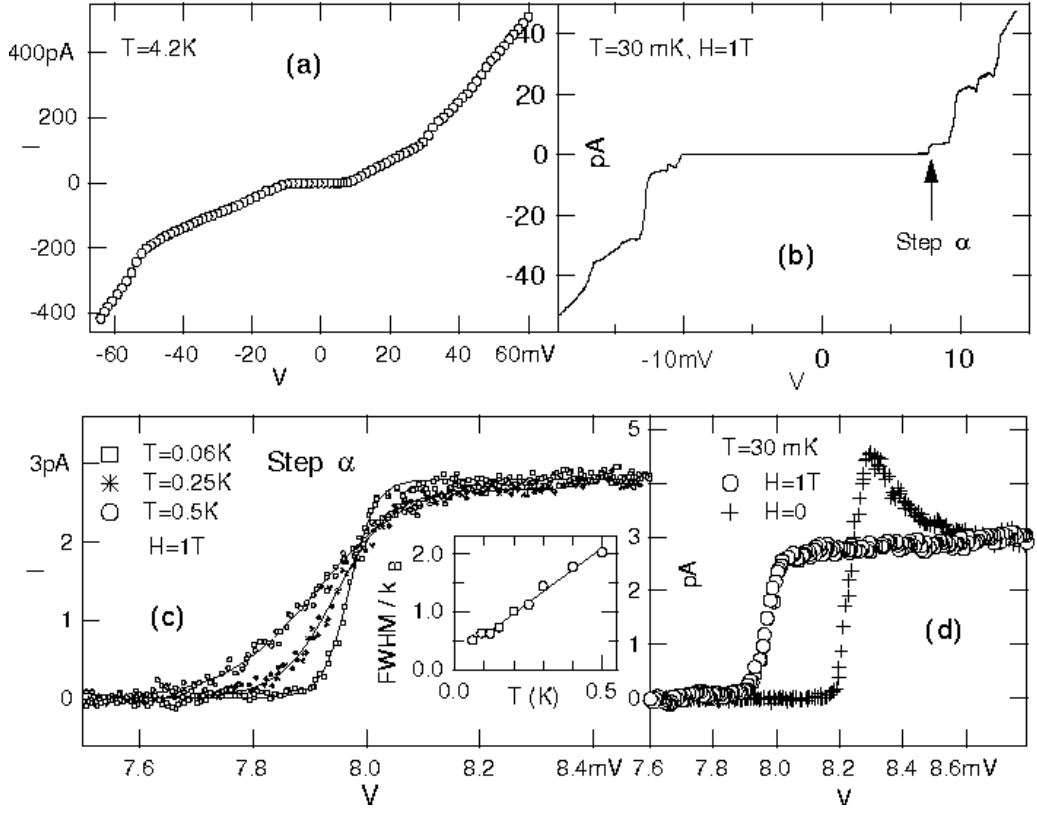,%
width=0.98\linewidth}}
\caption[$I$-$V$ characteristics of an
ultrasmall single-electron transistor]{Current-voltage characteristics
  of an Au grain with radius $r \approx 4.5$~nm, from
  \cite{davidovich99}. 
 (a) $I$-$V$ curve of a typical device
  at 4.2K.  (b). $I$-$V$ curve of the device at 30~mK. (c) First
  current step ($\alpha$) of (b), at three different refridgerator
  temperatures.  The curves between the points are fits to the Fermi
  distribution.  After correcting for the capacitive division of
  voltage, the full-width-half-maximum (FWHM) of the peak, shown in
  inset, is linear with the refridgerator temperature, with a slope of
  ${\rm FWHM}/\kB T= 3.8$, which is close to the expected slope of 3.5.
    (d) Lineshape of the level when the leads are superconducting
    ($H=0$) compared to the lineshape when the leads are normal
    ($H=1$~T).
\label{fig:drago-set}}
\end{figure}

Only those devices are studied in detail whose $I$-$V$ characteristics
display the large-scale Coulomb-staircase [\Fig{fig:generic-IV}]
expected for a SET, since its presence indicates that the tunneling
current flows through a single grain only (as opposed to several). The
basic parameters of the SET ($R_\ssL$, $R_\ssR$, $C_\ssL$, $C_\ssR$,
$C_\ssg$ and $\Ec$) can be determined by fitting this
Coulomb-staircase to the standard expressions for mesoscopic SETs
(with $C_\ssg =0$ for devices without a gate), as described in detail
by Hanna and Tinkham \cite{hanna91}.
To estimate the grain radius, RBT crudely assumed
\cite{rbt95,rbt96a,rbt96b,rbt97} the grain to be a hemisphere with
radius $r$ and $\Vol = 2 \pi r^3/3$ (actually their grains were more
pancake-shaped) and determined $r$ by equating the (larger of the)
grain-to-lead capacitances to $\pi r^2 C_\Box$; here $C_\Box$ is the
capacitance per unit area of the tunnel junctions.  In \Ref{rbt95},
RBT estimated that $C_\Box \simeq 0.075 {\rm aF/nm}^2$ for the ${\rm
  Al}_2{\rm O}_3$ layers insulating their grains from the leads and
gates.  More detailed subsequent studies \cite{Lu98} gave a slightly
revised value of $C_\Box \simeq 0.050 {\rm aF/nm}^2$.  Finally, the
grain's single-particle mean level spacing $d$ can be estimated using
\Eq{eq:d-estimate}.

\subsubsection{Superconducting leads}
\label{sec:sc-leads}

\begin{figure}[t]
\begin{center}
\epsfig{figure=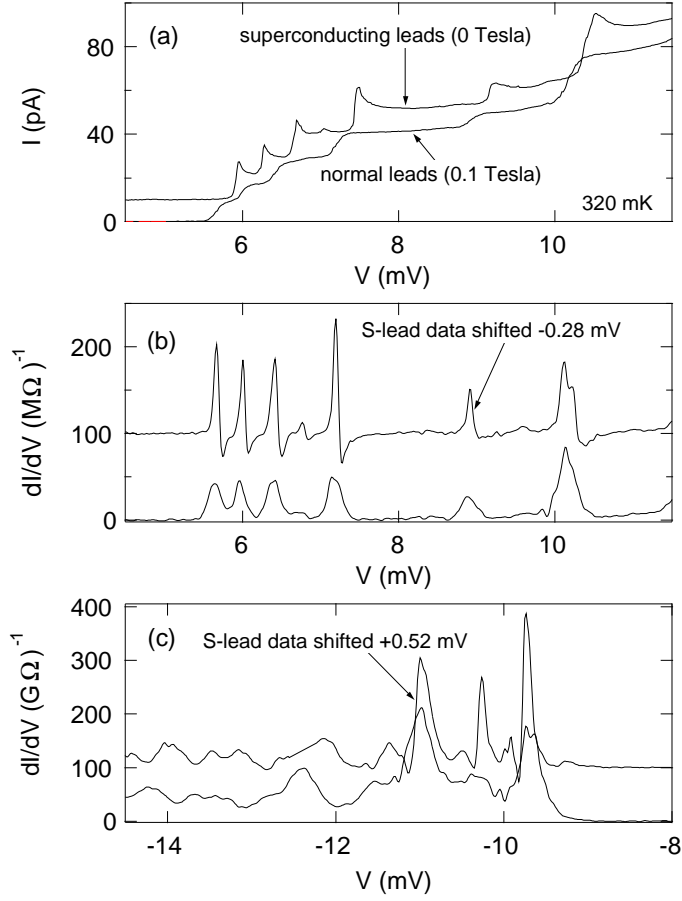,%
width=0.65\linewidth}
\end{center}
\caption[Effect of superconducting versus normal leads]{
  Current-Voltage characteristics for an Al grain connected to
  superconducting and normal leads \cite{rbt95}, without a gate
  electrode (\ie\ $C_\ssg = 0$). The device parameters were $C_\ssL =
  4.9 \pm 0.5$~aF, $C_\ssR = 8 \pm 1$~aF, $R_\ssL / R_\ssR = 8 \pm 2$,
  $R_\ssL + R_\ssR = 9 \pm 1$~M$\Omega$, $Q_0/e = 0.20 \pm 0.03$, and
  $\Ec \sim 6$~meV.  (a) $I$ vs.\ $V$ curves; the $N$-lead curve shows
  steps, the $S$-lead curve (displaced 10~pA in $I$) shows spikes,
  reflecting the BCS quasiparticle density of states in the leads [cf.
  \Eq{eq:sc-densityofstates}].  (b) and (c) ${\d} I/{\d} V$ vs. $V$ for
  positive and negative $V$, with the $S$-lead data shifted in $V$, as
  labeled, so as to align the maxima of ${\d} I/{\d} V$ with the $N$-lead
  data. For ease of comparison, the amplitude of the $S$-lead data is
  reduced by a factor of 2 and offset on the ${\d} I/{\d} V$ axis in (b)
  and (c).  }
\label{fig:sc-normal-IV-steps}
\end{figure}

In many of RBT's samples, the leads are made from Al. They thus
become superconducting below $\Tc = 1.21$K, but can  be driven normal
by turning on a magnetic field (the critical
field is very small, about $H \simeq 20$~mT, 
since RBT's leads are in effect disordered thin films). 
\Figs{fig:sc-normal-IV-steps} 
[and also \ref{fig:drago-set}(d)] show how the current and conductance
curves change when the leads  switch from normal to superconducting
upon removal of this magnetic field: each current step is shifted to a
higher $|V|$, relative to the normal-lead data, and takes the form of
a spike, with, correspondingly, a region of negative ${\d} I/{\d} V$. This
can readily be understood by redoing, for superconducting leads, the
golden-rule calculations of \Sec{sec:tunnelingcurrent} (see
\App{sec:superconductingleads}): at $H=0$ and temperatures $T \ll
\Delta_r$ (the BCS gap in lead $r$), \Eqs{eq:sigmar+short} and
(\ref{eq:sigmar-short}) for the rates $\Sigma^{r\pm}$ are modified to
\begin{eqnarray}
  \label{eq:sigmarplusminusshort.sc}
  \Sigma_{\alpha \alpha'}^{r\pm} &=&  
f (\E_\alpha - \E_{\alpha'} \mp e \bar V_r) \, 
\N_{r, \qp} (\E_\alpha - \E_{\alpha'} \mp e \bar V_r) \, 
  \Gamma_{\alpha \alpha'}^{r\pm} \; ,
\\
  \label{eq:sc-densityofstates}
  \N_{r,\qp} (E) &=& \left\{ 
\begin{array}{ll} {| E | /\sqrt{ E^2 - \Delta^2_r}} &
\qquad \mbox{for} \quad |E| > \Delta_r \, ,
\\
  0 & \qquad \mbox{for} \quad |E| \le \Delta_r \, .
\end{array} \right.
\end{eqnarray}
The novel feature in \Eq{eq:sigmarplusminusshort.sc} is the appearance
of $\N_{r,\qp}(E)$, the BCS quasiparticle density of states in
lead $r$, which reflects the fact that a quasiparticle is created each
time an electron enters or leaves a lead.  A detailed analysis
\cite{rbt95} showed that the shape of the spikes in the
superconducting-lead data in \Fig{fig:sc-normal-IV-steps}(a) [and also
\ref{fig:drago-set}(d)] quantitatively agrees rather well with the
prediction of \Eq{eq:sigmarplusminusshort.sc}, up to a slight amount
of broadening near the threshold (cf.\ Fig.~3 of \cite{rbt95}).  A
possible reason for this broadening was suggested by Levy-Yeyati
\etalia\ \cite{levy-yeyati}: as a discrete state is swept past the
threshold, its life-time broadening becomes anomalously large, since
this broadening is proportional to the leads' density of states and
hence reflects the BCS peak in the latter.

Despite this slight amount of anomalous broadening, the conductance
peaks in the ${\d} I / {\d} V$ curve still are much sharper for
superconducting than for normal leads, as is evident in
\Fig{fig:sc-normal-IV-steps}(b). This is a very useful feature that
can be exploited to enhance spectroscopic resolution.

The shift in $V$-thresholds for superconducting versus normal leads
can be used to determine which barrier ($r=$ L or R) acts as
bottleneck for a given conductance peak: For superconducting leads,
the threshold value for $e \bar V_r$ at which a rate $\Sigma^{r\pm}$
is switched on, namely $ \pm e \bar V_r \ge (\E_\alpha - 
\E_{\alpha'}) + \Delta_r$ [by
\Eq{eq:sigmarplusminusshort.sc}], is shifted by $\delta (e \bar
V_r) = \pm \Delta_r$ relative to the threshold of \Eq{eq:Vrthresholds}
for normal leads, reflecting the added energy cost for removing or
adding an electron from or to a superconductor. By
(\ref{eq:barVr-explicitly}), this corresponds to a shift in bias
voltage of
\begin{equation}
  \label{eq:vshift-due-to-Delta}
  \phantom{.} \hspace{-0.5cm} 
  \delta V = { \mp  C \Delta_r  \over |e| (C_\ssR + C_\ssg/2) } \quad
  \mbox{if} \; \; r = {\rm L}, \quad
  \delta V = { \pm C \Delta_r \over |e| (C_\ssL + C_\ssg/2) } \quad
  \mbox{if} \; \; r = {\rm R} .
\end{equation}
Whereas the sign of $\delta V$ always equals that of $V$ [as follows
by combining \Eqs{eq:vshift-due-to-Delta} and (\ref{eq:eV><0}), with
$e<0$], its magnitude evidently depends on $r$, allowing the
bottleneck barrier $r$ to be identified.  Incidentally, once $r$ is
known, \Eq{eq:eV><0} can be used to determine whether bottleneck
transitions for a given sign of bias voltage involve tunneling onto or
off the grain.

\Figs{fig:sc-normal-IV-steps}(b) and (c) illustrate these ideas: they
show two kinds of shifts, whose magnitudes 0.28 and 0.52~mV agree well
with the predictions of \Eq{eq:vshift-due-to-Delta} for L and R, using
the sample parameters listed in the figure caption. In
\Figs{fig:sc-normal-IV-steps}(b), {\em all}\/ peaks are shifted by the
same amount of $0.28$~mV, thus L is the bottleneck barrier, with
$\Sigma^{\ssL -}_{\alpha_N G_{N+1}}$ as bottleneck rates [by
\Eq{eq:eV><0}, since $eV < 0$ for \Figs{fig:sc-normal-IV-steps}(b)].
Thus, this is an example of the ``one-bottleneck'' situation described
in \Sec{sec:extractingspectrum}(a), from which fixed-$N$ excitation
spectra can be extracted.  In \Figs{fig:sc-normal-IV-steps}(c), both
types of shifts occur, implying that this is the ``two-bottleneck''
situation described in \Sec{sec:extractingspectrum}(b): the peaks
between $-9$ and $-12$~mV are shifted by $-0.52$~mV and hence have
bottleneck rates $\Sigma^{\ssR-}_{\alpha_N G_{N+1}}$, whereas the
peaks between $-12$ and $-14$~mV with shifts of $-0.28$~mV have
bottleneck rates $\Sigma^{\ssL +}_{\alpha_{N+1} G_{N}}$.  Finally, the
fact that {\em only two\/} values of $\delta V$ shifts were observed
confirms that all current steps are due to tunneling through the {\em
  same\/} grain.

\subsubsection{Effects of gate voltage on tunneling spectrum}
\label{sec:effects-of-gate}

\begin{figure}
\centerline{
  \epsfig{figure=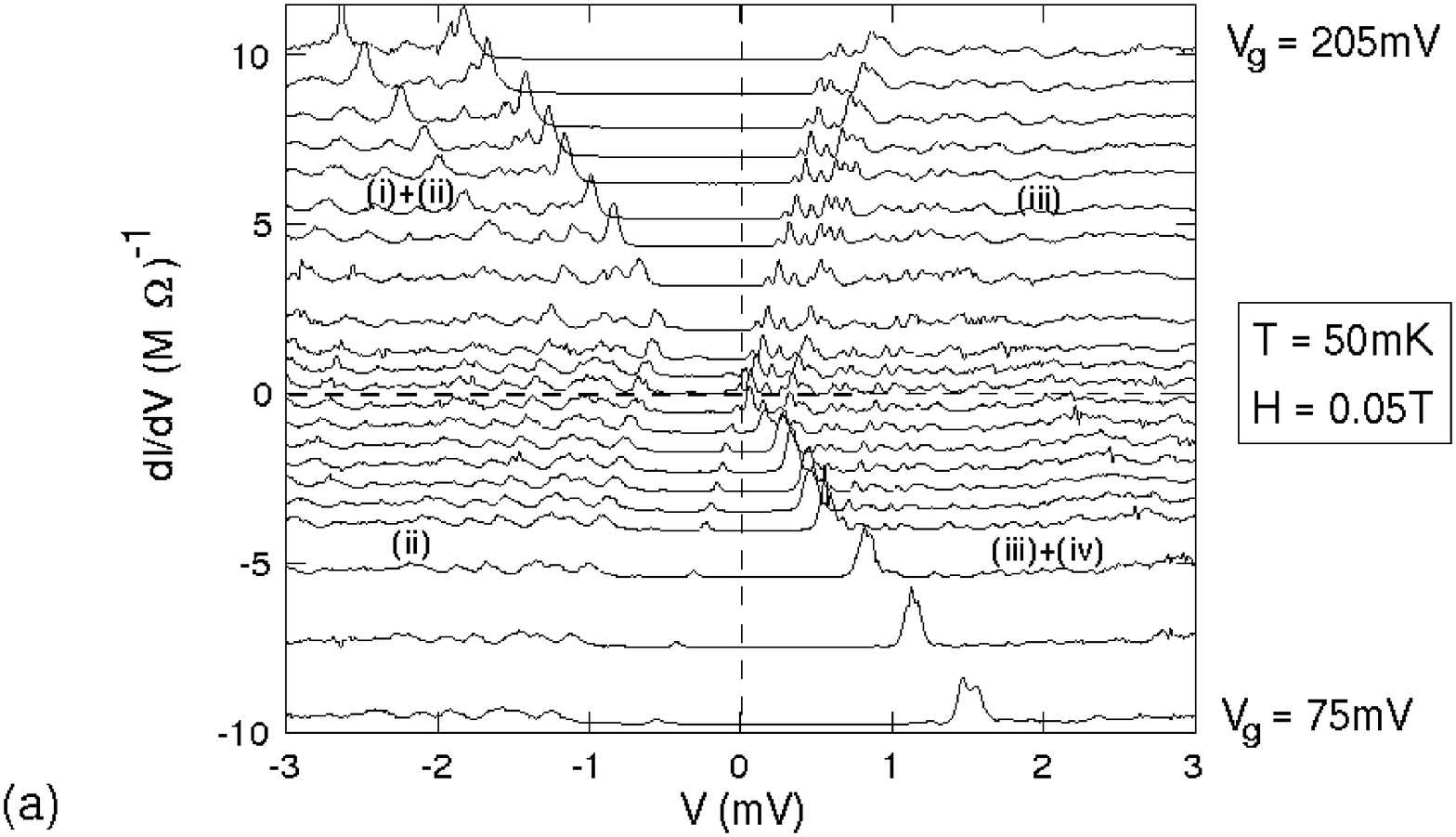,width=0.85\linewidth}}
\centerline{
  \epsfig{figure=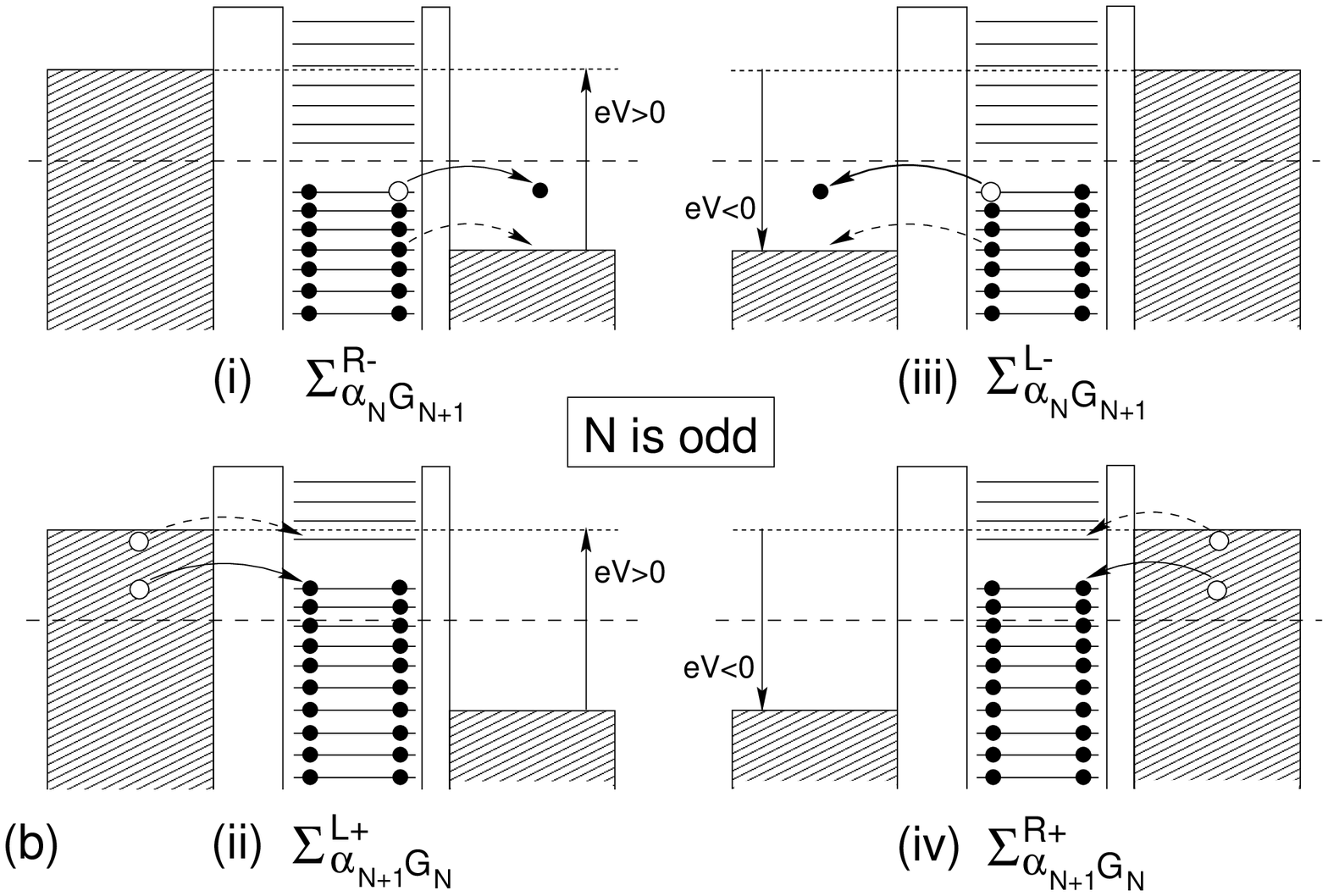,width=0.85\linewidth}}
  \caption[$V$ and $\Vg$ dependence  of a gated device]{
    (a) ${\rm d}I/ {\rm d}V$ vs.\ bias voltage $V$ for the same sample
    \cite{rbt97} as in Fig.~\ref{fig:generic-IV}, plotted with
    different vertical offsets for a set of $\Vg$-values ranging from
    75~mV (bottom) to 205~mV (top). All data are taken at $T=50$~mK
    and $H=0.05$~T (to drive the Al leads normal).  (b) Schematic
    depiction (using the conventions of \Fig{fig:energydiagram}) of
    the processes acting as bottlenecks in the four quadrants of (a),
    for $N$ being odd, with rates: (i)
    $\Sigma^{R-}_{\alpha_N,G_{N+1}}$ (upper left quadrant), (ii)
    $\Sigma^{L+}_{\alpha_{N+1},G_N}$ (upper and lower left quadrants),
    (iii) $\Sigma^{L-}_{\alpha_N,G_{N+1}}$ (upper and lower right
    quadrants), (iv) $\Sigma^{R+}_{\alpha_{N+1},G_N}$ (lower right
    quadrant).  Solid (dashed) arrows depict bottleneck tunneling
    transitions into the lowest- (highest) energy {\em final\/} states
    accessible for the chosen value of $V$, and filled circles
    represent the electron configuration of the lowest-energy {\em
      final\/} state.  Increasing $\Vg$ decreases $e \VD$ (recall:
    $e<0$) and shifts the drawn grain levels down; the degeneracy
    point between the $N$- and $(N+1)$-electron ground states, at
    which the Coulomb blockade regime has shrunk to zero, is reached
    when the topmost non-empty level of the final ground state
    coincides with the equilibrium, $V=0$ chemical potential of the L
    and R leads, indicated by the horizontal long-dashed line.  }
\label{fig:Vg-V}
\label{fig:explainV-Vg}
\end{figure}

In devices with a gate electrode, $\Vg$ can be used to controllably
change the average number of electrons on the grain, and to maximize
or minimize nonequilibrium effects, as desired.  This is illustrated in
\Fig{fig:Vg-V}(a), which shows a series of differential conductance
curves for a range of different gate voltages, for the same device as
in \Fig{fig:generic-IV} (whose caption gives the device parameters).
The bottleneck processes in its various regimes are depicted
schematically in \Fig{fig:explainV-Vg}(b), which one can construct by
the following arguments: First note in \Fig{fig:Vg-V}(a) that as $\Vg$
is increased, the extent of the Coulomb blockade region at low $|V|$,
in which there are no conductance peaks, decreases, goes to 0, and
then increases.  At this zero crossing the SET is at a Coulomb
blockade degeneracy point, at which, say, the $N$ and $(N+1)$-electron
ground states are degenerate.  When passing this point while
increasing $\Vg$, which lowers $e \VD$ (since $e<0$) and hence [by
\Eq{eq:Epot}] favors larger $\Nex$, the grain's average electron
number changes from $N$ to $N+1$.  Moreover, nonequilibrium effects
(discussed in detail in \Sec{sec:nonequilibrium}) are weak close the
the zero crossing but become stronger as $\Vg$ is tuned away from it,
since their strength is governed, roughly speaking, by the size of the
threshold voltage $|V|$ at which the first peak occurs.  If one wants
to make an ``ideal'' measurement of the ``isolated'' grain's
properties in which the grain is disturbed as little as possible by
the measuring process, one has to minimize nonequilibrium effects by
tuning $\Vg$ close to a zero crossing.

Next, note from \Eq{eq:eV><0} that the left half of \Fig{fig:Vg-V}(a)
(where $eV > 0$) is governed by rates of the type
$\Sigma^{R -}_{\alpha_{N} \alpha'_{N+1}}$ 
and $\Sigma^{L +}_{\alpha_{N+1} \alpha'_N}$ , and
the right half by $\Sigma^{L -}_{\alpha_{N} ,\alpha'_{N+1}}$ 
and $\Sigma^{R +}_{\alpha_{N+1}
  , \alpha'_N}$, as illustrated in the left and right halves of
\Fig{fig:explainV-Vg}(b), respectively.  To find out which of these
rates ($r=$ L or R) are bottlenecks to transport for a given $\Vg$,
consider the change $\delta V$ in the position of a conductance peak
upon a small change $\delta \Vg$ in gate voltage; these changes are
related by the condition that the corresponding threshold voltage
$\bar V_r$ in (\ref{eq:Vrthresholds}) remain unchanged [by
\Eq{eq:barVr-explicitly}]:
\begin{equation}
  \label{eq:DeltaVgDeltaV}
 {\delta V \over \delta V_\ssg} = { C_\ssg \over C_{\ssR} + C_\ssg/2} 
\quad \mbox{if} \; \; r = {\rm L} , \qquad
 {\delta V \over \delta V_\ssg} = {-  C_\ssg \over C_{\ssL} 
+ C_\ssg/2} 
\quad \mbox{if} \; \; r = {\rm R}  \, .
\end{equation}
A peak that has barrier L or R as bottleneck will thus move with slope
$\delta V/ \delta \Vg > 0$ or $<0$, respectively.  In
\Fig{fig:Vg-V}(a), all peaks indeed do shift uniformly with $\Vg$,
which is another proof that all are due to tunneling through the {\em
  same\/} grain.  In the upper left and lower right quadrants of
\Fig{fig:Vg-V}(a), slopes of both signs occur (with magnitudes 0.009
and $-0.025$, consistent with the capacitances listed in the caption
of \Fig{fig:generic-IV}); each of these quadrants thus corresponds to
a ``two-bottleneck'' situation [see \Sec{sec:extractingspectrum}(b)],
in which \emph{both} of the rates shown in the corresponding halves of
\Fig{fig:explainV-Vg}(b) act as bottlenecks.  In contrast, in the
upper right and lower left quadrants \emph{all} peaks move with slope
$\delta V/ \delta \Vg > 0$, implying that only the left barrier acts
as bottleneck (consistent with the fact that $R_\ssL/R_\ssR \simeq 18
\gg 1 $); each of these quadrants thus corresponds to a
``one-bottleneck'' situation [see \Sec{sec:extractingspectrum}(a)],
from which excitation spectra can be extracted.  The upper right
quadrant has $\Sigma^{\ssL -}_{\alpha_N ,\alpha'_{N+1}}$ as bottleneck
rate and hence yields the fixed-$N$ excitation spectrum [cf.\ 
\Eq{eq:fixed-N-excitation-spectrum2}], while the lower left quadrant
has $\Sigma^{\ssL +}_{\alpha_{N+1},\alpha'_N}$ as bottleneck rate and
hence yields the fixed-$(N+1)$ excitation spectrum [cf.\ 
\Eq{eq:fixed-N-excitation-spectrum1}].

\emph{Excitation spectra such as the latter two are the central
  measurement results of SET tunneling spectroscopy.} One can proceed
to extract from them all sorts of interesting information about the
grain and the correlations which its electrons experience. For
example, a very remarkable property of the fixed-$(N+1)$ excitation
spectrum in the lower left quadrant of \Fig{fig:Vg-V}(a) is the
sizeable distance between the first and second peaks (counted from
$|V|=0$).  As will be discussed in more detail in
\Sec{sec:gap-in-spectrum}, this ``spectroscopic gap'' is evidence for
superconducting pairing correlations on the grain: roughly speaking,
it reflects the gap of $2 \Delta_\grain$ (the pairbreaking energy
cost) in the excitation spectrum of a superconducting grain with an
{\em even\/} number of electrons.  This also implies that $N$ is odd.
The behavior of these excitation spectra in a magnetic field,
discussed in \Secs{sec:sc-measuredgap},
\ref{sec:paramagnetic-breakdown} and \ref{sec:magfield}, confirms
these conclusions.

\subsubsection{Capacitance ratios}

The accuracy of tunnel-spectroscopic measurements depends on the
accuracy with which one knows the capacitance ratios $(C_{\ssR/\ssL} +
C_\ssg/2)/C$ [or $C_{\ssR/\ssL}/(C_\ssL + C_\ssR)$ if no gate is
present]. These occur, \eg, in the voltages $\bar V_{\ssL/\ssR}$
[\Eq{eq:barVr-explicitly}] that enter in the threshold condition
(\ref{eq:Vrthresholds}), or as voltage-to-energy conversion factors in
\Eqs{eq:fixed-N-excitation-spectrum1} and
(\ref{eq:fixed-N-excitation-spectrum2}). These capacitance ratios can
of course be determined from the large-scale Coulomb-blockade $I$-$V$
curves of the grain. However, two additional methods are available,
which are more accurate (typically to within 1\% \cite{gueron99}),
since they exploit the properties of the discrete tunneling spectrum
itself:

(i) If the leads are made from a superconducting material, then the
ratios $(C_{\ssR/\ssL} + C_\ssg/2)/C$ can be read off directly from
the $\delta V$ shifts [\Eq{eq:vshift-due-to-Delta}] in peak positions
that occur when superconductivity in the leads is switched off by
switching on a magnetic field.

(ii) One can compare the positions of two tunneling peaks occurring for
opposite signs of the bias voltage but involving the same transition,
\eg\ $|G_N \rangle\to |\alpha_{N+1} \rangle$
(the case $|G_{N+1} \rangle\to |\alpha_{N} \rangle$ is
analogous). If $eV > 0$, this
transition will  [by \Eq{eq:eV><0}]
occur by tunneling across barrier L with rate
$\Sigma^{L+}_{\alpha G}$, and if $eV < 0$ across barrier R with rate
$\Sigma^{R+}_{\alpha G}$. Denote the bias voltages
at which the two conductance peaks occur by $V^{\alpha,>}$ and
$V^{\alpha,<}$, respectively, and use \Eq{eq:barVr-explicitly} to
define from these two corresponding $\bar V_r$-voltages, denoted by
$\bar V_\ssL^{\alpha,>}$ and $\bar V_\ssR^{\alpha,<}$. 
By \Eq{eq:Vrthresholds}, these in fact are equal, $\bar V_\ssL^{\alpha,>}
= \bar V_\ssR^{\alpha,<} (=\E^{N+1}_\alpha - \E^N_G)$.  Using
\Eq{eq:barVr-explicitly} to rewrite this equality in terms of
$V^{\alpha,>}$ and $V^{\alpha ,<}$, one immediately finds that $ 
|V^{\alpha,>} / V^{\alpha,<}| = {(C_\ssL + C_\ssg/2)/ (C_\ssR + C_\ssg /
  2)}$. Adding unity to this or to its inverse
gives the desired capacitance ratios.

\subsubsection{Ground state energy differences are 
currently not measurable}
\label{sec:ground-state-energies-not-measurable}

According to \Eq{eq:Vrthresholds}, the threshold values of $e \bar
V_r^\alpha$ and $e V_\alpha$ at which the {\em first\/} conductance
peak occurs as $|V|$ is increased from 0 should allow one, in
principle, to also determine the ground state energy difference
$E_{G_{N \pm 1}} - E_{G_N}$ between grains with adjacent electron
numbers.  In practice, however, it has so far {\em not\/} been
possible to do this accurately: by \Eq{eq:barVr-explicitly}, $e \bar
V_r^\alpha$ depends through $e (\Vg C_\ssg + Q_0)/C$ on the gate
voltage and the random offset charge $Q_0$, and in the devices studied
so far, $e Q_0/C$ could not be determined with sufficiently high
accuracy, \ie\ with an uncertainty smaller than the scale of the
grain's mean level spacing $d$, typically $\simeq$ 0.1~meV, for the
following reason:

$Q_0$ can of course be determined reasonably  accurately 
by studying the large-scale Coulomb oscillations of the $I$-$V$ curve
that occur as functions of $\Vg$ at fixed $V$, a procedure that is
well-established for mesoscopic SETs, for which indeed it has been
possible to measure the ground state energy difference between a
superconducting island with an even or odd number of electrons
\cite{hanna91,Tuominen-92,Tuominen-93,Tinkham-95,Saclay,eiles93}.
However, a complication arises for the nanoscopic grains of present
interest, due to the smallness of their gate capacitances (typically
$\simeq 0.1$~aF): to sweep $\Vg$ through one Coulomb oscillation, the
gate voltage $\Vg$ must be swept through a range so large (namely
$e/C_g \simeq 1 \mbox{V}$) that during the sweep, RBT routinely
observed small ``rigid'' shifts of the entire tunneling spectrum at
random values of $\Vg$. (Similar spontaneous shifts are discussed,
\eg, in \cite{eiles93}.) Likewise, shifts are also observed when $eV$
is swept over a range of order $\Ec$.  These shifts presumably are due
to single-electron changes in the charges contained in \emph{other}
metal grains\footnote{ This interpretation is bolstered by the fact
  that the shifts usually occur at approximately periodic values of
  $\Vg$.}  in the neighborhood of the grain of interest.  These
changes produce sudden shifts in the electrostatic potential energy of
the grain, \ie\ shifts in $e Q_0/C$ by a few \%, which are comparable
in magnitude to the mean level spacing $d$ and hence spoil any
attempts to determine $e Q_0/C$ with an uncertainty smaller than $d$.
--- The $\Vg$ and $Q_0$ dependencies are subtracted out, however, when
one considers the distances $V_{\alpha \tilde \alpha}$ \emph{between}
peaks, as in \Eqs{eq:fixed-N-excitation-spectrum1} and
(\ref{eq:fixed-N-excitation-spectrum2}). Moreover, to measure only the
first few discrete states, the range over which $V$ needs to be varied
typically is small enough that (with some luck) no shifts occur.

\newpage \section{Normal grains in an applied magnetic field}
\label{sec:normal-grain-magfield}

In this section we describe how 
the ability to resolve individual eigenenergies allowed RBT to
determine the parity (even or odd) of the
number of electrons, say $N$, in the $V=0$ ground state of a
normal-state metal grain \cite{rbt95}
[\Sec{sec:kramers}]. We also
explain why orbital magnetism is negligible in ultrasmall grains
[\Sec{sec:orbitalmagnetism}].

\subsection{Breaking of Kramers-degeneracy by applied magnetic field}
\label{sec:kramers}

In the
absence of an applied magnetic field $(H=0)$, a normal-state grain 
will have time-reversal symmetry.  For an even-$N$ grain, the
many-electron wave function for the ground state will be a spin
singlet, in order that the orbital energy be minimized.  In contrast,
the ground state of an odd-$N$ grain for $H=0$ necessarily is 
two-fold degenerate, by Kramers' theorem,  forming a Kramers doublet.
When $H$ is turned on, this doublet is Zeeman-split by $\pm \half \muB
\, g \, H$.  Therefore, for an {\em even\/}-$N$ grain at small $H$,
the lowest-lying tunneling excitations correspond to transitions from
the even-grain ground state singlet to the odd-grain ground state
doublet, \ie\ to two states split by $H$, so that the lowest-$V$
conductance peak should exhibit Zeeman splitting in an applied field.
On the other hand, for an {\em odd\/}-$N$ grain with $T \ll \muB \, g
\, H/ \kB$, the odd-grain ground state will be the lower-energy state
of the Kramers doublet; the lowest-lying tunneling excitation will
thus consist only of a single transition from this odd-grain ground
state to the even-grain ground state singlet, so that the lowest-$V$
conductance peak should {\em not\/} split into two as a function of
$H$.

\begin{figure}
\centerline{\epsfig{figure=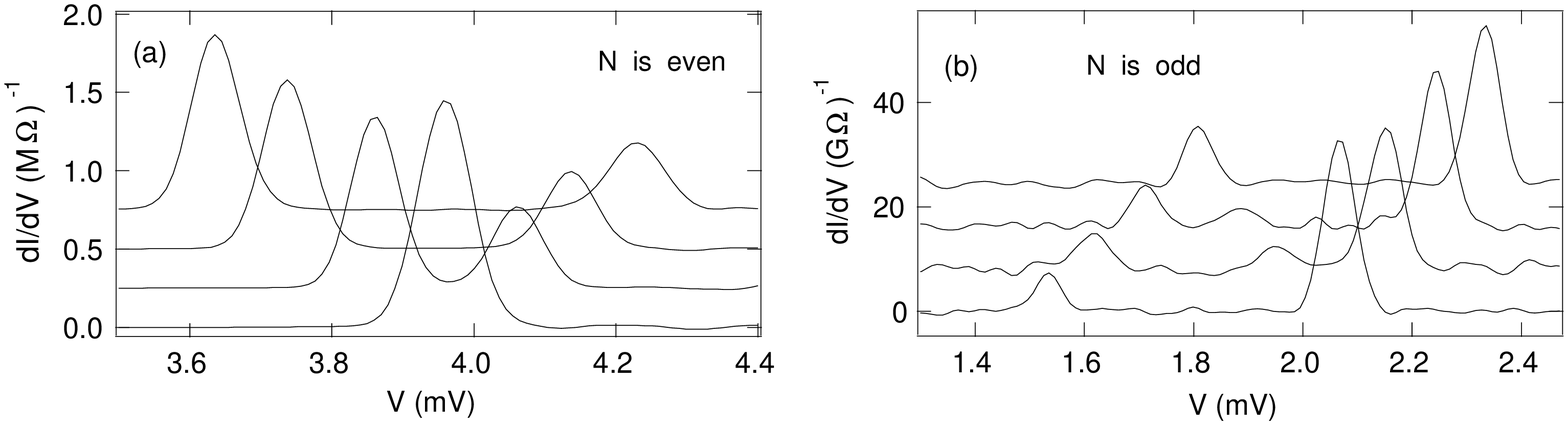,%
width=0.98\linewidth}}
\caption[Kramers splitting of odd ground state]{
 $\d I / \d V$
  vs.\ $V$ at 50~mK and $H=0.03$, 1, 2, 3~T from bottom to top, for
  two different samples \cite{rbt95}.  The first transition above the
  Coulomb-blockade threshold exhibits Zeeman splitting in (a) but not
  in (b), hence the electron number $N$ of the $V=0$ ground state is
  even for grain (a) and odd for grain (b). }
\label{fig:kramers}
\end{figure}

RBT observed both kinds of behavior, first in several different
ungated Al grains \cite{rbt95,rbt96b}, in each of which $N$ has a fixed,
random value, and subsequently also in a given gated Al grain, in which
$N$ could be tuned via $\Vg$ \cite{rbt97,ralph-curacao}.
Bulk Al superconducts, but the present grains were so small ($r
\lesssim 3$~nm) that superconducting pairing correlations on the
grain were negligible, since the mean level spacings ($\simeq
0.5$~meV) were substantially larger than the bulk gap $\tilde \Delta$
($\simeq 0.18$~meV for Al).  \Figs{fig:kramers}(a) and (b) show the
lowest-$V$ conductance peaks for an even-$N$ and an odd-$N$ grain,
respectively. The lack of splitting of {\em both\/} of the prominent
peaks in \Fig{fig:kramers}(b) indicates that the first two even-grain
excited states for this grain are both spin singlets. The small peak
in \Fig{fig:kramers}(b) (visible below the second large peak), which
moves to lower $V$ with increasing $H$, is attributable to the
nonequilibrium occupation of the higher-energy level of the odd-grain
initial-state Kramers doublet.

By measuring the difference in $V$ between the Zeeman-split peaks and
converting them to energy using \Eqs{eq:fixed-N-excitation-spectrum1} or
(\ref{eq:fixed-N-excitation-spectrum2}), RBT determined the $g$-factors for
the states in \Fig{fig:kramers}(a) and (b) to be $g=1.87 \pm 0.04$ and $g=1.96
\pm 0.05$, respectively.  In the former case the deviation from the value
expected for pure Al, namely $g^{\rm pure}=2$, 
is significant, and can be attributed to
spin-orbit scattering from the surface or impurities \cite{halperin,averin90},
a subject that will be discussed in detail in \Sec{sec:spin-orbit}.

\subsection{Why orbital diamagnetism is negligible in ultrasmall grains}
\label{sec:orbitalmagnetism}

Of course, a magnetic field in principle also couples to the orbital
motion of the conduction electrons. Orbital effects in spherical and
cylindrical superconductors whose dimensions are smaller than the
penetration depth were first considered by Larkin \cite{Larkin65}.
However, in grains as small as those of RBT, orbital diagmagnetic
effects are negligible \cite{salinas99}, as can be seen from the
following argument \cite{glazman-priv}: Orbital diamagnetism becomes
important when the splitting which it produces in the eigenenergies of
two time-reversed states (analogous to a Zeeman splitting) becomes
comparable to the mean level spacing.  This can be shown to occur [see
\Sec{sec:estimate-g-eff}] when the flux enclosed by an electron
traversing a ``closed'' trajectory corresponding to a discrete quantum
state exceeds one flux quantum $\Phi_0 $ ($= hc/2e$), \ie\ when the
magnetic field exceeds the scale $H_\orb = \Phi_0 / A_\typ$, where $
A_\typ$ is the typical (directed) area covered by the electron's
trajectory during the ``period'' of its motion, $\hbar / d$. Since the
number of bounces off the grain's boundaries during this time is
roughly $ \gdc = \ETh/d$, where $\hbar / \ETh$ is the time to cross
the grain once, the directed area is $A_\typ \approx r^2 \sqrt \gdc $,
where the square root accounts for the fact that the direction of
motion after each bounce is random \cite{matveev00}. It follows that
\begin{eqnarray}
  \label{eq:Horbital}
  H_\orb \approx { \Phi_0 \over r^2 \sqrt{ \gdc } } 
  = { 2067  \, \mbox{T} . \mbox{nm}^2 \over r^2 \sqrt{ \gdc }} \; . 
\end{eqnarray}
Thus, $H_\orb$ grows with decreasing grain size.  Using $d$ from
\Eq{eq:d-estimate} and the ballistic estimate of
\Eqs{eq:g-dimensionlessconductance} and (\ref{eq:define-Thouless})
for $\gdc$, with $a = 3$, we find that
hemispherical Al grains with radii of (say) $r \approx 3$ or 5~nm
have  $H_\orb \approx 19$ or 7~T, respectively.
If larger values are used for $a$, as  would be
appropriate for more pancake-shaped grains
[cf.\ footnotes \ref{f:Ethouless} and \ref{f:estimate-gdim}],
$H_\orb$ would be even larger.

We may thus conclude that orbital diagmagnetic effects only begin
to play a role for largish grains
($\gtrsim 5$~nm), and then only for the highest fields (of
7~T) studied by RBT.

\newpage \section{Superconductivity: experiment and phenomenological
theory}
\label{sec:superconductivity}

Among RBT's most striking experimental results are those for Al
grains: they found a significant spectroscopic gap indicative of
superconducting pairing correlations in the fixed-$N$ excitation
spectra of largish Al grains ($r  \gtrsim 5$ nm) if $N$ is even,
but not if it is odd (see \Figs{fig:sc-spectra(h=0)} and
\ref{fig:sc-magneticfield} below); however, in their smallest grains
($r \lesssim 3$ nm) no such parity-dependent gap could be discerned.

These results are of rather general interest and significance, since
they invite reconsideration of an old but fundamental question: {\em
  what is the lower size limit for the existence of superconductivity
  in small grains?}\/ Anderson \cite{anderson59} addressed this
question already in 1959: he argued that if the sample is so small
that its electronic eigenspectrum becomes discrete,
``superconductivity would no longer be possible'' when its mean level
spacing $d$ becomes larger than the bulk gap, which we shall denote by
$\tilde \Delta$.  Heuristically, this is obvious (see
\Fig{fig:v2u2-prb97} below): $\tilde \Delta / d$ is the number of
free-electron states that pair-correlate (those with energies within
$\tilde \Delta $ of $\eF$), i.e. the ``number of Cooper pairs'' in the
system; when this becomes $\lesssim 1$, it clearly no longer makes
sense to call the system ``superconducting''.

Although Anderson's answer is correct in general, it generates further
questions: What, precisely, does ``superconductivity'' mean in
ultrasmall grains, for which many of the standard criteria such as
zero resistivity, Meissner effect and Josephson effect, are not
relevant\footnote{For an isolated nm-scale grain, (i) its resistivity
  is not defined, since electron motion is ballistic and the mean free
  path is boundary-limited; (ii) the grain radius is smaller than the
  penetration depth, so that no Meissner effect occurs; and (iii) the
  electron number is fixed, so that the order parameter cannot have a
  well-defined phase.}?  How is one to modify the grand-canonical BCS
theory to obtain a fixed-$N$ theory appropriate for ultrasmall grains,
whose charging energy suppresses number fluctuations?  What happens in
the regime $d \gtrsim \tilde \Delta$ in which superconductivity has
broken down? Is the breakdown parity dependent?  How is it influenced
by a magnetic field?  \Secs{sec:superconductivity} and
\ref{sec:crossover} are concerned with providing detailed answers to
these and related questions.

\Sec{sec:superconductivity} 
is devoted to the experiments themselves. We 
analyze and qualitatively explain them
in the framework of a  phenomenological
theory by Braun \etalia\ \cite{braun97,braun99,braun-thesis},
which offers a simple intuitive picture for visualizing
the pairing correlations and the changes these
incur when the grain size is decreased.
Further theoretical developments, inspired by RBT's experiments 
but not directly concerned with their interpretation, will be
discussed in \Sec{sec:crossover}.

\Sec{sec:superconductivity} is organized as follows: subsection
\begin{itemize}
\item[(\ref{sec:gap-in-spectrum})]
presents the experimental results of RBT;
\item[(\ref{sec:model})]  proposes a phenomenological model
for an isolated ultrasmall grain;
\item[(\ref{sec:canonical-pair-mixing})] discusses how pairing
  correlations can be visualized in a fixed-$N$ system and explains
  when and in what sense it can be called ``superconducting'';
\item[(\ref{sec:generalBCS})] presents  a generalized variational BCS
  approach for calculating the eigenenergies of various variational
  eigenstates of general spin $|s\rangle$;
\item[(\ref{sec:CC-transition})] discusses how an increasing magnetic
  field induces a transition from a pair-correlated to a normal
  paramagnetic state;
\item[(\ref{sec:tunneling-spectra-prb97})]
  presents theoretical tunneling spectra of the RBT type,
  which are in qualitative agreement with RBT's measurements;
\item[(\ref{sec:time-reversed})] explains how RBT's experiments give
  direct evidence for the dominance of purely time-reversed states in
  the pairing interaction;
\item[(\ref{sec:BCS-parity})] discusses various parity effects that
  are expected to occur in ultrasmall grains.
\end{itemize}

\subsection{A gap in the excitation spectrum}
\label{sec:gap-in-spectrum}
\label{sec:sc-introduction}
\label{sec:sc-measuredgap}

\begin{figure}[t]
  \centerline{\epsfig{figure=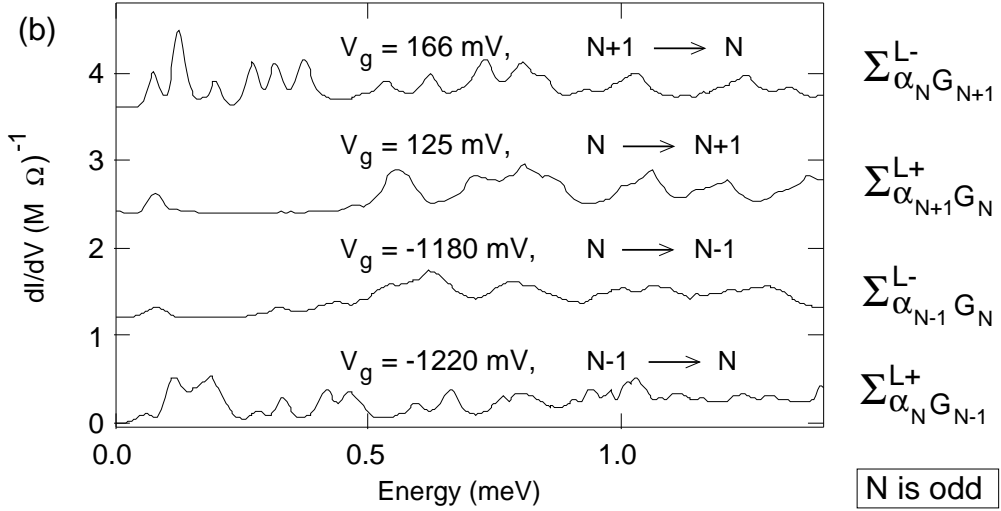,%
width=0.95\linewidth}}
\caption[Excitation Spectra of a Superconducting
Grain with an Even or Odd Number of Electrons]{
 Excitation spectra for the same sample \cite{rbt97} as
  \Figs{fig:generic-IV} and \ref{fig:Vg-V}(a), measured at $T=50$~mK and
  $H=0.05$~T (to drive the Al leads normal), for four different $\Vg$-values,
  corresponding to different values for the grain's average electron number
  (from top to bottom: $N+1,N,N,N-1$).  The curves are artificially offset on
  the vertical axis and each is labeled by the associated bottleneck tunneling
  rate $\Sigma^{L \pm}$ depicted in \Fig{fig:Vg-V}(b),
  the bottleneck barrier being $r=\ssL$ in this case.  Plotted is $\d
  I/\d V$ vs.\ energy, where the latter is given by $|eV| (C_\ssR + 
  C_\ssg/2)/C = 0.73 |e V|$ [see \Eqs{eq:fixed-N-excitation-spectrum1} and
  (\ref{eq:fixed-N-excitation-spectrum2})]; the voltage-to-energy conversion
  factor reflects the voltage drop across barrier L.
  The sizeable spectroscopic gap between the first two peaks in the
  middle two curves, and its absence in the top and bottom curves, reflects
  the pairbreaking energy cost in the excitation spectrum of a superconducting
  grain with an {\em even\/} number of electrons, and implies that $N$ is odd.
  }
\label{fig:sc-spectra(h=0)}
\end{figure}

\begin{figure}[t]
\centerline{\epsfig{figure=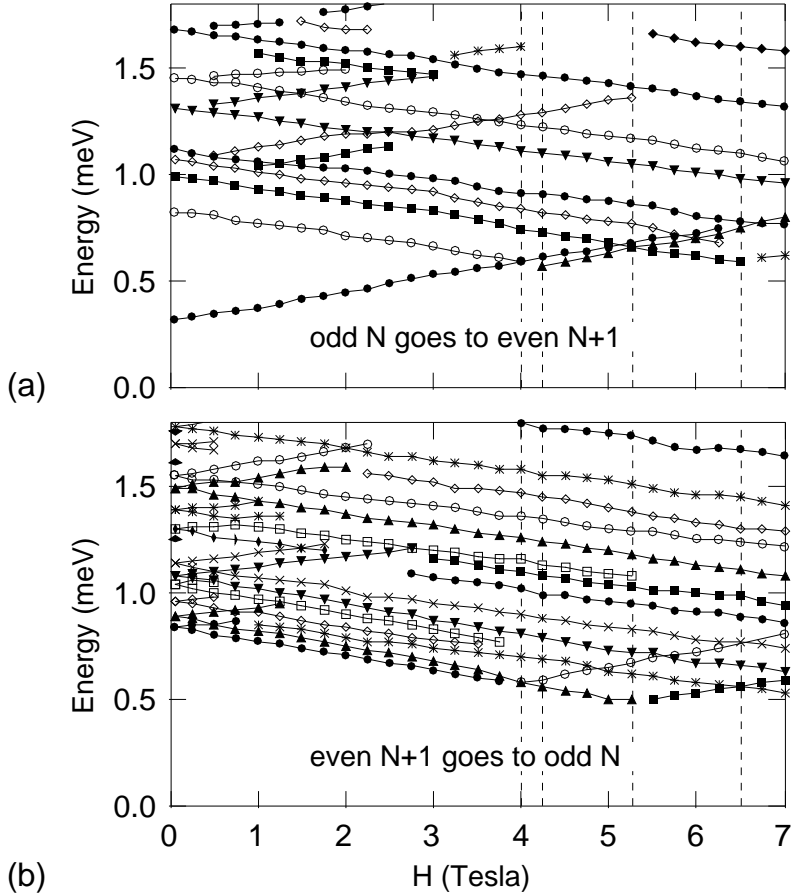,width=0.75\linewidth}}
  \caption[Experimental excitation spectra for superconducting
grains]%
{ Magnetic field dependence \protect\cite{rbt97} of excitation spectra
  such as those of \Figs{fig:Vg-V} and \ref{fig:sc-spectra(h=0)} and
  taken for the same grain, at (a) $\Vg \approx 110$~mV and (b) $\Vg
  \simeq180$~mV [lower left and upper right quadrants of
  \Figs{fig:Vg-V}(a), respectively]. Each line 
  represents a distinct conductance peak in the $\d I / \d V$ curve
  and traces how its energy changes with $H$. 
  Upward-moving peaks are broader and less distinct than
  downward-moving ones (for reasons poorly understood) and can be
  followed only for a limited range of $H$ before they are lost in the
  background.  The distances between lines directly give the grain's
  (a) fixed-($N+1$) and (b) fixed-$N$ excitation spectrum, where $N$
  is odd; the corresponding rate-limiting processes are schematically
  illustrated in \Fig{fig:Vg-V}(b,ii) for (a) and in
  \Fig{fig:Vg-V}(b,iii) for (b).  The vertical dashed
  lines indicate the first four level-crossing fields $H_{s,s'}$
  (defined in \Eq{eq:hcrit} and assigned by comparison with
  \Fig{fig:spectra}, see \Sec{sec:tunneling-spectra}),
  namely $H_{0,1} \!=\!4$T, $H_{1/2,3/2}\!=\!4.25$T,
  $H_{1,2}\!=\!5.25$T and $H_{3/2,5/2}\!=\!6.5$T with uncertainty $\pm
  0.13$T (half the $H$-resolution of 0.25T). }
\label{fig:sc-magneticfield} 
\label{fig:experimental-spectra}
\end{figure}

RBT's results for largish Al grains ($r \! \gtrsim \!5$ nm) are
exemplified by \Figs{fig:sc-spectra(h=0)} and
\ref{fig:sc-magneticfield}: if the number of electrons on the grain in
the final state after the bottleneck tunneling process is even (middle
two curves of \Fig{fig:sc-spectra(h=0)}), the excitation spectra
display a spectroscopic gap between the first two conductance peaks
that is significantly larger than the mean spacing between subsequent
peaks, whereas no such gap occurs if the final-state electron number
is odd [top and bottom curves of \Fig{fig:sc-spectra(h=0)}).  In other
words, even-$N$ excitation spectra (number parity $p =0$) are gapped,
whereas odd-$N$ excitation spectra ($p = 1$) are not.  This is even
more clearly apparent when comparing \Figs{fig:sc-magneticfield}(a)
and \ref{fig:sc-magneticfield}(b), which give the magnetic-field ($H$)
dependence of an even-$N$ and odd-$N$ excitation spectrum,
respectively.  However, in their smallest grains ($r \lesssim 3$ nm)
no such clear parity-dependent spectroscopic gap could be discerned.

BRT interpreted these observations as evidence for {\em
  superconducting pairing correlations} in their larger grains, using
notions from the BCS theory of superconductivity: in an even grain,
all \emph{excited} states involve at least one broken Cooper pair, \ie\ two
BCS quasiparticles, and hence lie at least $2 \Delta$ above the
fully-paired BCS ground state; in contrast, in an odd grain {\em
  all}\/ states have at least one unpaired electron, \ie\ at least one
quasiparticle, and hence no significant gap exists between ground- and
excited states.  \Fig{fig:explainV-Vg}(b) is a cartoon illustration of
this interpretation, if one imagines two electrons drawn on the same
energy level to represent a ``Cooper pair'' (making this cartoon
precise will be one of the main goals of this chapter): in
\Figs{fig:explainV-Vg}(b,ii) and \ref{fig:explainV-Vg}(b,iv) the final
electron number is even, and all final excited states (reached via
dashed arrows) have one less ``Cooper pair'' than the final ground
state (reached via the solid arrow); in contrast, in
\Figs{fig:explainV-Vg}(b,i) and \ref{fig:explainV-Vg}(b,iii) the final
electron number is odd, and the final ground and excited states have
the same number of ``Cooper pairs''.

The approximately linear $H$-dependence of the excitation spectra in
\Fig{fig:sc-magneticfield} was attributed by RBT to the Zeeman energy
shifts of discrete levels in a magnetic field, as discussed in
\Sec{sec:kramers}.  Indeed, the method described there for determining
the grain's number parity applies here too: the fact that the lowest
state in \Figs{fig:sc-magneticfield}(a) or (b) does not or does
display Zeeman splitting, respectively, confirms that $N$ is odd.  The
reduction of the spectroscopic gap in \Figs{fig:sc-magneticfield}(a)
therefore is purely due to Zeeman energy shifts and has nothing to do
with the reduction of the BCS gap parameter due to pair-breaking that
occurs in bulk samples in a magnetic field \cite{tinkham-book}.  A
detailed discussion of the spectra's magnetic field dependence will be
given in \Secs{sec:CC-transition} and 
\ref{sec:tunneling-spectra-prb97}.

For completeness, it should be remarked that a spectral gap in
ultrasmall superconducting grains was observed as long ago as 1968 by
Giaever and Zeller \cite{GiaeverZeller-68,ZellerGiaever-69}, who
studied tunneling through granular thin films containing electrically
insulated Sn grains. They found gaps for grain sizes right down to the
critical size estimated by Anderson (radii of 2.5~nm in this case),
but were unable to prove that smaller particles are always ``normal''.
RBT's experiments are similar in spirit to this pioneering work, but
their ability to focus on \emph{individual} grains makes a much more
detailed study possible.

\subsection{A model for ultrasmall grains with pairing correlations}
\label{sec:model}

In this section we construct a model for an isolated ultrasmall grain
with pairing correlations, using phenomenological arguments valid for
the regime $d \lesssim \tilde \Delta$. The model,
which we shall call ``\dbcsm'', allows a rather
detailed qualitative understanding of the measurements of RBT
\cite{braun97,braun99} and hence is ``phenomenologically successful''.
For $d\gg \tilde \Delta$ it is unrealistically simple, however, and
should rather be viewed as a toy model for studying how pairing
correlations change as the grain is made smaller and smaller.

\subsubsection{A simple reduced BCS interaction plus a Zeeman term}
\label{sec:Hamiltonian}

Following the philosophy of the ``orthodox theory'' for Coulomb
blockade phenomena mentioned in \Sec{sec:chargingenergy}, we assume
that the only effect of the Coulomb interaction is to contribute an
amount $\Ec \Nexsq$ [cf.\ \Eq{eq:Epot}] to the eigenenergy of each
eigenstate of the grain.  Since the charging energy is huge (5 to 50
meV) in ultrasmall grains, this term strongly supresses electron
number fluctuations, so that, to an excellent approximation, all
energy eigenstates will also be number eigenstates.\footnote{ An
  exception occurs \emph{at} a degeneracy point, where $ \Epot(\Nex) =
  \Epot(\Nex+1)$ [cf.\ \Eq{eq:Epot}]; the resulting complications
  will not be considered here.} Since $\Ec \Nexsq$ is constant within
every fixed-$N$ Hilbert space, we shall henceforth ignore it, with the
understanding that the model we are about to construct should always
be solve in a fixed-$N$ Hilbert space (and that the use of
grand-canonical approaches below, after having dropped $\Ec \Nexsq$,
simply serves as a first approximation to the desired canonical
solution).

The only symmetry expected to hold in realistic, irregularly-shaped
ultrasmall grains at zero magnetic field is time-reversal symmetry. We
therefore adopt a single-particle basis of pairs of time-reversed
states $|j \pm \rangle$, enumerated by a discrete index $j$. Their
discrete energies $\varepsilon_j$ are assumed to already incorporate
the effects of impurity scattering and the average of
electron-electron interactions, etc.  As simplest conceivable model
that incorporates pairing interactions and a Zeeman coupling to a
magnetic field, we adopt a Hamiltonian $\op H = \op H_0 + \op H_\red$
of the following reduced BCS form:
\begin{equation} 
  \label{eq:hamiltonian}
  \label{eq:hamilton-1}
 \op  H_0 = \sum_{j, \sigma= \pm} (\varepsilon_j -  \mu  - \sigma h) 
    c^\dagger_{j\sigma}c^\ds_{j\sigma} \; , \qquad
  \op H_\red = 
    - \lambda d \sum_{ij}
    c^\dagger_{i+}c^\dagger_{i-}c^\ds_{j-}c^\ds_{j+} \; .
\end{equation}
Here $ - \sigma h \equiv \sigma \frac12 \mu_Bg H$ is the Zeeman energy
of a spin $\sigma$ electron in a magnetic field $H$, and we shall take
$h>0$ below.  Models of this kind had previously been studied by
Richardson \cite{richardson63a,richardson63b,richardson64,%
  richardson65a,richardson65b,richardson66,richardson66-b,richardson67,%
  richardson77},
Strongin \etalia\ \cite{Strongin-70}, M\"uhlschlegel \etalia\ 
\cite{Muehlschlegel-72,muehlschlegel94} and Kawataba
\cite{kawabata-80,kawabata-81}. The first application to RBT's grains
for $h=0$ was by von Delft \etalia\ \cite{vondelft96} and for $h \neq
0$ by Braun \etalia\ \cite{vondelft96,braun97,braun99}. 

Due to level repulsion the $\varepsilon_j$'s will, to first
approximation, be uniformly spaced. Unless otherwise specified, we
shall for simplicity always (except in \Sec{sec:sc-level-statistics})
take a completely uniform spectrum with level spacing $d$.
Fluctuations in the level spacings have been studied with methods of
random matrix theory \cite{smith96,sierra99}, with qualitatively similar
results (see \Sec{sec:sc-level-statistics}).  For a system with a
total of $N$ electrons, where the \emph{electron number parity} $p
\equiv N \mbox{mod} 2$ is equal to $0$ for even $N$ and $1$ for odd
$N$, we use the label $j=0$ for the lowest-lying non-doubly-occupied
level (with occupation number $p$) in the $T=0$ Fermi sea, which we
shall denote by $|\F_N\rangle$. We choose the Fermi energy at $\eF
\equiv 0$ write\footnote{ This convention differs slightly from that
  used in \cite{vondelft96,braun97,braun99}, namely $\varepsilon_j = j
  d + \varepsilon_0$. The latter is a little less convenient,
  resulting, \eg, in a $p$-dependent chemical potential for the
  variational BCS ground states discussed below, $\mu_p^\BCS =
  \varepsilon_0 + (p-1)d/2$, whereas (\ref{eq:ejs}) results simply in
  $\mu_p^\BCS = 0$. \label{f:mu-value}}
\begin{equation}
  \label{eq:ejs}
\varepsilon_j = j d + (1-p)d/2 \; ,
\end{equation}
thereby taking the doubly-occupied and empty levels of $|\F_N\rangle$
to lie symmetrically above and below $\eF$
(see \eg\ \Fig{fig:alpha-states} below).  The parameter $\mu$ 
in \Eq{eq:hamiltonian} is, in \gc\ theories, 
the chemical potential, whose
value${}^{\ref{f:mu-value}}$ determines the average particle number.
For canonical theories, which make no reference to a chemical
potential, $\mu$ is not needed and can be dropped (\ie\ set equal to
0).


The pairing interaction is of the reduced BCS form, in that it
scatters a pair of electrons from one pair of time-reversed states
into another.  It is taken to include only states whose energy
separation from the Fermi energy lies within the cutoff given by the
Debye frequency: $|\varepsilon_j | < \omegaD$.  Experimental support
for the sufficiency of neglecting couplings between non-time-reversed
pairs of states,\footnote{ A theoretical motivation for the simple
  form assumed for the pairing interaction, based on random matrix
  theory, may be found in \Ref{agam98} and is briefly mentioned at the
  beginning of \Sec{sec:beyound-orthodox-model}.}  \ie\ of using only
a \emph{reduced} BCS-Hamiltonian, is given in \Sec{sec:time-reversed}.
The pair-coupling constant in \Eq{eq:hamilton-1} is written as
$\lambda d$, where $\lambda$ is a dimensionless parameter independent
of the grain's volume, to make it explicit that both $\hat H_0$ and
$\hat H_\red$ make extensive ($\propto \Vol$) contributions to the
ground state energy (since the number of terms in each sum $\sum_j$ in
\Eq{eq:hamilton-1} scales with $N$, and $d \propto \eF /N$).  The
``bulk gap'' of the model, obtained by solving the standard
BCS gap equation [\Eq{eq:gap-bulk}] at $T=0$ in the bulk limit,
 thus is
\begin{equation}
  \label{eq:lambda-definition}
  \tilde \Delta =\omegaD/\sinh(1/\lambda) \; .
\end{equation}
To be precise, by ``bulk limit'' we shall always mean $d/ \tilde
\Delta \to 0$ and $N \to \infty$ while the product $N  d $ is kept
fixed, and use $d \sum_j \to \int \! {\d} \varepsilon_j$.

An applied magnetic field will completely penetrate an ultrasmall
grain, since its radius (typically $r \lesssim 5$nm) is much smaller
than the penetration length of 50 nm for bulk Al. The Zeeman term in
Eq.~(\ref{eq:hamiltonian}) models the fact that RBT's measured tunnel
spectra of \Fig{fig:experimental-spectra} evolve approximately
linearly as a function of magnetic field, with $g$ factors
between\footnote{\label{f:wrong-g} Claims of smaller $g$ factors made
  in \Ref{rbt96a} are wrong, the result of confusing different orbital
  states as Zeeman-split spin states.  This was made clear in
  \Ref{rbt97}, where it was observed that upward-trending Zeeman
  states can have significantly smaller amplitude than
  downward-trending states, making them difficult to observe.}  $1.95$
and $2$ (determined from the differences between measured slopes of
up- and downward-moving lines).  Deviations from $g = 2$ probably
result from spin-orbit scattering, known to be small but nonzero in
thin Al films \cite{Meservey-70,Meservey-94}, but neglected below
(where $g=2$ is used).  Furthermore, orbital diamagnetism is also
negligible, just as for thin films in a parallel magnetic field
\cite{Meservey-70,Meservey-94} but in marked contrast to bulk samples
where it causes the Meissner effect: as explained in
\Sec{sec:orbitalmagnetism}, a magnetic field only begins to
significantly affect the orbital motion of the electrons once it
exceeds the scale $H_\orb \simeq \Phi_0 /r^2 \sqrt{\gdc}$
\cite{glazman-priv,matveev00};
but according to the estimates mentioned after \Eq{eq:Horbital}, RBT's
ballistic grains with $r \lesssim 5$~nm have $H_\orb \gtrsim 7$~T,
hence orbital effects should set in only near the highest fields of
7~T studied by RBT.  Indeed, some larger grains do show slight
deviations from $H$-linearity \cite{rbt96a} for large fields, which
probably reflect the onset of such orbital effects \cite{Bahcall-priv};
however, these are much smaller than Zeeman effects in the grains of
present interest, and will be neglected here.  Thus, the model assumes
that Pauli paramagnetism due to the Zeeman energy completely dominates
orbital diamagnetism, similarly to the case of thin films in parallel
magnetic fields \cite{Meservey-70,Meservey-94}.

Intuitively speaking, it is clear that the \dbcsm\ introduced above
contains all ingredients necessary to make contact with the spectra of
\Fig{fig:sc-magneticfield}: it is formulated in terms of discrete
levels, it contains a pairing interaction which is known, from bulk
BCS theory \cite{BCS-57,tinkham-book}, to cause a gap in the
excitation spectrum, and it contains a Zeeman term that will cause
eigenenergies to linearly depend on an applied magnetic
field.  Indeed, we shall see in \Sec{sec:tunneling-spectra-prb97}
that it can be used to obtain a rather detailed qualitative
understanding of the spectra of \Fig{fig:sc-magneticfield}.

\subsubsection{Choice of numerical values for model parameters}

When doing numerical calculations for this model, some choices must be
made for the numerical values of its parameters (though slight changes
in their values will not change the results qualitatively).  We shall
follow the choices made by Braun \etalia\ \cite{braun99}, since these
led to  reasonable agreement between experimental and theoretical
excitation spectra. For the Debye frequency they used the textbook
value \cite{ashcroft-mermin} for Al of $\omegaD=34$meV.  Making an
appropriate choice for the ``bulk gap'' $\tilde \Delta$ is less
straightforward, since its experimental value for systems of reduced
dimensionality often differs from that of a truly bulk system,
presumably due to (poorly-understood) changes in the phonon spectrum
and the effective electron-phonon coupling.  For example, for thin Al
films \cite{Strongin-70,Garland-68} it is known that $\tilde
\Delta_{\rm thin \, film} \simeq 0.38$~meV, which is about twice as
large as the gap of a truly bulk system, $\tilde \Delta_\bulk =
0.18$~meV.  (This increase in $\tilde \Delta$ is not universal,
though; \eg, for Nb $\tilde \Delta$ is smaller in thin films than in
the bulk.)  Since ultrasmall grains are in many ways analogous to thin
films in a parallel magnetic field [see \Sec{sec:CC-transition}],
Braun \etalia\ adopted the thin-film value for grains too, \ie\ used
$\tilde\Delta \simeq 0.38$meV.  These choices imply that the
dimensionless pair-coupling constant $\lambda= [\sinh^{-1}(\omegaD/
\tilde \Delta)]^{-1} $ [cf.\Eq{eq:lambda-definition}] has the value
$\lambda = 0.194$.  (In \Sec{sec:tunneling-spectra-prb97} we shall
see, \emph{a posteriori}, that the choices $\tilde \Delta = 0.34$ and
$\lambda = 0.189$ would have been slightly more appropriate.)
Finally, for those numerical calculations that are explicitly cut-off
dependent, Braun \etalia\ smeared the cutoff of the BCS interaction
over two single-electron levels; this smooths out small
discontinuities that would otherwise occur in $d$-dependent quantities
each time the energy $|\varepsilon_j|$ of some large-$|j|$ level moves
beyond the cutoff $\omegaD$ when $d$ is increased.

Note that the above way of choosing $\lambda$ lumps into a single
phenomenological constant all the poorly-understood effects of reduced
dimensionality \cite{Strongin-70} on the phonons that mediate the
attractive electron-electron interaction. 
Studying these effects in detail would be interesting in its own
right, but would require systematic investigations with grains of
well-controlled shapes and sizes.  For the case of RBT's
irregularly-shaped grains, using a phenomenological coupling constant
seems the best one can do. Note, though, that the precise value of
$\lambda$ is not very important as long as all energies are measured
in units of $\tilde \Delta$ (as we shall do for all numerical
calculations), since most of the $\lambda$-dependence is thereby
 normalized away.  Therefore, the slight difference between the
$\lambda$-values proposed above and those used in
\cite{braun98,braun-thesis,sierra99} (namely 0.224) hardly matters.

\subsubsection{Some general properties of the eigenstates
-- the blocking effect}
\label{sec:generalproperties}

The eigenstates of the \dbcsm\ of \Eq{eq:hamiltonian} have some
simple but general properties that are worth stating at the outset.

Firstly, every eigenstate of $\hat H$ will also be an eigenstate of
the number operator $\hat N = \sum_{j \sigma} c^\dagger_{j \sigma}
c^\ds_{j \sigma}$, since $[\hat H, \hat N] = 0$.

Secondly, since the interaction only involves levels within the cutoff
energy $\omegaD$ of $\eF$, the dynamics of those lying outside this
range is trivial. We shall thus ignore them henceforth and focus only
on the remaining set of \emph{interacting} levels, denoting this set
by $\I$. 

\begin{figure}[t]
  \centerline{\epsfig{figure=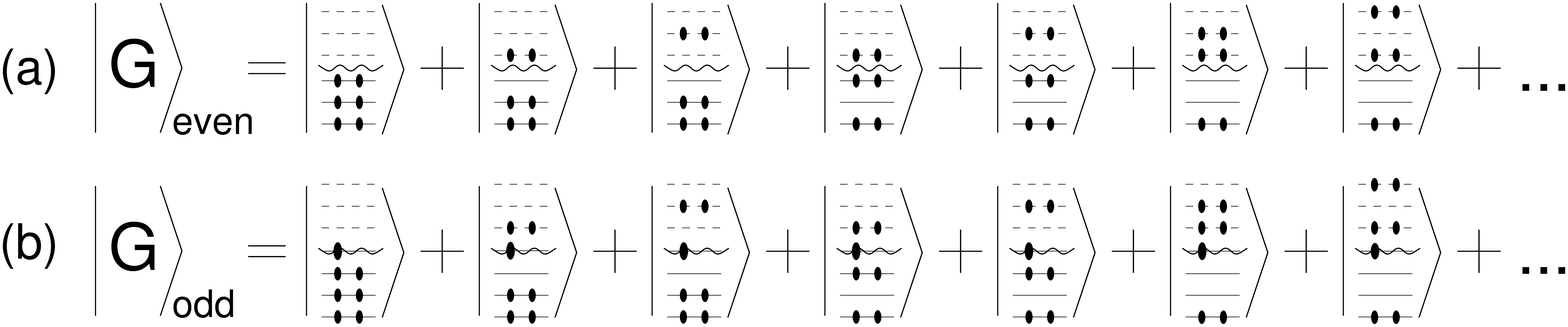,%
width=0.95\linewidth}}
  \caption[Cartoon of the exact even and odd ground states of a 
  reduced BCS Hamiltonian]{A cartoon depiction of the exact ground
    state for a reduced BCS Hamiltonian, for $N$ being even (a) or odd
    (b): they are coherent superpositions of eigenstates of $\hat H_0$
    (whose respective amplitudes are not depicted) that all have the
    same $N$; the leftmost of these is (a) the even or (b) the odd Fermi
    ground state $|\F_N \rangle$, whose Fermi energy is indicated
    by the wavy line.  }
    \label{fig:exactgroundstate}
  \end{figure}

Thirdly, {\em singly-occupied}\/ levels do not participate in the
pair scattering described by $\hat H$: ``unpaired'' electrons in such
levels are not scattered to other levels, hence the labels of
singly-occupied levels are good quantum numbers.  Moreover, every
unpaired electron Pauli-blocks the scattering of other pairs into its
own singly-occupied level, \ie\ it restricts the phase space available
to pair scattering and thereby weakens the amount of pairing
correlations, as we shall see in detail later.  This was called the
\emph{``blocking effect''} by Soloviev \cite{Soloviev-61}, who
discussed it extensively in the early 1960's in the context of nuclear
physics.  The eigenstates $|\alpha \rangle$ and corresponding
eigenenergies $\E_\alpha$ of $\hat H$ thus have the following general
forms:
\begin{eqnarray}
|\alpha \rangle &= & |\Psi_n,\B\rangle = 
\prod_{i \in \B} c_{i \sigma_i}^\dagger |\Psi_n\rangle ,
  \label{eq:generaleigenstate} 
\\
  \label{eq:generaleigenstate-2} 
 |\Psi_n \rangle & = & \sum_{j_1, \dots, j_n}^\U
\psi (j_1, \dots , j_n) \prod_{\nu=1}^n b_{j_\nu}^\dagger  
|\Vac \rangle
\; , 
\\
\label{eq:generaleigenenergy}
\E_{\alpha} &=& \E_n + \E_\B (h) \, , 
\qquad   \E_\B  (h) = \sum_{i \in \B} 
(\varepsilon_i - \mu - \sigma_i h) \, .
\end{eqnarray}
This describes $N = 2n + b$ electrons, $b$ of which are unpaired and
sit in a set $\B$ of singly-occupied, blocked levels, making a
contribution $\E_\B (h)$ to the total eigenenergy.
The remaining $n$ pairs of electrons, created by the pair operators
$b_j^\dagger = c^\dagger_{j +} c^\dagger_{j -}$, are distributed among
the remaining set $\U= \I \backslash \B$ of \emph{unblocked} levels,
with wavefunction $\psi (j_1, \dots , j_n)$ ($\sum_j^\U \equiv
\sum_{j \in \I\backslash \B}$ denotes a sum over all {\em unblocked\/}
levels in $\I$).  The corresponding state $|\Psi_n \rangle$ is an
eigenstate of the pair number operator and a Hamiltonian $\hat H_\U$
involving only pair operators:
\begin{eqnarray}
  \label{eq:eigenpsi}
  &&  \sum_j^\U b_j^\dagger b^\ds_j |\Psi_n\rangle
  = n |\Psi_n\rangle 
   , \qquad    \hat H_\U |\Psi_n\rangle = \E_n |\Psi_n\rangle \; , 
\\
 && \hat H_\U = \sum_{ij}^\U \left[ 2 (\varepsilon_j - \mu) \delta_{ij} -
 \; \lambda \, d \right]  b_i^\dagger b^\ds_j \; .
 \label{1}
 \end{eqnarray}
 Each eigenstate $|\Psi_n ,\B\rangle$ may 
  be visualized as a coherent superposition of eigenstates of
 $\hat H_0$ that all lie in the same fixed-$N$ Hilbert space,
 and in all of  which each pair of unblocked $(j \in \U$),
 time-reversed levels $|j \pm\rangle $  is either
 doubly occupied or empty. This is illustrated in
 \Figs{fig:exactgroundstate}(a) and (b), which schematically depict
 the exact ground states for even and odd $N$, respectively.
The odd ground state has a 
single blocked level, at the Fermi energy, 
containing an unpaired electron. The latter somewhat 
weakens pairing correlations relative to the even ground state
and hence leads to parity effects, which will 
be extensively discussed in later sections.

  A useful measure for the amount of energy gained by $|\alpha
  \rangle$ via its correlations is its ``condensation energy''
  relative to the uncorrelated state $|\alpha \rangle_0$,
\begin{equation}
  \label{eq:define-condensation-energy}
  E^\cond_\alpha = \E_\alpha -
 {}_0 \langle \alpha | \hat H | \alpha \rangle_0 , \qquad
\mbox{where} \quad | \alpha \rangle_0 = \prod_{i \in \B} c^\dagger_{i
\, \sigma_i} |\U  \rangle_0 \, ,
\end{equation}
and $|\U \rangle_0$ is the ``Fermi ground state'' in $\U$, for which
the $n$ pairs occupy the $n$ lowest-lying levels in $\U$.
 
 Note that $\hat H_\U$ is $h$-independent, since the total Zeeman
 energy of any \emph{pair} of electrons is zero. Hence the full
 $h$-dependence of the eigenenergies resides in the rather trivial
 contribution $\E_\B(h)$ of the blocked levels, which is a very
 important and useful simplification.
 
 Diagonalizing $\hat H_\U$ would be trivial if the $b$'s were true
 bosons.  However, they are not, and in the subspace spanned by the
 set $\U$ of all non-singly-occupied levels instead satisfy the
 ``hard-core boson'' relations,
\begin{eqnarray}
\label{hard-core-boson-1}
  b^{\dagger 2}_j = 0, \qquad
\label{hard-core-boson-2}
\mbox{[} b^\ds_j, b^\dagger_{j^{\prime}}\mbox{]} =
\delta_{j j^{\prime}} (1 - 2 b^\dagger_j b^\ds_j), \qquad 
\label{hard-core-boson-3}
\mbox{[} b^\dagger_{j} b^\ds_{j},  
b^\dagger_{j'} \mbox{]} &=& \delta_{j j'} b^\dagger_j \; ,
\end{eqnarray}
which reflect the Pauli principle for the fermions from which the
$b$'s are constructed. In particular, $b^{\dagger 2}_j=0$ implies that
only those terms in (\ref{eq:generaleigenstate-2}) are non-zero for
which the indices $j_1, \dots j_n$ are all distinct.

The task of finding the eigenstates $|\Psi_n \rangle$ is thus highly
non-trivial. Nevertheless, an exact solution does exist. Unbeknownst
to most of the condensed-matter physics community, it was found and
studied extensively by Richardson in the 1960's and will be presented
in \Sec{sec:sc-richardson}. Throughout the present section
\ref{sec:superconductivity}, however, we shall use more well-known
approaches based on the variational wavefunctions introduced
by BCS \cite{BCS-57}, and that had been used to study the \dbcsm\ 
before Richardson's solution was revived towards the end of 1998.

\subsection{Canonical characterization of pairing correlations}
\label{sec:canonical-pair-mixing}
\label{sec:meaningofsc}
\label{chap:fixedN}

Since the \dbcsm\ has the standard reduced BCS form, the most natural
first step toward understanding its $T=0$ properties is to use
BCS-like variational wavefunctions (or equivalently Bogoljubov's
mean-field approach), and indeed this will be done in
\Secs{chap:generalBCS} to \ref{sec:BCS-parity}.  However, the
limitations of such an approach should be realized at the outset: the
spectra measured by RBT are excitation spectra for a grain with a {\em
  fixed electron number\/} $N$, and hence should be calculated for a
grain with definite electron number $N$ (\ie\ completely isolated from
the rest of the world, \eg\ by infinitely thick oxide barriers).  In
contrast, the variational wavefunctions of BCS [\Eq{eq:BCSground}
below] do not have the fixed-$N$ form [\Eq{eq:generaleigenstate-2}]
which any true eigenstate should have, but instead are formulated in a
{\em grand-canonical\/} (\gc) framework (as is the Bogoljubov
mean-field approach to which they are intimately related).

When considering a truly isolated superconductor such as a perfectly
insulated grain (another example would be a superconductor levitating
in a magnetic field due to the Meissner effect), one therefore needs
to address the following question, which will be the main theme of the
present section: \emph{how is one to incorporate the fixed-$N$
  condition into BCS theory, and how important is it to do so?}  This
issue is well understood and was discussed at length in the early days
of BCS theory (Rickayzen's book gives a beautiful discussion
\cite{rickayzen-book}), in particular in its application to pairing
correlations in nuclei \cite[p.~439]{RingSchuck-80} (see also the
general remarks in \cite{Lipkin-60}). Nevertheless, for pedagogical
reasons the arguments are well worth recapitulating in the present
context.

We shall first remind the reader that the use of a \gc\ framework is
only a matter of convenience, since the essence of the pairing
correlations that lie at the heart of BCS theory is by no means
inherently \gc\ and can easily be formulated in canonical language
\cite{vondelft96}. We then show how standard BCS theory fits into this
scheme, point out that the differences between results obtained using
\gc\ and canonical wavefunctions are negligible for $d \ll \tilde
\Delta$, and conclude that for the purposes of gaining a
phenomenological understanding of the experimental data, standard
grand-canonical BCS theory should be sufficient.  Nevertheless, the
fundamental question of how to improve on this theory, in order to
achieve a truly canonical description and to properly treat
fluctuation effects, which become important for $d \gtrsim \tilde
\Delta$ \cite{muehlschlegel62,Janko-94,matveev97}, is interesting and
important in its own right and will be addressed at length in
\Sec{sec:sc-canonical}.

For simplicity, throughout the present subsection
\ref{sec:canonical-pair-mixing} we shall consider only the even ground
state in the thermodynamic limit (in which even-odd differences are
negligible), so that $\U=\I$ and blocking effects need not be worried
about.

\subsubsection{The grand-canonical BCS wavefunction}

Conventional BCS theory describes the pairing correlations induced by
an attractive pairing interaction such as $\hat H_\red$ of
\Eq{eq:hamiltonian} within a \gc\ ensemble, formulated on a Fock space
of states in which the total particle number $N$ is not fixed.  This
is illustrated by BCS's famous variational ground state Ansatz
\begin{eqnarray}
  \label{eq:BCSground}
    |\BCS \rangle = \prod_j 
    (u_j + \e^{\ii\phi_j} 
    v_j b^\dagger_j )\,|\Vac\rangle \; , 
    \qquad \mbox{with} \quad u_j^2 + v_j^2 = 1,
\end{eqnarray}            
where the variational parameters $u_j$ and $v_j$ are real and $\phi_j$
is a phase (which, it turns out, must be $j$-independent, for reasons
discussed below).  $|\BCS \rangle$ is not an eigenstate of $\hat N$
and its particle number is fixed only on the average by the condition
$ \langle \hat N \rangle_\BCS = N$, which determines the \gc\ chemical
potential $\mu$.  Likewise, the commonly used \gc\ definition
\begin{equation}
  \label{eq:BCS-gap}
  \Delta_{\MF} \equiv \lambda \, d \sum_j \langle b_j
\rangle_\BCS = \lambda \, d  \sum_{j} u_j v_j \e^{\ii\phi_j}
 \; 
\end{equation}
for the superconducting pairing parameter only makes sense in a \gc\ 
ensemble, since $\langle b_j \rangle$ would trivially give
zero when evaluated in a canonical ensemble, formulated on a strictly
fixed-$N$ Hilbert space of states.  (We shall use the term ``pairing
parameter'' instead of ``order parameter'', since the latter carries
the connotation of a phase transition, which would require the
thermodynamic limit $N\to \infty$, which is not applicable for
ultrasmall grains).

\subsubsection{Canonically meaningful definition
for the pairing parameter}
\label{sec:meaningfulDelta}

A theory of strictly fixed-$N$ superconductivity must evidently entail
modifications of conventional BCS theory. However, these are only of
technical, not of conceptual nature, since the essence of the pairing
correlations discovered by BCS can easily be formulated in a
canonically meaningful way, including a definition for the pairing
parameter.  We shall now attempt to explain, in intuitive,
non-technical terms, how this may be done (our discussion is indebted
to that of Rickayzen \cite{rickayzen-book}).  Readers with a
preference for rigor may consult \Sec{sec:sc-canonical} for a
corroboration, using Richardson's exact solution, of the arguments
presented below.

Let $|G \rangle$ be the exact even ground state of the system,
depicted in \Fig{fig:exactgroundstate}(a). As explained in
\Sec{sec:generalproperties}, it is a coherent superposition of
eigenstates of $\hat H_0$ that all have the same $N$ and in all of which
each pair of time-reversed levels $|j \pm\rangle $ is either doubly
occupied or empty.  Due to this coherent superposition, $|\G \rangle$
entails strong pairing correlations, whose essential properties may be
understood by investigating how they modify the correlators
\begin{eqnarray}
  \label{eq:defineu2v2}
\label{eq:C_ij}
 C_{ij}  \equiv  \langle 
    b_i^\dagger b^\ds_j \rangle \, , \qquad 
 \bar v_j^2 \equiv C_{jj} = \langle b_j^\dagger b^\ds_{j}
 \rangle \, ,  \qquad 
  \bar u_j^2  \equiv \; \langle  b^\ds_{j} b_j^\dagger \rangle \,  , 
\qquad 
\end{eqnarray}
relative to the form these take on for the Fermi ground state
$|\F_N\rangle$:
\begin{eqnarray}
(C_{ij})_\F  = \delta_{ij} (\bar v_j^2)_\F \, , \qquad
(\bar v_j^2)_\F
= \theta (  -
  \varepsilon_j) \, ,  \qquad
 (\bar u_j^2)_\F  = \theta ( \varepsilon_j  ) \, .
\end{eqnarray}
$C_{ij} (= C_{ji}^\ast)$
is the matrix element for the interaction to be able to scatter a pair
of electrons from level $j$ to $i$, and 
$\bar v_j^2$ and $\bar u_j^2$ are the probabilities to find level $j$
doubly occupied or empty, respectively.
The pairing correlations in $|\G\rangle$ must be such that $\hat
H_\red$ lowers the ground state energy below that of the uncorrelated
Fermi sea $|\F_N\rangle$ by an amount that is extensive ($\propto N
\propto d^{-1}$)
in the thermodynamic limit.  Clearly, this requires that $ \langle \op
H_\red \rangle_\G - \langle \op H_\red \rangle_\F $ is negative and
extensive, \ie\ that
\begin{eqnarray}
  \label{eq:exptation-Hred}
 \lambda \, d  \sum_{ij}  
\mbox{[} C_{ij} - (C_{ij})_\F \mbox{]}  \simeq 
 \lambda \, d  \sum_i \sum_{j < i} 2  \re ( C_{ij} ) 
\, \propto \, N \;  \quad \mbox{(and positive}).
\end{eqnarray}
In the second expression we neglected the diagonal terms, since their
number is so small (only $\propto N $) that $\lambda d \sum_j [\bar
v_j^2 - (\bar v_j^2)_\F ] $ is at best of order unity in the
thermodynamic limit.  For \Eq{eq:exptation-Hred} to hold, $|\G\rangle$
must have two properties:
\begin{enumerate}
\item[(i)] the number of $C_{ij}$'s that differ significantly from
  zero (\ie\ are of order unity) should scale like $N^2$, \ie\ one
  power of $N$ per index \cite[p.~167]{rickayzen-book};
\item[(ii)] most or all of the $C_{ij}$ for $i<j$ should have the same
  phase, since a sum over random phases would average out to zero.
\end{enumerate}
Since a suitable pairing parameter should vanish in the thermodynamic
limit unless both these conditions hold, the definition
\begin{eqnarray}
  \label{eq:canonical-order-parameter}
  \Delta^2_{\can} \equiv  (\lambda \, d )^2 
\sum_{ij} (C_{ij} - 
\langle c^\dag_{i+} c^\ds_{j+} \rangle
 \langle c^\dag_{i-} c^\ds_{j-} \rangle )
\end{eqnarray}
(or its square root) suggests itself, where the subscript emphasizes
that (\ref{eq:canonical-order-parameter}) is meaningful in a canonical
ensemble too, and we subtracted\footnote{This subtraction was
  suggested to us by Moshe Schechter, who pointed out that then
  \Eq{eq:canonical-order-parameter} has a natural generalization to
  position space: it is the spatial average,
  $  \Delta^2_\can  \equiv  (\lambda \, d)^2
 \int \! \! \d \vec r_1  \d \vec r_2 \, {\cal F}(\vec r_1,
 \vec r_2)$, 
of the two-point function
\begin{eqnarray}
\nonumber 
{\cal F}(\vec r_1, \vec r_2) \equiv
 \langle \psi^\dag_+ (\vec r_1) \psi_-^\dag (\vec r_1) 
          \psi_-^\ds (\vec r_2) \psi_+^\ds (\vec r_2) \rangle 
  \;  - \; \langle \psi^\dag_+ (\vec r_1) \psi_+^\ds (\vec r_2) \rangle
     \langle \psi^\dag_- (\vec r_1) \psi_-^\ds (\vec r_2) \rangle
 \,    \label{eq:Delta_can-realspace}
\end{eqnarray}
(with $\psi_\sigma (\vec r) \equiv \Vol^{-1/2} \sum_{\vec k} \e^{\ii
  \vec k \cdot \vec r} c_{\vec k \sigma}$), which evidently measures
the amplitude for the propagation of \emph{pairs} as opposed to
uncorrelated electrons.  Other definitions for a canonically
meaningful pairing parameter have been suggested
\protect\cite{vondelft96,braun98,braun99}, such as $\lambda d \sum_j
\bar u_j \bar v_j$ or $ \lambda d \sum_j [ \langle b^\dagger_{j}
b^\ds_{j}\rangle - \langle c^\dagger_{j+}c^\ds_{j+} \rangle \langle
c^\dagger_{j-}c^\ds_{j-} \rangle ]^{1/2} \; $, but these focus only on
requirement (i) and fail to incorporate requirement (ii).  A quantity
very similar to \Eq{eq:canonical-order-parameter} was recently
proposed in Eq.~(55) of \Ref{Rossignoli-99a}, namely $(\lambda d)^2
\sum_{ij} [C_{ij} - (C_{ij})_{\lambda=0}]$. }  the ``normal-state
contribution to  $C_{ij}$.''  If (i) and (ii) hold, $\Delta_\can$ will
take on a finite value; its relation to a gap in the spectrum will
become clear below.  In the bulk limit,
$\Delta_\can$ can be shown [see \Sec{sec:bulk-few-n-differences}]
to reduce to the ``bulk pairing parameter'' $\tilde \Delta$
of \Eq{eq:lambda-definition}.

\subsubsection{Redistribution of occupation probability
across $\eF$}

\begin{figure}
   \centerline{\epsfig{figure=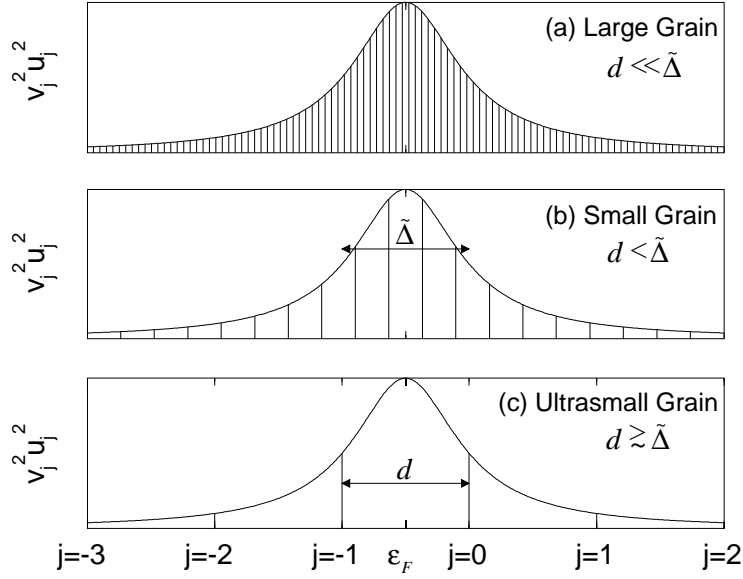,%
width=0.7\linewidth}}
  \caption[Cartoon of breakdown of superconductivity]{A cartoon
    depiction of why ``superconductivity breaks down'' when the sample
    becomes sufficiently small. Vertical lines are drawn at each
    single-particle energy $\varepsilon_j$, spaced with a mean level
    spacing $d$ corresponding to (a) a ``large'' grain
    ($d\ll\tilde\Delta$); (b) a ``small'' grain ($d \simeq 0.25 \tilde
    \Delta$); (c) an ``ultrasmall'' grain ($d \simeq \tilde \Delta$).
    The lines' height represents the function $u_j^2 v_j^2 = \frac14
    \tilde \Delta^2/(\varepsilon_j^2 + \tilde \Delta^2)$ of standard
    bulk BCS theory, to illustrate the energy range (of width $\tilde
    \Delta$ around $\eF$) within which pairing correlations are
    strongest.  Loosely speaking, the number of single-particle levels
    in this regime, namely $ \tilde \Delta /d$, corresponds to ``the
    number of Cooper pairs'' of the system.  Evidently, this number
    becomes less than one when $d \gtrsim \tilde \Delta$ as in (c), so
    that it then no longer makes sense to call the system
    ``superconducting'' [cf.\ \Sec{sec:qual-disc}].}
    \label{fig:v2u2-prb97}
  \end{figure}

Now, property (i) can be realized if all $C_{ij}$ in a finite
($d$-independent) range of $\varepsilon_i$'s and $\varepsilon_j$'s
around the Fermi energy  $\eF$ differ significantly from
zero; the width of this range will evidently determine the magnitude
of $\Delta_\can$ (provided (ii) also holds), which conversely can be
viewed as a measure of this width. But a nonzero $C_{ij}$ evidently
requires \emph{both} $b^\dagger_i b^\ds_j | \G \rangle \neq 0$,
implying $(\bar v_{j})_\G \neq 0$ and $(\bar u_i)_\G \neq 0$,
\emph{and also} $\langle \G | b^\dagger_i b^\ds_j \neq 0$, implying
$(\bar v_i)_\G \neq 0$ and $(\bar u_{j})_\G \neq 0$.  The product
$(\bar u_j \bar v_j)_\G$ must thus be different from zero [in contrast
to $(\bar u_j \bar v_j)_\F = 0$] for all $\varepsilon_j$ within a
finite range around $\eF$ (cf.\ \Fig{fig:v2u2-prb97}).  This can be
achieved by smearing out the sharp steps of the $\theta$-functions of
$(\bar v_j)_\F $ and $(\bar u_j)_\F$, so that $(\bar v_j)_\G$ [or
$(\bar u_j)_\G$] is nonzero also for a finite range of $\varepsilon_j$
above [or below] $\eF$.  In other words, for $|\G\rangle$ some
occupation probability must be redistributed (relative to
$|\F_N\rangle$) from below to above $\eF$, as illustrated in \Fig
{fig:exactgroundstate}.  This redistribution, which was called
pair-mixing in \cite{vondelft96,braun99}, frees up phase space for
pair scattering and so achieves a gain in interaction energy (provided
(ii) also holds) that more than compensates for the kinetic energy
cost incurred thereby.

  Furthermore, note that properties (i) and (ii) imply, even without
  detailed calculations, that the spectrum will be gapped.  Consider,
  for example, a ``blocking excitation'' that disrupts pairing
  correlations by having $|j +\rangle$ definitely occupied and $|j
  -\rangle$ definitely empty; since pair-scattering involving level
  $j$ is blocked, the energy cost is \label{p:blocking}
\begin{eqnarray}
&& (\varepsilon_j - \mu) - [
(\varepsilon_j - \mu)  2\langle 
b^\dagger_j  b^\ds_j \rangle) - \lambda \, d  
\sum_{i (\neq j)} 
\langle b^\dagger_i b_j^\ds + b^\dagger_j b^\ds_i  \rangle ]
\\ & &
\label{eq:excitationenergy}
= (\varepsilon_j - \mu) (1- 2 \bar v_j^2)
 +  \lambda \, d  
\sum_{i (\neq j)} ( C_{ij} + C_{ji} ) \, ,
\end{eqnarray}
in which the restriction on the sum reflects the blocking of
scattering events involving level $j$.  Since the first term of
(\ref{eq:excitationenergy}) is positive definite (particle-hole
symmetry ensures that $(\half - \bar v_j^2) \gol 0$ if $\varepsilon_j
- \mu \gol 0$) and the second of order $\Delta_\can$, the excitation
energy will be \emph{finite} even for $d\to 0$, implying the existence
of a gap of order $\Delta_\can$. Similarly, ``phase-breaking
excitations'' that violate the fixed-phase condition (ii) are gapped
too: for example, if $(C_{i j})_{\rm excited} = - (C_{ij})_{\rm
  ground}$ for a given $j$ and all $i (\neq j)$, the energy cost is 
$- \lambda d \sum_{i(\neq j)} [
(C^{\phantom{\ast}}_{ij} + C^\ast_{j i})_{\rm excited} 
- (C^{\phantom{\ast}}_{ij} + C_{j i})_{\rm ground} ]
$,  which is at least of order $ 2 \Delta_\can$.

We see, therefore, that the essence of pairing correlations can
readily be formulated in a canonical framework: (i) a redistribution
of occupation probability across $\eF$ occurs, such that each level
$j$ in a finite range around $\eF$ has a finite probability of both
being doubly occupied or empty, and (ii) any two components of the
ground state wavefunction that differ only by the exchange of a pair
of electrons between two levels $i$ and $j$ have the same phase.

Since pairing correlations with these properties are the
microscopic property at the heart of all manifestations of
``superconductivity'', it seems reasonable to call a sample
``superconducting'' as long as it exhibits pairing correlations with
measurable consequences.  And by this criterion the gap observed in
the even grains of RBT certainly qualifies.

\subsubsection{Gauge symmetry breaking}

Note that property (ii) will be preserved under the gauge
transformation $c_{j\sigma} \to \e^{\ii\phi^\prime_j} c_{j \sigma}$,
\ie\ $C_{ij} \to \e^{-2 \ii ( \phi^\prime_i -\phi^\prime_j)} C_{ij}$,
only if all $\phi^\prime_j $ are equal, say $\phi^\prime_j =
\phi^\prime$.  Property (ii), and likewise the pairing parameter
$\Delta_\can$, therefore (a) are not gauge invariant ``locally'' in
$j$-space, but (b) are gauge invariant globally. These are obvious
consequences of the facts that (a) a \emph{correlated fixed-$N$} state
consists of a \emph{phase-coherent} superposition of many different
components, and hence cannot be invariant under arbitrary changes of
the phases of individual components; and that (b) all of these
components contain the \emph{same} number of electrons $N$ and hence
under a global gauge transformation all pick up the \emph{same} phase
factor $\e^{\ii N \phi'}$. Obviously, global gauge symmetry can
therefore never be broken in a canonical ensemble.  In contrast, the
breaking of global gauge symmetry by the \gc\ pairing parameter
$\Delta_\MF$ of Eq.~(\ref{eq:BCS-gap}), which transforms as
$\Delta_\MF \to \e^{2 \ii \phi^\prime} \Delta_\MF$, is an inevitable
consequence or artefact of its \gc\ definition
\cite[p.~142]{rickayzen-book}.

\subsubsection{Making contact with standard BCS theory}
\label{sec:contact-with-BCS}

One of the breakthrough achievements of BCS was, of course, to propose
a simple variational ground state which has precisely the properties
(i) and (ii) described above: when evaluating the correlators of
\Eq{eq:defineu2v2} using $|\BCS\rangle$ of \Eq{eq:BCSground}, one
finds
\begin{eqnarray}
  \label{eq:uvC-BCS}
 (\bar u_j)_\BCS^2 = u_j^2,  \qquad (\bar v_j)_\BCS^2 = v_j^2,
\qquad (C_{ij})_\BCS = u_i v_i u_j v_j \e^{- \ii(\phi_i - \phi_j)}\; ,
\end{eqnarray}
and  also $(\Delta^2_\can)_\BCS = |\Delta_\MF|^2$.  The definite-phase
requirement (ii) can thus be implemented by choosing all the phases
$\phi_j$ to be the same, say $\phi_j = \phi$ for all $j$, thereby
breaking local gauge invariance (usually one simply takes $\phi = 0$);
and requirement (i) is fulfilled automatically when minimizing the
expectation value $\langle \hat H \rangle_\BCS$ w.r.t.\ $u_j$ and
$v_j$, since this does yield smeared-out step functions,
namely  \cite{BCS-57,tinkham-book}
\begin{equation}
\label{vj-bulk}
v_j^2 = \half  \left[1 - (\varepsilon_j - \mu)/E_j \right] \, ,
\qquad E_j \equiv \sqrt{(\varepsilon_j - \mu)^2 
+ |\Delta_\MF|^2} \; .
\end{equation} 
Here we neglected terms that vanish for $d \to 0$, and 
 $\Delta_\MF$ is determined by the famous gap equation
(for $T=0$), 
\begin{eqnarray}  \label{eq:gap-bulk}
  \frac1\lambda & = & 
d \sum_{|\varepsilon_j| < \omegaD} \frac1{2 E_j} \; . 
\end{eqnarray}
The BCS wavefunction instructively illustrates some of
the general properties discussed above.
Firstly, the product $u_j^2v_j^2$, shown in \Fig{fig:v2u2-prb97}, has a
bell-shaped form with a well-developed peak around $\eF$ of width
$\simeq |\Delta_{\MF}|$, illustrating that pairing correlations are
strongest within a region of width $|\Delta_{\MF}|$ around the Fermi
surface.  Secondly,  the energy
of a blocking excitation  [\Eq{eq:excitationenergy}]
reduces to $(\varepsilon_j - \mu) (1 - 2 v_j^2) + 2 u_j v_j |\Delta_{\MF}| =
E_j$, which is just the well-known energy of the 
Bogoljubov quasiparticle state $\gamma^\dag_{j +} | \BCS \rangle$,
where 
\begin{eqnarray}
  \label{eq:Bogoljubov}
\gamma_{j \sigma} = u_j c_{j \sigma} - \sigma v_j e^{\ii \phi}
  c^\dagger_{j-\sigma} \; . 
\end{eqnarray}
Thirdly, an example of a phase-breaking excitation is 
\begin{eqnarray}
  \label{eq:phase-breaking-excitation}
\gamma^\dag_{j +} 
\gamma^\dag_{j -} | \BCS \rangle
 = (- v_{j} e^{- \ii \phi} + u_{j} 
b^\dag_{j})
\prod_{i (\neq {j})} 
(u_i + v_j e^{\ii \phi} b^\dag_i) | \Vac \rangle
 \; ,   
\end{eqnarray}
which has $(C_{i j})_{\rm excited}
 = - u_i v_i u_{j} v_{j}$
and energy $2 E_{j}$.

It should be appreciated, however, that BCS chose a
\emph{grand-canonical} construction purely for calculational
convenience (as is made clear on p.~1180 of their original paper
\cite{BCS-57}): the trick of using a factorized form of
\emph{commuting} products in (\ref{eq:BCSground}), at the cost of
$N$-indefiniteness, makes it brilliantly easy to determine the
variational parameters $u_j$ and $v_j$. In fact, BCS proposed
themselves to use the projection of $|\BCS \rangle$ to fixed $N$ as
the actual variational ground state, namely \cite{rickayzen-book}
\begin{eqnarray}
  \label{eq:BCSground-N}
    | {\PBCS} \rangle & \equiv & \int_0^{2 \pi} \!\! {\rm d}
    \phi\, \e^{- \ii \phi N} \! \prod_j 
    (u_j \! + \! \e^{2 \ii \phi}
    v_j b^\dagger_j )\,| \Vac \rangle
\\
  \label{eq:BCSground-N-2}
 & = & {1 \over (N/2)!} \Bigl( \prod_j u_j \Bigr)
\Bigl( \sum_j {v_j \over u_j} b^\dagger_j
\Bigr)^{N/2} | \Vac \rangle \; 
\end{eqnarray}      
(PBCS for \underline{P}rojected BCS), which \emph{is} of the general
form of \Eq{eq:generaleigenstate-2}.  In the bulk limit ($d/\tilde
\Delta \ll 1$), however, it is completely adequate to use $| \BCS
\rangle$: firstly, the relative error which its factorized form
causes, by taking the occupation amplitude of level $j$ to be
independent of that of level $i$, scales like $1/N$
\cite[pp.~150,163]{rickayzen-book}; and secondly, the fluctuations in
its particle number, $(\Delta N^2)_\BCS \equiv \langle N^2
\rangle_\BCS - N^2 = \sum_j (2 u_j v_j)^2$, are equal to $\pi \tilde
\Delta / d$ in the bulk limit, in which the relative fluctuations
$(\Delta N^2)_\BCS / N^2 \propto d \tilde \Delta / \eF^2 $ therefore
vanish.  Thus, bulk results obtained from $ | \PBCS \rangle$ or $ |
\BCS \rangle$ are essentially identical.  In fact, Braun
\cite{braun98,braun-thesis} checked by explicit calculation that the
functions $(\bar v_j^2)_\G$, $(\bar v_j^2)_\PBCS$ and $v_j^2$ are
practically indistinguishable even for $d/ \tilde \Delta$ as large as
0.5 [see \Sec{sec:bulk-few-n-differences}]. Significant differences
\emph{do} develop between them once $d / \tilde \Delta$ increases past
0.5, however, as will be discussed in \Sec{sec:sc-canonical}.

To end this section, note that \Fig{fig:v2u2-prb97} offers a very
simple intuitive picture for why pairing correlations weaken with
increasing level spacing until, in Anderson's words \cite{anderson59},
``superconductivity is no longer is possible'' when $d \gtrsim \tilde
\Delta$: an increase in level spacing implies a decrease in the number
of levels within $\tilde \Delta$ of $\eF$ for which $u^2_j v^2_j$
differs significantly from zero, \ie\ a decrease in the number of
pairs with significant pairing correlations.  This number, namely
$\tilde \Delta/d$, can roughly speaking be viewed as the ``number of
Cooper pairs'' of the system, and when it becomes less than one, as in
\Fig{fig:v2u2-prb97}(c), it no longer makes sense to call the system
``superconducting''. However, this should not be taken to imply that
pairing correlations cease altogether in this regime; remnants of them
do persist, in the form of fluctuations, up to arbitrarily large
$d/\tilde \Delta$, as will be discussed in detail in
\Sec{sec:bulk-few-n-differences}.

\subsection{Generalized variational BCS approach}
\label{chap:generalBCS}
\label{sec:generalBCS}
\label{excited}

 
In the next several sections we review the generalized variational BCS
approach used by Braun \etalia\ \cite{braun98,braun99,braun-thesis} to
describe the paramagnetic breakdown of superconductivity in nm-scale
grains in a magnetic field.  This theory produces theoretical
excitation spectra that are in good qualitative agreement with the
measurements of BRT shown in \Fig{fig:sc-magneticfield} and thereby
yields the most direct confirmation available of the relevance to
experiment of the \dbcsm.  Moreover, it sheds considerable light on
how ``superconductivity breaks down'' (more precisely, how pairing
correlations weaken) with increasing $d$ and $h$: As mentioned in the
previous paragraph, in grains with $d \simeq \tilde \Delta$ (bulk
gap), near the lower size limit \cite{anderson59} of observable
superconductivity, the number of free-electron states with strong
pairing correlations (those within $\tilde \Delta $ of
$\eF$) is of order one. Thus, even in grains in which a
spectral gap can still be observed, pairing correlations are expected
to become so weak that they might be destroyed by the presence of a
single unpaired electron \cite{vondelft96}.  This can be probed
directly by turning on a magnetic field, since its Zeeman energy
favors paramagnetic states with nonzero total spin.

The theory reviewed below exploits analogies to thin films in a
parallel magnetic field \cite{Meservey-70,Meservey-94}, but explicitly
takes account of the discreteness of the grain's spectrum.  Since in
RBT's experiments the temperature $T= 50\mbox{mK}$ is much smaller
than all other energy scales ($d, \tilde \Delta$), we shall neglect
finite-temperature effects and set $T=0$.  In \Sec{chap:generalBCS}
the eigenenergies $\E_\alpha$ of the grain's lowest-lying eigenstates
$|\alpha \rangle$ are calculated approximately using a
\emph{generalized \gc\ variational BCS approach} that goes beyond
standard mean-field theory by using a different pairing parameter
$\Delta_\alpha$ for each $|\alpha \rangle$.  The $\E_\alpha$ are then
used to discuss various observable quantities, such as $h$-dependent
excitation spectra (\Sec{sec:tunneling-spectra-prb97}), direct
experimental evidence for the dominance of purely \emph{time-reversed}
states in the pairing interaction (\Sec{sec:time-reversed}), and
various parity effects (\Sec{sec:BCS-parity}).

The reasons for deciding to calculate the excitation spectra, despite
their \emph{fixed-$N$} nature, within a \emph{grand-canonical}
framework are as follows: Firstly, its simplicity.  Secondly and
perhaps most importantly, the exact eigenenergies have the general
form $\E_\alpha= \E_n + \E_\B (h)$ [\Eq{eq:generaleigenenergy}], in
which \emph{all $h$-dependence resides in the exactly known
  contribution $\E_\B (h)$ from the blocked levels.}  The choice of
approximation scheme therefore only affects $\E_n$, which determines
the $h=0$ properties of the spectrum, such as the size of the
zero-field spectral gap, etc., but not the qualitative features of the
$h$-dependence. In particular, this means that all of the analysis
below could easily be ``made exact'' by simply replacing the \gc\ 
approximations for $\E_n$ by the exact values from Richardson's
solution. However, this is expected to cause only slight quantitative
differences, since, thirdly, canonical calculations (mentioned after
\Eq{eq:BCSground-N-2} and discussed in \Sec{sec:sc-canonical}) yield
very similar results to \gc\ ones as long as $d/\tilde \Delta \lesssim
0.5$, which, by inspection of \Fig{fig:sc-magneticfield}, does seem to
be the case for the grain in question (the analysis of
\Sec{sec:tunneling-spectra-prb97} yields $d / \tilde \Delta \simeq
0.67$).

\subsubsection{The generalized variational Ansatz}
\label{sec:BCS-ansatz}

\begin{figure}[t]
  \centerline{\epsfig{figure=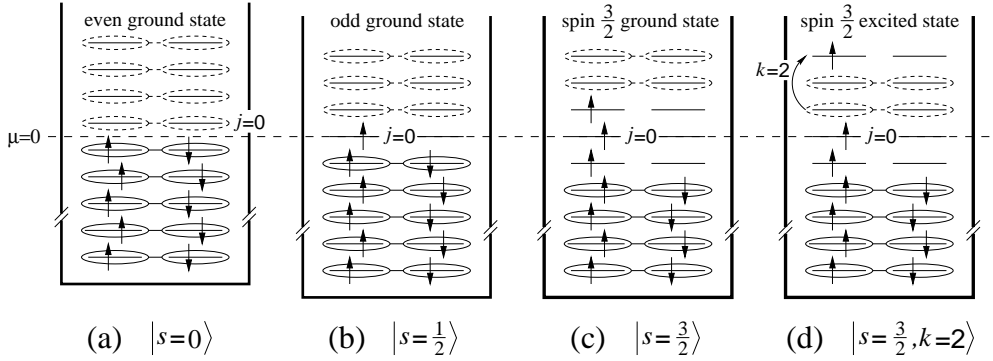,%
width=0.95\linewidth}}
  \caption[Generalized BCS wavefunctions]{Cartoon of 
    four typical variational states, 
    labeled using the notation of \Eq{eq:ansatzs} for (a-c) and
    \Eq{eq:ansatzs12} for (d).  They represent (a) the even ground
    state $|0\rangle$; (b) the odd ground state $|\half\rangle$; (c)
    the spin-${3 \over 2}$ ground state $|{3\over 2}\rangle$; (d) a
    spin-${3 \over 2}$ excited state $|{3\over 2},2\rangle$.  The
    single-particle levels are drawn for $h=0$, with the chemical
    potential half-way between levels 0 and 1 for even systems (a),
    but exactly on level 0 for odd ones (b,c,d).  The ellipses joining
    states on the same level are meant to represent a ``Cooper pair'',
    and signify its being empty or doubly occupied with amplitude
    $(u_j + v_j b_j^\dag)$; solid (dashed)
    ellipses are used for levels that would be completely filled
    (empty) in the absence of pairing correlations.  }
    \label{fig:alpha-states}
\end{figure}

The Zeeman term in the Hamiltonian of \Eq{eq:hamiltonian} favors
states with a nonzero total $z$-component of the total spin, 
$s=\half \sum_{j\sigma} \sigma c^\dagger_{j \sigma}
c^\ds_{j \sigma}$ 
(henceforth simply called ``spin''). Increasing $h$ will thus
eventually lead to a series of ground state changes to states with
successively larger spins.  In general, therefore, we are interested
in pair-correlated states with nonzero spin, and in particular in
their eigenenergies.  Following Braun \etalia\ 
\cite{braun97,braun99,braun-thesis}, we now show how this can be
calculated variationally, using the following general BCS Ansatz for a
state $|s,\bbalpha \rangle$ with $N=2n + 2s$
electrons and a definite total spin $s \ge 0$
(first introduced by Soloviev for application in nuclei
\cite{Soloviev-61}):
\begin{eqnarray}
  \label{eq:ansatz}
    |s, \bbalpha\rangle = \prod_{i \in \B} c^\dagger_{i+} 
                \prod^\U_j (u^{(s,\bbalpha)}_j + v^{(s,\bbalpha)}_j
       b^\dagger_j )\,|\Vac\rangle.
\end{eqnarray}                                               
If the spin is nonzero, it is built up by placing $2s$ unpaired spin-up
electrons in a set $\B$ of $b=2s$ single-particle levels [cf.\ 
\Eq{eq:generaleigenstate}] while the remaining single-particle levels have
BCS-like amplitudes to be either empty $(u^{(s,\bbalpha)}_j)$ or doubly
occupied by a pair $(v^{(s,\bbalpha)}_j)$, with
$(u^{(s,\bbalpha)}_j)^2+(v^{(s,\bbalpha)}_j)^2=1$.  The product $\prod_j^\U$
thus constitutes a \gc\ approximation to the state $|\Psi_n\rangle$ of
\Eq{eq:generaleigenstate-2}.  The superscript $\U$ over products (and over
sums below) indicates exclusion of the singly occupied levels in $\B$, for
which $u^{(s,\bbalpha)}$, $v^{(s,\bbalpha)}$ are not defined.

More specifically, in a given spin-$s$ sector
of Hilbert space the following two types of
specializations of \Eq{eq:ansatz} were studied in detail
($p = 2s\,{\rm mod}\,2$):
\begin{eqnarray}
  \label{eq:ansatzs}
    |s\rangle &=& 
    \prod_{i=-s + p/2}^{s-1+p/2}  c^\dagger_{i +} 
                \prod^\U_j (u^{s}_j + v^{s}_j
       b^\dagger_j)\,|\Vac\rangle.
\\
  \label{eq:ansatzs12}
    |s,k \rangle &= & 
    c^\dagger_{( s-1 +  p/2 + k ) +} c_{( s-1 +  p/2) +} 
     |s\rangle \, . 
\end{eqnarray}
$|s\rangle$ is the spin-$s$ state with the lowest energy,
\ie\ the ``variational spin-$s$ ground state'', obtained by placing
the $2s$ unpaired electrons as close as possible to $\eF$
[\Fig{fig:alpha-states}(b,c)], in order to minimize the kinetic
energy cost of having more spin ups than downs. $|s,k\rangle$ is a
particular type of excited spin-$s$ state, obtained from $|s\rangle$
by moving one electron from its topmost occupied level ($s-1+p/2$)
upwards by $k$ units of $d$ into a higher level ($ s-1+p/2 + k$).
These constructions are
illustrated in \Fig{fig:alpha-states}, of which 
 (a) and (b) represent the
variational ground states of a grain with an even or odd number of
electrons, respectively.

The orthogonality of the wavefunctions, 
 $\langle s, \bbalpha | s', \bbalpha'\rangle
= \delta_{ss'}\delta_{\bbalpha\bbalpha'}$, implies
that the variational
parameters $v_j^{(s,\bbalpha)}$ and $u_j^{(s,\bbalpha)}$ must be found
\emph{anew} for each $(s, \bbalpha)$ (hence the superscript),
by minimizing the variational ``eigenenergies'' 
\begin{eqnarray}
  \label{eq:Esalpha}
\lefteqn{  
\E^\GC_{s,\bbalpha} (h,d)  \equiv  
    \langle s,\bbalpha | \hat H | s,\bbalpha \rangle }
\\   &=& 
  -2sh + \sum_{i \in \B} (\varepsilon_{i} - \mu)
  + \sum^\U_j \left[2 (\varepsilon_j - \mu) (v^{(s,\bbalpha)}_j)^2 
 + \lambda d (v^{(s,\bbalpha)}_j)^4 \right] 
\\
 & &
  - \lambda d\Big(\sum^\U_{j}
  u^{(s,\bbalpha)}_jv^{(s,\bbalpha)}_j\Big)^2 , \nonumber
\end{eqnarray}
which we use as approximations to the exact  eigenenergies
$\E^\ex_{s,\bbalpha}(h,d)$. The 
$v_j^4$ term is not extensive and hence neglected in
the bulk case where only effects proportional to the system volume are
of interest.  Here it is retained, since in ultrasmall systems it is
non-negligible (but not dominant either) \cite{braun97,braun99}.
Solving the variational conditions 
$ 
 \frac{\partial \E^\GC_{s,\bbalpha}}{\partial v^{(s,\bbalpha)}_j} = 0
$
in standard BCS fashion yields 
\begin{equation}
\label{vj}
(v_j^{(s,\bbalpha)})^2 = \half ( 1 - \xi_j /
 [\xi_j^2 + \Delta_{s,\bbalpha}^2]^{1/2}) \; , \qquad
\xi_j \equiv \varepsilon_j-\mu - \lambda d
(v_j^{(s,\bbalpha)})^2 \, ,
\end{equation} 
where  the ``pairing parameter'' 
$\Delta_{s,\bbalpha}$ is defined by the relation
\begin{eqnarray}
  \label{eq:gap1}
  \Delta_{s,\bbalpha} & \equiv & \lambda d \sum^\U_j u_j^{(s,\bbalpha)}
  v_j^{(s,\bbalpha)} \; ,\qquad\mbox{or}  \qquad 
  \label{eq:gap}
  \frac1\lambda  =  
d\sum^\U_j \frac1{2 \sqrt{\xi_j^2+\Delta_{s,\bbalpha}^2}} \; ,
\end{eqnarray}
which in the limit $d/\tilde \Delta \to 0$ reduces to the standard
bulk $T=0$ gap equation.  Note that it is $h$-independent, because it
involves only unblocked levels $j\in \U$, which are populated by pairs
with zero total Zeeman energy. Note also that in \Eq{vj} the $ \lambda
d (v_j^{(s,\bbalpha)})^2$ shift in $\xi_j$, usually neglected because
it simply renormalizes the bare energies, is retained, since for large
$d$ it somewhat increases the effective level spacing near $\eF$ (and
its neglect turns out to produce a significant upward shift in the
$\E^\GC_{s,\bbalpha} (h,d)$'s, which one is trying to minimize).

The chemical potential $\mu$ is fixed by requiring that
\begin{eqnarray}
  \label{eq:mu}
   2n + 2s =  \langle s,\bbalpha | \hat N | s,\bbalpha \rangle
  = 2s + 2\sum_j^\U (v_j^{(s,\bbalpha)})^2.
\end{eqnarray} 
In contrast to conventional BCS theory, the pairing parameter
$\Delta_{s,\bbalpha}$ can in general not be interpreted as an energy
gap and is \emph{not} an observable. It should be viewed simply as a
mathematical auxiliary quantity which was introduced to conveniently
solve the variational conditions. 
However, by parameterizing  $v_j^{(s,\bbalpha)}$ and $u_j^{(s,\bbalpha)}$,
$\Delta_{s,\bbalpha}$ does serve as a measure of the pairing
correlations present in $|s,\bbalpha\rangle$: for
vanishing $\Delta_{s,\bbalpha}$ the latter reduces to an uncorrelated
paramagnetic state $|s,\bbalpha\rangle_0$ with spin $s$
and energy $\E^0_{s,\bbalpha}$, namely
\begin{eqnarray}
  \label{eq:param}
  |s,\bbalpha\rangle_0 \equiv 
\prod_{i \in \B} c_{i +}^\dagger 
                          \prod_{j< 0}^\U b^\dagger_j |0\rangle \; ,
\qquad \mbox{with} \quad \E^0_{s,\bbalpha} \equiv  
{}_0\langle s,\bbalpha| \hat H |s,\bbalpha\rangle_0 \; ,
\end{eqnarray}
and the condensation energy $E_{s,
  \bbalpha}^\cond \equiv \E_{s,\bbalpha}^\GC-\E^0_{s,\bbalpha}$ of
$|s,\bbalpha\rangle$ reduces to zero.

\subsubsection{General numerical solution -- illustration
of the blocking effect}
\label{sec:generalnumerics}
\label{sec:qual-disc}

\begin{figure}
  \centerline{\epsfig{figure=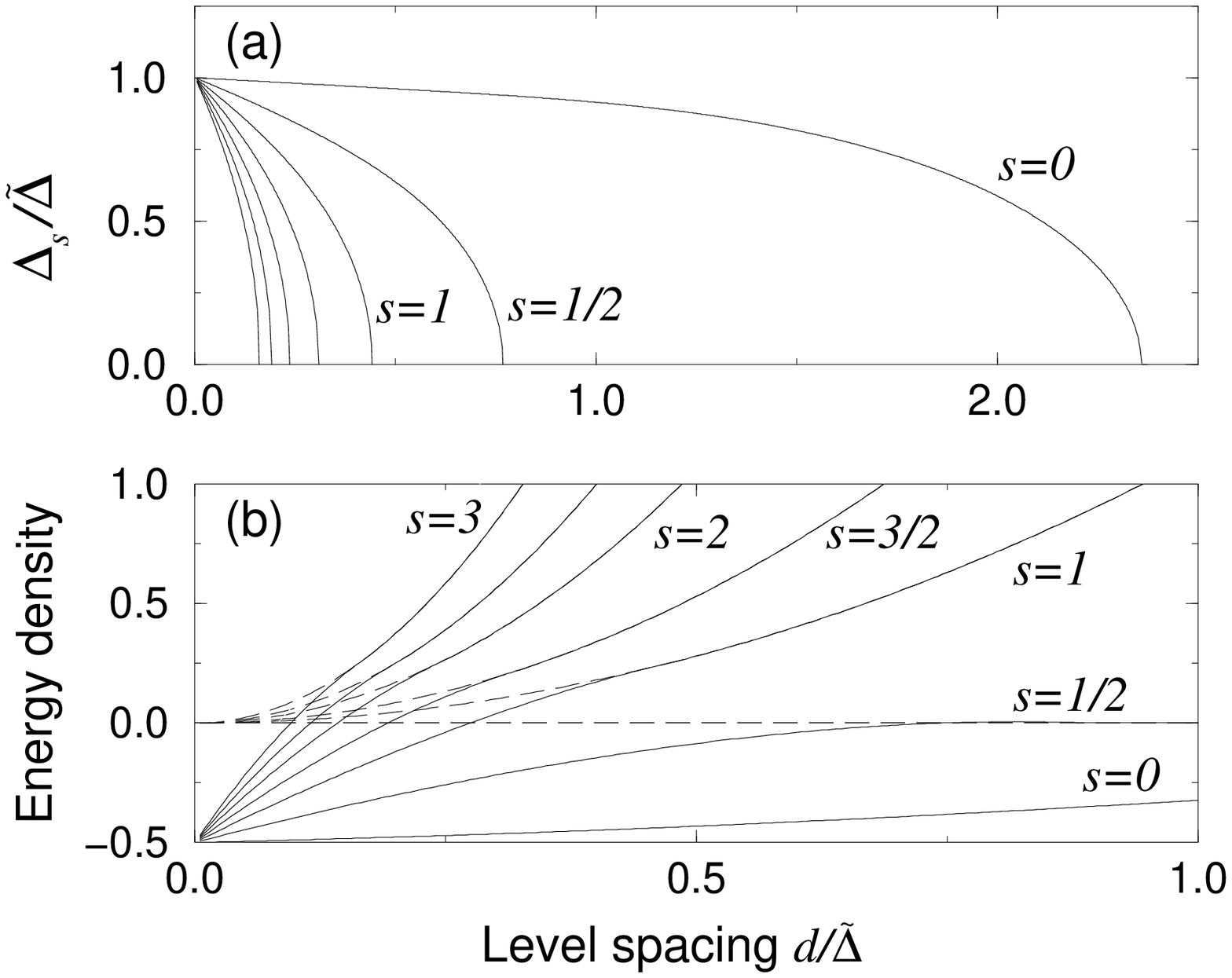,%
height=0.40\linewidth}
\epsfig{figure=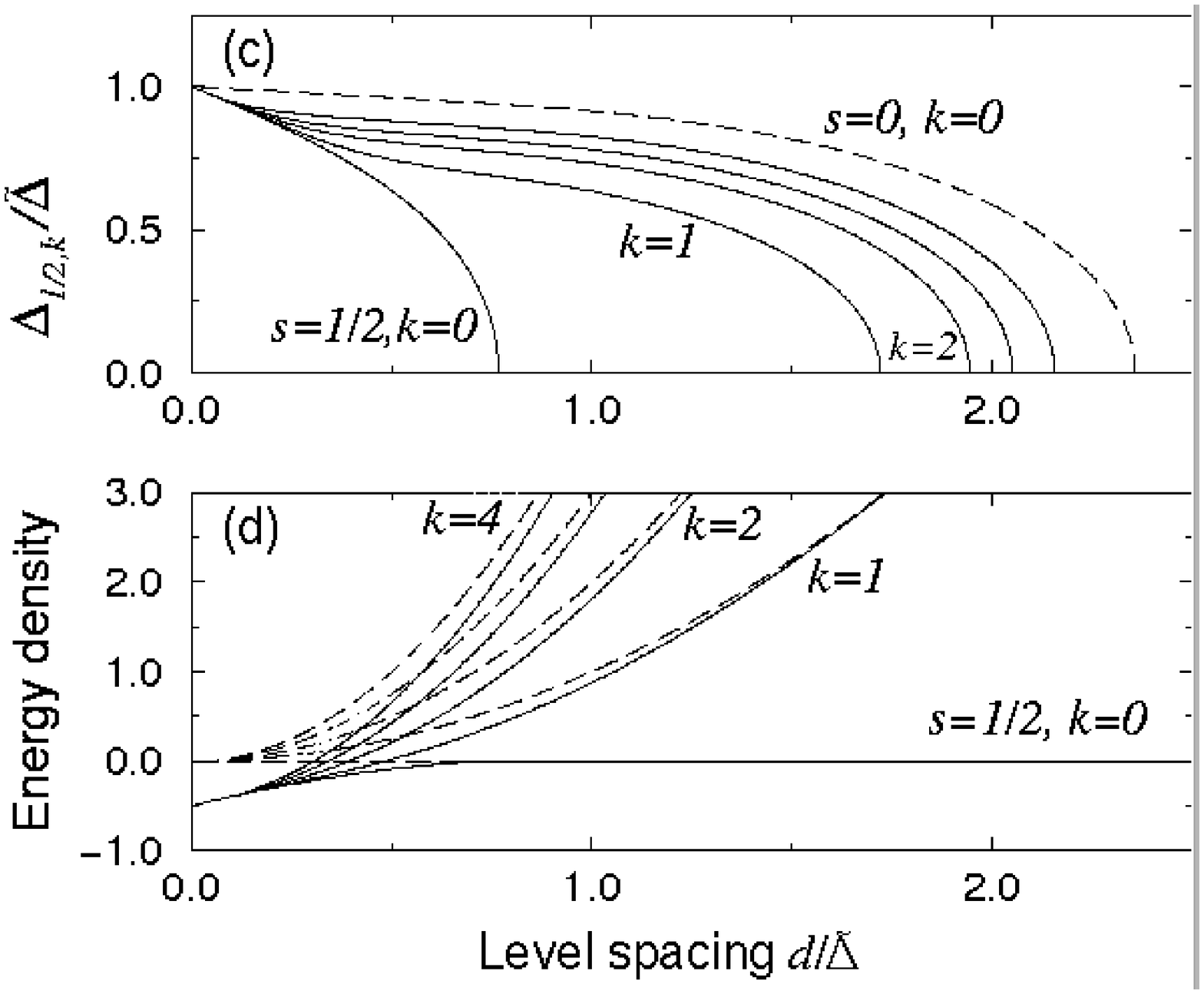,%
height=0.40\linewidth}}
   \caption[Behavior of spin-$s$ ground states and spin-$\half$
  excited states]{Properties of (a,b) spin-$s$ ground states
    $|s\rangle = |s,0\rangle$ [Eq.~(\protect\ref{eq:ansatzs})] and
    (c,d) spin-$\half$ excited states $|\half, k \rangle$ (for $k = 0,
    \dots, 4$) [Eq.~(\ref{eq:ansatzs12})], as functions of $d/\tilde
    \Delta$ (\ie\ decreasing grain size), calculated for $\lambda =
    0.194$.  (a) The pairing parameters $\Delta_s(d)/ \tilde \Delta$,
    which vanish at critical level spacings $d^\GC_{s}$ of 2.36, 0.77,
    0.44, 0.31,$\ldots$ for $s = 0, 1/2, 1, 3/2, \dots$, respectively.
    (c) The pairing parameters $\Delta_{1/2, k}$, together with their
    $k \to \infty$ limit, which equals $\Delta_0$ (dashed line).
    (b,d) show relative energy \emph{densities} (since normalized by
    $d/\tilde \Delta^2 \propto \Vol^{-1}$) at $h=0$ for both
    correlated (solid) and uncorrelated (dashed) states, the latter
    obtained by setting $\Delta_{s, \B} = 0$ in the former.  (b) shows
    $(\E^\GC_s - {\E}^0_{p/2})d / \tilde \Delta^2$ (solid) and
    $({\E}^0_s - {\E}^0_{p/2}) d / \tilde \Delta^2$ (dashed), the
    energy differences of $|s\rangle$ and $|s\rangle_0$ relative to
    the uncorrelated spin-$p/2$ Fermi sea $|p/2\rangle_0$.  (d) shows
    $(\E^\GC_{1/2,k} - {\E}^0_{1/2}) d / \tilde \Delta^2$ (solid) and
    $(\E^0_{1/2,k} - {\E}^0_{1/2}) d / \tilde \Delta^2$ (dashed), the
    energy differences of $|\frac12,k\rangle$ and
    $|\frac12,k\rangle_0$ relative to the uncorrelated spin-$\half$
    ground state $|\frac12,0\rangle_0 = |\frac12\rangle_0$.  Solid and
    dashed lines meet at the critical level spacing $d^\BCS_{s,k}$ at
    which $\Delta_{s , k}$ becomes 0 and the condensation energy
    $E^\cond_{s,k} = {\cal E}^\BCS_{s,k} - {\cal E}_{s,k}^0$
    vanishes.}
    \label{fig:pairing-parameter}
\end{figure}

The simultaneous solution of \Eqs{vj}, (\ref{eq:gap}) and
(\ref{eq:mu}) is a straightforward numerical exercise which Braun and
von Delft performed \cite{braun97,braun99}, for the sake of
``numerical consistency'', without further approximations. (Analytical
solutions can be found only in the limits $d\ll\tilde\Delta$ and
$d\gg\Delta_{s,B}$, see App.~A of \cite{braun99}.) The numerical
results are summarized in \Fig{fig:pairing-parameter}, which shows the
pairing parameters $\Delta_{s,\bbalpha}$
[\Figs{fig:pairing-parameter}(a,c)] and energies
$\E^{\BCS}_{s,\bbalpha}$ [\Figs{fig:pairing-parameter}(b,d), solid
lines] of some selected variational states $|s,\bbalpha \rangle$, as
well as the energies $\E^0_{s,\bbalpha}$ of the corresponding
uncorrelated states $|s,\bbalpha \rangle_0$
[\Figs{fig:pairing-parameter}(b,d), dashed lines]; both
$\E^{\BCS}_{s,\bbalpha}$ and $\E^0_{s,\bbalpha}$ are plotted relative
to the energy $\E^0_{p/2}$ of the uncorrelated spin-$p$ Fermi sea
$|p/2\rangle$.  The results have a number of salient features:

(i) In the bulk limit $d/\tilde \Delta \to 0$, all of the pairing
parameters $\Delta_{s,\B}$ reduce to $\tilde \Delta$, as expected, and
the energy differences $\E^{\BCS}_{s,\B} - \E^0_{p/2}$ between the
correlated states $|s,\B\rangle$ and the uncorrelated Fermi sea $|p/2
\rangle$ reduce to $- \half \tilde \Delta^2 /d = - \half \N(\eF)
\tilde \Delta^2 $, which is the standard bulk result for the
condensation energy.

(ii) Each $\Delta_{s,\bbalpha}$ in \Figs{fig:pairing-parameter}(a,c)
decreases with increasing $d$.  This reflects the fact that with
increasing $d$, the number of pair-correlated states within $\tilde
\Delta$ of $\eF$ decreases [cf.\ \Fig{fig:v2u2-prb97} and the last
paragraph of \Sec{sec:contact-with-BCS}], so that the amount of
pairing correlations, for which $\Delta_{s,\bbalpha}$ is a measure,
decreases too.

(iii) Each $\Delta_{s, \B}$ vanishes abruptly at a critical level
spacing $ d^\GC_{s,\bbalpha}$ (whose precise numerical value depends
sensitively on model assumptions such as the value of $\lambda$ and
the use of uniformly-spaced levels \cite{smith96}).  For $d >
d^\GC_{s,\bbalpha}$ no pairing correlations exist at this level of
approximation, so that that the condensation energy
$E^\cond_{s,\bbalpha} $ (difference between solid and dashed lines)
vanishes and the solid and dashed lines in
\Figs{fig:pairing-parameter}(b,d) meet.

(iv) In \Figs{fig:pairing-parameter}(a), the pairing parameters
$\Delta_s$ for the spin-$s$ ground states decrease rapidly with
increasing $s$ at fixed $d$ (and $d^\GC_{s} < d^\GC_{s'}$ if $s>s'$).
[This is a generalization of a parity effect discussed by von Delft
\etalia\ \cite{vondelft96}, who studied only ground state pairing
correlations and found that these are weaker in odd $(s=1/2)$ grains
than in even $(s=0)$ grains, $\Delta_{1/2} < \Delta_0$, cf.\ 
\Sec{sec:BCS-parity}.]  This tendency is a direct consequence of the
\emph{blocking effect} described in \Sec{sec:generalproperties} and
is independent of model details: larger $s$ means more unpaired
electrons, more terms missing from the sum $\sum_j^\U$, less
correlated pairs and hence smaller $\Delta_{s , \bbalpha}$.

(v) As $d$ increases the blocking effect described in (iv) becomes
stronger, \ie\ the difference between the various $\Delta_{s}$ for
different $s$ becomes more pronounced, since then the relative weight
of each term missing in the sum $\sum_j^\U$ increases.
The blocking effect is most dramatic in the
regime $d/\tilde\Delta\in[0.77, 2.36]$ in which $\Delta_{0} \neq 0 $
but $\Delta_{s \neq 0 } = 0$. This is a regime of ``minimal
superconductivity'' \cite{braun97,braun99}, in the sense that all
pairing correlations that still exist in the \emph{even} 
variational ground state $|0\rangle$ 
(since $\Delta_0 \neq 0$) are completely destroyed by the addition of
a single electron or the flipping of a single spin (since $\Delta_{s
  \neq 0} = 0$).

(vi) Considering the spin-$\half$ excited states $|\half, k \rangle$ of
\Figs{fig:pairing-parameter}(b,d), one finds that the larger $k$, the
longer the pairing correlations survive with increasing $d$: the
critical spacings $d^\GC_{1/2,k} $ increase with $k$, approaching the
value $d^\GC_{0}$ of the spin-0 case as $k \to \infty$;
correspondingly, the larger $k$, the larger the $d$-value at which the
condensation energies $E^\cond_{1/2,k}$ [differences between solid and
dashed lines in \Fig{fig:pairing-parameter}(d)] vanish. The intuitive
reason why the amount of pairing correlations in an excited
$|s,k\rangle$ increases with $k$ is of course quite simple: the
further the unpaired electron sits from the Fermi surface where
pairing correlations are strongest, the less it disrupts the latter
(since $u_k v_k$ becomes very small for large $k$, see
\Fig{fig:v2u2-prb97}).  In fact, the state $|\frac12,k \to \infty
\rangle$ will have just about the same amount of pairing correlations
as the even ground state $|0 \rangle$ ($\Delta_{1/2,k \to \infty}
\simeq \Delta_0$).

(vii) Similar effects hold for excited states in other spin sectors
 (not shown): The higher the excitation, the larger the
pairing parameter $\Delta_{s,\bbalpha}$. However, the concomittant
gain in correlation energy is always less than the
kinetic-energy cost of having an
unpaired electron far from $\eF$. 

(viii) The strong dependence of $\Delta_{s,\bbalpha}$ on $s$ and $d$
for $d \gtrsim \tilde \Delta$ illustrates why in this regime a
conventional mean-field treatment is no longer sufficient: \emph{the
  system cannot be characterized by a {\bf single} pairing parameter,
  since the amount of pairing correlations vary from state to state,
  each of which is characterized by its own pairing parameter.}
Instead, the present variational approach is, roughly speaking,
equivalent to a doing a \emph{separate} mean-field calculation for
each new choice $\U$ of unblocked levels within the Fock space spanned
by them (\ie\ replacing $b_j \to \{ b_j - \langle b_j \rangle \} +
\langle b_j \rangle $ 
and neglecting terms quadratic in the fluctuations $\{b_j -
\langle b_j \rangle \}$).  Indeed, the behavior of $\Delta_{
  s,\bbalpha} (d)$ near $d^\GC_{s,\bbalpha}$ has the standard
mean-field form $\sqrt{1-d/d^\GC_{s,\bbalpha}}$, as can be shown
analytically \cite[App.~A]{braun99}.

To summarize: pairing correlations decrease with increasing $d$ and
$s$ and decreasing $k$. These features survive also in more accurate
canonical calculations. This is not the case, however, for the abrupt
vanishing of $\Delta_{s, \B}$ at $d^\BCS_{s, \B}$, which signals the
breakdown of the \gc\ approach once $d$ becomes of order $\tilde
\Delta$: canonical methods show that, regardless how large $d$
becomes, some remnants of pairing correlations survive and the pairing
parameters $(\Delta_{s,\B})_\can$ do not vanish
[\Sec{sec:sc-canonical}], in accordance with the rule of thumb that
``in a \emph{finite system} no abrupt phase transition can occur
between a zero and nonzero order parameter.''

\subsection{Softening of the $H$-induced 
transition to a paramagnetic state}
\label{sec:CC-transition}
\label{sec:paramagnetic-breakdown}

Since  states with nonzero spin are favored
by the Zeeman energy but have smaller correlation energy due
to the blocking effect, a competition arises between Zeeman energy and
correlation energy.  The manifestations of the blocking effect can
thus be probed by turning on a magnetic field; if it becomes large
enough to enforce a large spin, excessive blocking will destroy all
pairing correlations.

The situation is analogous to ultra-thin films in a parallel magnetic
field \cite{Meservey-70,Meservey-94}, where orbital diamagnetism is
negligible for geometrical reasons and \emph{superconductivity is
  destroyed at sufficiently large $h$ by Pauli paramagnetism.}  This
occurs via a first order transition to a paramagnetic state, as
predicted by Clogston and Chandrasekhar (CC)
\cite{Clogston-62,Chandrasekhar-62} by the following argument (for
bulk systems): A pure Pauli paramagnet chooses its spin $s$ 
such that the
sum of the kinetic and Zeeman energies, $s^2 /\N(\eF) - 2
h s$, is minimized, and hence has spin $s = h \N(\eF)$ and ground
state energy $-h^2 \N (\eF)$.  When this energy drops below the bulk
correlation energy $-\frac12\tilde\Delta^2 \N(\eF)$ of the
superconducting ground state, which happens at the critical field
$h_{\CC} =\tilde\Delta/\sqrt{2}$, a transition will occur from the
superconducting to the paramagnetic ground state. The transition is
first-order, since the change in spin, from 0 to $ s_{\CC} = h_{\CC}
\N(\eF) = \tilde \Delta / (d \sqrt 2)$, is macroscopically large
($\N(\eF) = 1/d \propto {\rm Vol}$).

\begin{figure}[t]
  \centerline{\epsfig{figure=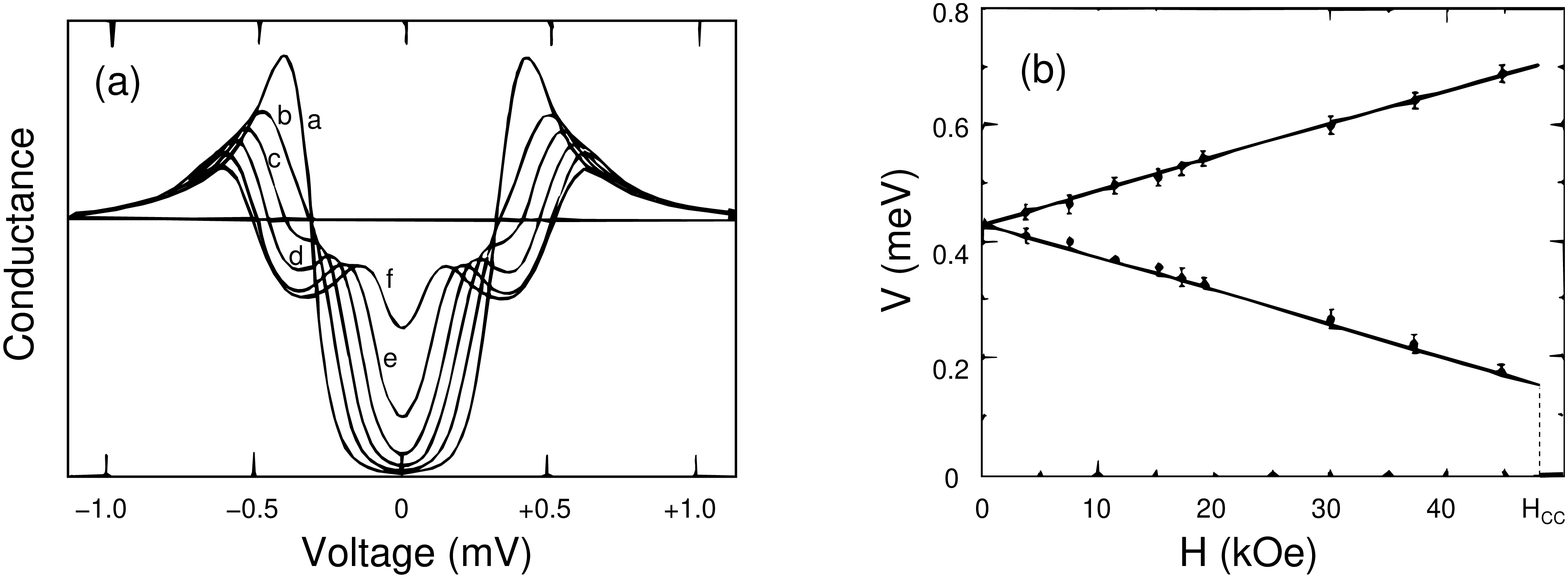,%
width=0.95\linewidth}}
    \caption[Superconductivity in thin Al films]{Thin 
      films in a magnetic field (Figs.~11 and 12 of
      \protect\cite{Meservey-70}). (a) Tunneling conductance from a
      normal metal through a tunnel barrier into a thin
      superconducting Al film, as function of voltage, for several
      magnetic fields labeled in increasing order ``a'' to ``f''. The
      conductance reflects the BCS quasiparticle density of states,
      whose single peak (for a given sign of $V$) for $H=0$ splits
      into two separate peaks for $H\neq0$, corresponding to the
      Zeeman energy difference between quasiparticles with spin up and
      down. (b) Voltage corresponding to the maxima of spin-up and
      spin-down density of states as a function of magnetic field. At
      the critical field $H_\CC$ superconductivity is destroyed and the
      tunneling threshold drops abruptly to zero.}
    \label{fig:MT}
\end{figure}

This transition has been directly observed by Meservey and Tedrow
\cite{Meservey-70,Meservey-94} in ultra-thin (5nm) superconducting Al
films ($\tilde\Delta = 0.38$meV), whose density of states
[\Fig{fig:MT}(a)] they measured via the tunnel conductance through an
oxide layer between a normal metal and the film.  They found that in a
magnetic field the BCS quasiparticle peak splits up into two subpeaks,
separated in energy by $2 \muB H$ [\Fig{fig:MT}(b)], which simply
reflects the Zeeman splitting of quasiparticles states\footnote{Recall
  that the BCS quasiparticles $\gamma^\dag_{j,\sigma} = u_j
  c^\dag_{j,\sigma} - \sigma v_j c^\ds_{j,-\sigma}$ have well-defined
  spins.} with spin up or down (and $g=2$). Remarkably, the tunneling
threshold abruptly dropped to zero at a field of 4.7~T
[\Fig{fig:MT}(b)], which they associated with the field $H_\CC$ at
which the phase transition from the superconducting to the
paramagnetic ground state occurs.  Indeed, \Fig{fig:MT}(b)
demonstrates clearly that the transition to the normal state is first
order: \emph{the mean of the spin-up and spin-down spectral gaps, \ie\ 
  the pairing parameter $\tilde\Delta$, is constant until the critical
  field $H_{\CC}$ is reached, at which it abruptly drops to zero.}

For the case of isolated ultrasmall grains, the above picture of the
transition needs to be rethought in two respects due to the
discreteness of the electronic spectrum: Firstly, the spin must be
treated as a discrete (instead of continuous) variable, whose changes
with increasing $h$ can only take on (parity-conserving) integer
values.  Secondly, one needs to consider more carefully the
possibility of $h$-induced transitions to nonzero spin states that are
still \emph{pair-correlated} (instead of being purely paramagnetic),
such as the variational states $ |s, \bbalpha\rangle $ discussed
above.  (In the bulk case, it is obvious that such states play no
role: the lowest pair-correlated state with nonzero spin obtainable
from the ground state by spin flips is a two-quasiparticle state,
costing energy $2\tilde \Delta - 2h$; when $h$ is increased from 0,
the paramagnetic transition at $h_{\CC} = \tilde \Delta/\sqrt 2$ thus
occurs before a transition to this state, which would require $h =
\tilde \Delta$, can occur.)

\begin{figure}
  \centerline{\epsfig{figure=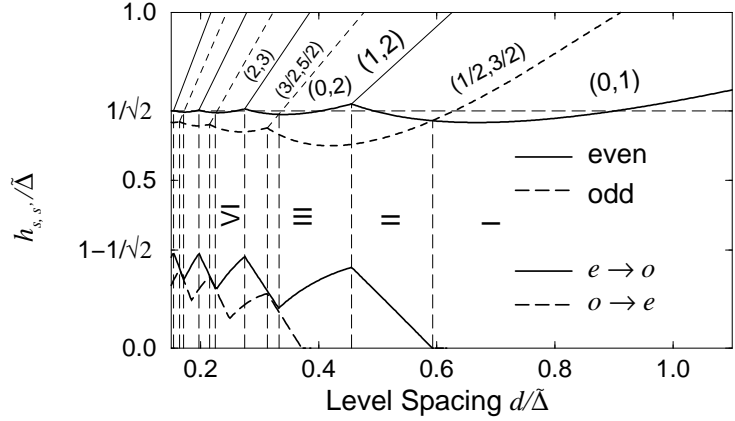,%
width=0.65\linewidth}}
  \caption[$d$-dependence of the 
  level-crossing fields $h_{s,s'} (d)$.]{$d$-dependence of the
    level-crossing fields $h_{s,s'} (d)/ \tilde \Delta$
    [Eq.~(\ref{eq:hcrit})] at which $\E^\BCS_{s'}$ drops below
    $\E^\BCS_s$ with increasing $h$. Only those level crossing fields
    are shown that belong to the cascade of (\emph{fixed}-$N$) ground
    state changes (GSCs) $(s_0,s_1)$; $(s_1,s_2)$; \dots that occur as
    $h$ increases from 0 at given $d$.  Solid (dashed) lines are used
    for even (odd) grains with integer (half-integer) spins, and some
    are labeled by the associated GSC $(s,s')$. (In contrast, in
    \Fig{fig:spectra} the $N$-\emph{changing} tunneling transitions in
    are labeled by $|s_i\rangle \to |s_f\rangle$.)  The size $|\Delta
    \E_{s_1,f'} - \Delta \E_{s_0,f}|$ of the first jump (occuring at
    the level-crossing field $h_{\CC}(p,d) = h_{s_0,s_1}$) in the
    lowest line of the tunneling spectra of \Fig{fig:spectra} is shown
    by the lowest two (jagged) curves (solid for $e\to o$ and dashed
    for $o \to e$ tunneling spectra), which both approach the CC value
    $1 - 1/\sqrt2$ as $d \to 0$.}
      \label{fig:h-crit}
\end{figure}

Quite generally, the effect of increasing $h$ from 0
can be analyzed as follows: At given $d$ and $h$, the grain's ground
state is the lowest-energy state among all possible spin-$s$ ground
states $|s \rangle$ having the correct parity $p = 2s\,{\rm mod}\,2$.
Since $\E_{s}(h,d) = \E_{s}(0,d) - 2hs$, level crossings occur with
increasing $h$, with $ \E_{s'}$ dropping below $\E_s$ at the
\emph{level crossing field}
\begin{eqnarray}
  \label{eq:hcrit}
  h_{s,s'} (d) = \frac{\E_{s'}(0,d)-\E_s(0,d)}{2(s'-s)}. 
\end{eqnarray}
Therefore, as $h$ is slowly turned on from zero with initial ground
state $|s_0 = p/2\rangle$, a cascade of successive ground-state
changes (GSCs) to new ground states $|s_1 \rangle$, $|s_2 \rangle$,
\dots will occur at the fields $h_{s_0, s_1}$, $h_{s_1, s_2}$, \dots
Let us denote this cascade by $(s_0,s_1); (s_1,s_2);\ldots$; for each
of its GSCs the corresponding level-crossing field $h_{s,s'} (d)$ is
shown in \Fig{fig:h-crit}.  Generalizing CC's critical field to
nonzero $d$, let us denote the (parity-dependent) field at which the
\emph{first} transition $(s_0, s_1)$ occurs by $h_{\CC} (d,p) \equiv
h_{s_0, s_1} (d) $, which simply is the lower envelope of the
level-crossing fields $h_{s_0, s_1}$ in \Fig{fig:h-crit} (shown as
bold solid and dashed lines for $s_0=0$ and $s_0=\frac12$,
respectively).  In the limit $d\to0$ it is numerically found
to reduce to the Clogston-Chandrasekhar value, \ie\ $h_{\CC}(0,p) =
\tilde\Delta/\sqrt{2}$, as expected. 

In general, the order in which the GSCs occur with increasing $h$
within a cascade (\ie\ the order of $h_{s,s'}$ lines encountered when
moving vertically upward in \Fig{fig:h-crit}) depends sensitively on
$d$, and an infinite number of distinct regimes (cascades) I, II, III,
\dots can be distinguished: Starting at large $d$ we find the typical
normal-grain behavior $(0,1); (1,2); (2,3); \ldots$ for even grains
and $(\frac12,\frac32); (\frac32,\frac52); \ldots$ for odd grains,
with $h_{0,1} < $ (or $>$) $ h_{\frac12,\frac32}$ in regimes I (or
II).  In regimes III and IV of somewhat smaller $d$, the order of GSCs
is $(0,2); (2,3); \ldots$ and $(\frac12,\frac32); (\frac32,\frac52);
\ldots$, etc, \ie\ the spin $s_1$ attained after the first GSC
$(s_0,s_1)$ has increased to 2 in the even case.  This illustrates a
general trend: \emph{the spin $s_1 (d)$ after the first transition
  increases with decreasing $d$ and becomes macroscopically large in
  the $d\to0$ limit}, where $s_1 = h_{\CC} / d = \tilde \Delta / (d
\sqrt 2)$, as explained in recounting CC's argument above.

Furthermore, it turns out that $\Delta_{s_1} (d)= 0$ and therefore
$\E^\GC_{s_1}=0$ for \emph{all} $d$, implying that after the first GSC
the new ground state $|s_1\rangle$ is \emph{always} (not only in CC's
bulk limit) a purely paramagnetic state, \ie\ without any pairing
correlations in the \gc\ framework (canonical calculations would yield
some weak remnant pairing correlations in the form of fluctuations).
In this regard, CC's picture of the transition remains valid
throughout as $d$ is increased: at $h_{\CC} (d,p)$, a transition
occurs from the superconducting ground state to a paramagnetic,
\emph{uncorrelated} state $|s_1 \rangle_0$, the transition being
first-order in the sense that $\Delta_{s_1} (d)= 0$; however,
\emph{the first-order transition is ``softened'' with increasing $d$,
  in the sense that the size of the spin change, $s_1 - s_0$,
  decreases from being macroscopically large in the bulk to being
  equal 1 at $d \gg \tilde \Delta$ (regimes I and II)}.

To conclude this section, we mention that the above analysis of the
paramagnetic breakdown of superconductivity has recently been
generalized to finite temperatures
\cite{Rossignoli-00}, using the so-called static path approximation
[explained in \Sec{sec:SPA}] to treat fluctuation effects properly. 

\subsection{Excitation spectrum in a magnetic field}
\label{sec:tunneling-spectra-prb97}
\label{sec:tunneling-spectra}
\label{sec:magfield}

In this section we compare the theoretical tunneling spectra for a
grain coupled to leads, calculated as functions of $h$ and $d$
\cite{braun97,braun99,braun-thesis}, and compare these to RBT's
measurements of \Fig{fig:sc-magneticfield}.

The form of the tunneling spectrum depends in a distinct way on the
specific choice of level spacing $d$ and on the electron number parity
$p$ of the final states $|f\rangle$ of the bottleneck tunneling
processes $|i\rangle \to |f \rangle$ (or $|\alpha' \rangle \to |\alpha
\rangle$ in the notation of \Sec{sec:theoryofultrasmallSET}).
However, for the uniformly spaced $\varepsilon_j$-levels used here,
particle-hole symmetry ensures that there is no difference between
electron addition or removal spectra $|i_{N \mp 1} \rangle \to |f_N
\rangle$.  To calculate the spectrum \emph{for given $d$ and $p$},
Braun \etalia\ \cite{braun97,braun99,braun-thesis} proceeded as
follows: they first analyzed at each magnetic field $h$ which
tunneling processes $|i \rangle \to |f \rangle$ are possible, then
calculated the corresponding tunneling energy thresholds $\Delta
\E_{if} (h) \equiv \E_f(h) - \E_i(h)$ [cf.\ \Eq{eq:Vrthresholds}] and
plotted $\Delta \E_{if} (h) - \Delta\E_{{\rm min}} (0)$ as functions
of $h$ for various combinations of $|i\rangle$ and $|f\rangle$, each
of which gives a line in the spectrum.  Since the selection rule $s_f
- s_i = \pm 1/2$ holds, only slopes of $\pm 1$ can occur.  The reason
for subtracting $\Delta\E_{{\rm min}} (0)$, the $h=0$ threshold energy
cost for the \emph{first} (lowest-lying) transition, is that in
experiment, this energy depends on $V_g$ and hence yields no
significant information, as explained in
\Sec{sec:ground-state-energies-not-measurable}.  Neglecting
nonequilibrium effects \cite{rbt97,agam97a,agam97b,agam98} (which were
minimized in the present experiment by tuning $\Vg$, cf.\ 
\Sec{sec:effects-of-gate}, and which are discussed in
\Sec{sec:sub-gap-structures}), the initial state is always taken to be
the ground state of a given spin-$s$ sector.  The appropriate
$s_i(h,d)$ must be determined from Fig.~\ref{fig:h-crit}.

\begin{figure}
  \centerline{\epsfig{figure=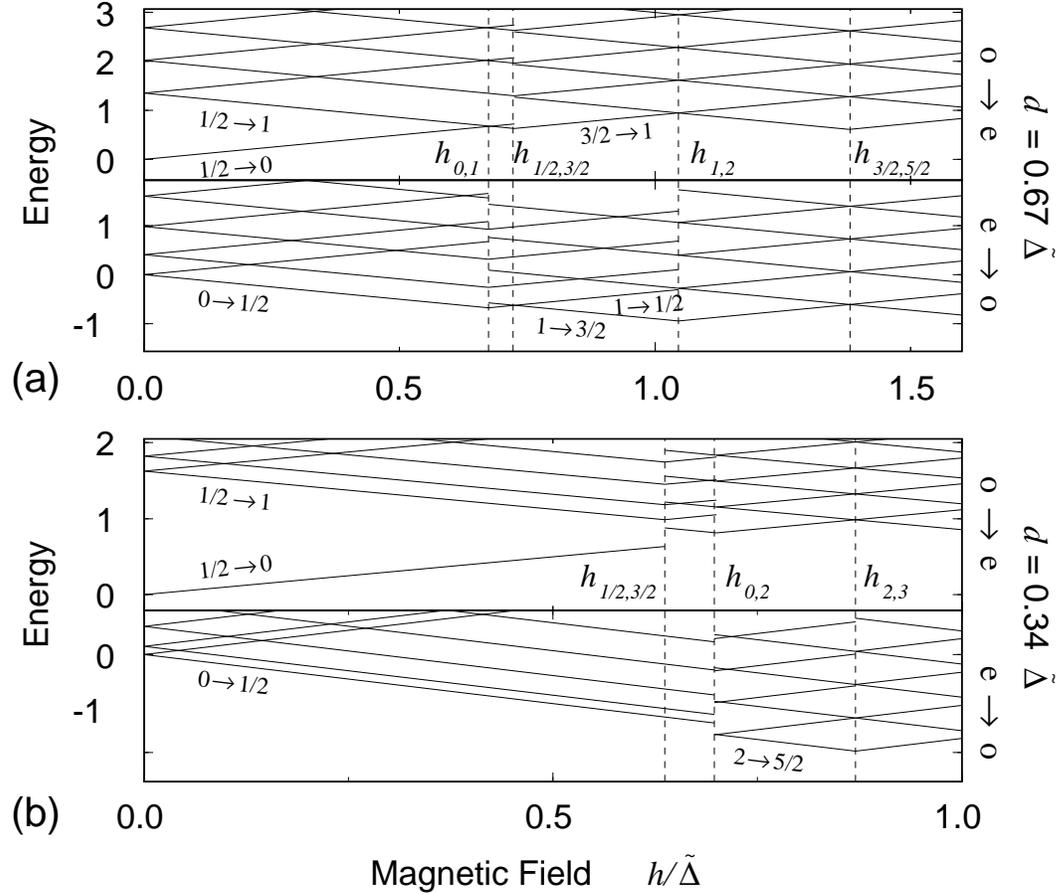,%
width=\linewidth}}
  \caption[Theoretical tunneling spectra]{The theoretical 
    odd-to-even and even-to-odd tunneling spectra $(\Delta \E_{if} -
    \Delta\E_{{\rm min}} (0))/ \tilde \Delta $ predicted for an
    ultrasmall superconducting grain as a function of magnetic field
    $h$, for two different level spacings: (a) $d=0.67\tilde\Delta$
    and (b) $d=0.34\tilde\Delta$ (corresponding to regimes I and III
    of Fig.~\protect\ref{fig:h-crit}, respectively).  Some lines are
    labeled by the corresponding $s_i\to s'_i$ tunneling transitions.
    Not all possible higher lines (corresponding to excited final
    states $|s,j\rangle$) are shown.  Vertical dashed lines indicate
    those level-crossing fields $h_{s,s'}$ [\Eq{eq:hcrit}]
    at which kinks or jumps occur, with $h_{0,1}<h_{1/2,3/2}
    < h_{1,2} < h_{3/2,5/2}$ in (a) and 
    $h_{1/2,3/2}<h_{0,2} < h_{2,3}$ in (b). }
      \label{fig:spectra}
\end{figure}

\Fig{fig:spectra} shows four typical examples of such theoretical
tunneling spectra, with some lines labeled by the corresponding $|i
\rangle \to |f \rangle$ transitions.  Whenever $h$ passes through one
of the level-crossing fields $h_{s_i, s_{i'}}$ of \Eq{eq:hcrit},
the grain experiences a ground state change $(s_i, s_{i'})$, at which
the set of allowed tunneling transitions changes from $|s_{i} \rangle
\to \{ |s_{f} \rangle \}$ to $|s_{i'} \rangle \to \{ |s_{f'} \rangle
\}$. Therefore, at $h_{ s_i, s_{i'}}$ one set of lines in the
tunneling spectrum ends and another begins, producing kinks or
discontinuities.  A \emph{kink} occurs if one of the new final states
coincides with one of the old ones, $|f'\rangle = |f\rangle$, meaning
that it can be reached from both $|s_i \rangle $ and $|s_{i'} \rangle$
[\ie\ $s_f - s_i = - (s_f - s_{i'})$], in which case $\Delta \E_{if}
(h)$ and $\Delta\E_{if'} (h)$ have slopes of opposite sign.  However,
for most lines this is not the case, so that at $h_{ s_i, s_i'}$ the
line $|s_{i}\rangle \to |f \rangle$ simply ends while new lines
$|s_{i'} \rangle \to |f' \rangle$ begin.  This results in
\emph{discontinuities} (or ``jumps'') in the spectrum at $h_{ s_i, s_i'}$ of
size $(\Delta\E_{i'f'} - \Delta \E_{if})(h_{ s_i, s_i'})$, unless by
chance some other final state $|f '\rangle $ happens to exist for
which this difference equals zero.

Since the order in which the GSCs $(s_i, s_{i'})$ occur as functions
of increasing $h$ depend on $d$ and $p$, as indicated by the distinct
regimes I, II, III, \ldots in \Fig{fig:h-crit}, one finds a
distinct kind of tunneling spectrum for each regime, differing from
the others in the positions of its jumps and kinks.  In regime~I,
where the order of occurrence of GSCs with increasing $h$ is $(0,1);
(\frac12,\frac32); (1,2); (\frac32,\frac52);\ldots$, there are no
discontinuities in the evolution of the lowest line [see
\Fig{fig:spectra}(a)].  For example, for the $e\to o$ spectrum,
the lowest $|0 \rangle \to |1/2\rangle$ line changes
\emph{continuously} to $|1\rangle \to |1/2\rangle$ at $h_{0,1}$, since
$|s_f - s'_i | = 1/2$.  However, in all other regimes the first change
in ground state spin (at $h_{0, s_1}$ from 0 to $s_1$) is $ > 1$,
implying a \emph{jump} (though possibly small) in all $e\to o$ lines,
as illustrated by \Fig{fig:spectra}(b).

The jump's magnitude for the tunneling thresholds, \ie\ the
\emph{lowest} $e\to o$ and $o\to e$ lines, is shown as function of $d$
in the lower part of \Fig{fig:h-crit}.  It starts at $d=0$ from the CC
value $\tilde \Delta (1 - 1/\sqrt2)$ measured for thin Al films
\cite{Meservey-70,Meservey-94}, and with increasing $d$ decreases to 0
(non-monotonically, due to the discrete spectrum).  This
\emph{decrease of the size of the jump in the tunneling threshold}
reflects the fact, discussed in \Sec{sec:CC-transition}, that the
change in spin at the first ground state change $(s_0,s_1)$ decreases
with increasing $d$ (as $s_1 \!-\! s_0 \sim h_{\CC}/d$), and signals
the softening of the first-order superconducting-to-paramagnetic
transition.
 
The fact that the measured tunneling thresholds in
\Fig{fig:experimental-spectra} show no jumps at all, which might at
first seem surprising when contrasted to the threshold jumps seen at
$h_{\CC}$ in \Fig{fig:MT} for thin films in a parallel field
\cite{Meservey-70,Meservey-94}, can therefore naturally be explained
\cite{braun97,braun99} by assuming the grain to lie in the ``minimal
superconductivity'' regime I of \Fig{fig:h-crit} (where the jump size
predicted in \Fig{fig:h-crit} is zero).  Indeed, \emph{the overall
  evolution (\ie\ order and position of kinks, etc.)  of the lowest
  lines of \Fig{fig:experimental-spectra} qualitatively agrees with
  those of a regime I tunneling spectrum, \Fig{fig:spectra}(a)}. This
important result rather convincingly establishes the phenomenological
success of the \dbcsm. It also allows one to deduce the following
values for the level-crossing fields $H_{s_i,s'_i}$ (indicated by
vertical dashed lines in Figs.~\ref{fig:experimental-spectra}
and~\ref{fig:spectra}): $H_{0,1} = 4$T, $H_{1/2,3/2} = 4.25$T,
$H_{1,2} = 5.25$T and $H_{3/2,5/2} = 6.5$T. \label{HSS} As
corresponding uncertainties we take $\Delta H_{s_i,s'_i} = 0.13$T,
which is half the $H$ resolution of 0.25T used in experiment.

By combining the above $H_{s_i,s'_i}$ values with \Fig{fig:h-crit},
some of the grain's less-well-known parameters can be determined
somewhat more precisely: 
\begin{itemize}
\item[(i)]
To estimate the grain's ``bulk $H_{\CC}$'',
note that since $H_{1/2,3/2} / H_{0,1} \simeq 1.06$, this grain lies
just to the right of the boundary between regions II and I in
\Fig{fig:h-crit} where $d/ \tilde \Delta \simeq 0.63$,
at which we
have $h_{0,1} / h_{\CC} \simeq 0.95$, so that $H_{\CC} = H_{0,1}/ 0.95
\simeq 4.2$~T.  This is quite close to the value $H_{\CC} \simeq 4.7$~T
found experimentally \cite{Meservey-70,Meservey-94} in thin films in a
parallel field, confirming our expectation that these correspond to
the ``bulk limit'' of ultrasmall grains as far as paramagnetism is
concerned.
\item[(ii)] The grain's corresponding bulk gap is $\tilde \Delta =
  \sqrt 2 \mu_B H_{\CC} \simeq 0.34$~meV, implying a coupling constant
  of $\lambda = 0.189$ [by \Eq{eq:lambda-definition}].  \emph{A
    posteriori}, these values can be regarded as being more
  appropriate for the present grain than the choices $\tilde \Delta =
  0.38$~meV and $\lambda = 0.194$ made in \Sec{sec:generalnumerics},
  though the differences are evidently not significant (12\% for
  $\tilde \Delta$ and 3\% for $\lambda$).
\item[(iii)] 
The mean level spacing implied
by $d/ \tilde \Delta \simeq 0.63$ is $d \simeq 0.21$~meV.
\label{p:estimate-d}
The crude volume-based value $d \simeq 0.45$~meV cited in
the caption of \Fig{fig:generic-IV} thus seems to have been an
overestimate.  It would be useful if this determination of $d$
could be checked via an independent accurate experimental
determination of $d$ directly from the spacing of lines in the
excitation spectrum. Regrettably, this is not possible: the measured
levels are shifted together by pairing interactions, implying that
their spacing does not reflect the mean \emph{independent}-electron
level spacing $d$. Nevertheless, note that the
measured spacing of $0.05$~meV 
between the lowest two states
of the  odd grain agrees quite well with the crude
BCS estimate $\sqrt{\tilde \Delta^2 + d^2} - \tilde \Delta$
[cf.\ \Eq{vj-bulk}], which gives $0.06$~meV when evaluated for
$d=0.21$~meV and $\tilde \Delta = 0.34$~meV. 
\end{itemize}

The higher lines plotted in \Fig{fig:spectra} correspond to
transitions into spin-$s_f$ state of the form $|s_f, k \rangle$ [cf.\ 
\Eq{eq:ansatzs12} and \Fig{fig:alpha-states}(d)] (for simplicity
these were the only ones considered in
\cite{braun97,braun99,braun-thesis}, though in general others are
expected to occur too). The jumps in these lines, \eg\ in
\Fig{fig:spectra}(a) at $h_{1,2}$, occur whenever the two final
excited states $|s_f, k_f \rangle$ and $|s_{f'}, k_{f'} \rangle$
before and after the GSC at $h_{s_i, s'_i}$ have different correlation
energies.  (Recall that the correlation energy of an excited state
$|s_f, \bbalpha_f \rangle$ can be nonzero even if that of the
corresponding ground state $|s_f \rangle$ is zero, since the former's
unpaired electrons are further away from $\eF$, so that
$\Delta_{s_f,\bbalpha_f}>\Delta_{s_f}$, see point (vi) of
\Sec{excited}.)  Experimentally, these jumps have not been observed.
This may be because up-moving resonances lose amplitude and are
difficult to follow \cite{rbt97} with increasing $h$, or because the
widths of the excited resonances ($\simeq 0.13\tilde\Delta$) limit
energy resolution \cite{agam97a,agam97b,agam98}.

For somewhat larger grains, the present theory predicts jumps even in
the lowest line, as illustrated in \Fig{fig:spectra}(b).  It remains
to be investigated, though, whether orbital effects, which rapidly
increase with the grain size, would not smooth out such jumps.

To conclude this section, we emphasize once again that more than
qualitative agreement between theory and experiment can not be
expected, since both the model and our variational treatment thereof
are very crude: the model neglects, for instance, fluctuations in
level spacing and in pair-coupling constants, and the \gc\ wave
functions become unreliable for $d /\tilde \Delta \gtrsim 0.5$.
Furthermore, we neglected nonequilibrium effects in the tunneling
process and assumed equal tunneling matrix elements for all processes.
In reality, though, random variations of tunneling matrix elements
could suppress some tunneling processes which would otherwise be
expected theoretically.

\subsection{Time-reversal symmetry} 
\label{sec:time-reversed}

In this section we argue that RBT's spectra give direct support for
the dominance of purely time-reversed states in the pairing
interaction, implying that the sufficiency of using only a reduced BCS
Hamiltonian, well established for bulk systems and dirty
superconductors, holds for ultrasmall grains, too.

\begin{figure}
  \centerline{\epsfig{figure=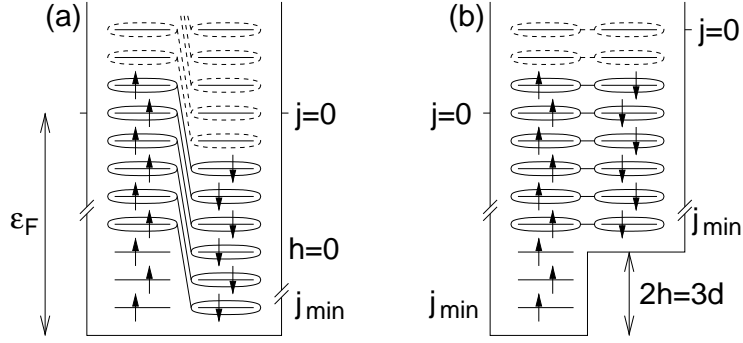,%
width=0.7\linewidth}}
  \caption[Wavefunction of non-time-reversed state]{Schematic 
representations
    of the non-time-rever\-sed-pairing state $|3/2\rangle'$ defined in
    Eq.~(\protect\ref{sprime}).  The energies $\varepsilon_{j} \mp h$
    of the single-particle states $|j, \pm \rangle$ are drawn (a) for
    $h=0$ and (b) for $2h = 3d$. We indicated schematically how
    non-time-reversed states are paired according to $(u_i + v_i
    c^\dagger_{(i+3)+} c^\dagger_{i-})$ in the BCS-like
    Ansatz~(\ref{sprime}), with solid or dashed ellipses encircling
    states that would be completely filled or empty in the absence of
    pairing correlations.}
  \label{fig:repaired-state} 
\end{figure}

When defining the \dbcsm\ in Eq.~(\ref{eq:hamiltonian}), we adopted a
{\em reduced\/} BCS Hamiltonian, in analogy to that conventionally
used for macroscopic systems. In doing so, we neglected interaction
terms of the form
\begin{equation}
  \label{eq:nontimereversed}
  -d \sum_{iji'j'} \lambda (i,j,i',j') c^\dagger_{i+}c^\dagger_{j-} 
  c_{i'-}c_{j'+} 
\end{equation}
between non-time-reversed pairs $c^\dagger_{i+}c^\dagger_{j-}$,
following Anderson's argument \cite{anderson59} that for a
short-ranged interaction, the matrix elements involving time-reversed
states $c^\dagger_{j+} c^\dagger_{j-}$ are much larger than all
others, since their orbital wavefunctions interfere constructively.
(This argument can be substantiated using $1/\gdc$ as small parameter
\cite{agam98}, 
where $\gdc$ of \Eq{eq:g-dimensionlessconductance} is the grain's
dimensionless conductance, see \Sec{sec:beyound-orthodox-model}.)
Interestingly, the experimental results of RBT provide strikingly
direct support for the correctness of neglecting interactions between
non-time-reversed pairs of the from (\ref{eq:nontimereversed}) at
$h=0$: Suppose the opposite, namely that the matrix elements $\lambda
(j+k,j, j'+ k', j')$ were all roughly equal to $\lambda$ for a finite
range of $k$- and $k'$-values [instead of being negligible for $k$ or
$k' \neq 0$, as assumed in the reduced BCS Hamiltonian
(\ref{eq:hamiltonian})].  Then for $2s < k$, one could construct a
spin-$s$ state $|s \rangle'$ with manifestly lower energy ($\E'$) than
that ($\E$) of the state $|s\rangle$ of Eq.~(\ref{eq:ansatzs}), namely
\begin{eqnarray}
\label{sprime}
    |s\rangle' =  \prod_{i=j_{\rm min}}^{j_{\rm min}+2s-1}  
    c^\dagger_{i+} 
                \prod_{j= j_{\rm min}}^{\infty} (u^{(s)}_j +  v^{(s)}_j
       c^\dagger_{(j+2s)+}c^\dagger_{j-})\,|\mbox{Vac}\rangle,
\end{eqnarray}
where $j_{\rm min}$ labels the lowest-lying of the interacting
levels (\ie\ the bottom of the interacting band).
Whereas in $|s\rangle$ pairing correlations involve 
only time-reversed
partners, in $|s\rangle'$ we have allowed correlations between
\emph{non}-time-reversed partners, while choosing the $2s$ unpaired
spin-up electrons that occupy their levels with unit amplitude to sit
at the band's \emph{bottom} (see \Fig{fig:repaired-state}).  To
see that $|s\rangle'$ has lower energy than $|s\rangle$,
\begin{equation}
  \label{eq:Esalphaprimed}
  \E'_s \;   = \; E^{\prime \, \cond}_s \! +  \E^{\prime \, 0}_s
  \; < \; E^{\cond}_s \! + {\E}^0_s \;= \; {\E}_s \; ,
\end{equation}
we argue as follows: Firstly, $ \E^{\prime \, 0}_s = \E^0_s $, since
the corresponding uncorrelated states $|s\rangle'_0$ and $|s
\rangle_0$ are identical (and given by \Eq{eq:param}, where $\B$ is
the set of levels $i= - s+p/2 , \dots, s-1+p/2$).  Secondly,
$\Delta'_s = \Delta_0 $ (implying $E^{\prime \, \cond}_s =
E^\cond_0$), because the $2s$ unpaired electrons in $|s\rangle'$ sit
at the band's bottom, \ie\ so far away from $\eF$ that their blocking
effect is negligible [cf.\ point (vi) of \Sec{sec:generalnumerics}].
In contrast, $\Delta_s < \Delta_0$ (implying $E^{\cond}_s >
E^\cond_0$), because the $2s$ unpaired electrons in $|s\rangle$ sit
around $\eF $ and cause significant blocking.  Thus the condensation
energies satisfy $E^{\prime \, \cond}_s < E^\cond_s ( \le 0)$, so that
Eq.~(\ref{eq:Esalphaprimed}) holds, implying that $|s \rangle'$ would
be a better variational ground state for the
interaction~(\ref{eq:nontimereversed}) than $|s \rangle$.

Now, the fact that $E^{\prime \, \cond}_s =E^\cond_0$ is
\emph{independent} of $s$ means that flipping spins in $|s \rangle'$
does not cost correlation energy.  Thus, the energy cost for turning
$|0 \rangle'$ into $|1 \rangle'$ by flipping one spin is simply the
kinetic energy cost $d$, implying a threshold field $h'_{0,1} = d/2$
[see Eq.~(\ref{eq:hcrit})]; in contrast, the cost for turning $|0
\rangle$ into $|1 \rangle$, namely $(\E_1 - \E_0)$, implies a
threshold field $h_{0,1} = \half(\E_1 - \E_0)$, which (in the regime
$d \lesssim \tilde \Delta$) is rather larger than $d/2$. [$(\E_1 -
\E_0)$ is the $h=0$ ``spectral gap'' between the first and second
lines of \Figs{fig:sc-magneticfield}(a) and \ref{fig:spectra}(a)].
The fact that RBT's experiments [\Fig{fig:sc-magneticfield}(b)]
clearly show a threshold field $h_{0,1}$ significantly larger than
$d/2$ shows that the actual spin-$1$ ground state which nature chooses
is better approximated by $|1 \rangle$ than by $|1 \rangle'$, in spite
of the fact that $\E'_1 < \E_1 $.  Thus the premise of the above
argument was wrong, and we can conclude that those terms in
Eq.~(\ref{eq:nontimereversed}) not contained in the reduced BCS
Hamiltonian can indeed be neglected, as done throughout in this
review.

\subsection{Measurable consequences of the 
 blocking effect: parity effects}
\label{sec:BCS-parity}

This section is devoted to various measurable manifestations of
the blocking effect, in the form of parity effects,
\ie\ differences between a grain with an even or
odd number of electrons.

\subsubsection{Bulk consequences of blocking}
\label{sec:two-qp}

The most obvious measurable manifestation of the blocking effect is 
the very existence of a spectral gap: ``breaking a pair'' and
placing the two newly unpaired electrons in two singly-occupied levels
costs a significant amount of correlation energy, because the unpaired
electrons loose pairing energy themselves and also 
disrupt the pairing correlations of the other pairs. This,
of course, is already present in standard bulk mean-field BCS theory
via the energy cost of at least $2 \tilde \Delta$ involved in creating
two quasiparticles, and is one of the hallmarks of superconductivity.

In the context of ultrasmall grains, let 
us denote the \emph{pair-breaking energies}
for an even (odd) grain, \ie\ the  minimum
energy cost per electron for breaking a pair by flipping a
single spin at $h=0$, by $\Omega_e$ ($\Omega_o$):
\begin{equation}
  \label{eq:spectralgaps}
      \Omega_e \equiv \half (\E_1 - \E_0)_{h=0}
, \qquad   
\Omega_o \equiv \half (\E_{3/2} - \E_{1/2})_{h=0} \, . 
\end{equation}
The even pair-breaking gap $\Omega_e$ is of course strikingly visible
in RBT's $h=0$ spectra as a large spectral gap for even grains [cf.\ 
\Figs{fig:sc-spectra(h=0)} and \ref{fig:sc-magneticfield}; the latter
gives $\Omega_e = 0.26$~meV].  Its presence is direct evidence for the
existence of pairing correlations in the grain, which in that sense
can still be called ``superconducting''. 

In contrast, the odd pair-breaking gap $\Omega_o$ can not be obtained
from $h=0$ spectra, since in an odd grain the lowest excitation does
not involve breaking a pair, but simply exciting the unpaired
electron, which does not require a correlation-induced gap to be
overcome.  To measure $\Omega_o$, a finite field is needed: by
\Eq{eq:hcrit}, $ \Omega_e = h_{0,1}$ and $\Omega_o = h_{1/2,3/2}$,
hence both spin-flip gaps are equal to level-crossing fields that can
be deduced from $h \neq 0$ data, as explained in
\Sec{sec:tunneling-spectra-prb97}.  For \Fig{fig:sc-magneticfield}
this yields $ \Omega_e = 0.23\pm 0.01$~meV and $\Omega_o = 0.24 \pm
0.01$~meV [a result further discussed in \Sec{sec:parity-pairbreaking}].
The reason that the $\Omega_e$-value determined in this way
is somewhat {\em smaller}\/ than the above-mentioned 0.26~meV
determined at $h=0$ is presumably that the experimental
spectral lines are not perfectly linear in $h$ (having a small
$h^2$-contribution due to orbital diamagnetism,
 which should cause the spectroscopic gap to
decrease faster with $h$ than in our model).

Another consequence of the blocking effect is that the condensation
energies $E^\cond_{p/2} = \E_{p/2} - \E_{p/2}^0$ for an even and odd grain
differ: the unpaired electron of an odd grain weakens its pairing
correlations relative to an even grain, so that $E^\cond_{1/2}$ is less
negative than $E^\cond_0$.  In the bulk limit their difference approaches $
E^\cond_{1/2} - E^\cond_0 \to \tilde \Delta$, the energy of a single
quasiparticle.  For large mesoscopic islands (with $d/\tilde \Delta
\ll 1$) this energy difference has indeed been directly observed: it
causes a change from $e$- to $2e$-periodicity in the gate-voltage
dependence of Coulomb oscillations
\cite{hanna91,Tuominen-92,Tuominen-93,Tinkham-95,Saclay,eiles93}.
 For ultrasmall grains, however,
ground state energy differences are currently not directly measurable,
due to experimental difficulties explained in
\Sec{sec:ground-state-energies-not-measurable}.

The parity effects discussed above survive in the bulk limit. Let us
now turn to parity effects that result from even-odd differences in
the \emph{$d$-dependence} of various quantities.

\subsubsection{Parity-dependent pairing parameters}

As is evident from
\Fig{fig:pairing-parameter}(a,b), not only the condensation energies
$E_{p/2}$ are parity dependent; as soon as one leaves the bulk regime,
the pairing parameters $\Delta_{p/2}$ become parity-dependent too,
with $\Delta_{0} > \Delta_{1/2}$.  In the context of ultrasmall grains
this was first emphasized by von Delft \etalia\ 
\cite{vondelft96}, but it had been anticipated before by Janko, Smith
and Ambegaokar \cite{Janko-94} and Golubev and Zaikin \cite{Golubev-94},
who had studied the first correction to the bulk limit, finding
$\Delta_0 - \Delta_{1/2} = d/2$ to leading order in $d/\tilde
\Delta$; and this result, in turn, had already been published by
Soloviev in the nuclear physics literature as long ago as 1961
\cite{Soloviev-61}.

The \gc\ results of \Fig{fig:pairing-parameter}(a), in particular the
fact that the critical level spacing $d^\BCS_{p/2}$ at which
$\Delta_{p/2}$ vanishes is smaller for odd than even grains
($d_{1/2}^\BCS < d_{0}^\BCS$), suggest that ``pairing correlations
break down sooner in odd than even grains'' \cite{vondelft96}.
However, it should be remembered that the vanishing of $\Delta_{p/2}$
signals the breakdown of the \gc\ approach. A more accurate statement,
that is born out by the canonical calculations reviewed in
\Sec{sec:sc-canonical}, is that the inequality $\E_{1/2} > \E_0$
persists for arbitrarily large $d$ (see \Fig{fig:exact-gse} in
\Sec{comparison}), \ie\ pairing correlations are always weaker for odd
than even grains, although they never vanish altogether in either.

\subsubsection{Matveev-Larkin parity parameter}
\label{sec:ML-parameter}

\begin{figure}
  \centerline{\epsfig{figure=%
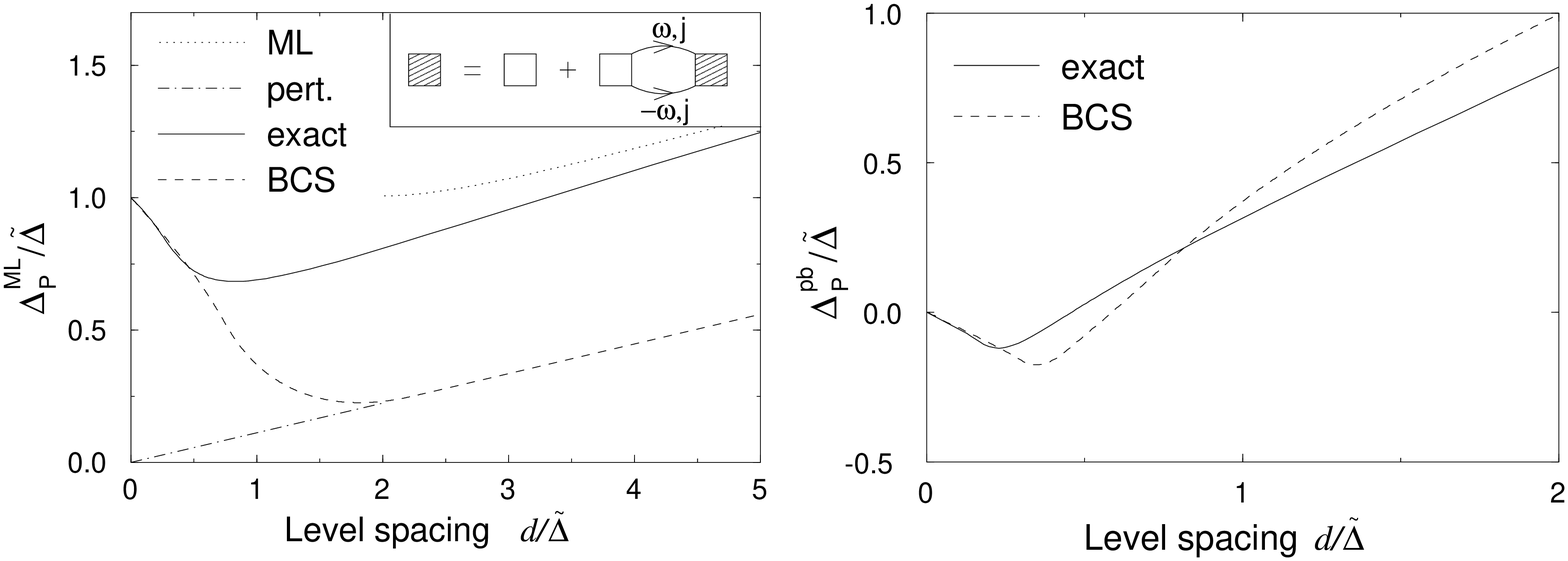,height=0.34\linewidth}}
  \caption[Two parity parameters]{The 
    parity parameters (a) $\Delta^\ML_\PP$ of Matveev-Larkin
    [\Eq{eq:def-even-odd-effects}] and (b) $\Delta^\pb_\PP$ for the
    pair-breaking energies [\Eq {eq:Delta-pairbreaking}], as functions
    of $d/\tilde \Delta$, calculated using the \gc\ variational BCS
    approach of \Sec{sec:BCS-ansatz} (dashed lines), and Richardson's
    exact solution of \Sec{sec:richardson} (solid lines).  In (a), we
    also show the perturbative result for the uncorrelated Fermi sea,
    $(\Delta_\PP^\ML)_\pert= \frac12 \lambda d$ (straight dash-dotted
    line); and the renormalized result $(\Delta_\PP^{\ML})_\ren \simeq
    d/[2\ln(ad/\tilde\Delta)]$ of \Eq{MLren}, in its range of validity
    $d / \tilde \Delta \gg 1$ (dotted line).  The parameter $a=1.35$
    is chosen to ensure quantitative agreement with the exact result in
    the limit $d/\tilde \Delta \gg 1$.  For a summary of the results
    of various other canonical calculations of $\Delta_\PP^{\ML}$, see
    Fig.~\protect\ref{fig:exact-gse}. The inset of (a) shows the Dyson
    equation used to calculated the renormalized coupling $\tilde
    \lambda$ in \Eq{eq:lambda-Dyson}.}
  \label{fig:spectral-gap}
  \label{fig:Matveev-parity}
\end{figure}
To capture the difference between correlations in even and odd grains
in terms of measurable quantities (which $\Delta_{p/2}$ are not),
Matveev and Larkin \cite{matveev97} proposed the parameter (sometimes
called ``pairing energy'' in nuclear physics \cite{richardson65a})
\begin{eqnarray}
  \label{eq:def-even-odd-effects}
  \Delta^\ML_\PP & \equiv & 
\E_{1/2}^{N+1} - \half (\E_0^{N}  + \E_0^{N+2} )
  \qquad \mbox{(where $N$ is even)} \, ,
\end{eqnarray}
\ie\ the difference between the ground state energy of an odd grain
and the mean of the ground state energies of the even grains obtained
by removing or adding one electron.  \Fig{fig:Matveev-parity}(a) shows
its behavior as function of $d/ \tilde \Delta$.  In the bulk limit we
have $\E_0^{N} \simeq \E_0^{N+2} $ and $\Delta^\ML_\PP \simeq \tilde
\Delta$, which is simply the energy cost for having an unpaired
electron on the odd grain.  With increasing $d/\tilde \Delta$, this
energy cost decreases since pairing correlations get weaker, hence
$\Delta_\PP^\ML$ initially decreases.  It begins to increase again for
$d \gtrsim \tilde \Delta$, since then pairing correlations are so weak
that the behavior of $\Delta_\PP^\ML$ is governed by the
``self-energy'' of the one extra pair in $\E_0^{N+2}$ relative to
$\E_0^N$. For example, in the \gc\ variational BCS result for
$\Delta^\ML_\PP$, namely
\begin{equation}
  \label{eq:ML-BCS}
(\Delta^{\ML}_\PP)_\BCS = \E^\BCS_{1/2}
- \E^\BCS_0 + \lambda d/2 \, ,   
\end{equation}
it is this self-energy which produces the $ \lambda d/2$ contribution.

A more careful calculation for the regime $d \gg \tilde \Delta$ was
performed by Matveev and Larkin \cite{matveev97}, whose considered the
renormalization of $\lambda$ due to ``pairing fluctuations'' about the
uncorrelated Fermi ground state $|p/2 \rangle_0$. Summing up the
leading logarithmic vertex corrections \cite{AGD63} [see inset of
\Fig{fig:Matveev-parity}(a)], they obtained a renormalized coupling
$\tilde \lambda$ given, with logarithmic accuracy, by
\begin{eqnarray}
  \label{eq:lambda-Dyson}
  \tilde \lambda & = & \lambda + \lambda d \left[ \sum_j^U \int
   { \d \omega \over 2 \pi} 
   { 1 \over [ \ii \omega - (\varepsilon_j -    \mu) ] }
   { 1 \over [ - \ii \omega - (\varepsilon_j -    \mu) ] } \right] \tilde
   \lambda\\
  \label{eq:renormalized-lambda}
\tilde \lambda & = & {\lambda \over 1 - \lambda d \sum_j^U  {1 \over
2 | \varepsilon_j - \mu|}}  \; \simeq \; 
{ \lambda \over 1 - \lambda \log (\omegaD/d)} \; .
\end{eqnarray}
This result evidently is valid only if $d \gg \omegaD \e^{-1/\lambda}
\simeq \tilde \Delta/2$ (which, incidentally, is another way of seeing
that $d \simeq \tilde \Delta$ defines the crossover between the
fluctuation-dominated and bulk regimes).  Matveev and Larkin concluded
that
\begin{equation}
  \label{MLren}
  (\Delta_\PP^{\ML})_\ren \simeq \tilde \lambda d / 2 =
 d /(2 \log d/ \tilde \Delta)
\,   \qquad \mbox{for} \quad d \gg \tilde \Delta  \;  .
\end{equation}
This logarithmic renormalization is beyond the reach of the \gc\ 
variational BCS method, but was confirmed using exact methods
\cite{mastellone98,braun-thesis,sierra99} (see \Sec{comparison}).  Its
occurrence, in a regime that in \gc\ variational calculations appears
to be ``uncorrelated'', can be regarded as the ``first sign of pairing
correlations'', in particular since, by \Eq{eq:renormalized-lambda},
the interaction strength \emph{increases} upon renormalization only if
the interaction is attractive ($\lambda <0$ would imply $|\tilde
\lambda | < | \lambda |$).  The pairing fluctuations responsible for
this renormalization will be discussed in more detail in
\Secs{comparison} and \ref{sec:bulk-few-n-differences}.

Unfortunately, $\Delta_\PP^{\ML}$ is at present not measurable in
ultrasmall grains, for the same experimental reasons as apply to
$\E_{1/2} - \E_0$, see
Section~\ref{sec:ground-state-energies-not-measurable}.

\subsubsection{Parity effect for pairbreaking energies}
\label{sec:parity-pairbreaking}

Braun and von Delft \cite{braun98,braun99,braun-thesis} discussed yet
another parity effect, based on 
\begin{equation}
  \label{eq:Delta-pairbreaking}
\Delta^\pb_\PP =
\Omega_o- \Omega_e  \; , 
\end{equation}
the difference between the \emph{pair-breaking energies} of an even
and an odd grain [see \Eq{eq:spectralgaps}].
\Fig{fig:Matveev-parity}(b) shows its behavior as function of $d$.  In
the bulk limit $\Omega_e \simeq \Omega_o \simeq \tilde \Delta$ and
$\Delta^\pb_\PP \simeq 0$.  The most interesting feature of
$\Delta^{\pb}_\PP$ is that it initially becomes negative as $d/\tilde
\Delta$ increases; this occurs because in an odd grain pairing
correlations are weaker and hence breaking a pair costs less energy
than in an even grain.  $\Delta^{\pb}_\PP$ becomes positive again for
$d / \tilde \Delta \gtrsim 0.5$, since then pairing correlations are
so weak that $\Delta^{\pb}_\PP$ is governed by the kinetic energy cost
of flipping a spin, which is $2d$ for an odd grain but only $d$ for an
even grain.

$\Delta^\pb_\PP$ \emph{is} directly measurable in RBT's grains, via
the level-crossing fields $h_{0,1} = \Omega_e$ and $h_{1/2,3/2} =
\Omega_o $ [\Eq{eq:hcrit}].  The measured values $ \Omega_e = 0.23\pm
0.01$~meV and $\Omega_o = 0.24 \pm 0.01$~meV cited in \Sec{sec:two-qp}
give a positive value of $\Delta^\pb_\PP = 0.1$~meV, implying that the
grain under study was too small to fall in the most interesting regime
where $\Delta^\pb_\PP$ is negative.  Braun and von Delft suggested
that the latter should be observable in a somewhat larger grain with
$h_{1/2,3/2} < h_{0,1}$, i.e.\ in Regime~II of Fig.~\ref{fig:h-crit}.
(This suggestion assumes that despite the increased grain size, the
complicating effect of orbital diamagnetism is still non-dominant in
Regime~II.)  \emph{To look for negative $\Delta^\pb_\PP$
  experimentally would thus require good control of the ratio $d /
  \tilde \Delta$, i.e.\ grain size.}  This might be achievable if
recently-reported new fabrication methods, which allow systematic
control of grain sizes by using colloidal chemistry techniques
\cite{Klein-97,sunmurray,Schmid97}, could be applied to Al grains.

\newpage \section{Superconductivity: crossover from the bulk to
the limit of a few electrons}
\label{sec:crossover}
\label{sec:sc-overview}
\label{sec:sc-canonical}


This section is devoted to the question: \emph{How do pairing
correlations change when the size of a superconductor is decreased
from the bulk to the limit of only a few electrons?} In particular,
we shall attempt to refine the answer given by Anderson
\cite{anderson59}, namely that superconductivity as we know it breaks
down for $d \gtrsim \tilde \Delta$.

First steps towards a more detailed answer were taken in the early
1970s by Strongin \etalia\ \cite{Strongin-70} and by M\"uhlschlegel
\etalia\ \cite{Muehlschlegel-72}, who calculated the thermodynamic
properties of ensembles small superconducting grains.  Experimental
realizations of such ensembles were, \eg, the granular films studied
by Giaver and Zeller \cite{GiaeverZeller-68,ZellerGiaever-69}.
The interest of theorists was rekindled in 1995 by RBT's success in
probing \emph{individual} superconducting grains. Apart from
motivating the phenomenological theory of Braun \etalia\ reviewed in
\Sec{sec:superconductivity}, these experiments also inspired a
substantial and still growing number of
theoretical studies \cite{vondelft96,braun97,braun99,smith96,%
balian-short,balian-long,%
bonsager98,matveev97,Rossignoli-98,Rossignoli-99a,Rossignoli-99b,%
Rossignoli-00,mastellone98,berger98,braun98,dukelsky99a,dukelsky99b,%
braun-vieweg,sierra99,vondelft-ankara99,dukelsky99c,%
tian99,tanaka99,dilorenzo99} of how
superconducting pairing correlations in such grains are affected by reducing
the grains' size, or equivalently by increasing its mean level spacing $d
\propto {\rm Vol}^{-1}$ until it exceeds the bulk gap $\tilde \Delta$.

In the earliest of these, von Delft {\em et al.\/} studied the \dbcsm\ 
of \Sec{sec:model} within a parity-projected \gc\ BCS approach
\cite{vondelft96} closely related to the variational BCS method of
\Sec{sec:generalBCS}.  Their \gc\ results suggested that pairing
correlations, as measured by the pairing parameter or the condensation
energy, vanish abruptly once $d$ exceeds a critical level spacing
$d^\BCS_{p/2}$ \emph{that depends on the parity ($p= 0$ or $1$) of the
  number of electrons on the grain}, being smaller for odd grains
($d^\BCS_{1/2} \simeq 0.89 \tilde \Delta$) than even grains $(d^\BCS_0
\simeq 3.6 \tilde \Delta$).  Parity effects were also found in a
number of subsequent papers that used parity-projected \gc\ methods to
study the behavior of the BCS mean-field gap parameter $\Delta_\MF$
and related quantities as functions of level spacing
\cite{braun97,braun99,smith96,balian-short,balian-long,%
bonsager98,matveev97,Rossignoli-98,Rossignoli-99a,Rossignoli-99b,%
Rossignoli-00},
temperature and magnetic field.  All these parity effects are
consequences of the blocking effect (cf.\ 
\Sec{sec:generalproperties}): for odd grains, the unpaired electron
somewhat disrupts the pairing correlations of the remaining paired
ones, by reducing the phase space available for pair scattering.

A series of more sophisticated canonical approaches
\cite{mastellone98,berger98,braun98,%
dukelsky99a,dukelsky99b,braun-vieweg,sierra99,%
vondelft-ankara99,dukelsky99c}
(summarized in \Sec{comparison}) {\em confirmed the parity dependence
  of pairing correlations,\/} but established that the abrupt
vanishing of pairing correlations at $d^\BCS_{p/2}$ is an artifact of
\gc\ treatments: \emph{pairing correlations do persist, in the form of
  so-called fluctuations, to arbitrarily large level spacings}
  \cite{matveev97}, and the crossover between the bulk
superconducting (SC) regime $(d \ll \tilde \Delta)$ and the
fluctuation-dominated (FD) regime $(d \gg \tilde \Delta)$ is
completely smooth \cite{dukelsky99a,dukelsky99b,braun-vieweg,sierra99,%
vondelft-ankara99}.  Nevertheless, these two regimes are
qualitatively very different
\cite{braun98,dukelsky99a,dukelsky99b,braun-vieweg,%
sierra99,vondelft-ankara99}: the
condensation energy, \eg, is an extensive function of volume in the
former and almost intensive in the latter, and pairing correlations
are quite strongly localized around the Fermi energy $\eF$, or more
spread out in energy, respectively. Very recently, Di Lorenzo \etalia\ 
\cite{dilorenzo99} suggested that the remnant pairing correlations in
the FD regime might be detectable via susceptibility measurements.

Toward the end of 1998 and after the appearance of most of these
works, R.W.  Richardson pointed out \cite{richardson-private-98} to
their various authors that the discrete BCS Hamiltonian on which they
are based actually has an exact solution, discovered by him in 1963
\cite{richardson63a} (and independently by Gaudin in 1968
\cite{gaudin}). Richardson published his solution in the context of
nuclear physics in a series of papers between 1963 and
1977 \cite{richardson63a,richardson63b,richardson64,%
  richardson65a,richardson65b,richardson66,%
  richardson66-b,richardson67,richardson77} which seem to have
completely escaped the attention of the condensed matter community.
Very recently, the model was also shown to be integrable
\cite{cambiaggio,sierra99b}.  The revival of this remarkably simple
exact solution after such a long and undeserved period of neglect is
perhaps one of the most important consequences of RBT's experimental
breakthrough:
%
%
Richardson's solution 
allows the elucidation and illustration by {\em exact\/} means of
many important conceptual ingredients of the standard BCS theory of
superconductivity, such as the nature of pairing correlations, the
importance of phase coherence, the validity of using a mean-field
approximation and a grand-canonical formulation for bulk systems, and
the limitations of the latter approaches for ultrasmall systems.
Moreover, it allows the exact calculation of 
essentially all quantities of interest for ultrasmall grains. 

We shall therefore start this section by discussing the exact solution
[\Sec{sec:richardson}]. We then summarize the other canonical
approaches somewhat more briefly than they perhaps would have deserved
had an exact solution not existed, and compare their results to those
of the exact solution [\Sec{comparison}].  Next we analyze the
qualitative differences between the bulk and FD regimes
[\Sec{sec:bulk-few-n-differences}], then discuss the case of randomly
(as opposed to uniformly) spaced energy levels $\varepsilon_j$
[\Sec{sec:sc-level-statistics}], and finally discuss finite
temperature parity effects [\Sec{sec:finite-T}].
Throughout this section we set $\mu = 0$, since canonical
treatments make no reference to a chemical potential.

\subsection{Richardson's exact solution}
\label{sec:sc-richardson}
\label{sec:richardson}

In this section we summarize some of the central results of Richardson's
exact solution of the \dbcsm.

\subsubsection{General eigenstates}
Consider $N= 2n + b$ electrons, $b$ of which are unpaired, as
in \Sec{sec:generalproperties}.  According to the general
discussion there, the nontrivial aspect of solving the model is
finding the eigenenergies $\E_n$ and corresponding eigenstates
$|\Psi_n\rangle$ [\Eq{eq:eigenpsi}] of the pair Hamiltonian [\Eq{1},
in which we set $\mu=0$ below]
\begin{eqnarray}
 && \hat H_\U = \sum_{ij}^\U \left( 2 \varepsilon_j \delta_{ij} -
 \; \lambda d \right)  b_i^\dagger b^\ds_j \; ,  
 \label{pairH}
 \end{eqnarray}
 in the Hilbert space of all states containing exactly $n$ pairs
 $b^\dagger_j = c^\dagger_{j+} c^\dagger_{j-}$ of electrons, where $j$
 runs over the set of all unblocked single-particle levels, $\U = \I
 \backslash \B$ [$I$ is the set of all interacting levels, $B$ the set
 of all blocked levels].  In general, degenerate levels are allowed in
 $\I$, but are to be distinguished by distinct $j$-labels, i.e.\ they
 have $\varepsilon_i = \varepsilon_{j}$ for $i \neq j$.

Richardson showed that the sought-after eigenstates (with
normalization \linebreak
\mbox{$\langle \Psi_n | \Psi_n \rangle = 1$}) and
eigenenergies have the general form
\begin{eqnarray}
  \label{eq:truebosoneigenstates-cc}
  | \Psi_n \rangle = {\cal N}\prod_{\nu=1}^n 
B_{\nu}^\dagger |0\rangle \, , \quad 
\E_n = \sum_{\nu = 1}^n E_{\nu}, 
\quad \mbox{with} \quad 
  B_{\nu}^\dagger = \sum_j^\U 
{b_j^\dagger \over 2 \varepsilon_j - E_{\nu}} \; .
\end{eqnarray} 
Here ${\cal N}$ is a  normalization constant and 
the $n$ parameters $E_{\nu}$ ($\nu = 1, \dots , n$) are a
solution of the set of $n$ coupled algebraic equations
\begin{eqnarray}
  \label{eq:richardson-eigenvalues-cc}
  {1\over \lambda d }  - \sum_j^\U 
{1  \over 2 \varepsilon_j - E_{\nu}}    
+ \sum_{\mu = 1(\neq \nu)}^n {2 \over E_{{\mu}} - E_{{\nu}}} = 0 \; ,
\qquad \mbox{for}\quad \nu = 1, \dots, n \; ,
\end{eqnarray}
which are to be solved (numerically, see
\App{app:electrostatic-analogy}) subject to the restrictions $E_\mu
\neq E_\nu$ if $\mu \neq \nu$.
Richardson originally derived this remarkably simple result by solving
the Schr\"odinger equation for the wave-function $\psi (j_1, \dots ,
j_n)$ of \Eq{eq:generaleigenstate-2}.  A simpler proof, also due to
Richardson \cite{richardson-private-99}, may be found in
\Refs{vondelft-ankara99,vondelft-annalen}
and in \App{app:richardson};  
its strategy is to
verify that $(\hat H_\U - \E_n) |\Psi_n \rangle = 0$ by simply
commuting $\hat H_\U$ past the $B_{\nu}^\dagger$ operators in
(\ref{eq:truebosoneigenstates-cc}).

Below we shall always assume the $\varepsilon_j$'s to be all distinct
(the more general case that degeneracies are present is discussed by
Gaudin \cite{gaudin}).  Then it can be shown explicitly \cite{gaudin}
that (i) the number of distinct solutions of
Eq.~(\ref{eq:richardson-eigenvalues-cc}) is equal to the dimension of
the $n$-pair Hilbert space defined on the set of unblocked levels
$\U$, namely ${N_U \choose n}$, where $N_U$ is the number of unblocked
levels; and (ii) that the corresponding eigenstates
(\ref{eq:truebosoneigenstates-cc}) are mutually orthogonal to each
other, thus forming an eigenbasis for this Hilbert space. This can
easily be understood intuitively, since there exists a simple relation
between the bare pair energies $2\varepsilon_j$ and the solutions of
\Eqs{eq:richardson-eigenvalues-cc}: as $\lambda$ is reduced to 0, it
follows by inspection that each solution $\{ E_{1}, \dots,$ $E_{n}\}$
reduces smoothly to a certain set of $n$ bare pair energies, say $\{2
\varepsilon_{j_1}, \dots, 2\varepsilon_{j_n} \}$; this particular
solution may thus be labeled by the indices $j_1, \dots, j_n$, and the
corresponding eigenstate (\ref{eq:truebosoneigenstates-cc}) written as
$|\Psi_n \rangle \equiv |j_1, \dots j_n \rangle_\U$.  By inspection, its
$\lambda \to 0 $ limit is the state $|j_1, \dots j_n \rangle_{\U,0}
\equiv \prod_{\nu=1}^n b_{j_\nu}^\dagger |0 \rangle $, thus there is a
one-to-one correspondence between the sets of all states $\{ |j_1,
\dots, j_n \rangle_\U \} $ and $\{ | j_1, \dots j_n \rangle_{\U,0}
\}$.  But the latter constitute a complete eigenbasis for the $n$-pair
Hilbert space defined on the set of unblocked levels $\U$, thus the
former do too.

\subsubsection{Ground state}
\label{sec:richardson-ground-state}

For a given set of blocked levels $B$, the lowest-lying of all states
$|\Psi_n,\B\rangle$ of the form (\ref{eq:generaleigenstate}), say
$|\Psi_n,\B\rangle_{\G}$, is obtained by using that particular
solution $|j_1, \dots j_n \rangle_\U$ for which the total ``pair energy''
${\cal E}_n$ takes its lowest possible value.  The lowest-lying of all
eigenstates with $n$ pairs, $b$ blocked levels and total spin $s =
b/2$, say $|n,s \rangle_\G$ with energy ${\cal E}^\G_s (n)$, is that
$|\Psi_n, \B \rangle_\G$ for which the blocked levels in $\B$ all
contain spin-up electrons and are all as close as possible to $\eF$,
the Fermi energy of the uncorrelated $N$-electron Fermi sea
$|\F_N\rangle$. The $E_{\nu}$ for the ground state $|n,s\rangle_\G$
coincide at $\lambda =0$ with the lowest $n$ energies $2
\varepsilon_{j_\nu}$ ($\nu = 1, \dots, n$), and smoothly evolve toward
(initially) lower values when $\lambda$ is turned on, a fact that can
be exploited during the numerical solution of
\Eq{eq:richardson-eigenvalues-cc}. As $\lambda$ is increased further,
some of the $E_\nu$'s become complex; however, they always occur in
complex conjugate pairs, so that ${\cal E}_n$ remains real
\cite{richardson66}. For details, see \Ref{richardson66} and our
\App{app:electrostatic-analogy}, where some algebraic transformations
are introduced that render the equations less singular and hence
simplify their numerical solution considerably.

\subsubsection{General comments}
Since the exact solution provides us with wave functions, it is in
principle straightforward to calculate arbitrary correlation functions
of the form $\langle \Psi_n | b^\dagger_{i} b^\dag_j \dots b^\ds_{i'}
b^\ds_{j'} | \Psi_n \rangle$, by simply commuting all $b$'s to the
right of all $b^\dag$'s.  However, due to the hard-core boson
commutation relations (\ref{hard-core-boson-2}) of the $b$'s, the
combinatorics is rather involved. Nevertheless, Richardson succeeded
to derive \cite{richardson65b} explicit results for the normalization
constant ${\cal N}$ of (\ref{eq:truebosoneigenstates-cc}) and the
occupation probabilities $\bar v_j^2$ and correlators $C_{ij}$ of
\Eq{eq:C_ij}, which we summarize in \App{app:correlators}. 
The exact result for the $C_{ij}$'s show that they are 
\emph{all} positive, in agreement with
the requirement (ii) formulated in \Sec{sec:meaningfulDelta}.
 It is also
natural to ask whether in the bulk limit ($d \to 0$ at fixed $n \,
d$), the standard BCS results can be extracted from the exact
solution.  Indeed they can, as Richardson showed in
\cite{richardson77} (following unpublished work by Gaudin
\cite{gaudin}), by interpreting the problem of solving the eigenvalue
equations (\ref{eq:richardson-eigenvalues-cc}) for the $E_\nu$ as
a problem in two-dimensional electrostatics (see
\App{app:electrostatic-analogy}).  Exploiting this analogy, he showed
that in the bulk limit, Eqs.~(\ref{eq:richardson-eigenvalues-cc})
reduce to the well-known BCS equations determining the gap and
chemical potential at $T=0$ [\Eqs{eq:gap} and (\ref{eq:mu})], and the
ground state condensation energy ${\cal E}^\cond_0(n)$
[\Eq{eq:define-condensation-energy}] to its BCS result, namely $-
\tilde \Delta^2/2d$.

Finally, let us mention that Cambiaggio, Rivas and Saraceno have
recently shown that the \dbcsm\ is integrable and have constructed
explicit expressions for all its constants of the motion
\cite{cambiaggio}. The latter's relation to Richardson's solution was
clarified by Sierra \cite{sierra99b}, who has also explored possible
connections between the exact solution and conformal field theory.  It
would be an interesting challenge for mathematical physicists to try
to exploit this integrability to calculate finite-temperature
properties exactly --- although these can in principle be obtained
from Richardson's solution by ``simply'' computing the partition
function over all states, this is forbiddingly tedious in practice for
large temperatures, since the eigenenergy of \emph{each} state
requires a separate (non-trivial) numerical calculation.

\subsection{Comparison of other canonical methods with the
exact solution}
\label{sec:fixedN}
\label{sec:sc-exact-diagonalization}
\label{sec:sc-dmrg}
\label{comparison}

\begin{figure}
\centerline{\epsfig{figure=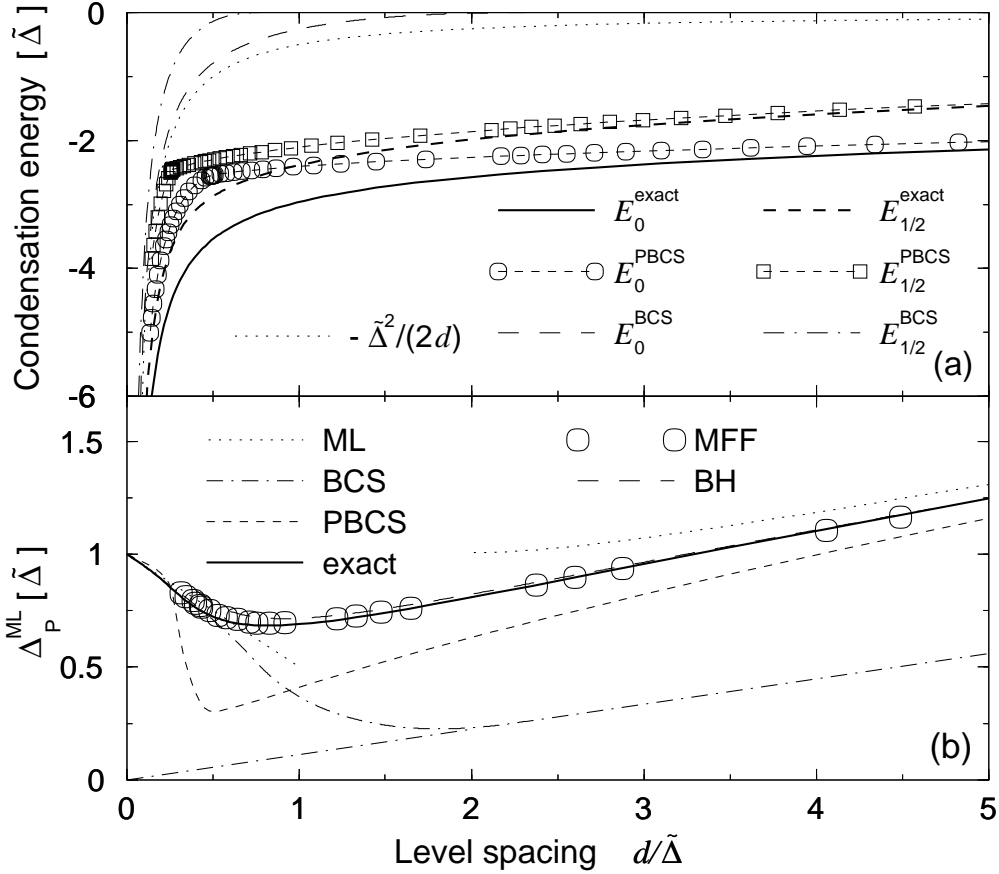,%
width=0.95\linewidth}} 
  \caption[Exact ground state condensation energies]{
    (a) The even and odd $(s=0,1/2)$ condensation energies
    ${E}^\cond_{s}$ of Eq.~(\ref{eq:condensation-again}) [in units of
    $\tilde \Delta$], calculated with BCS, PBCS and exact wave
    functions \cite{sierra99}, as functions of $d/ \tilde \Delta = 2
    \sinh (1/ \lambda) / (2n + 2s)$, for $\lambda = 0.224$.  For
    comparison, the dotted line gives the ``bulk'' result $E_0^{\rm
      bulk} = -\tilde \Delta^2/(2d)$.  (b) Comparison \cite{sierra99}
    of the parity parameters $\Delta^\ML_\PP $
    \protect\cite{matveev97} of Eq.~(\ref{eq:ML-again}) [in units of
    $\tilde \Delta$] obtained by various authors: ML's analytical
    result (dotted lines) [$\tilde \Delta(1-d/2 \tilde \Delta)$ for
    $d\ll \tilde \Delta$, and $d/2\log(a d/\tilde \Delta)$ for $d\gg
    \tilde \Delta$, with $a=1.35$ adjusted to give asymptotic
    agreement with the exact result]; grand-canonical BCS approach
    (dash-dotted line) [the naive perturbative result $\frac12\lambda
    d$ is continued to the origin]; PBCS approach (short-dashed line);
    Richardson's exact solution (solid line); exact
    diagonalization and scaling by MFF (open circles) and BH
    (long-dashed line).  }
  \label{fig:exact-gse}
\end{figure}

In this section we briefly mention the various canonical methods by
which the \dbcsm\ had been investigated prior to the revival of
Richardson's exact solution in 1999. All of these studies used a
half-filled band with fixed width $ 2\omegaD$ of 
$N=2n + p$ uniformly-spaced
levels [i.e.\ $\varepsilon_j = j \, d + (1-p)d/2$, 
with $p = 0$ or 1, as in \Eq{eq:ejs}],
 containing $N$ electrons.  
Then the level spacing is $d= 2 \omegaD /N$ and the bulk
gap is $\tilde \Delta = \omegaD / \sinh (1/\lambda)$. Following
\cite{braun98}, we take $\lambda = 0.224$ throughout this section.  To
judge the quality of the various approaches, we compare in
\Fig{fig:exact-gse} the results which they yield with those from
Richardson's solution, for the even and odd $(s=0,1/2)$ condensation
energies $E_{s}^\cond $ and the Matveev-Larkin parity parameter
$\Delta^\ML_\PP$ [cf.\ \Sec{sec:ML-parameter}]. In the notation of
\Sec{sec:richardson-ground-state}, these are given by
\begin{eqnarray}
  \label{eq:condensation-again}
E^\cond_{s} (n) & = & {\cal E}^\G_{s} (n) -
\langle \F_N| \hat H | \F_N\rangle \; ,
\\
  \label{eq:ML-again}
\Delta^\ML_\PP (n) & = & {\cal E}^\G_{1/2} (n) - 
[ {\cal E}^\G_0(n) + {\cal E}^\G_0 (n+1)]/2 \, .
\end{eqnarray}

Following the initial g.c.\ studies
\cite{vondelft96,braun97,braun99,smith96,matveev97} of the \dbcsm, the
first purely canonical study was that of Mastellone, Falci and Fazio
(MFF) \cite{mastellone98}, who used Lanczos exact diagonalization.
Despite being limited to $n \le 12$, they managed to reach reasonably
small ratios of $d/\tilde\Delta$ by using an ingenious scaling
approach: for a given level spacing $d$, they increased the coupling
constant $\lambda$ to about $0.5$, thereby decreasing $d/\tilde\Delta
= d/\omegaD \sinh(1/\lambda)$ to values as small as 0.5.  This allowed
them to probe, coming from the few-electron side, a remarkably large
part of the crossover to bulk limit. They found, \ia, that the
condensation energies are negative for \emph{all} $d$, showing that
the system can \emph{always} gain energy by allowing pairing
correlations, even for arbitrarily large $d$.

Berger and Halperin (BH) \cite{berger98} showed that almost identical
results can be achieved with less than 6 pairs, thus significantly
reducing the calculational effort involved, by first performing a
``poor man's scaling'' renormalization: they reduce the bandwidth from
$\omegaD \approx n d$ to, say, $\baromegaD \approx \bar n d$
(with $\bar n \le 6$) and incorporate the effect of the removed levels
by using a renormalized coupling constant,
\begin{eqnarray}
  \label{eq:lambda-ren}
  \bar \lambda = \lambda \left[ {1-  \sum_{ \baromegaD
 < |\varepsilon_j | < \omegaD} {\lambda  \over 2  |\varepsilon_j |
}}\right]^{-1}
 \; .
\end{eqnarray}
The reduced system is then diagonalized exactly.  Note that the
renormalization of Matveev and Larkin [\Eq{eq:renormalized-lambda}]
corresponds to taking $\baromegaD \simeq d$ in \Eq{eq:lambda-ren},
\ie\ to integrating out the entire band. Also note that the
renormalization prescription of (\ref{eq:lambda-ren}) has the property
that it would leave the bulk gap invariant in the limit $d / \tilde
\Delta \to 0$, for which \Eq{eq:lambda-ren} would imply $\baromegaD
e^{-1/\bar \lambda} \simeq \omegaD e^{-1/\lambda} \simeq \tilde
\Delta$.

To access larger values of $n$, Braun and von Delft \cite{braun98}
used a \emph{fixed-$n$ projected} BCS approach (PBCS), in which
BCS-like variational wavefunctions are projected to fixed particle
number, as in \Eq{eq:BCSground-N}.  The projection integrals occurring
in \Eq{eq:BCSground-N} were evaluated numerically for $n \le 600$,
using tricks developed in the nuclear physics literature by Bayman
\cite{Bayman-60}, Dietrich, Mang and Pradal \cite{Dietrich-64} and Ma
and Rasmussen \cite{Ma-77}, and summarized in part in the book of Ring
and Schuck \cite{RingSchuck-80}.  (A much simpler way of dealing with
the projection, using recursion relations, was recently found by
Dukelsky and Sierra \cite{dukelsky99b}.) The PBCS method gives
condensation energies that (i) are significantly lower than the
grand-canonical ones [see \Fig{fig:exact-gse}], thus the projection much
improves the variational Ansatz, and that (ii) are negative for
\emph{all} $d$, confirming that the abrupt vanishing of the \gc\ 
condensation energies is indeed an artifact of the \gc\ treatment.
The PBCS method is able to fully recover the bulk limit, but the
crossover is not completely smooth and shows a remnant of the \gc\ 
breakdown of pairing correlations: the $d$-dependence of the
condensation energy $({ E}_{s}^\cond)^\PBCS$ changes rather abruptly
[kinks in the short-dashed lines in \Fig{fig:exact-gse}(a)] from being
\emph{extensive} ($\sim 1/d$) to being practically \emph{intensive}
(almost $d$ independent).

It should be mentioned here that a generalization of the PBCS
method to finite temperatures has been worked out by Essebag
and Egido in the context of nuclear physics \cite{Essebag-93}.
The PBCS method has recently also been applied to the attractive
Hubbard model in one dimension by Tanaka and Marsiglio
\cite{tanaka99}, who found even-odd and super-even effects.  The
latter consist of differences between the number of pairs being equal
to $n=2m$ or $2m+1$, and arise if boundary conditions are used that
produce doubly-degenerate levels ($\varepsilon_{\vec k} =
\varepsilon_{- \vec k}$) near the Fermi surface.

Dukelsky and Sierra \cite{dukelsky99a,dukelsky99b} used the density
matrix renormalization group (DMRG) (with $n \le 400$) to achieve
significant improvements over the PBCS results for the \dbcsm, in
particular in the regime of the crossover, which they found to be
\emph{completely smooth}.  In general, the DMRG approach is applicable
to systems that can be divided into two pieces, called block and
environment, which interact via a preferably rather small number of
states. One starts with a small block and environment, computes their
combined density matrix, then enlarges both and recomputes the density
matrix, etc, until a large part of the system has been treated.
Dukelsky and Sierra chose the block and environment to consist,
respectively, of all particle or hole states relative to the Fermi sea
(for a detailed description of the method, see \cite{dukelsky99b}).
Since the pairing correlations involving coherent superpositions of
particle and hole states are peaked in a rather small regime of width
$\tilde \Delta$ around the Fermi energy [compare \Figs{fig:v2u2-prb97}
or \ref{fig:exact-wf}], the ``interaction'' between block and
environment is ``localized'', so that the DMRG can \emph{a priori} be
expected to work rather well for this problem.

Finally, Dukelsky and Schuck \cite{dukelsky99c} showed that a
self-consistent RPA approach, which in principle can be extended to
finite temperatures, describes the FD regime rather well (though
not as well as the DMRG).

To check the quality of the above methods, Braun
\cite{braun-thesis,sierra99} computed $E_{s}^\cond$ (for $s=0,1/2$)
and $\Delta^\ML_\PP$ using Richardson's solution
(Fig.~\ref{fig:exact-gse}).  The exact results
\begin{enumerate}
\item[(i)] quantitatively agree, for $d \to 0$, with the leading $-
  \tilde \Delta^2/2d$ behavior for $E^\cond_{s}$ obtained in the \gc\ 
  BCS approach \cite{vondelft96,braun98,braun99}, which in this sense
  is exact in the bulk limit, corrections being of order $d^0$;
\item[(ii)] confirm that the even ground state energy \emph{always}
lies below the odd one (this had independently been proven
rigoroulsy by Tian and Tang \cite{tian99});
\item[(iii)] confirm that a completely smooth
  \cite{dukelsky99a,dukelsky99b} crossover occurs around the scale $d
  \simeq \tilde \Delta$ at which the g.c.\ BCS approach breaks down;
\item[(iv)] show that the PBCS crossover \cite{braun98} is
qualitatively correct, but not quantitatively, being somewhat too
abrupt; 
\item[(v)] are reproduced remarkably well by the approaches of MFF
\cite{mastellone98} and BH \cite{berger98}; 
\item[(vi)] are fully reproduced by the DMRG of
  \cite{dukelsky99a,dukelsky99b} with a relative error of $< 10^{-4}$
  for $n \le 400$; our figures don't show DMRG curves, since they are
  indistinghuishable from the exact ones and are discussed in detail
  in \cite{dukelsky99a,dukelsky99b}.
\end{enumerate}

The main conclusion we can draw from these comparisons is that the two
approaches based on renormalization group ideas work very well: the
DMRG is essentially exact for this model, but the band-width rescaling
method of BH also gives remarkably (though not quite as) good results
with rather less effort.  In contrast, the PBCS approach is rather
unreliable in the crossover region. To study generalizations of the
\dbcsm, \eg\ using state-dependent couplings of the form $d \sum_{ij}
\lambda_{ij} b^\dagger_i b^\ds_j$, the DMRG would thus be the method
of choice.

\subsection{Qualitative differences between
the bulk and the few-electron regimes}
\label{sec:bulk-few-n-differences}

\begin{figure}
  \begin{center}
  \epsfig{figure=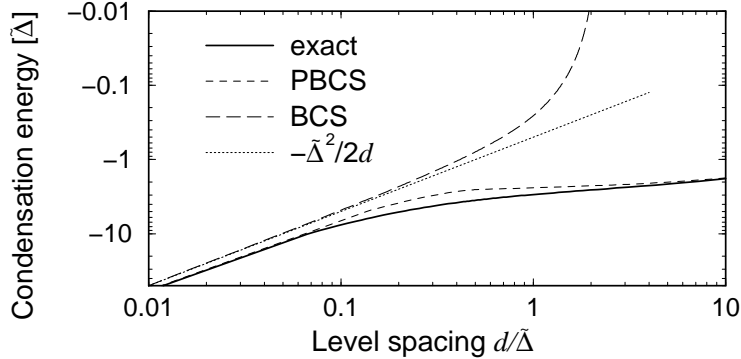,%
width=0.7\linewidth}
    \caption[Condensation energies: log-log plot and
    $\lambda$-dependence]{Log-log plot \cite{braun-thesis} of some of
      the curves of Fig.~\protect\ref{fig:exact-gse}(a) for the even
      condensation energy $E^\cond_0$ [in units of $\tilde \Delta$],
      for $\lambda = 0.224$; its asymptotic $-\tilde\Delta^2/(2d)$
      behavior for $d / \tilde \Delta \to 0$ is shown by the dotted
      line.
}
    \label{fig:exact-log-log}
    \label{fig:likharev-crossover}
  \end{center}
\end{figure}

Does the fact that the exact condensation energy $E_{s}^\cond$ is
always negative, even for arbitrarily large $d/\tilde \Delta$, mean
that the system stays ``superconducting'' even if it is arbitrarily
small? The answer is certainly no, since in the fluctuation-dominated
(FD) regime, the pairing correlations are qualitatively different than
in the bulk, superconducting regime. In this section we shall try to
make this statement more precise by analyzing the qualitative
differences between the two regimes, with regard to the $\lambda$- and
$d$-dependence of $E_{s}^\cond$, and the behavior of the occupation
probabilities $\bar v_j^2$.

\Fig{fig:exact-log-log}(a) shows, on a log-log plot, the
$d$-dependence of the even condensation energy $E_{0}^\cond (d)$.
Note that even on the log-log plot, the crossover of the exact
$E^\cond_{0}$ from the bulk to the FD regime is completely smooth.
According to Sierra and Dukelsky \cite{dukelsky99b}, the exact result
for $E_{0}^\cond (d)$ can be fitted very well to the form
\begin{eqnarray}
  \label{eq:E-cond-d-dependence}
  E_0^\cond (d) = -\tilde\Delta^2/(2d) - 
\eta_0 (\ln 2) \omegaD \lambda^2  + \gamma_0
 (\tilde\Delta d/2 \omegaD)   \log (2\omegaD/d) \; ,
\end{eqnarray}
where $\eta_0$ and $\gamma_0$ are constants of order unity \cite{dukelsky99b}.
The first term is \emph{extensive} ($\propto \Vol$) and dominates in the bulk
limit; its standard heuristic interpretation \cite{tinkham-book} is that
roughly $\tilde \Delta / d$ levels (those within $\tilde \Delta$ of $\eF$) are
strongly affected by pairing, with an average energy gain per level of
$-\tilde\Delta /2$.  The second term, which is \emph{intensive} and dominates
in the FD limit, is equal (up to the numerical factor $\eta_0$) to the result
from second-order perturbation theory \cite{dukelsky99b}, namely $(\lambda
d)^2 \sum_{i = 1}^n \sum_{j = n+1}^{2n} (2 \varepsilon_i- 2
\varepsilon_j)^{-1}$.  This subleading term's $d$-independence (which was
anticipated in \cite{anderson59,muehlschlegel62}) may be interpreted by
arguing that in the FD regime, the number of levels that contribute
significantly to $E_0^\cond$ is no longer of order $\tilde \Delta/d$: instead,
fluctuations affect \emph{all} $n \simeq 2\omegaD/d$ unblocked levels within
$\omegaD$ of $\eF$ (this is made more precise below), and each of these levels
contributes an amount of order $ -(\lambda d)^2/d$ (corresponding, in a way,
to its selfenergy). Finally, the third term
 contains the small parameter $\tilde \Delta/ \omegaD$
and thus represents a very small correction.

The $\lambda$- and volume-dependencies of $E_0^\cond$ in
\Eq{eq:E-cond-d-dependence} strikingly illustrate the qualitative
differences between the bulk and FD regimes: in the bulk regime,
dominated by the first term, $E_0^\cond$ is nonperturbative in
$\lambda$ (since $\tilde \Delta \simeq 2 \omegaD e^{- 1/\lambda}$) and
extensive, as expected for a strongly-correlated state; in constrast,
in the FD regime, dominated by the second term, $E_0^\cond$ is
perturbative in $\lambda$ and practically intensive (up to the weak
$\log d$ dependence of the third term).

\label{sec:dmp-wavefunctions}

\begin{figure}
  \begin{center}
  \epsfig{figure=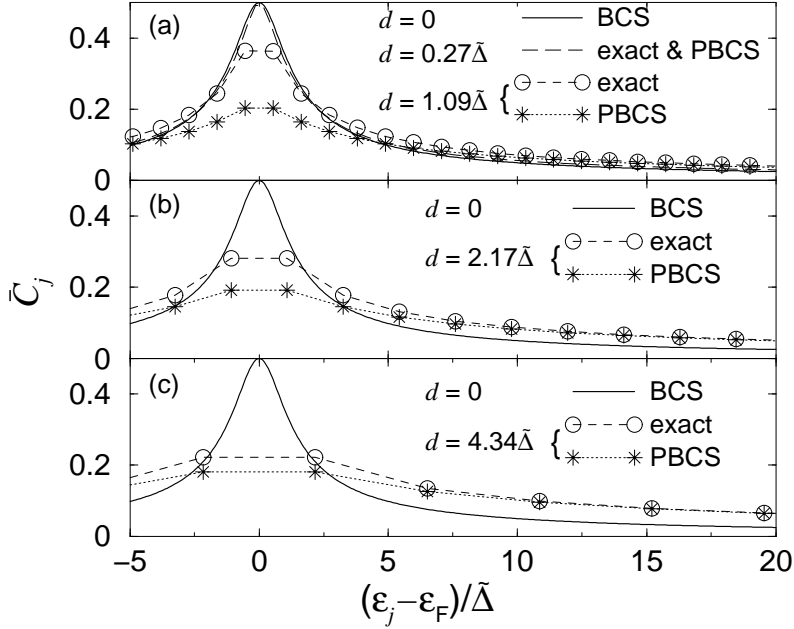,width=0.75\linewidth}
    \caption[Pairing amplitudes $\bar C_j$]{The occupation probabilities
      $\bar C_j$ of Eq.~(\ref{eq:C2_j}) for $d/ \tilde \Delta = 0$,
      0.27, 1.09, 2.17 and 4.34 \cite{braun-thesis}.  In all three
      figures, the thick solid lines give the $d=0$ bulk BCS result,
      whereas circles and stars represent $\bar C_j$-values evaluated
      for discrete $j$'s using the exact solution and PBCS method,
      respectively.  For $d=0.27\tilde\Delta$, the PBCS and exact
      results are indistinguishable, and are shown in (a) as a single
      long-dashed line, which is also virtually identical to the bulk
      curve. For small $d$, pairing correlations are evidently
      localized within a few $\tilde\Delta$ of $\eF$.  With increasing
      $d$ more and more weight is shifted away from $\eF$ into the
      tails; compared to the exact results, the PBCS method somewhat
      overemphasizes this delocalization, which is one of the reasons
      why it produces a somewhat too abrupt crossover.}
    \label{fig:exact-wf}
  \end{center}
\end{figure}

Perhaps the most vivid way of illustrating the
qualitative difference between the bulk and FD regimes
is to study properties of the ground state wavefunction. 
We shall consider here the correlators
  \cite{braun98}
\begin{equation}
  \label{eq:C2_j}
\bar   C^2_j (d) =
\langle b^\dagger_j b^\ds_j \rangle
\langle b^\ds_j b^\dag_j \rangle \; , 
\end{equation}
which measure the probability that a level can be ``both occupied and
empty'', and vanish identically for states without pairing
correlations.  For the \dbcsm\ $\bar C_j^2$ identically equals
$\langle b^\dagger_j b^\ds_j \rangle - \langle b^\dagger_j b^\ds_j
\rangle^2 = \bar v_j^2 - \bar v_j^4$ [by \Eqs{hard-core-boson-2} and
(\ref{eq:defineu2v2})], which measures the fluctuations
in the pair occupation number of level  $j$, 
and it vanishes for any blocked single-particle
level.  Note that $\bar C^2_j $ also equals $\langle b^\dagger_j
b^\ds_j \rangle -\langle c^\dagger_{j+} c^\ds_{j+} \rangle\langle
c^\dagger_{j-} c^\ds_{j-}\rangle$; this form, which was used in
\cite{braun98} and corresponds to the diagonal terms under the sum in
\Eq{eq:canonical-order-parameter} for $\Delta_\can$, can be
interpreted as the probability enhancement for finding a \emph{pair}
of electrons instead of two uncorrelated electrons in a
single-particle level $|j,\pm\rangle$.

When evaluated using the grand-canonical BCS wavefunction, $(\bar
C_j^2)_\BCS$ is equal to $ u_j^2 v_j^2 = {1 \over 4} \tilde \Delta^2/
(\varepsilon_j^2+\tilde \Delta^2)$ [thick solid lines in
\Fig{fig:exact-wf}, the same function as that plotted in
\Fig{fig:v2u2-prb97}]. The $(\bar C_j)_\BCS$'s thus have a
characteristic peak of width $ \propto \tilde \Delta$ around $\eF$,
implying that pairing correlations are ``localized around $\eF$ in
energy space'', which may be taken to be the defining property of
``BCS-like correlations''.  Moreover, in the bulk regime $d \ll
\tilde\Delta$, the $(\bar C_j)_\BCS$ are virtually identical to $(\bar
C_j)_\ex$ [long-dashed line of \Fig{fig:exact-wf}(a)], \emph{vividly
  illustrating why the grand-canonical BCS approximation is so
  successful: not performing a canonical projection hardly affects the
  parameters $\bar u_j$ and $\bar v_j$ if $d \ll \tilde\Delta$, but
  tremendously simplifies their calculation}.

As one enters the FD regime $d\gtrsim \tilde \Delta$, the character of
the correlator $(\bar C_j)_\ex$ changes [\Fig{fig:exact-wf}(b),
circles]: weight is shifted into the tails far from $\eF$ at the
expense of the vicinity of the Fermi energy.  Thus \emph{pairing
  correlations become delocalized in energy space} (as also found in
\cite{mastellone98,dukelsky99a,dukelsky99b}), so that referring to
them as mere ``fluctuations'' is quite appropriate.  In the extreme
case $d \gg \tilde \Delta$, the $(\bar C_j)_\ex$ for all interacting
levels are roughly equal.

Richardson's solution can also be used to calculate,
for a given set $B$ of blocked levels, the $d$-dependence
of the canonical order parameter $\Delta^B_\can (d)$ of
\Eq{eq:canonical-order-parameter}.  Schechter has found
\cite{moshe-priv} that it can be fit to the form $\Delta^B_\can (d) =
\tilde \Delta (1 + \tilde \gamma_B d/\tilde \Delta)$, 
where $\tilde \gamma_B$ is a
positive numerical constant, and the linear term essentially reflects
the factor of $d$ in the definition of $\Delta^B_\can$. The fact that
$\Delta^B_\can $ is a strictly increasing function of $d$ is in very
striking contrast to the behavior of the grand-canonical pairing
parameters $\Delta_s (d)$ shown in \Fig{fig:pairing-parameter}(a).


\subsection{Effect of level statistics}
\label{sec:sc-level-statistics}

\begin{figure}
\centerline{\epsfig{figure=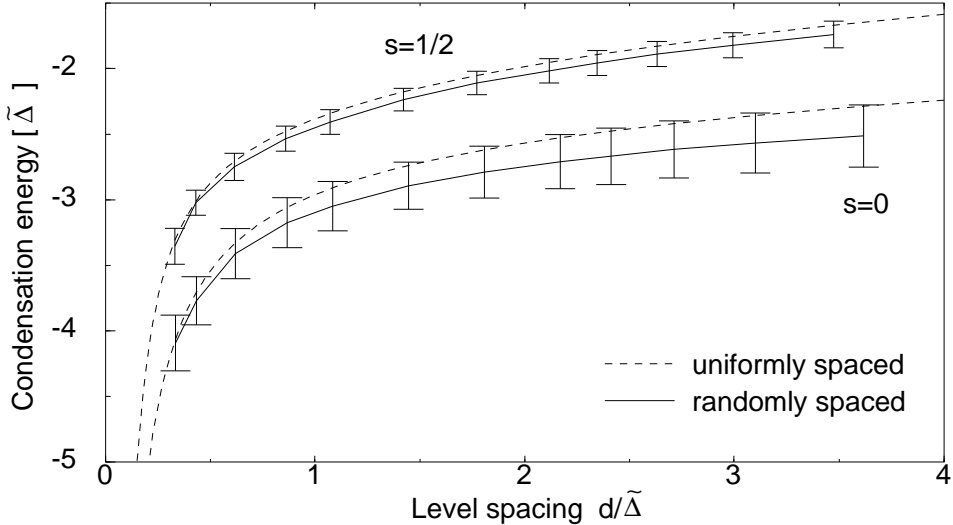,width=0.9\linewidth}} 
\caption{Exact even and odd condensation energies
  $E^\cond_s$ [in units of $\tilde \Delta$] for equally spaced levels
  (dashed line), and the ensemble-average $\langle E^\cond_s \rangle$
  for randomly-spaced levels (solid line) \cite{sierra99}. The height
  of the fluctuation bars gives the variances $\delta E^\cond_s$.}
\label{fig:randomly-spaced-levels}
\end{figure} 

Smith and Ambegaokar investigated the effect of level statistics on
the crossover between the bulk and FD regimes \cite{smith96}.  In
contrast to the uniform level spacing used in previous works, they
employed a random spacing of levels, distributed according to the
gaussian orthogonal ensemble.  Using a \gc\ mean-field BCS approach,
they found, interestingly, that \emph{randomness enhances pairing
  correlations:} compared to uniform spacings (u.s.)
\cite{vondelft96}, it (i) on average {\em lowers}\/ the condensation
energy $E^\cond_{s}$ to more negative values, $\langle E_s^\cond
\rangle < E_s^\cond \mbox{(u.s.)}$, but (ii) these still are parity
dependent, $\langle E_0^\cond \rangle < \langle E_{1/2}^\cond
\rangle$. These results can readily be understood intuitively: pairing
correlations become stronger the higher the density of levels around
$\eF$, where pair-mixing costs the least energy.  When determining the
amount of pairing correlations for a set of randomly-spaced levels,
fluctuations that increase the level density near $\eF$ are thus
weighted more than those that decrease it, so that randomness enhances
pairing correlations.

Although the \gc\ mean-field treatment of Smith and Ambegaokar breaks
down for mean level spacings much larger than $\tilde \Delta$, just as
was the case in \cite{vondelft96,braun97,braun99}, their main
conclusions (i) and (ii) are robust. Indeed, these were recently
confirmed by Sierra \etalia\ \cite{sierra99}, who used Richardson's
exact solution to calculate $E^\cond_{s}$ for ensembles of random
levels [\Fig{fig:randomly-spaced-levels}].  Moreover, they found that
the blocking effect responsible for (ii) manifests itself in the
fluctuations too, which likewise are parity dependent: for example,
\Fig{fig:randomly-spaced-levels} shows that both the variances $\delta
E^\cond_s \equiv [{\langle(E^\cond_s)^2\rangle - \langle
E^\cond_s\rangle ^2 } ]^{1/2}$ and the randomness-induced changes in
condensation energies $|\langle E^\cond_s \rangle - E^\cond_s {\rm
(u.c)}|$ were larger for even than for odd grains.

\subsection{Finite temperature parity effects} 
\label{sec:finite-T}

Although finite-temperature studies of the \dbcsm\ are not of direct
relevance for spectroscopic measurements of the BRT-type (a finite $T$
would simply smear out the discrete spectra, thereby blurring their
most interesting features), they are important in their own right for
extending our understanding of superconductivity in ultrasmall grains.
We hence review several recent finite-$T$ developments below.

To begin, let us note that parity effects are of course not restricted to the
$T = 0$ limit discussed so far. To be observable
\cite{Tuominen-92,Tuominen-93,Tinkham-95,Saclay,eiles93}, they only require
the temperature to be smaller than the free energy difference $\delta {\cal F}
\simeq \tilde \Delta - \kB T \ln[ N_{\rm eff} (T)]$ between an odd and even
grain. Here $N_{\rm eff} (T)$ is the effective number of states available for
quasiparticle excitations at temperature $T$, and for $d \ll \tilde \Delta$ is
given by $N_{\rm eff} (T) = \sqrt{8 \pi T \tilde \Delta /d^2}$
\cite{Tuominen-92}.  Below the corresponding crossover temperature where
$\delta {\cal F} = 0$, determined by $\kB T^\ast_{\rm cr} = \tilde \Delta /
\ln[ N_{\rm eff} (T^\ast_{\rm cr})]$ and roughly equal to $\tilde \Delta / \ln
\sqrt{8 \pi \tilde \Delta^2 /d^2} $, the single unpaired electron begins to
matter: it causes a crossover from $e$-periodicty to $2e$-periodicity in the
$I$-$V$ characteristics of mesoscopic superconducing SET's
\cite{Tuominen-92,Tuominen-93,Tinkham-95,Saclay,eiles93}, due to the ground
state energy difference ${\cal E}_{1/2} - {\cal E}_{0} \simeq \tilde \Delta$.
Since $T^\ast_{\rm cr}$ becomes of order $\tilde \Delta$ in nanoscopic grains
with $d \simeq \tilde \Delta$, parity effects should survive to temperatures
as high as the (bulk) superconducting transition temperature $T_{\rm c}$
itself.

Regrettably, the canonical methods discussed in the preceding sections
become impractical at finite temperatures, since the number of states
that need to be considered increases rapidly for $T \gtrsim d, \tilde
\Delta$. On the other hand, \gc\ finite-$T$ methods, some of which we
review below, are, in principle, inherently unreliable for $d \gtrsim
\tilde \Delta$.  This applies in particular to the simplest of these,
parity-projected mean-field theory \cite{Janko-94,Golubev-94}
(\Sec{sec:sc-parity-projection}) and certain variational
generalizations thereof \cite{balian-short,balian-long}
(\Sec{sec:balian}): they yield the same sharp phase transition as
function of temperature for finite systems as for bulk systems,
whereas on general grounds no sharp transitions are possible in finite
systems. The reason for this problem is that they neglect fluctuations
in the order parameter, which become very important in the transition
region.  The sharp transition is smoothed out once fluctuations are
included. A rather efficient way of doing this is the so-called static
path approximation (\Sec{sec:SPA}). Its use is illustrated in
\Sec{sec:sc-susceptibility} for a calculation of the spin
susceptibility, which shows an interesting parity effect that should
be measurable in ensembles of ultrasmall grains.

\subsubsection{Parity-projected mean-field theory}
\label{sec:sc-parity-projection}

The simplest finite-$T$ approach that is able to keep track of parity
effects is parity-projected mean-field theory, first used in nuclear
physics by Tanabe, Tanabe and Mang \cite{Tanabe-81}, and,
indepedently, introduced to the condensed-matter community by Jank\'o,
Smith and Ambegaokar \cite{Janko-94} and Golubev and Zaikin
\cite{Golubev-94}.
One
projects the \gc\ partition function exactly onto a subspace of Fock
space containing only even or odd $(p=0,1)$ numbers of particles,
using the parity-projector $\hat P_p$:
\begin{eqnarray}
  \label{eq:parity-projection}
      Z_p^{\MF}  \equiv  \mbox{Tr}^\MF \hat P_p
      e^{- \beta (\hat H  - \mu_p \hat N)} \; , \qquad
      \hat P_{0,1} \equiv {\textstyle {1 \over 2}} [1 \pm (-1)^{\hat N}]
      \; .
\end{eqnarray}
One then makes the mean-field replacement $ b_j \to \{ b_j - \langle b_j
\rangle_p \} + \langle b_j \rangle_p \; , $ neglects terms quadratic in the
fluctuations represented by $\{ \quad \}$, and diagonalizes $\hat H$ in terms
of the Bogoljubov quasiparticle operators $\gamma_{j \sigma}$ of
\Eq{eq:Bogoljubov}. The self-consistency condition $\Delta_{p/2} \equiv
\lambda d \sum_j \langle b^\ds_j \rangle_p$, evaluated in a parity-projected
\gc\ ensemble according to \Eq{eq:parity-projection}, leads to a gap equation
of the standard form,
\begin{equation}
\label{gap}
   {1 \over \lambda} = d  \sum_{|\varepsilon_j| < \omegaD }
   {1 \over 2 E_{ j}} \left( 1 - \sum_\sigma f_{p j \sigma} \right) \; ,
 \quad E_{j} \equiv \sqrt{(\varepsilon_j - \mu)^2 
+ |\Delta_{p/2}|^2} \; , 
\end{equation}
which is parity-dependent, via the occupation function $ f_{p j\sigma}
= \langle \gamma_{j \sigma}^\dagger \gamma_{j \sigma} \rangle_p $ for
quasiparticles. Since their number parity is restricted to be $p$,
$f_{pj\sigma}$ differs from the usual Fermi function $f^0_{j\sigma}$.
The condition $2n + p = \langle \hat N\rangle_p$ fixes the chemical
potential $\mu$ to lie exactly half-way between the last filled and
first empty levels if $p=0$, and exactly on the singly-occupied level
if $p=1$, implying $\mu = 0$ in both cases [by \Eq{eq:ejs}].

 \begin{figure}
\centerline{\epsfig{figure=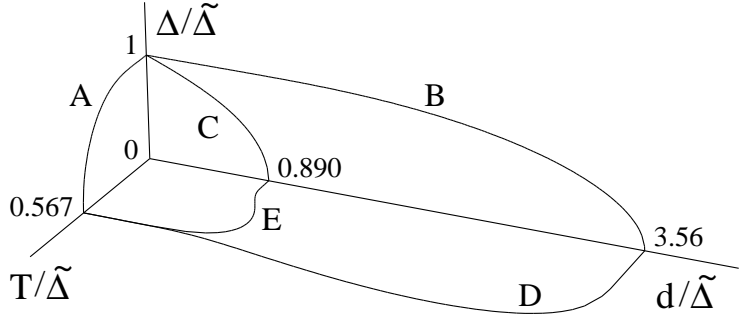,width=0.7\linewidth}}
\caption{\label{fig:T=0}
  $d$-and $T$-dependence of the pairing parameter $\Delta_{p/2}
  (d,T)$, calculated using parity-projected mean-field theory
  \cite{vondelft96}.  Curve A gives the bulk gap $\Delta (0,T)$, with
  $\Delta(0,0) \equiv \tilde \Delta$; curves B-E give $\Delta_{p/2}
  (d, T) / \tilde \Delta$ as a function of $d/\tilde \Delta$ and
  $T/\tilde \Delta$ for $p=0$ (B, D) and $p=1$ (C, E).  The critical
  spacings $d^\BCS_0 = 3.56 \tilde \Delta$ and $d^\BCS_{1/2} = 0.890
  \tilde \Delta$ given here differ somewhat from those in
  \Fig{fig:pairing-parameter}(a), because the present mean-field
  approach differs in minor details (via terms that vanish when $d\to
  0$) from the variational approach of \Sec{sec:generalnumerics}.
  \label{fig:3D}}
\end{figure}

von Delft \etalia\ \cite{vondelft96} applied this approach to the
\dbcsm\ with uniformly-spaced levels, and solved \Eq{gap} for the
parity parameter $\Delta_{p/2} (d,T)$ as function of both level
spacing and temperature.  \Fig{fig:3D} summarizes their results. At
zero temperature, $\Delta_{p/2} (d,0)$ corresponds to the spin-dependent
parity parameters $\Delta_{s= p/2}$ discussed in
\Sec{sec:generalnumerics} [cf. \Fig{fig:pairing-parameter}(a)], and
drops to zero at a critical level spacing $d^\BCS_{s}$.  The
$\Delta_{p/2} \to 0$ limit of Eq.~(\ref{gap}) defines the
parity-dependent ``critical temperature'' $T_{c,p} (d)$, which can be
viewed as another measure of how rapidly pair-mixing correlations
break down as function of level spacing (although ultrasmall grains of
course cannot undergo a sharp thermodynamic phase transition, which
can only occur if $n \to \infty$).  In both the even and odd cases,
the behavior of $T_{c,P} (d)$ shows direct traces of the parity
projection:

In the even case, $T_{c,0} (d)$ 
[\Fig{fig:3D}, curve D] is non-monotonic as function of
increasing $d$, initially increasing
slightly before dropping to zero very rapidly as $d \to d^\BCS_{0}$.
The intuitive reason for the initial increase is that the difference
between the actual and usual quasiparticle occupation functions is
$f_{pj\sigma} - f_{j \sigma}^0 < 0$ for an even grain (becoming
significant when $d \simeq \tilde \Delta$), reflecting the fact that
exciting quasiparticles two at a time is more difficult than one at a
time. Therefore the quasiparticle-induced weakening
of pairing correlations with increasing $T$ will set in at slightly higher
$T$ if $d \simeq \tilde \Delta$.

In the odd case, the critical level spacing $d_{1/2}^\BCS (T)$ 
[\Fig{fig:3D}, curve E] is
non-monotonic as a function of increasing $T$, first increasing to a
maximum before beginning to decrease toward $d^\BCS_{1/2} (T_c) = 0$.
The intuitive reason for this is that for $0 < \Delta_{1/2} \ll T,d$,
the odd $j=0$ function $f_{p0\sigma} (T)$ becomes somewhat smaller
than its $T=0$ value of $1/2$, because with increasing $T$
some of the probability for finding a quasiparticle in state $j$
``leaks'' from $j=0$ to higher states with $j \neq 0$, for which
$E^{-1}_{j} < E^{-1}_{0}$ in Eq.~(\ref{gap}). Thus, the
blocking-of-pair-scattering effect of the odd quasiparticle becomes
slightly less dramatic as $T$ is increased, so that $d^\BCS_{1/2}$
increases slightly.

It should be noted, however, that although the non-monotonicities of
$T_{c,0} (d)$ and $d_{1/2}^\BCS (T)$ are intuitively plausible within the
\gc\ framework in which they were derived, their physical significance
is doubtful, since they fall in the regime where $d / \Delta_s \gtrsim
1$ and the \gc\ approach is unreliable, due to its neglect of
fluctuations.

\subsubsection{Variational extensions of BCS theory}
\label{sec:balian}

The above-mentioned results of von Delft \etalia\ \cite{vondelft96}
were reproduced and extended to finite magnetic fields by Balian,
Flocard and V\'en\'eroni, using a more general \gc\ variational BCS
approach \cite{balian-short,balian-long}.  It is designed to optimize
the characteristic function $\varphi (\xi) \equiv \ln \mbox{Tr} \hat
P_p \e^{-\beta (\hat H - \mu \hat N)} \hat A (\xi)$, where $\hat
P_p$ is the parity projector of \Eq{eq:parity-projection} and $\hat
A(\xi) \equiv \exp (- \sum_\gamma \xi_\gamma \hat Q_\gamma)$, with $\hat
Q_\gamma$ being observables of interest (\eg\ the total spin) and
$\xi_\gamma$ the associated sources (\eg\ the magnetic field).  This
approach goes beyond the usual minimization of the free energy
\cite{BCS-57}, since it optimizes not only thermodynamic quantities
but also equilibrium correlation functions, which can be obtained by
differentiating $\e^{\varphi (\xi)}$ with respect to $\xi_\gamma$.
However, its \gc\ version also suffers from the drawback of yielding
abrupt, spurious phase transitions even though the systems are finite.
Presumably this problem would be cured if an exact projection to fixed
particle number were incorporated into this approach, but this is
technically difficult and has not yet been worked out.

\subsubsection{Static path approximation}
\label{sec:SPA}

\begin{figure}
\centerline{\epsfig{figure=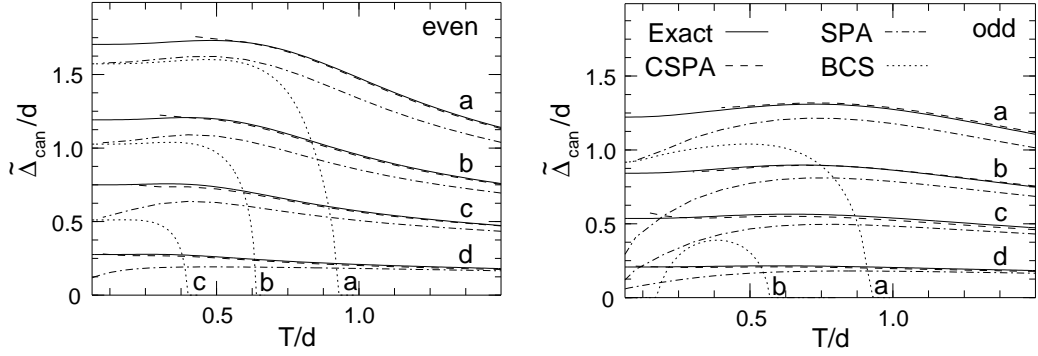,%
width=0.98\linewidth}}  
\caption[Pairing correlation energy at finite temperatures.]%
{Temperature dependence of the pairing correlation energy $\tilde
  \Delta_\can$ of \Eq{eq:Delta-rossignoli} [in units of level spacing
  $d$], as calculated in \cite{Rossignoli-99a} using parity-projected
  mean-field BCS theory (dotted lines), SPA (dash-dotted), CSPA (dashed)
  and exact diagonalization (solid lines).  A system of 10
  equally-spaced, doubly-degenerate, levels was studied, containing 10
  (left panel) or 11 (right panel) electrons.  $\tilde \Delta_\can
  (T)$ is shown at four fixed values of $d/\tilde \Delta$ (thus this
  figure elaborates \Fig{fig:3D}), namely 0.60, 0.91, 1.7, and 15,
  labeled by a,b,c and d, respectively (calculated using $\lambda =$
  0.55, 0.45, 0.35 and 0.2). CSPA data are shown only above the CPSA
  breakdown, which occurs at $T_{\rm CSPA} < \tilde \Delta/4$ for the
    cases considered.  The absence of dotted lines for the cases d
    (even) and c,d (odd) means that for these no nontrivial mean-field
    BCS solution exists. Of course, $(\tilde \Delta_\can)_{\rm exact}$ is
    nonzero nevertheless.  The abrupt BCS transition is completely
    smeared out for the SPA, CSPA and exact results, for which the
    asymptotic decay at $T \gg \tilde \Delta$ can be shown
    \cite{Rossignoli-99a} to be $\tilde \Delta_\can \sim \frac14
    (\lambda^3 d/T)^{1/2}$.}
\label{fig:rossignoli}
\end{figure}

For finite systems, in contrast to infinite ones, \emph{fluctuations}
of the order parameter about its mean-field value are very important
in the critical regime, causing the phase transition to be smeared
out; conversely, the spurious sharp transition found in the \gc\ 
approaches above is a direct consequence of the neglect of such
fluctuations.
A rather successful way of including fluctuations is the so-called
\emph{static path approximation} (SPA), pioneered by M{\"u}hlschlegel,
Scalapino and Denton \cite{Muehlschlegel-72} and developed 
by various nuclear theorists 
\cite{Alhassid-84,Alhassid-92,Arve-88,Lauritzen-88,%
Rossignoli-92,Rossignoli-93,Rossignoli-94,Puddu-90,%
Lauritzen-90,Puddu-91,Puddu-92,Attias97,Rossignoli-97a,Rossignoli-97b,%
Rossignoli-95,Rossignoli-96a,Rossignoli-96b},
while recently an exact parity projection has also been incorporated
\cite{Rossignoli-98,Rossignoli-99a,Rossignoli-99b,Rossignoli-00}.  A
detailed and general discussion, including a complete list of relevant
references, was given very recently by Rossignoli, Canosa and Ring
\cite{Rossignoli-99a}. We therefore confine ourselves below to stating
the main strategies of the SPA and illustrating its capabilities by
showing its results [\Fig{fig:rossignoli}] for the quantity
\begin{eqnarray}
  \label{eq:Delta-rossignoli}
  \tilde \Delta_\can^2 = (\lambda d)^2
\sum_{ij} \left[ C_{ij} - (C_{ij})_{\lambda=0}\right] .
\end{eqnarray}
$\tilde \Delta_\can$ is reminiscent\footnote{The definitions of
  $\tilde \Delta_\can$ and $\Delta_\can$ differ by terms of order
  $(d/\omega_D)^{1/2}$; for example, when evaluating both
  using $|\BCS\rangle$ of \Eq{eq:BCSground} and comparing to
  $\Delta_\MF$ of \Eq{eq:BCS-gap}, one finds $( \Delta_\can)_\BCS =
  \Delta_\MF = (\tilde \Delta_\can)_\BCS + {\cal
    O}[(d/\omega_D)^{1/2}]$.}
of $\Delta_\can$ of \Eq{eq:canonical-order-parameter}, and 
measures the increase in pairing correlation energy due to a nonzero
coupling strength $\lambda$.

One starts by decoupling the quartic interaction $\hat H_\red$ of
\Eq{eq:hamiltonian} into a quadratic form using a Hubbard-Stratonovich
transformation with a complex auxiliary field $\Delta (\tau) =
\Delta^+ (\tau) + \ii \Delta^- (\tau)$, with  Matsubara-expansion 
$\Delta^\pm (\tau) = \sum_{n} \Delta^\pm_{n} \e^{\ii 2 \pi n \tau /
  \beta}$ in the interval $\tau \in (0,\beta)$. The parity-projected
partition function of \Eq{eq:parity-projection}  then 
has the following path integral representation
(our notation is deliberately schematic; see \cite{Rossignoli-99a}
for a precise version): 
\begin{eqnarray}
  \label{eq:hubbard-stratonovich}
  Z_p & \propto &  \int \! \! \prod_n { \d} \Delta^+_n
{\d} \Delta^-_n \; {\cal Z}_p [\Delta] ,  \quad
{\cal Z}_p [\Delta] = 
\mbox{Tr} \left\{ \hat P_p \hat {\cal T}
    \e^{- \int_0^\beta {\d} \tau \, h[ \Delta(\tau)] } \right\}  , 
\\
  h[\Delta]  & = & 
  \sum_{j \sigma} (\varepsilon_j - \mu - \lambda d/2)
  c^\dagger_{j\sigma} c^\ds_{j \sigma}
  - \sum_j (b^\dagger_j \Delta + \Delta^\ast b^\ds_j) + 
  {|\Delta|^2 \over \lambda d} \; . 
\end{eqnarray}
The path integral can be treated at several levels of
sophistication:

(i) In the simplest, one uses a ``fixed-phase saddle-point
approximation'' for the ``static'' $n=0$ modes and neglects all
$n \neq 0$ modes, 
\ie\ one fixes the phase of $\Delta_0^+ + \ii \Delta_0^- =
|\Delta_0| \e^{\ii \phi_0} $ by, say, setting $\phi_0=0$, so that 
$\int \!  {\d} \Delta^+_{0} {\d} \Delta^-_{0}$ is replaced by
$\int {\d} |\Delta_0|$,
and approximates this integral by its saddle-point value.  The
saddle-point condition for maximizing ${\cal Z}_p[|\Delta_0|]$ then
yields the gap equation (\ref{gap}), thus this approach simply
reproduces the \emph{parity-projected mean-field} approach of
\Sec{sec:sc-parity-projection}, including its sharp phase transition
(\Fig{fig:rossignoli}, dotted lines).

(ii) The next-best approximation is obtained if one writes $\int {\d}
\Delta^+_{0} {\d} \Delta^-_{0} = \int_0^\infty |\Delta_{0}| {\d}
|\Delta_{0}| \int_0^{2 \pi} {\d} \phi_{0}$ and performs the phase
integral fully. Remarkably, ``liberating'' the phase degree of freedom
in this way already suffices to smooth out the phase transition
\cite{Rossignoli-95,Rossignoli-99a}, even if the $\int \d |\Delta_0 |$
integral is again replaced by its saddle-point value, provided that
the latter is found by now maximizing $|\Delta_0| {\cal Z}_p
[|\Delta_0|]$ (\ie\ including the factor $|\Delta_0|$ from the
integration measure).  This yields a modified gap equation with a
nontrivial solution for arbitrarily large $T$, \ie\ no abrupt
transition.

(iii) For finite systems, fluctuations about the saddle become large
in critical regions. To obtain an improved description of the latter
(\Fig{fig:rossignoli}, dash-dotted lines), the \emph{static path
  approximation} (SPA) 
\cite{Muehlschlegel-72,Alhassid-84,Alhassid-92,Arve-88,Lauritzen-88,%
Rossignoli-92,Rossignoli-93,Rossignoli-94} incorporates all
\emph{static} fluctuations exactly, via a (numerical) evaluation of
the full integral $\int_0^\infty |\Delta_{0}| \d |\Delta_{0}|$ $
\int_0^{2 \pi} \d \phi_{0}$ over all ``static paths''.

(iv) In the so-called \emph{correlated static path approximation}
(CSPA) (also called SPA+RPA), small-amplitude quantum fluctuations
around each static path are included too, by performing the remaining
$\int \!  \d \Delta^\pm_{n \neq 0}$ integrals in
the gaussian approximation \cite{Puddu-90,Lauritzen-90,%
Puddu-91,Puddu-92,Attias97,Rossignoli-97a,%
  Rossignoli-97b,Rossignoli-98,Rossignoli-99a,Rossignoli-99b,%
Rossignoli-00}. The CSPA yields
qualitatively similar but quantitatively more reliable results 
(\Fig{fig:rossignoli}, dashed lines) than
the SPA, but breaks down below a temperature $T_{\rm CSPA}$, below
which the fluctuations of the $\Delta^\pm_{n \neq 0}$ modes become
large at unstable values of $|\Delta_{0}|$, causing the gaussian
approximation to fail.

(v) Finally, in the so-called \emph{canonical CSPA} one projects the
partition function not only to fixed number parity (as done throughout
above) but also to fixed particle number, by performing an integration
over the chemical potential (\emph{before} performing any of the
$\Delta_n^\pm$ integrals)
\cite{Essebag-93,Rossignoli-92,Rossignoli-93,Rossignoli-94}.  However,
this too is usually done only in the gaussian approximation (and would
produce negligible corrections to the CPSA results for the quantities
shown in \Fig{fig:rossignoli}).

Comparisons with exact diagonalization results \cite{Rossignoli-96a}
(\Fig{fig:rossignoli}, solid lines) show that in its regime of
validity $(T > T_{\rm CSPA}$), the CSPA produces results that are
qualitatively completely similar and also quantitatively very close to
the exact ones, whereas the quantitative agreement is significantly
worse if only the SPA is used.  Since the CSPA is conceptually simple,
well-documented \cite{Rossignoli-99a} and straightforward to
implement, it seems to be the method of choice for not too low
temperatures.  A possible alternative is a quantum Monte Carlo
evaluation of the path integral (\ref{eq:hubbard-stratonovich})
\cite{Lang93,koonin}, but the numerics is much more demanding than for
the CPSA, while the convergence at low $T$ is in general rather poor,
due to the familiar sign problem of Monte Carlo methods.

The development of canonical finite-$T$ methods that remain
quantitatively reliable for $d \gtrsim \tilde \Delta$ and arbitrarily
small $T$ is one of the open challenges in this field. It would be
very interesting if progress in this direction could be made by
exploiting the integrability \cite{cambiaggio,sierra99b} of the model,
using Bethe Ansatz techniques. For the FD regime, another possibility
would be to develop a finite-$T$ generalization of the self-consistent
RPA approach of Dukelsky and Schuck \cite{dukelsky99c}.

\subsubsection{Re-entrant spin susceptibility}
\label{sec:sc-susceptibility}

For grains so small that $d \gg \tilde \Delta$, the spectroscopic
\emph{transport} measurements of BRT are not able, in principle, to
reliably detect the effect of pairing correlations, since in this
regime these cause only small changes to the eigenspectrum of a normal
metallic grain, whose spectrum is, however, irregular to begin with.
In contrast, \emph{thermodynamic} quantities do have the potential to
measurably reveal the existence of pairing correlations for $d \gg
\tilde \Delta$. Since very recently parity effects for the \emph{spin
  susceptibility} have been observed experimentally for an ensemble of
small, normal metallic grains \cite{Volotikin-96}, it is an
interesting and experimentally relevant question to investigate how
pairing correlations affect its behavior in superconducting grains.

\begin{figure}
\centerline{\epsfig{figure=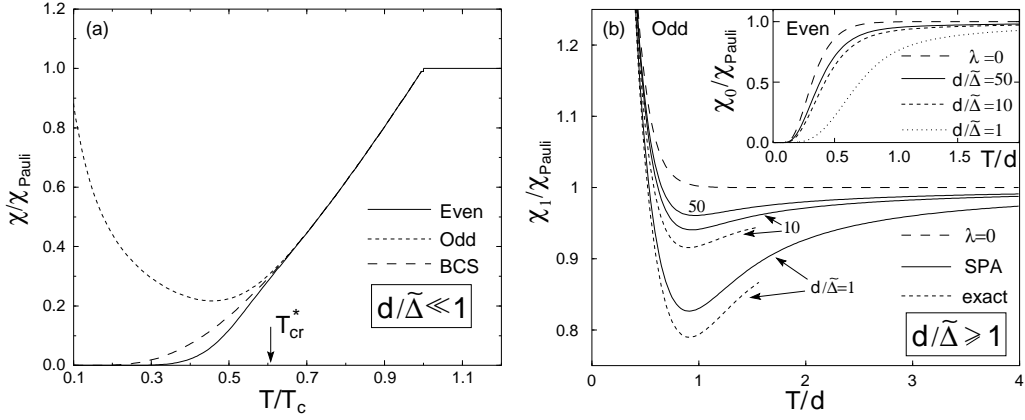,%
width=0.98\linewidth}}
\caption[Spin susceptibility of an even and odd superconducting grain]%
{Spin susceptibility $\chi_0$ ($\chi_1$) of an even (odd)
  superconducting grain as function of $T$, plotted in units of its
  bulk high-$T$ value $\chi_{\rm Pauli} = 2 \muB^2/d$.  (a) Bulk limit
  ($d/\tilde \Delta \ll 1$): the even (solid) and odd (short-dashed)
  curves were calculated using parity-projected mean-field theory, the
  long-dashed curve using standard  (unprojected) BCS theory.  (b)
  Fluctuation-dominated regime ($d/\tilde \Delta \gtrsim 1$) for
  $\chi_1$ (main figure) and $\chi_0$ (inset): All curves were
  calculated using the static path approximation, except the
  short-dashed curves in the main figure, calculated using
  Richardson's exact solution, and the long-dashed curves for the
  non-interacting case ($\lambda = 0$). 
\label{fig:susceptibilites}}
\end{figure}

This question was worked out in detail by Di Lorenzo \etalia\ 
\cite{dilorenzo99}, whose results are summarized in
\Fig{fig:susceptibilites}.  The spin susceptibility for an isolated
grain is defined as
\begin{equation}
  \label{eq:susceptibility}
  \chi_p (T) = - \left. {\partial^2 {\cal F}_p (T,H) \over \partial H^2}
  \right|_{H=0} \; , 
\end{equation}
where ${\cal F}_p = - \kB T \ln Z_p^\can$ is the free energy of a
grain with parity $p$ and $Z_p^\can$ is the canonical partition
function.

In the bulk limit [\Fig{fig:susceptibilites}(a)], it is well-known
\cite{Yosida-58} that the spin susceptibility decreases below its
Pauli value $\chi_\Pauli = 2 \muB^2/d$ once $T$ drops below the
superconducting transition temperature $T_{\rm c}$, since the
electrons tend to bind into Cooper pairs, which are spin singlets and
do not contribute to the spin susceptibility.  Interestingly, however,
the spin susceptibility becomes parity-dependent as $T$ is lowered
below the crossover temperature $T^\ast_{\rm cr}$ mentioned in the
opening paragraphs of \Sec{sec:finite-T}: In the even case, $\chi_0$
exponentially drops to zero for sufficiently small temperatures, $T
\ll \max(\tilde \Delta,d)$, for reasons that are intuitively obvious
in the two limits $\tilde \Delta \gg d$ (all electrons bound into
Cooper pairs) and $\tilde \Delta \ll d$ (no Cooper pairs, but all
levels doubly occupied). In contrast, in the odd case $\chi_1$ shows a
\emph{re-entrant} behavior, in that it increases as $\muB^2/T$ for low
temperatures, due to a Curie-like contribution from the unpaired odd
electron. As a result, $\chi_1 (T)$ has a \emph{minimum} somewhat
below $T^\ast_{\rm cr}$, which can be viewed as a ``smoking gun'' for
pairing correlations, since it is absent for odd normal grains.  For
the latter, $\chi_1 (T)$ [long-dashed $\lambda=0$ curve in
\Fig{fig:susceptibilites}(b)] also has the Curie-like increase at
very low $T$, but lacks the initial pairing-induced decrease as $T$ is
reduced below $T_{\rm c}$.

Remarkably, Di Lorenzo \etalia\ found that \emph{this re-entrance of
  $\chi_1$ survives also for $d \gtrsim \tilde \Delta$}
[\Fig{fig:susceptibilites}(b)]: although pairing correlations survive
here only as fluctuations, these are evidently sufficiently strong to
still significantly reduce $\chi_1 (T)$ relative to $\chi_\Pauli$ [by
several percent even for $d /\tilde \Delta \simeq 50$ (!)], before the
Curie-like increase sets in at low $T$.  Di Lorenzo \etalia\ 
established this result by considering the limits $T \gg d$ and $T \ll
2 d$ analytically, using a static path approximation to capture the
crossover numerically, and checking the results for $T \lesssim d$
using Richardson's exact solution (they considered all eigenstates
with excitation energy up to a cutoff $\Lambda \sim 40 d$, for grains
with $N \le 100$ electrons).  This check shows that the static path
approximation somewhat underestimates the amount of pairing
correlations (its minima for $\chi_1 (T)$ are too shallow), but in
general is in good qualitative agreement with the exact results,
confirming that it is a useful and qualitatively reliable tool for
describing the crossover regime.

\newpage \section{Nonequilibrium effects}
\label{sec:nonequilibrium}

So far, we have always assumed that the gate voltage has been tuned
such that the system is close to a Coulomb-blockade degeneracy point,
where the Coulomb-blockade barrier $\mbox{min}[ \delta E_{\rm
  pot}^\pm]$ of \Eq{eq:U+-} is small (of order $d$) and nonequilibrium
effects can be neglected. However, since this requires fine-tuning
$\Vg$, it is a rather non-generic situation.  In this section, we
discuss the opposite case in which this Coulomb barrier is so large that
transport occurs only in the nonequilibrium regime, for which several
interesting new phenomena have been observed
\cite{agam97a,rbt97,agam97b,agam98,davidovich99}.  Among these, we
focus 
\begin{enumerate} 
\item[(i)]  in subsection 
(\ref{subsec:neq-experiment})
on   the observation of \emph{clusters} of closely-spaced
levels in normal grains; 
\item[(ii)] in subsection (\ref{sec:sub-gap-structures}) on the occurrence of
  \emph{prethreshold structures} in the excitation spectra of
even superconducting grains; and 
\item[(iii)] in subsection (\ref{sec:estimating-relaxation}) on a direct
  observation of the crossover from a discrete to a continuous excitation
  spectrum at energies of order $\ETh$.
\end{enumerate}
  All of these can be understood
only by going beyond the simple theoretical framework presented in
\Sec{sec:theoryofultrasmallSET}: (i) and (iii) require consideration
of corrections to the ``orthodox'' model for treating the Coulomb
interaction, which are of order $d/\gdc$ and hence become important
for sufficiently small grains; and (ii) requires consideration of
cotunneling processes, whereas we had hitherto restricted our
attention to sequential tunneling.

The general significance of the results presented below is that they
provide a very direct confirmation and illustration of the general
theoretical picture \cite{Sivan94,AGKL97} (summarized in
\Sec{sec:estimating-relaxation}) for the nature of excitations in
disordered interacting systems that has emerged during the last
decade.  The development of this picture was initiated by studies of
semiconductor quantum dots \cite{Sivan94}, but is here beautifully
confirmed for metallic grains too.

\subsection{Clusters of resonances} 
\label{subsec:neq-experiment}

\subsubsection{Experimental results}
\label{sec:clusters-experiments}

\begin{figure}
\centerline{\epsfig{figure=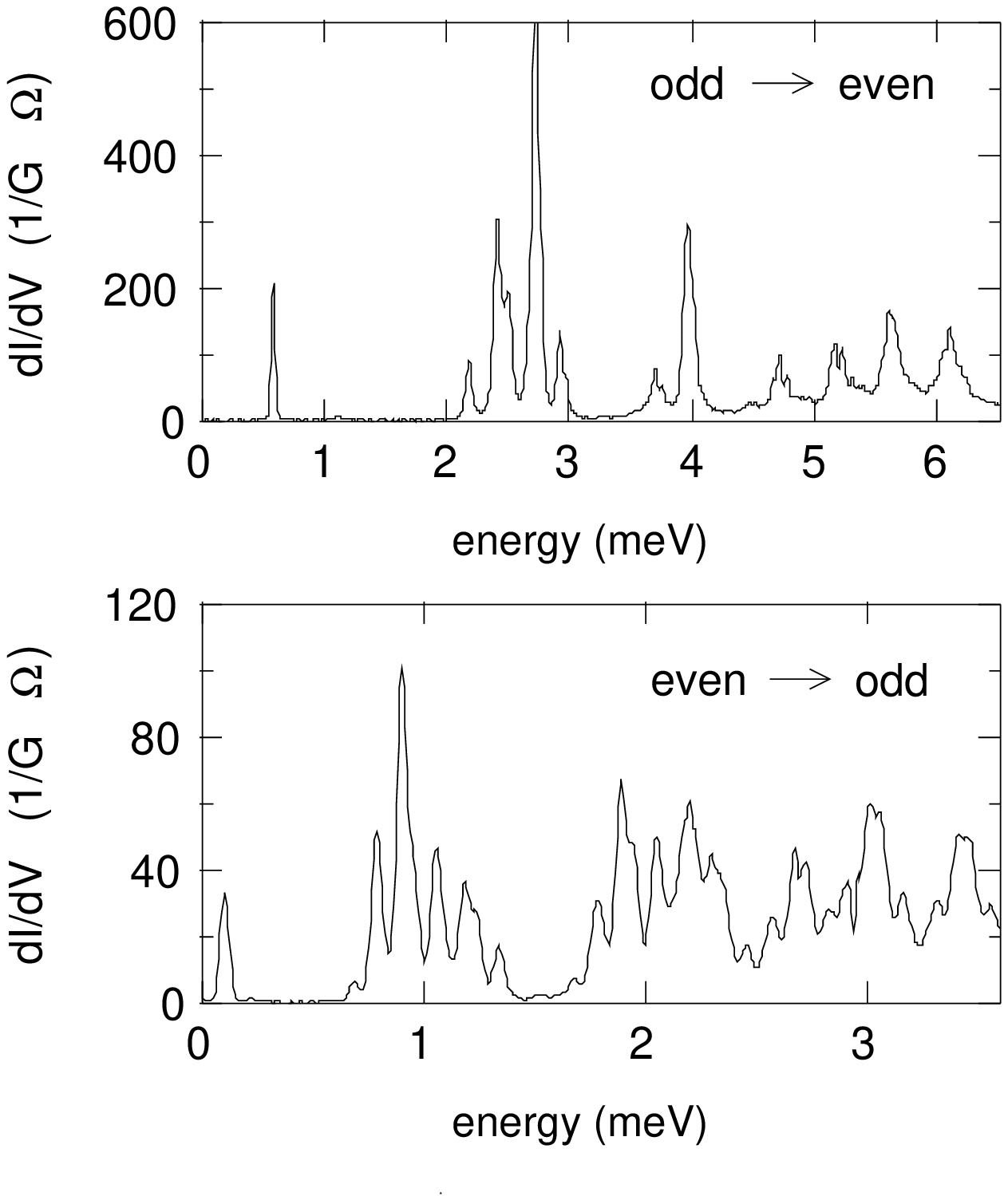,%
width=0.7\linewidth}}
\centerline{\epsfig{figure=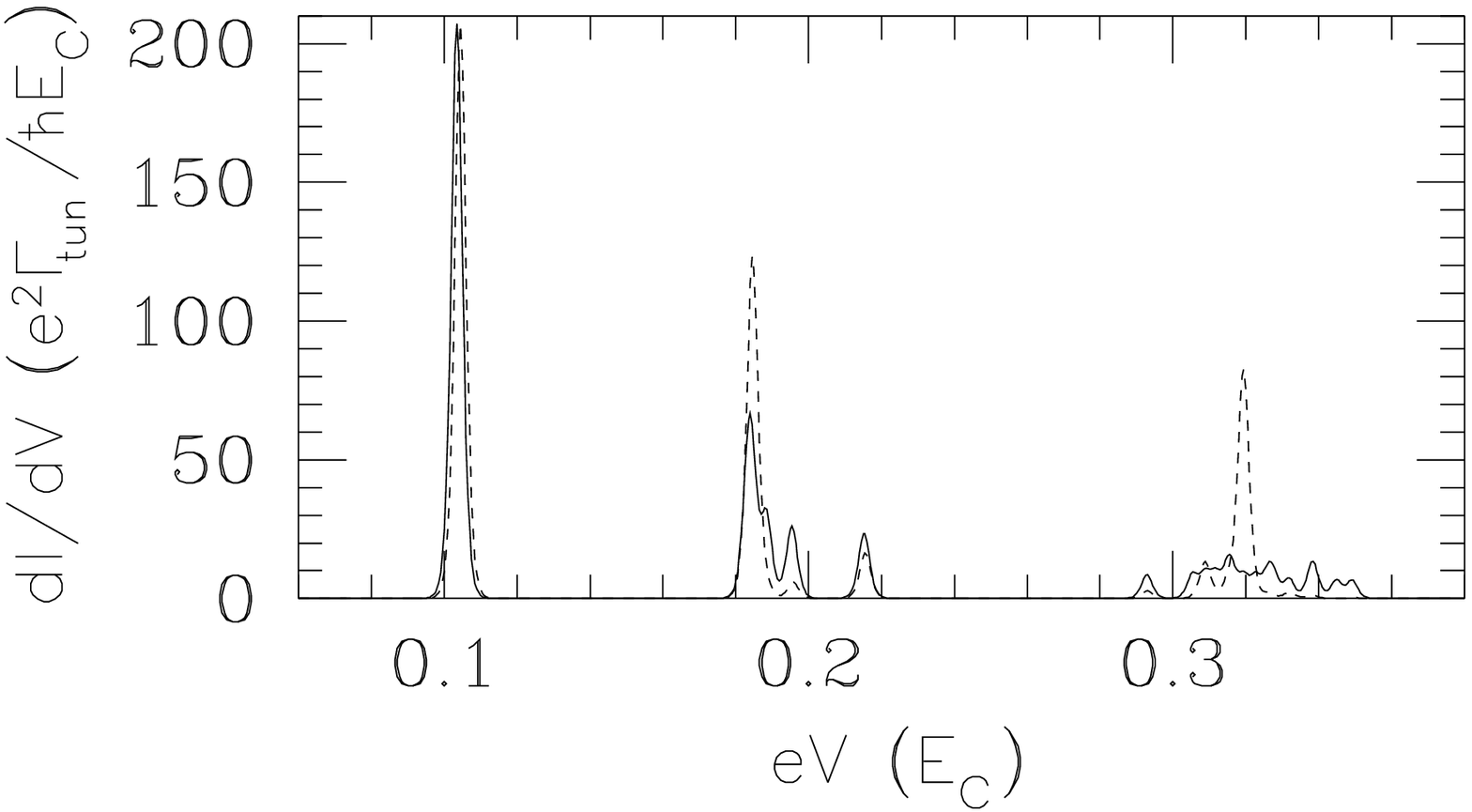,%
width=0.7\linewidth}}
\caption[Clusters of resonances in nonequilibrium 
transport]{(a,b) The excitation spectra (at $T=$ 30 mK, $H=0$) of two
  ultrasmall Al grains with volumes $\approx$ 40 nm$^3$ (a) and
  $\approx$ 100 nm$^3$ (b) \cite{agam97a}. The first resonance is
  isolated while subsequent resonances are clustered in groups. The
  distance between nearby groups of resonances is approximately the
  single-particle mean level spacing $d$.  (c) Differential
  conductance for a model system of 7 equally spaced levels, occupied
  alternately by 4 or 5 electrons in a current-carrying steady state,
  calculated \cite{agam97a} from \Eq{eq:currentexpectation} by solving
  the master equation (\ref{generalmasterequation}) numerically.  The
  tunneling rate $\Gamma^r_\alpha$ into each level was chosen to be
  uniform, $\Gamma^r_\alpha \equiv \Gamma_\tun$, and a temperature $T
  = d/100$ and $\delta U = d/5$ were used, where $\delta U$ is the
  variance of the fluctuations [Eq.~(\ref {eq:interaction-fluctuation})] in
  the interaction energy, see text.  Dashed line: inelastic relaxation
  rate $\Gamma^\inel $ of each level was chosen to be
  larger than the tunneling rate into it
  ($\Gamma^\inel = 5 \Gamma_\tun$); solid line: no inelastic processes
  ($\Gamma^\inel = 0$).
  \label{fig:clusters}}
\end{figure}
\Figs{fig:clusters}(a,b) display the excitation spectrum, ${\d}I/{\d}V$
versus energy, of two different {\em normal} metallic grains, of radii
roughly $2.7$~nm and $3.6$~nm (if assumed hemispherical).  The
spectra display four clear features:
\begin{enumerate}
\item[(1)]
The low resonances of the differential conductance
 are grouped in \emph{clusters}. 
\item[(2)] The distance between 
nearby clusters is of order the mean level spacing $d$ of the 
noninteracting electrons in the dot. 
\item[(3)] The first cluster contains only a single resonance.  
\item[(4)] Higher clusters consist of several 
resonances \emph{spaced much more closely than $d$}.
\end{enumerate}
Agam, Wingreen, Altshuler, Ralph and Tinkham 
\cite{agam97a} have shown that these features are
manifestations of the interplay between nonequilibrium effects and
electron-electron interactions, where the latter must be treated more
carefully than in the usual orthodox model for charging effects.  The
general idea is simple: if the threshold voltage for the onset of
current through the grain is large, the transfer of one electron
through the grain by sequential tunneling can leave behind several
possible particle-hole excitations on the grain. If sufficiently
long-lived, each of these excitations will modify the threshold energy
for the next electron that tunnels into the grain in a slightly
different way, and thus cause what would have been a single
conductance resonance to split up into a cluster of subresonances.


\subsubsection{Reaching nonequilibrium via sequential tunneling}
\label{sec:nonequilibrium-sequential}

To explain the interpretation of Agam \etalia\ in more detail, we
shall begin our analysis within the framework of the orthodox model
defined by \Eqs{eq:generalhamiltonian-leads} to
(\ref{eq:generalhamiltonian-tunnel-l}) of
\Sec{sec:theoryofultrasmallSET}, where the notation used below is
introduced and explained.  The present grains are so small that the
influence of superconducting pairing may be neglected, since the
single-particle mean level spacing, $d \simeq 1 $~meV, is much larger
than the BCS superconducting gap for Al, which is 0.18~meV. We thus
describe the grain by the normal-state single-particle Hamiltonian of
\Eq{eq:H-dot-normal} and (\ref{eq:H-dot-normal-eigenstates}),
\begin{eqnarray}
  \label{eq:H-dot-normal-nonequib}
 && \hat H^\normal_{\ssD, \orth}  =  
  \Epot (\hNex) + \sum_p
\varepsilon_{p} c^\dagger_{p} c_{p}
= \sum_\alpha 
( e V_\ssD  N_{{\rm ex},\alpha} + \E_\alpha  ) 
|\alpha \rangle \langle \alpha | \, , 
\\
\label{eq:H-dot-normal-eigenstates-nonequib}
&& |\alpha \rangle  = | \{ n^{p} \} \rangle
\, , \qquad \quad {\cal E}_\alpha =  \Ec  N^2_{{\rm ex},\alpha} +
\sum_{p} \varepsilon_{p} \, n^{(\alpha)}_{p} 
 \; ,
\end{eqnarray} 
where, for notational simplicity, we have here used the roman label
$p$ for the combined labels $l \sigma$D of \Eq{eq:H-dot-normal}.
$\hat H^{\rm normal}_{\ssD,\orth}$ incorporates the effect of
electron-electron interactions only through the electrostatic
contribution $\Epot (\hNex)$ [defined in \Eq{eq:Epot}], which, as we
shall see below, is too simplistic to explain the clustering
properties (1-4).  Corrections to the orthodox model will be
introduced [\Sec{sec:beyound-orthodox-model} below] once its
insufficiency has become apparent.

\begin{figure}
\centerline{\epsfig{figure=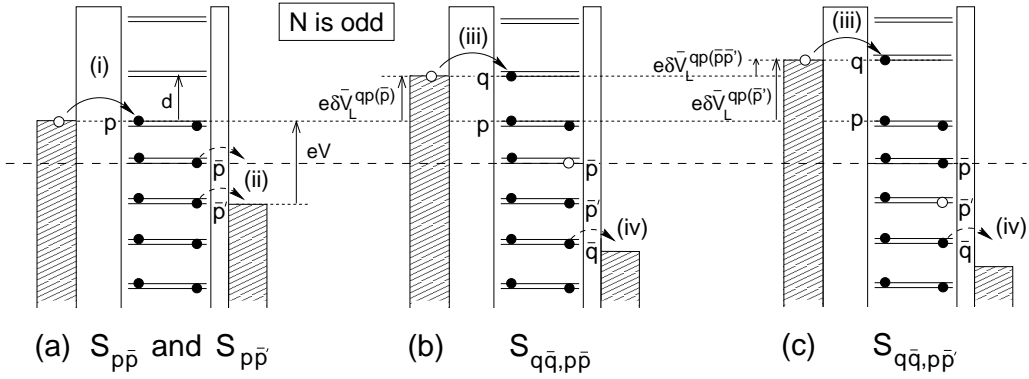,%
width=0.98\linewidth}}
\caption{
  Schematic depiction (using the conventions of
  \Fig{fig:energydiagram}) of the tunneling sequences, starting from
  the ground state $| \G\rangle_N$, that lead to clusters of resonances
  in nonequilibrium transport.  Filled circles depict the electron
  configuration of the intermediate $N+1$-particle states (with $N$
  odd), solid arrows the bottleneck tunneling transitions into them;
levels  drawn close together represent time-reversed pairs
that are degenerate in the absence of a magnetic field.
  (a) $S_{p \bar p}$ and $S_{p \bar p'}$ [\Eq{eq:sequential-pp}] are
  two possible sequences via level $p$ that leave the grain in an
  excited final state, $|p \bar p\rangle_N$ or $|p \bar p' \rangle_N$.
  (b) $S_{q \bar q, p \bar p}$ or (c) $S_{q \bar q, p \bar p'}$
  [\Eq{eq:sequential-qqpp}] are two possibilities, differing only in
  the hole position ($\bar p$ vs.\ $\bar p'$), for the next tunneling
  sequence via level $q$; their thresholds differ by an amount $e
  \delta V_\ssL^{pq(\bar p \bar p')}$
  [\Eq{eq:voltage-differenceqqpp-cluster}], since the interaction
  energies of two electrons in levels $q$ and $\bar p'$ or $q$ and
  $\bar p$ differ [\Eq{eq:interaction-fluctuation}].  The figure is
  drawn such that $p=0$ and $q=1$ are the two lowest-lying empty
  levels of $|\G \rangle_N$.  }
\label{fig:nonequilibrium}
\end{figure}

For definiteness, suppose that the overall ground state of the grain
has $N$ electrons and that the measured excitation spectra arise via
tunneling transitions, with the left barrier acting as bottleneck,
between $N$ and $(N+1)$ electron states [see \Fig{fig:nonequilibrium}].
Denoting their respective ground states by $|\G\rangle_N$ and
$|\G\rangle_{N+1} \equiv c_0^\dag |\G\rangle_N$, where $c_0^\dagger$ is
the creation operator for the lowest unfilled level of $|\G\rangle_N$,
the corresponding ground state energy difference is
\begin{eqnarray}
  \label{eq:EN+1-ENgroundstates}
  \E_\G^{N+1} - \E_\G^{N} + e V_\ssD  = 
\delta E_{\rm pot}^+ + \varepsilon_0 \;   ,
\end{eqnarray}
where $\delta E_{\rm pot}^+ = \Epot (\Nex + 1 ) - \Epot (\Nex)  $
is the electrostatic
energy cost for entering the grain. All $N$-electron excited states
are particle-hole excitations w.r.t.\ $|\G\rangle_N$.  We shall denote
them by $ |p \bar p \rangle_N \equiv c_p^\dag c_{\bar p}^\ds | \G
\rangle_N$ and $|q \bar q, p \bar p \rangle_N \equiv c_q^\dag c_{\bar
  q}^\ds c_p^\dag c_{\bar p}^\ds | \G\rangle_N$, etc.,
and within the orthodox model of \Eqs{eq:H-dot-normal-nonequib}
and (\ref{eq:H-dot-normal-eigenstates-nonequib}), 
their excitation energies are 
\begin{eqnarray}
\label{eq:particle-holes-1}
(\E^N_{p \bar p} - \E_\G^N)_\orth = \varepsilon_p -
  \varepsilon_{\bar p} \; , \qquad 
(\E^N_{q \bar q, p \bar p} - \E^N_{\G})_\orth = \varepsilon_q - 
  \varepsilon_{\bar q} + \varepsilon_p -
  \varepsilon_{\bar p} \; ,
\end{eqnarray}
etc.\ Analogous definitions hold for $N+1$. Now, sequential tunneling
through the grain occurs via sequences of transitions of the type
[\Fig{fig:nonequilibrium}(a)] 
\begin{eqnarray}
  \label{eq:sequential-pp}
&& S_{p \bar p}: 
\qquad | \G \rangle_N \quad 
\stackrel{{\rm (i)}}{\longrightarrow}  \quad
|p 0 \rangle_{N+1} \quad \stackrel{{\rm (ii)}}{\longrightarrow} \quad
| p \bar p \rangle_N \, .
\end{eqnarray}
In step (i), an electron tunnels via the left (bottleneck) barrier
into an empty level $p$ (with $\varepsilon_p \ge \varepsilon_0$),
leaving the grain in an intermediate state $|p 0 \rangle_{N+1}$, which
we shall call a ``quasiparticle state'' (following \cite{AGKL97}); and in
step (ii), one tunnels out via the right barrier from a filled level
$\bar p$, leaving the grain in the excited (if $\bar p \neq p$) final
particle-hole state $|p \bar p \rangle_N$.  According to
\Eq{eq:Vrthresholds}, steps (i) and (ii) become possible only once the
following inequalities hold, respectively
(with $\bar V_r \equiv V_r - V_\ssD$):
\begin{eqnarray}
&& \label{eq:sequential-pp-inequalities}
\mbox{(i)} \quad e \bar  V_\ssL 
\ge \E_{p0}^{N+1} - \E_\G^N \; ,  \qquad 
\mbox{(ii)} \quad - e \bar  V_\ssR 
\ge \E_{p \bar p}^{N} - \E_{p 0}^{N+1} \;  .
\end{eqnarray}
Adding
Eqs.~(\ref{eq:sequential-pp-inequalities}.i)+(\ref{eq:EN+1-ENgroundstates})
and Eqs.~(\ref{eq:sequential-pp-inequalities}.i)+%
(\ref{eq:sequential-pp-inequalities}.ii) gives, respectively,
\begin{eqnarray}
  \label{eq:condition-for-nonequilibrium}
  {\rm (i)}^\prime \quad eV/2 
  = e V_\ssL \ge \delta E_{\rm pot}^+ + \varepsilon_p
  \; , \qquad
  {\rm (ii)}^\prime \quad e V  \ge \E_{p \bar p}^{N} - \E_\G^N \; ,
\end{eqnarray}
which have obvious physical interpretations: Condition ${\rm
  (i)}^\prime$ makes explicit that transport is possible only if the
bias voltage is large enough that an electron leaving the left lead
can overcome the Coulomb blockade barrier $ \delta E_{\rm pot}^+$; the
lowest-threshold 
conductance resonance  evidently occurs for $p=0$.  Condition
${\rm (ii)}^\prime$ states that an excited final state $|p \bar p
\rangle_N$ can be produced whenever the energy supplied
by the bias voltage exceeds the
corresponding excitation energy $\E_{p \bar p}^N - \E_\G^N$. Now, ${\rm
  (i)}^\prime$ implies that if the Coulomb blockade barrier is large
$(\delta E_{\rm pot}^+ \gg d)$, \ie\ for transport far away from a
degeneracy point, the threshold bias voltage for the current to begin
to flow will be large $(e V \gg d)$; but this would automatically
guarantee that ${\rm (ii)}^\prime$ is fulfilled, at least for
low-lying final-state excitations with an energy cost of only a few
$d$.  We can thus draw the following important conclusion: \emph{Since
  sequential tunneling transport away from a degeneracy point requires
  a large threshold bias voltage, it will always generate excited
  final states once this threshold is reached.}

Whether an excited final state $|p \bar p \rangle_N$ affects transport
or not depends on the ratio between $\Gamma^\inel_{p \bar p}$, its
inelastic decay rate, and $\Gamma^r_\tun$, the average tunneling rate
across the bottleneck barrier $r$ into an individual discrete level
near $\eF$. In all previous sections we had assumed that
$\Gamma^\inel_{p \bar p} \gg \Gamma^r_\tun$, \ie\ that $|p \bar p
\rangle_N$ will decay to $|\G\rangle_N$ long before the next electron
tunnels in across the left barrier. In this case, the distance between
two so-called ``quasiparticle conductance resonances'', associated
with the onset of sequential tunneling through two different
intermediate quasiparticle states $|q 0\rangle_{N+1}$ and $|p
0\rangle_{N+1}$, is determined by the bottleneck condition
(\ref{eq:sequential-pp-inequalities}.i) [or equivalently from
\Eq{eq:fixed-N-excitation-spectrum1}] to be
\begin{eqnarray}
  \label{eq:voltagepq}
e \delta \bar
V_\ssL^{qp} \equiv \E_{q0}^{N+1} - \E_{p 0}^{N+1} = \varepsilon_{q} -
\varepsilon_p \; .  
\end{eqnarray}
  However, estimates of $\Gamma^\inel_{p \bar p}$
(discussed in \Sec{sec:estimating-relaxation}) show that it is quite
possible for an ultrasmall grain to have have $\Gamma^\inel_{p \bar p}
\lesssim \Gamma^r_\tun$.  In this case, the grain will still be in
nonequilibrium when the next electron tunnels, so that new tunneling
sequences of the type [\Fig{fig:nonequilibrium}(b)]
\begin{eqnarray}
  \label{eq:sequential-qqpp}
&& S_{q \bar q, p \bar p}: 
\qquad | p \bar p \rangle_N \quad 
\stackrel{{\rm (iii)}}{\longrightarrow}  \quad
|q 0, p \bar p \rangle_{N+1} 
\quad \stackrel{{\rm (iv)}}{\longrightarrow} \quad
|q \bar q, p \bar p \rangle_N \, , \rule[-4mm]{0mm}{0mm} \\
&& \label{eq:sequential-qqpp-inequalities}
\mbox{(iii)} \quad  e \bar V_\ssL 
\ge \E_{q0, p \bar p}^{N+1} - \E_{p \bar p}^N  \; ,  \qquad 
\mbox{(iv)} \quad - e \bar V_\ssR 
\ge \E_{q \bar q, p\bar p}^{N} - \E_{q0, p \bar p}^{N+1}  \; , 
\end{eqnarray}
become possible once the corresponding conditions (iii) and (iv) are
met. These differ from $S_{q \bar q}$ by the presence
of additional particle-hole excitations $p \bar p$.
The change in $e \bar V_\ssL$ needed to reach threshold
(\ref{eq:sequential-qqpp-inequalities},iii) from threshold
(\ref{eq:sequential-pp-inequalities},i) is
\begin{eqnarray}
  \label{eq:voltage-differenceqqpp}
  e \delta \bar V^{q p (\bar p)}_\ssL \equiv
  (\E_{q 0, p \bar p}^{N+1} - \E_{p \bar p}^N) - 
  (\E_{p 0}^{N+1} - \E_{\G}^N)
  \; 
\end{eqnarray}
[see \Fig{fig:nonequilibrium}(b)], and each time this is positive, the
conductance should show another resonance (if $ e \delta \bar V^{q p
  (\bar p)}_\ssL <0$, the sequence $S_{q \bar q, p \bar p}$ simply
increases the peak height associated with the sequences $S_{p \bar p}$
through level $p$, without causing another peak, \ie\ these \emph{do}
contribute to the current, although they are spectroscopically
``hidden'').  But within the orthodox model, $ e \delta \bar V^{q p
  (\bar p)}_\ssL = \varepsilon_q - \varepsilon_p $ [by
\Eqs{eq:particle-holes-1}], which is the \emph{same} as $e \delta \bar
V_\ssL^{qp}$ of \Eq{eq:voltagepq}.  Thus, the orthodox model predicts
that even in the presence of nonequilibrium final states, the spacing
of conductance peaks will be of order $d$. It is therefore unable to
explain the observed occurrence in \Figs{fig:clusters}(a,b) of
\emph{clusters} with mean inter-cluster spacing $d$ and an
intra-cluster spacing much less than $ d$.

\subsubsection{Beyond the orthodox model: interaction fluctuations}
\label{sec:beyound-orthodox-model}

The  reason for this failure, identified by
Agam \etalia\ \cite{agam97a,agam98}, is that the  orthodox model
treats the electron-electron interaction  too simplistically.
Its general form is
\begin{eqnarray}
  \label{eq:H_int-el-el}
 \hat  H_\elel &=& 
  \frac{1}{2} \sum_{ ijkl} \sum_{ \sigma \sigma'} U_{ijkl}
c^\dagger_{i\sigma} c^\dagger_{j \sigma'} c_{k \sigma'} c_{l \sigma},
\\
U_{ijkl} &=& \int {\d} {\bf r}_1 {\d} {\bf r}_2  \,
U({\bf r}_1, {\bf r}_2) \psi_i^*({\bf r}_1)
\psi_j^*({\bf r}_2)\psi_l({\bf r}_1)\psi_k({\bf r}_2) \; , 
\end{eqnarray}
where $U({\bf r}_1, {\bf r}_2)$ is the (screened) interaction potential.
The orthodox model keeps only ``diagonal'' terms in which
two pairs of indices are identical ($U_{ijji}$, $U_{iijj}$,
$U_{ijij}$), namely 
\begin{eqnarray}
\nonumber
  \label{eq:H-el-el-orthodox}
  \hat H^\orth_\elel &=&
 \Ec \hat N_{\rm ex}^2 - \lambda d \sum_{ij} 
c^\dagger_{i +} c^\dagger_{i -} c_{j - } c_{j +}
+  \lambda_{\rm s} d \sum_{ij,\sigma \sigma'} 
c^\dagger_{i\sigma} c^\dagger_{j \sigma'} c_{i \sigma'} c_{j \sigma}
+ {\cal O}(d/\gdc) \;,
\end{eqnarray}
and neglects both ``\emph{off-diagonal} terms'' $U_{ijkl}$ (no equal
indices) and $ij$-dependent \emph{fluctuations} in diagonal terms.
($\lambda$ and $\lambda_s$ are dimensionless, volume-in\-de\-pen\-dent
interaction constants in the pair and spin channels; we took $\lambda
= \lambda_s = 0$ in this section, since we are not interested in
superconducting or magnetic correlations here).  Now, by studying the
correlations of eigenfunctions in chaotic systems (and an irregularly
shaped ultrasmall grain may be viewed as chaotic), the neglected terms
can be
shown \cite{agam98,Agam95,Blanter96,BMM-97,%
Blanter-Mirlin-97,Aleiner-Glazman-97,Blanter-Mirlin-96%
} 
to be small as $d/\gdc$, where $\gdc = \ETh /
d$ is the dimensionless conductance\footnote{\label{f:perron-frobenius}
Strictly speaking, in
  this context one should use $\gdc = \hbar \gamma_1/d$, where
  $\gamma_1$ is the first non-vanishing Perron-Frobenius eigenvalue
  \cite{Agam95,agam98}, which depends on the geometry and the amount
  of disorder present.  Eg.\, for a pancake-shaped grain with thickness $z$,
  radius $r$ and diffusive dynamics, Agam \cite{agam98} gives $\gdc
  \propto (\hbar v_\F /2 r d) (z/r)^2$.}  and $\ETh$ the Thouless
energy [cf.\ \Eqs{eq:g-dimensionlessconductance} and 
\ref{eq:define-Thouless}].  Intuitively, one may say that
the smaller $1/\gdc$, the more uniform are the wavefunctions and the
smaller the fluctuations in the interaction energy.  Indeed, the
neglect of these fluctuations is parametrically justified for bulk
samples ($1/\gdc \simeq 0$) and in mesoscopic samples ($1/\gdc \ll
1$). For ultrasmall grains, however, $1/\gdc$ can become of order
unity, so that interaction fluctuations are expected to become
measurable.

Agam \etalia\ \cite{agam97a,agam98} attributed the above-mentioned
clustering properties (1-4) to a combination of
precisely such interaction fluctuations and the presence of
nonequilibrium excitations on the grain: they argued that the energy
of the added electron in level $q$ of the intermediate state $|q0, p
\bar p \rangle_{N+1}$ depends, in general, on precisely which
particle-hole excitation $p \bar p$ is present.  In particular, two
intermediate states that differ only by the interchange of a hole
between levels $\bar p$ and $\bar p'$, say $|q0, p \bar p
\rangle_{N+1}$ and $|q0, p \bar p' \rangle_{N+1}$, will have
thresholds that differ, according to
Eq.~(\ref{eq:sequential-qqpp-inequalities},iii), by
[see \Fig{fig:nonequilibrium}(b,c)]
\begin{eqnarray}
  \label{eq:voltage-differenceqqpp-cluster}
  e \delta \bar V^{q p (\bar p \bar p')}_\ssL & \equiv &
  (\E_{q 0, p \bar p}^{N+1} - \E_{p \bar p}^N) - 
  (\E_{q 0, p \bar p'}^{N+1} - \E_{p \bar p'}^N) \; .
\end{eqnarray}
Agam \etalia\ argued that the magnitude of this expression may be
estimated using the Hartree term of the interaction
(\ref{eq:H_int-el-el}), which gives $U_{q \bar p' \bar p' q}
- U_{q \bar p \bar p q}$, \ie\
\begin{eqnarray}
\label{eq:interaction-fluctuation}
 e \delta \bar V^{q p (\bar p \bar p')}_\ssL 
& \simeq & \int {\d} {\bf r}_1 \int {\d} {\bf r}_2 \, 
 | \psi_q ({\bf r}_1 ) |^2 U({\bf r}_1 , {\bf r}_2) \left(
  | \psi_{\bar p'} ({\bf r}_2) |^2 -  
 | \psi_{\bar p} ({\bf r}_2 ) |^2  \right) \; .
\end{eqnarray}
Let $\delta U$ denote the variance of
(\ref{eq:interaction-fluctuation}), evaluated either by ensemble or
energy averaging. If diffusive dynamics is assumed, $\delta U$ can be
shown \cite{Blanter96} to be of order $c d/
\gdc$,${}^{\ref{f:perron-frobenius},}$\footnote{ This estimate does
  not take into account a change in the electrostatic potential due to
  the extra electron in level $q$.  Blanter, Mirlin and Muzykanskii
  \cite{BMM-97} showed that the latter effect will lead to even
  stronger fluctuations, namely $\delta U \sim d/\sqrt{\gdc}$.}  where
$c$ is a geometry-dependent constant of order unity. Thus, \emph{each
  quasiparticle resonance associated with the set of sequential
  tunneling sequences $S_{q0,p \bar p}$ through a level $q$ should
  split up into a cluster of subresonances, spaced by $d/\gdc$,
  associated with different nonequilibrium particle-hole
  configurations $p \bar p$.}  Since in \Fig{fig:clusters}(a,b) there
typically are about 5 subresonances per cluster, we can take $\gdc
\simeq 5$ to be an ``experimental estimate'' for the dimensionless
conductance, which is entirely reasonable for grains of the present
size.\footnote{
\label{f:pancake-estimate}
For comparison, note that a similar order
  of magnitude can obtained from \Eqs{eq:d-estimate},
  (\ref{eq:g-dimensionlessconductance}) and (\ref{eq:define-Thouless},ii), for
  a pancake-shaped grain with Volume $\approx 4 \pi r^2 z$ and $a \approx
  (r/z)^2$ [cf.\ footnotes~\ref{f:Ethouless} and \ref{f:estimate-gdim}], 
  for which $\gdc \approx 50 z^3/r (\mbox{nm})^{-2}$: for radius $r \approx
  3$~nm and thickness 
$z \approx 0.7$~nm this would give $\gdc \approx 6$. It is clear,
  though, that such estimates depend very sensitively on the relative
  magnitudes of $r$ and $z$.}

Whether the \emph{lowest}-lying quasiparticle resonance splits into
subresonances or not depends on the threshold voltage for the onset of
transport and on the grain's number parity.  Let $\varepsilon_0$ and
$\varepsilon_1$ denote the two lowest-lying empty levels in $|\G
\rangle_N$.  The lowest-lying resonance involves tunneling sequences
$S_{0 \bar p}$ via level $p=0$.  A necessary condition for it to split
into a cluster is that the level $\bar p$ from which the second
electron leaves the grain [step (ii) of $S_{0 \bar p}$] should come
from a Kramers doublet not involving the level $\varepsilon_0$ into
which the first electron tunneled [step (i) of $S_{0 \bar p}$],
implying $\varepsilon_{0} - \varepsilon_{\bar p} \gtrsim d$, which
requires a threshold [by
Eq.~(\ref{eq:condition-for-nonequilibrium},ii$)^\prime$] of $eV \ge
\E_{0 \bar p}^{N} - \E_\G^N \gtrsim d$.  This condition is evidently
not met in either of \Figs{fig:clusters}(a,b), for which the threshold
voltage at the first peak is significantly smaller than the
inter-cluster spacing, which explains in both figures why the first
peak not split into a cluster.

If this condition \emph{is} met, then the lowest-lying peak will or
will not split into subresonances, depending on whether
$\varepsilon_0$ and $\varepsilon_1$ are degenerate or not,
respectively, as can be seen by choosing $p=0$ and $q=1$: If $N$ is
even so that $\varepsilon_0 = \varepsilon_1$ (by Kramers degeneracy,
see \Sec{sec:kramers}), the lowest quasiparticle resonance will have a
cluster of sub-resonances on its upper shoulder, generated by those
tunneling sequences $S_{1 \bar q, 0 \bar p}$ through level $1$ for
which $ e \delta \bar V^{1 0 (\bar p)}_\ssL > 0$.  On the other hand,
if $N$ is odd so that $\varepsilon_1 - \varepsilon_0 \simeq d$, the
threshold difference $ e \delta \bar V^{1 0 (\bar p)}_\ssL$ is of
order $d$ too, hence the $S_{1 \bar q, 0 \bar p}$ cluster will be
separated from the lowest-lying resonance $S_{0 \bar p}$, which
remains unsplit, by an energy of order $d$.

In general, when $M$ available states below the highest accessible
energy level (including spin) are occupied by $M'<M$ electrons, there
are ${M \choose M'}$ different occupancy configurations.  The typical
width of a cluster of resonances in this case is of order $W^{1/2}
d/\gdc$, where $W=\mbox{min}(M-M', M')$. The width of a cluster of
resonances therefore {\em increases} with the bias voltage. The
distance between nearby resonances of the cluster, on the other hand,
{\em decreases} as $(W^{1/2}d/\gdc) / {M \choose M'}$.  This behavior
can be seen in \Fig{fig:clusters}(c), which shows the differential
conductance obtained in a model calculation by Agam \etalia\ 
\cite{agam97a} and explicitly demonstrates the splitting of resonances
induced by fixed fluctuations $\delta U$ in the interaction energy.

\subsection{Prethreshold structures in superconducting grains}
\label{sec:sub-gap-structures}

\begin{figure}
\centerline{\epsfig{figure=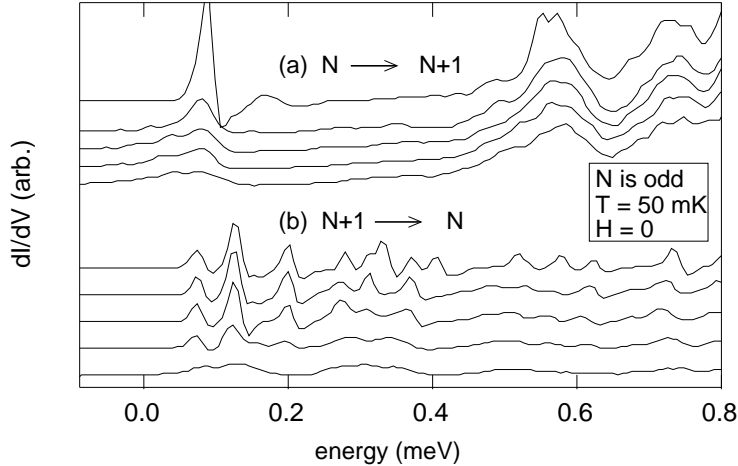,%
width=0.7\linewidth}}
\caption[Prethresold structures due to nonequilibrium in 
superconducting grains ]{The tunneling spectra \cite{rbt97}, for five
  different values of the gate voltage $\Vg$, of a largish Al grain
  with (a) an odd and (b) an even number of electrons in its $V=0$
  ground state.  The top curve in each case corresponds to $\Vg$ at a
  degeneracy point, \ie\ tuned such the peaks occur at the lowest
  possible values for the bias voltage $|V|$. As $\Vg$ is changed to
  shift them to larger $|V|$ values (by amounts $\delta |V| = 0.2$,
  0.4, 0.6 and 1.2~mV), the peaks broaden and develop substructure.
  The corresponding curves have been artificially shifted in energy to align
  peaks due to the same eigenstates.  The sample is the same as in
  \Figs{fig:generic-IV}, \ref{fig:Vg-V}, \ref{fig:sc-spectra(h=0)} and
  \ref{fig:sc-magneticfield}.  All curves were taken at $T= 50$~mK and,
  to improve spectroscopic resolution, at $H=0$, so that the Al leads
  were superconducting [this is the reason for the dip to the right of
  the first peak in the topmost curve, cf.\ \Sec{sec:sc-leads} and
  \Fig{fig:sc-normal-IV-steps}(b)].
 \label{fig:sugap-scclusters}}
\end{figure}

Nonequilibrium effects also cause interesting anomalies in Al grains
large enough for superconducting pairing correlations to be important
\cite{rbt97}.  Figs. \ref{fig:sugap-scclusters}(a) and
\ref{fig:sugap-scclusters}(b) show the evolution with gate voltage
$\Vg$ of the odd-to-even and even-to-odd tunneling spectra,
respectively, of the same grain as that discussed in
\Sec{sec:gap-in-spectrum}. The top curves in both (a) and (b)
correspond to $\Vg$ at a degeneracy point, and show features familiar
from \Sec{sec:gap-in-spectrum} and indicative of pairing correlations:
a distinct gap (of 0.55~meV) between the first and second main peaks
in the odd-to-even case, and its absence in the even-to-odd case [cf.\ 
\Figs{fig:sc-spectra(h=0)} and \ref{fig:sc-magneticfield}].  As $\Vg$
is changed away from the degeneracy point, we note the following
salient features:
\begin{itemize}
\item[(a)] \emph{Odd-to-even  spectrum}
  [\Fig{fig:sugap-scclusters}(a), $N \to N+1$ transitions, with $N$
  odd]:
\begin{itemize}
\item[(a1)] The height of the first peak
  rapidly decreases and a structure of ``\emph{prethreshold}''
  subresonances develops on its
  low-voltage shoulder, \ie\ \emph{below} the threshold of the first
  main peak.
\item[(a2)]
The  \emph{characteristic spacing} between these subresonances is 
comparable to that of the first few peaks of the
even-to-odd  spectrum in \Fig{fig:sugap-scclusters}(b).
\end{itemize}
\item[(b)] \emph{Even-to-odd spectrum} [\Fig{fig:sugap-scclusters}(b),
  $N+1 \to N$ transitions]:
\begin{itemize}
\item[(b1)] For $\Vg$ sufficiently close to the degeneracy point (top
  three curves), the first few peaks of the differential conductance
  are \emph{merely shifted} by increasing $V_g$, without changing their
  shape.
\item[(b2)] For sufficiently large changes of $\Vg$ (lowest two
  curves),
the resonances are rapidly \emph{smeared out} by increasing $\Vg$.
\end{itemize}
\end{itemize}
Properties (b1) and (b2) are easily understood: due to the large
spectral gap of 0.55~meV for creating an excitation on an even grain,
the even-to-odd spectrum will be affected by nonequilibrium
excitations only once the bias voltage $e|V|$ exceeds this value.
Therefore, the low-lying peaks of the even-to-odd spectrum are smeared
out by $\Vg$ only for those curves for which $\Vg$ has shifted the
bias threshold to $\delta |eV| > 0.55$~meV, namely the
lowest two of \Fig{fig:sugap-scclusters}(b).

Property (a1), however, stands in direct contrast to the predictions
of the sequential tunneling mechanism of
\Sec{sec:beyound-orthodox-model}, namely that for odd $N$, the lowest
peak of the spectrum does \emph{not} split into subresonances
[cf.\ third paragraph after \Eq{eq:interaction-fluctuation}].
Moreover, the existence of the \emph{prethreshold}
peaks mentioned in (a1) suggests that nonequilibrium
is reached even while  $|eV|$ still is \emph{below} the threshold
for the onset of equilibrium sequential tunneling through
the bottleneck barrier (say L), namely
\begin{eqnarray}
  \label{eq:sequential-tunneling-threshold}
  e \bar V_\ssL  \ge \E^{N+1}_\G - \E^N_\G \; .
\end{eqnarray}
Agam and Aleiner \cite{agam97b} have therefore proposed that in the
present case, nonequilibrium is established not by sequential
tunneling, but instead by \emph{cotunneling}. A cotunneling process
with $| \G\rangle_N$ as initial state
involves a coherent, second-order transition of the form 
 \begin{eqnarray}
   \label{eq:explain-cotunneling}
   |\G\rangle_N \stackrel{c^\dag_p}{\longrightarrow}
   |\beta\rangle_{N+1} 
   \stackrel{c^\ds_{\bar p}}{\longrightarrow}
   | \alpha \rangle_N \; , 
 \end{eqnarray}
 in which an electron tunnels into level $p$ via the bottleneck
 barrier (say L), producing a virtual intermediate state $
 |\beta\rangle_{N+1}$, and one tunnels out of level $\bar p$ via the
 other barrier (R), leaving the grain in the final state $| \alpha
 \rangle_N$.  (Since the grain is pair-correlated, $|\alpha\rangle_N$
 and $|\beta\rangle_{N+1}$ are,  in general, of the form
 of \Eq{eq:generaleigenstate}.)  The energy conditions for cotunneling
 are
\begin{eqnarray}
  \label{eq:conditions-for-cotunneling}
  {\rm (i)}  \quad e \bar  V_\ssL 
<  \E_{\beta}^{N+1} - \E_\G^N \; , \qquad
   {\rm (ii)} \quad e V  \ge \E_{\alpha}^{N} - \E_\G^N \; .
\end{eqnarray}
Here (ii) is the analogue of
Eq.~(\ref{eq:condition-for-nonequilibrium},ii$)^\prime$ for sequential
tunneling, and (i), which should be contrasted to
Eq.~(\ref{eq:sequential-pp-inequalities},i), is the condition that the
$|\G\rangle_N \to |\beta \rangle_{N+1}$ transition is virtual, not real
(once the inequality (\ref{eq:conditions-for-cotunneling},i) is
violated, cotunneling turns into sequential tunneling).  Since the
conditions (\ref{eq:conditions-for-cotunneling}) are much less
stringent than (\ref{eq:sequential-tunneling-threshold}), cotunneling sets
in long before equilibrium sequential tunneling.  

The current due to pure cotunneling is of course typically very small,
since the cotunneling rate $\Gamma_{p \bar p} \propto \Gamma_p^\ssL
\Gamma_{\bar p}^\ssR$ is smaller than that for either step of a
sequential tunneling sequence by one factor of $\Gamma$ [the latter
are defined in \Eq{eq:spectralGamma}]. However, if the excited state
$|\alpha \rangle_N$ produced by cotunneling is sufficiently
long-lived, it can enable the next electron to enter the grain via
\emph{sequential} tunneling whenever the  condition
\begin{eqnarray}
  \label{eq:cotunneling-reduced-threshold}
  e \bar V_\ssL \ge \E^{N+1}_\G - \E^{N}_{\alpha}
\end{eqnarray}
is satisfied, which will cause a step-like increase in current as $eV$
is tuned past this ``nonequilibrium sequential tunneling threshold''.
Now, since the latter lies \emph{below} the threshold of
\Eq{eq:sequential-tunneling-threshold} for equilibrium sequential
tunneling by an amount $\E^N_{\alpha} - \E^N_\G > 0$, a set of
subresonances will occur on the low-bias shoulder of the first main
resonance, precisely as observed in (a1). Moreover, the spacing
between these subresonances will reflect that between consecutive
excited states $\E_\alpha^N$ of the odd grain, \ie\ between the peaks
measured in the even-to-odd excitation spectrum, in accordance with
property (a2).

Agam and Aleiner \cite{agam97b,agam98} worked out these ideas
quantitatively and concluded that the rate of creation of nonequilibrium
excitations by cotunneling is indeed sufficient to explain the data of
\Fig{fig:sugap-scclusters}(a): they showed that the smallness of the
cotunneling rates due to their being quadratic in $\Gamma_p^r$ is offset
to a large extent by the increase in the density of states near $\eF$
of both the grain and the leads, due to superconducting pairing
correlations in both. 

\subsection{Discussion of relaxation rates}
\label{sec:estimating-relaxation}

As emphasized above, nonequilibrium can only be established
if relaxation processes are slower than tunneling processes, 
$\Gine \lesssim \Gamma_\tun$ for $\varepsilon
\simeq d$, hence it
is important to understand the mechanism contributing to 
$\Gine$. Agam \etalia\ \cite{agam97a} discussed
several possibilities: 
(i) The electron-phonon interaction; (ii) the electron-electron
interaction beyond the Hartree-Fock approximation; (iii) Auger processes
in which an electron in the dot relaxes while another one in the lead
is excited; (iv) the relaxation of an electron in the dot as another
electron tunnels out to the lead; and (v) thermalization with the
leads via tunneling.  Of these, it is easy to see that (iii) to (v)
are at best small corrections: (iii) is exponentially small in the
ratio $w/\chi$ of the barrier width $w$ to the screening length
$\chi$, and moreover the interaction between electrons on both sides of
the tunnel junction can take place only within a very limited volume;
and (iv) and (v) are small, since they clearly happen on time scales
larger than the tunneling time. (i) and (ii), however, require a
careful discussion.

\subsubsection{Electron-phonon interaction}

Since the temperature of $T \simeq  30$~mK at which
the experiments are being done is much smaller than the
mean level spacing,  the probability of phonon
absorption is negligible, and only emission may take place. 
The sound velocity in aluminum is $v_\sound=6420$ m/sec, 
therefore the wavelength of a phonon associated with relaxation
of energy $\varepsilon \sim d =1$ meV is approximately 5~nm,
the same  as the system size. In this regime, 
Agam \etalia\  estimate the phonon emission rate to be  
\begin{equation}
\Gamma^{\rm el-ph} \sim 
\left(\frac{2}{3}\eF \right)^2 \frac{\varepsilon^3 \tau_\el \, d}
{2\rho \hbar^4 v_\sound^5},
\end{equation}
where $\eF$ is the Fermi energy (11.7 eV in Al), and $\rho$ is the ion
mass density (2.7 g/cm$^3$ in Al). This rate is that of a clean metal
but reduced by a factor of $\tau_\el d/\hbar$, where $\tau_\el$ is the
elastic mean free time \cite{Reizer86} which, in ballistic systems,
equals the traversal time across the system of an electron at the
Fermi level. Assuming ballistic motion this factor is of order
10$^{-3}$.  The resulting relaxation rate for $\varepsilon \simeq d$
is therefore of order $\Gamma^{\rm el-ph}_{\varepsilon \simeq d}
\approx 10^8$ sec$^{-1}$ which is similar to the tunneling rate
$\Gamma_\tun \approx 6 \cdot 10^8$ sec$^{-1}$ (corresponding to an
increase in current of $10^{-10}$~A at the first current step).  Thus,
by increasing the resistance of the tunnel junctions, it should be
possible to cross over to the near-equilibrium regime shown by the
dashed line in \Fig{fig:clusters}(c).
  
\subsubsection{Electron-electron interaction beyond Hartree-Fock}

The effect of the electron-electron interaction on 
the resonance widths measured in a tunneling experiment,
or more generally on the life-time of quasiparticle excitations
in a finite, disordered system of interacting electrons,
has been elucidated by Altshuler,
Gefen, Kamenev and Levitov (AGKL) \cite{AGKL97}. Since their
conclusions are  directly relevant for the interpretation
of nonequilibrium tunneling spectra, we briefly review them here. 

Consider again tunneling transitions from the $N$ to $(N+1)$ particle
Hilbert spaces, which probe the excitations of the latter.  In
\Sec{sec:nonequilibrium-sequential}, we took these excitations to be
simply particle-hole excitations with respect to the Fermi ground
state $|\G\rangle_{N+1}$, such as $|p \bar p \rangle_{N+1}$, $|q \bar
q, p \bar p \rangle_{N+1}$, etc.; in the nomenclature of AGKL, these
are called Hartree-Fock states, $|p 0 \rangle_{N+1}$ is called a
``quasiparticle'', $|q 0 , p \bar p \rangle_{N+1}$ two quasiparticles
and a quasihole, etc. Now, while these are exact eigenstates of the
orthodox model Hamiltonian $H_{\ssD,\orth}^\normal$
[\Eq{eq:H-dot-normal-nonequib}], they will decay in the presence of
the more general interaction $H_\elel$ of \Eq{eq:H_int-el-el}, \eg,
a quasiparticle will decay into two quasiparticles and a quasihole,
etc.\ 

A quasiparticle state with energy $\varepsilon$ will thus acquire a
lifetime $1/ \Gine$, and in general should be viewed as a wave packet,
with packet width $\Gamma^\inel_\varepsilon$ (we set $\hbar = 1$ in
this section), constructed from the exact eigenstates of the full
Hamiltonian.  In a finite-sized system with a discrete spectrum, a
quasiparticle will be well-defined if $\Gamma^\inel_\varepsilon < d$
(note that this condition is more stringent than the one for infinite
systems, namely $\Gine \ll \varepsilon$).  Moreover, the quasiparticle
wave packet width $\Gine$ can be identified with the measured width of
a resonance in the single-particle tunneling density of states (DOS)
probed by an incident tunneling electron, since the mathematical
description of a single-electron tunneling process involves projecting
a single-particle state onto exact eigenstates of the system [this is
made explicit in \Eq{eq:Tchange-basis}].  In principle, in a finite
system \emph{every} exact eigenstate should produce an infinitely
sharp conductance resonance. However, in reality only a small fraction
of the exact eigenstates overlap sufficiently strongly with the
one-particle excitations produced by the incoming electron to be
detected as strong peaks by a measurement of finite sensitivity. The
nature and spacing of these strong peaks depends on the energy
$\varepsilon$ of the incoming electron; under certain conditions,
clarified by AGKL, they group into clusters of combined width $\Gine$
that can be interpreted as quasiparticle peaks.

The standard golden rule (GR) approach for estimating $\Gine$ gives
$\Gbulke \sim \varepsilon^2/\eF$ for bulk systems
\cite{Pines-Nozieres-89} and $\Gfinitee \approx d(\varepsilon/d
\gdc)^2$ for finite systems \cite{Sivan94}, but is valid only when the
quasiparticle effectively decays into a \emph{continuum} of final
states (otherwise the GR will not give the decay rate, but rather just
a first-order perturbation correction to the energy of a given
eigenstate). For sufficiently low $\varepsilon$,
the accessible final states will not form a continuum;
for example, the density of final states for the decay of a
quasiparticle into two quasiparticles and a quasihole, namely
$\varepsilon^2 / 2 d^3$, is much smaller than that of all many-body
states. To analyze how matters change with decreasing $\varepsilon$,
AGKL considered the decay of a quasiparticle into two quasiparticles
and a quasihole, which in turn can decay into five quasiparticles and
three quasiholes, etc., and thus mapped the problem onto one of
delocalization in Fock space, in which the ``distance'' between two
states is measured, roughly speaking, by the number of particle-hole
excitations by which they differ.  A state that is ``localized'' or
``delocalized'' in Fock space is associated with a quasiparticle peak
of zero and finite width, respectively.  Assuming the dimensionless
conductance to be large, $\gdc \gg 1$, AGKL exploited analogies to
Anderson localization to study the localization-delocalization
crossover quantitatively, and constructed the following picture.

The nature  of an excitation with energy $\varepsilon$ depends
on the relation of $\varepsilon$ to the hierarchy
$  E^{\ast \ast} < E^{\ast} < \ETh $ 
of energy scales, where 
\begin{eqnarray}
  \label{eq:AGKL-hierarchy}
 E^{\ast \ast} \equiv d \sqrt {\gdc/ \ln (\gdc)} \; , \qquad
 E^{\ast } \equiv d \sqrt {\gdc} \; , \qquad
\ETh \equiv d \, \gdc \; .
\end{eqnarray}
$ E^{\ast \ast} $ separates the localized from 
the delocalized regime, and altogether four cases be distinguished:
\begin{itemize}
\item[(i)] $\varepsilon < E^{\ast \ast}$ (localized phase): Here the
  true many-body excited states are just slightly perturbed
  Hartree-Fock states, \ie\ their overlap with simple particle-hole
  excitations of the Fermi ground state is very close to unity.  In
  other words, quasiparticles do not decay significantly, and the
  single-particle DOS will consist of a few $\delta$-function-like
  resolved peaks, which may have weak satellites due to coupling to
  many-particle states involving more particle-hole pairs.  If the
  latter are purposefully created by driving the grain into
  nonequilibrium, the satellites will become strong and each
  quasiparticle peak will appear as a cluster. 
 --- Since $\gdc \simeq
  5 $ for \Fig{fig:clusters}, at least its first few quasiparticle
  excitations fall into this regime; note that AGKL's conclusion that
  these are well-described by simple particle-hole states fully
  justifies using only such states in the analysis in
  \Secs{sec:nonequilibrium-sequential} and
  (\ref{sec:beyound-orthodox-model}).
\item[(ii)] $E^{\ast \ast} < \varepsilon < E^{\ast}$ (delocalized
  phase with non-Lorentzian peaks): A given quasiparticle state is
  connected to several others far away from it in Fock space, so that the
  number of satellites in a quasiparticle peak rapidly increases with
  $\varepsilon$. This causes the quasiparticle peaks in the DOS to
  have non-Lorentzian shapes, but their widths are still $\ll
  \Gfinitee$.
\item[(iii)] $E^\ast < \varepsilon < \ETh $ (delocalized phase
with Lorentzian peaks): A given quasiparticle state
is connected to so many others that it effectively
decays into a continuum, so its decay rate is
given by the GR result, $\Gfinitee \approx d(\varepsilon/d
\gdc)^2$. Since this is still $<d$ for
$\varepsilon < \ETh$, the peaks in the
DOS  can still be well-resolved, and have a  Lorentzian shape. 
\item[(iv)] $\ETh < \varepsilon$ (quasiparticles not defined):
In this regime $\Gfinitee > d$, hence the quasiparticles
are not well-defined, and the DOS is a featureless continuum.
\end{itemize}
\begin{figure}
\centerline{\epsfig{figure=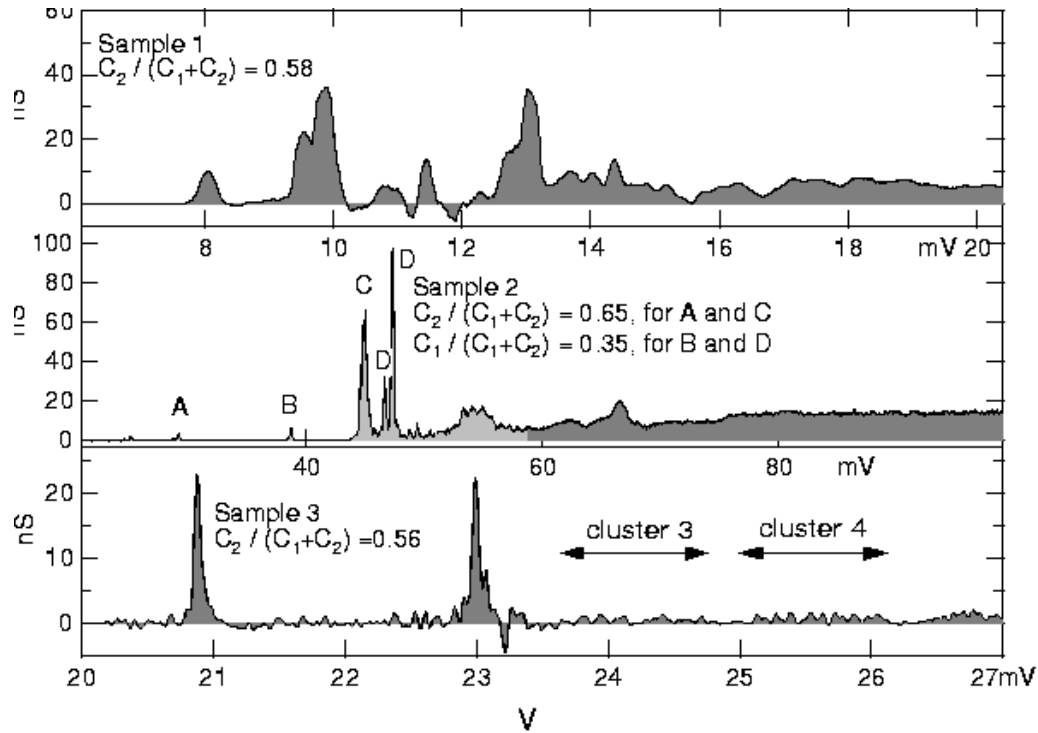,%
width=0.98\linewidth}}
\caption[Energy-evolution of tunneling spectra up to
and beyond $\ETh$]{Excitation spectra of three different nm-scale Au
  grains measured by Davidovi\'c and Tinkham \cite{davidovich99}, at
  $T=30$~mK and $H= 1$~T. In order to convert from bias  voltage to
  excitation energy, $V$ must be multiplied by $e$ times the 
  capacitative conversion
  factor [cf.\ \Eqs{eq:fixed-N-excitation-spectrum1} and
  (\ref{eq:fixed-N-excitation-spectrum2})] indicated for each graph.
  Assuming the grains to be hemispherical, the grains parameters were
  estimated to be as follows for samples 1, 2 and 3, respectively, :
  grain radius (estimated from $C_1 + C_2$ and a capacitance per unit
  area of 0.05~aF/$\mbox{nm}^2$): $r = 4.5$, 2.4, and 3~nm; estimated
  single-electron level spacing (from \Eq{eq:d-estimate}, using $r$ as
  input): $\bar d = 0.65$, 4.6, and 2.1~meV; measured
  single-electron level spacing $d$ (taken as distance between the two
  lowest energy peaks): $d = 1$, 7, and 1.2~meV; Thouless energy
  (estimated as $\ETh = \hbar v_{\rm F}/6 r'$ [\Eq{eq:define-Thouless}
  with $a=3$, cf.\ footnote~\ref{f:drago}], where $r'$ is the grain
  radius estimated from \Eq{eq:d-estimate}, using the \emph{measured}
  $d$-values as input): $\ETh = 37$, 75, and 40~meV.
  \label{fig:drago1}} 
\end{figure}
Experimental results strikingly reminiscent of this picture have
recently been obtained by Davidovi\'c and Tinkham \cite{davidovich99},
see \Fig{fig:drago1}.  \Fig{fig:drago1}(b), in particular, seems to
show the progression, with increasing energy, of the measured
tunneling spectrum through all the crossovers.  Estimates for $\ETh$,
cited in the caption of \Fig{fig:drago1}, agree within a factor of 2
with the voltage\footnote{Davidovi\'c and Tinkham argued that $\ETh$
  should be compared to the voltage $eV$ itself and not to the
  corresponding (capacitatively corrected) excitation energy, because
  the number of possible nonequilibrium excitations is controlled by
  $eV$, cf.\ Eq.~(\ref{eq:condition-for-nonequilibrium},ii$)^\prime$.}
$eV$ at which the measured spectra become continuous.  Note that in
order to observe the entire regime up to and beyond the Thouless
energy, all within the first step of Coulomb blockade, one needs to
have $\ETh \ll \Ec$. In this respect ultrasmall metallic grains, with
their huge charging energies, have an edge over disordered
semiconducting quantum dots, studies of which had initiated the
analysis of the role of $\ETh$ as the upper scale for the
observability of level discreteness \cite{Sivan94}.

\newpage \section{Spin-orbit interaction}
\label{sec:spin-orbit}

For several of the normal-state grains investigated so far, the
magnetic-field dependence of the excitation spectra showed marked
deviations from the simple behavior described in \Sec{sec:kramers}:
$g$ factors differing from 2 were measured
\cite{rbt96b,salinas99,davidovich99} (in Au grains, some were as small
as 0.28 \cite{davidovich99}), and deviations from linear
$H$-dependence, in the form of avoided level crossings, were observed
\cite{rbt96b,salinas99}. These features were attributed to the
spin-orbit interaction, which is the subject of this section.  
A summary of early work on the spin-orbit interaction in small metallic
grains, mainly with regard to thermodynamical properties, may be found
in the reviews by
Buttet \cite{Buttet}, Perenboom, Wyder and Meier \cite{wyder}, and
Halperin \cite{halperin}.

The information on spin-orbit scattering obtained from the analysis summarized
below is complementary to that obtained from studies of bulk systems.  There,
the quantity of primary interest is the average scattering rate ($\Gamma_\so$)
for an electron, assumed to be initially in a pure spin-up or spin-down state,
to be scattered into a \emph{continuum} of states with opposite spin. This
rate can be measured in studies of weak localization in disordered metals
\cite{Bergmann} or of tunneling between thin superconducting films in a
parallel magnetic field \cite{Meservey-94}, or it can be related to the
effective $g$ factors \cite{Monod-77} measured by electron spin resonance. In
ultrasmall grains, spin-orbit scattering can be studied at a more detailed
level via its effects on individual, discrete eigenstates of the grain, such
as the above-mentioned occurrences of anomalous $g$ factors and avoided
crossings.  Moreover, the effects of mesoscopic fluctuations \cite{matveev00}
on phenomena governed by spin-orbit scattering can be observed directly.

\begin{figure}
\centerline{\epsfig{figure=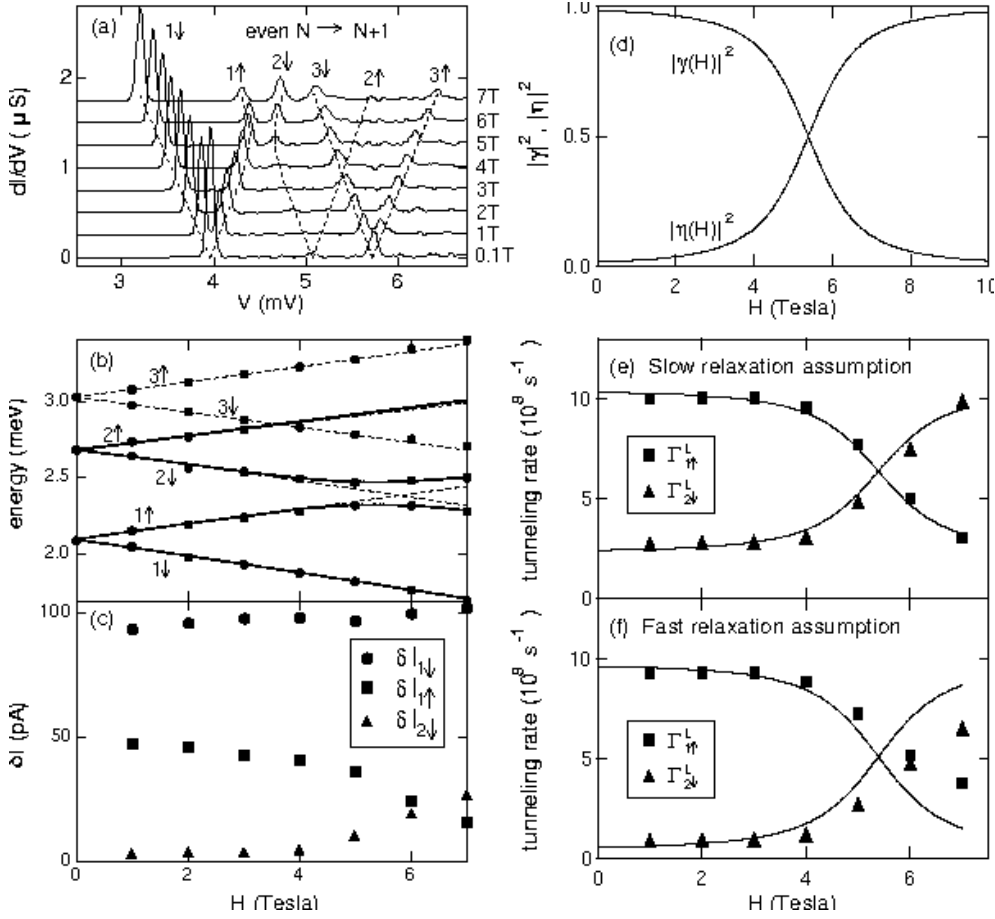,%
width=0.94\linewidth}}
\caption[The effect of spin-orbit scattering]{
  The effect of the spin-orbit interaction on the excitation spectrum
  of an Al grain of radius $r \simeq 3$~nm \cite{salinas99}: (a)
  Differential conductance for a range of applied magnetic fields, at
  $T=50$~mK. The curves are offset in ${\d} I / {\d} V$ for
  visibility. Orbital state no.\ 2 gives small but visible resonances
  at low $H$.  Small changes in offset charge ($Q_0$) occurred between
  the 0.1 and 1~T scans and between the 6 and 7~T scans, shifting peak
  positions. The 0.1 and 7~T scans have therefore been shifted along
  the voltage axis, to give the best fit to a linear dependence for
  peak $\onedown $. Dotted lines are guides to the eye. (b)
  Magnetic-field dependence of the peak positions of (a), which have
  been converted from voltage to energy by a conversion factor of $e
  C_\ssL/ (C_\ssL + C_\ssR) = e(0.53 \pm 0.01)$, R being the
  bottleneck barrier here [cf.\ 
  \Eqs{eq:fixed-N-excitation-spectrum1}].  Dots represent data points.
  Dashed lines are extensions of the low-field linear dependence of
  the energies on $H$.  Solid lines 
  show the result of the \som\
   of \Sec{subsec:spin-orbit-theory}, describing the
  avoided crossing between levels $| \oneup \rangle$ and 
  $|\twodown \rangle$.  (c) Areas  under the
  first three conductance peaks, which are equal  to 
  the current steps  $\delta I$ at the corresponding resonances. 
  Note the exchange in weight between $\delta I_{\oneup}$ and $\delta
  I_{\twodown}$, which occurs near the avoided crossing between the
  eigenstates $|\oneup\rangle $ and $|\twodown \rangle$ in (b). (d)
  Magnetic-field dependence of the coefficients $\chi (H)$ and $\eta
  (H)$ in \Eq{eq:true-eigenstates}, which relate the eigenstates
  $|\oneup \rangle$ and $|\twodown \rangle$ to the basis states
  $|\oneup \rangle_\tildezp$ and $|\twodown \rangle_\tildezp$.  (e)
  and (f): Markers: Tunneling rates $\Gamma^\ssL_\oneup$
  and $\Gamma^\ssL_\twodown$ for transitions out of the grain's
  eigenstates $|\oneup \rangle$ and $|\twodown \rangle$, extracted
  from the measured current as discussed in text, assuming that the
  relaxation rate of nonequilibrium excitations of the grain is either
  smaller (e) or larger (f) than the tunneling rates.  Lines: fits
  using the predictions of the \som. }
\label{fig:spin-orbit}
\end{figure}

\subsection{Experimental results for weak spin-orbit interaction}
\label{subsec:spin-orbit-exp}

\Fig{fig:spin-orbit} shows the results of measurements published by
Salinas, Gu\'eron, Ralph, Black and Tinkham \cite{rbt96b,salinas99} on
a grain of estimated radius $r \simeq 3$~nm, made from nominally pure
Al. It presumably contained an unintended defect or impurity, because
its behavior was similar to that observed in grains intentionally doped
with 4\% of Au impurities. The following salient features can be
identified:
\begin{itemize}
\item[(i)] This is an \emph{even-to-odd} excitation spectrum, since all three
  the lowest peaks in \Fig{fig:spin-orbit}(a) Zeeman-split in an
  applied magnetic field $H$.
\item[(ii)] \emph{Orbital effects are negligible:} For each pair of
  peaks the splitting is symmetric about their $H=0$ resonance energy,
  and the average of their resonance energies shows hardly any
  $H$-dependence (fitting the latter to a linear form, $g^\ave \muB
  H$, yields $g^\ave < 0.1$, which is negligible).  This was to be
  expected, since the largest value of $H$ used here (7T) is much
  smaller than the scale $H_\orb$ [\Eq{eq:Horbital}] beyond which
  orbital effects become significant.  The latter will thus be
  neglected below.
\item[(iii)] The \emph{effective  $g^\eff_{j}$ factor}
 (for low fields) of
  each Zeeman-split resonance is defined by writing its Zeeman splitting, to
  \emph{linear} order in $H$, as $\delta E_j = g^\eff_{j} \muB H$
  [cf.\ \Eq{g-effective}]. 
  Then \Fig{fig:spin-orbit}(b) gives $g^\eff_1 = 1.84 \pm 0.03$,
  $g^\eff_2 = 1.68 \pm 0.08$, and $g^\eff_3 = 1.76 \pm 0.05$.
\item[(iv)] An \emph{avoided crossing} occurs between resonances
  $\oneup $ and $\twodown$ in \Figs{fig:spin-orbit}(a,b);
\item[(v)] the \emph{areas} under the corresponding
peaks  $\oneup $ and $\twodown$  in  \Fig{fig:spin-orbit}(a),
plotted in \Fig{fig:spin-orbit}(c), 
are \emph{strongly $H$-dependent}: the smaller one ($\twodown$)
grows at the expense of the larger one ($\oneup$)
in the avoided-crossing region, while their sum remains
  approximately constant.
\end{itemize}
Properties (iii) to (v) contrast strongly with the behavior observed
for pure Al grains [cf.\ \Fig{fig:sc-magneticfield}]; the latter
typically have $g$-factors close to $g^\pure \equiv 2$ [see also
footnote \ref{f:wrong-g} in \Sec{sec:model}] that vary only weakly
from level to level (if at all), their spectra show no clear avoided
crossings, and they have essentially $H$-independent peak areas.

\subsection{A simple model for the spin-orbit interaction}
\label{subsec:spin-orbit-theory}

Salinas \etalia\ \cite{salinas99} have attributed anomalies (iii) to
(v) to spin-orbit scattering due to some defect in the grain, and have
constructed a simple model within which their measurements could be
analyzed quantitatively.  Using a somewhat more general notation than
they did and guided by the reviews of \Refs{wyder,halperin},
 we shall consider a Hamiltonian $\hat H_\tot = \hat H_0 + \hat H_\so$, to
be called the \emph{\som}, of the following simple form:
\begin{eqnarray}
  \label{eq:H-spin-orbit-0}
  \hat H_0 & = & \sum_{j \sigma} \varepsilon_{j \sigma,0}  \, 
  c^\dagger_{j\sigma,0} \, c^\ds_{j\sigma,0} \; , 
\qquad \varepsilon_{j \sigma,0} = \varepsilon_{j,0} + 
\half \sigma g^\pure \muB H  \; , 
\\
  \label{eq:H-spin-orbit-so}
  \hat H_\so & = & \sum_{i j \sigma} {v}_{ij}^{\sigma } \,
  c^\dagger_{i\sigma,0} \, c^\ds_{j \bar \sigma,0} \; ,
  \quad \mbox{with} \quad 
  {v}^{\sigma }_{ij} = 
  ({v}^{\bar \sigma}_{ji})^\ast = 
  - {v}^{\sigma}_{ji} \, , \quad
   {v}_{jj}^\sigma \equiv 0 \;  . 
\end{eqnarray}
Since time-reversal symmetry guarantees each orbital level to be
two-fold degenerate in the absence of an applied magnetic field
($H=0$), we have adopted a single-particle basis of pairs of
time-reversed states $|j \sigma \rangle_0$, labeled by a discrete
orbital index $j$ and a spin index $\sigma = (\uparrow, \downarrow) =
(+,-)$.  The orbital eigenenergies $\varepsilon_{j,0}$ are meant to
include the effects of all spin-independent forces and interactions,
and are taken to have a mean spacing of $d$.  A magnetic field
Zeeman-splits these levels by $\half \sigma g^\pure \muB H$. In
principle, it can also produce orbital diamagnetic effects, which
could cause $\varepsilon_{j,0}$ to be $H$-dependent and, via
spin-orbit coupling, contribute to reducing the effective $g$ factors
\cite{Kravtsov92,matveev00}.  We shall neglect these here, however, in
accordance with property (ii).  $\hat H_\so$ describes the spin-orbit
interaction and is taken to couple only states of \emph{opposite}
spin, $\sigma$ and $\bar \sigma \equiv - \sigma$.  The conditions
specified in \Eq{eq:H-spin-orbit-so} for the matrix elements $
{v}_{ij}^{\sigma }$ ensure that $\hat H_\so$ is hermitian and
invariant under time-reversal [which maps $c_{j \sigma}^\dagger \to
\sigma c_{j \bar \sigma}^\dagger$ and $ {v}_{ij}^{\sigma} \to
({v}_{ij}^{\sigma })^\ast$], \ie\ that $\hat H_\so$ represents a
symplectic matrix, as required on general grounds for the spin-orbit
interaction \cite{wyder,halperin}.

The average spin-orbit scattering rate in this model, expressed in
terms of the average (over $i,j$) of the matrix elements in $\hat
H_\so$, is $\Gamma_\so = 2 \pi \overline {|{v}_{ij}^{\sigma}|^2} /
(\hbar d)$.  The effects of spin-orbit scattering on an individual
eigenstate of $\hat H_\tot$ will be ``weak'' or ``strong'' depending
on whether the dimensionless parameter $\lambda_{\so} \equiv \hbar
\Gamma_\so / d$ is $\ll 1$ or $\gg 1$, respectively. Since the rate $
\Gamma_\so $ is an intensive (volume-independent) quantity (implying
that $v_{ij}^\sigma$ must scale as $\Vol^{-1/2}$), it should be
possible to reach the regime of weak spin-orbit scattering in grains
with sufficiently large $d$.  Indeed, the grain of
\Fig{fig:spin-orbit} is an example of this case, since the minimum
splitting at the avoided crossing, which is a measure of $\overline
{|{v}^{\sigma}_{ij}|}$, is much less than the mean level separation,
$d$.

Since $\hat H_\tot$ is quadratic, it can in principle be brought into
the diagonal form $\hat H_\tot = \sum_{j \sigma} \varepsilon_{j
\sigma} c^\dagger_{j \sigma} c^\ds_{j \sigma}$ by a linear
transformation. The time-reversal symmetry of $\hat H_\tot$ for $H=0$
implies that the exact eigenstates again come in time-reversed pairs,
say $|j \sigma \rangle$, degenerate at $H=0$.  Since this degeneracy
is lifted by the magnetic field,
\begin{eqnarray}
\label{g-effective}
  \varepsilon_{j \pm } = \varepsilon_j \pm \half g^\eff_j
   \muB H \; + {\cal O}(H^2) \; ,
\end{eqnarray}
the labels $(+,-) = (\uparrow$, $\downarrow)$ identify states with
``predominantly spin-up or down'' character (though no longer
eigenstates of $\hat \sigma_z$).  The linear term in \Eq{g-effective}
defines the effective $g$ factor $g^\eff_j$ for the pair of states $|j
\sigma \rangle$, which, being superpositions of spin-up and spin-down
states, will have\footnote{In situations where orbital effects are
  important, such as semiconductor quantum dots, they may reverse this
  trend and give $g^\eff_j > g^\pure$, see \cite{matveev00}.}
$g^\eff_j < g^\pure$.  For weak spin-orbit scattering ($\lambda_\so
\ll 1$), this reduction can be calculated perturbatively: using the
lowest-order perturbative approximations (designated by the subscript
$\tildez$) to $|j \sigma \rangle$ and $\varepsilon_{j \sigma}$ to
evaluate $g^\eff_j$, one finds \cite{halperin}
%
\begin{eqnarray}
  \label{eq:spin-orbit-eigenstates-pert}
   |j \sigma \rangle_{\tildez} & = & \left[ 1 - \half \sum_{i}
    { |{v}^{\bar \sigma}_{i j}|^2
\over 
(\varepsilon_{j \sigma, 0} - \varepsilon_{i \bar \sigma , 0})^2}
\right] |j \sigma \rangle_0 
+ \sum_{i} 
{ {v}^{\bar \sigma}_{i j}
\over 
\varepsilon_{j \sigma, 0} - \varepsilon_{i \bar \sigma , 0} }
|i \bar \sigma \rangle_0 \; , 
\\
\label{eq:spin-orbit-eigenvalues-pert}
\varepsilon_{j \sigma, \tildez} &=& \varepsilon_{j \sigma, 0}
+ \sum_{i} {| {v}_{i j}^{\bar \sigma}|^2 \over 
\varepsilon_{j \sigma, 0} - \varepsilon_{i \bar \sigma , 0}} \; , 
\\
\label{g-effective-pert}
  g^\eff_j & =  &  g^\pure \left[ 1 -  \sum_{i} 
{2 | {v}^{\bar \sigma}_{i j}|^2
\over 
(\varepsilon_{j , 0} - \varepsilon_{i , 0})^2}
\right] \; . 
\end{eqnarray}
(\Eq{g-effective-pert} follows by expanding
(\ref{eq:spin-orbit-eigenvalues-pert}) to linear order in $H$.)  Since
the bare energies $\varepsilon_{j,0}$ and matrix elements
${v}_{ij}^\sigma$ are subject to mesoscopic fluctuations, $g^\eff_j$
will be too.  This is in accord with the observation of
$j$-\emph{dependent}, \emph{reduced} $g^\eff_j$ values cited in
property (iii) above.

\subsection{Detailed analysis of an avoided crossing}
\label{sec:avoided-crossing}

In order to analyze the avoided crossing in \Figs{fig:spin-orbit}(a,b)
quantitatively, Salinas \etalia\ truncated the problem to an effective
four-state system, associated with the four lowest levels in
\Fig{fig:spin-orbit}(b). The idea is to treat the effect of all other
levels on the lowest four using perturbative expressions evaluated in
the low-field limit, while treating the spin-orbit interaction among
the lowest four fully, for arbitrary $H$.  To this end, one introduces
four ``bare'' basis states for the truncated problem, $|\onedown
\rangle_\tildezp$, $|\oneup \rangle_\tildezp$, $|\twodown
\rangle_\tildezp$, $|\twoup \rangle_\tildezp$, defined by the $H = 0$
limit of \Eqs{eq:spin-orbit-eigenstates-pert}, with ``bare''
eigenenergies $\varepsilon_{j \sigma, \tildezp}$ given by
(\ref{eq:spin-orbit-eigenvalues-pert}), expanded to linear order in
$H$. In order not to overcount the interaction among these states, the
$\sum_i$ sums in \Eqs{eq:spin-orbit-eigenstates-pert} and
(\ref{eq:spin-orbit-eigenvalues-pert}) should now \emph{exclude} $i=1$
and 2 (the tilde on the label $\tildezp$ indicates this restriction).
The effective Hamiltonian for the truncated system therefore
is\footnote{ When deriving $\hat H_\tot^\tildezp$ by reexpressing
  $\hat H_\tot$ in terms of $c_{j\sigma, \tildezp}$-operators,
  off-diagonal terms beyond linear order in ${v}^{\bar
    \sigma}_{ij}$ were dropped, since they contain further factors of
  the small quantity ${v}^{\bar \sigma}_{ij} /(\varepsilon_{j
    \sigma,0} - \varepsilon_{i \bar \sigma, 0})$, with $i= 1,2$ and $j
  \neq 1,2$.}
\begin{eqnarray}
  \label{eq:H-spin-orbit-effective}
  \hat H^{\tildezp}_\tot &=& \sum_{j = 1,2} \sum_\sigma 
  \varepsilon_{j \sigma, \tildezp}
   c^\dagger_{j\sigma,\tildezp} \, c^\ds_{j\sigma,\tildezp} \; + \; 
  \sum_{\sigma} 
 {v}^\sigma_{12} c^\dagger_{1 \sigma,\tildezp} 
c^\ds_{2 \bar \sigma,\tildezp}
 + \mbox{h.c.} 
\\
\label{eq:pert-g}
  \varepsilon_{j \sigma, \tildezp}
  & = & \varepsilon_{j,\tildezp} + (g^\ave_j + \half \sigma
   \tilde g^\eff_j ) \muB H \; , 
\qquad 
  \tilde g^\eff_j \equiv   g^\eff_j + 
{g^\pure \, 2 |{v}^{\downarrow}_{12}|^2  \over 
     (\varepsilon_{2,0} - \varepsilon_{1,0})^2} \; .
\end{eqnarray}
Note that $\tilde g^\eff_j$ is the contribution of the $i \neq 1,2$
levels to the total effective $g^\eff_j$ factor of state $|j
\sigma\rangle_{\tildez}$.  Moreover, simply for convenience in the
fitting, a linear term $g^{\ave}_j \mu_{B} H$ has been allowed too, to
model any shift in the average energy of the Zeeman-split pairs [the
fits turn out to give $g_j^\ave \!\approx\!0$, cf.\ property (ii)].

It is now straightforward to solve this truncated problem explicitly
and to find the eigenstates $|\onedown \rangle$, $|\oneup \rangle$,
$|\twodown \rangle$, $|\twoup \rangle$ of $\hat H_\tot^\tildezp$.
Fitting the corresponding eigenenergies to the lowest four lines of
\Fig{fig:spin-orbit}(b), the level repulsion between $|\oneup
\rangle$, $|\twodown \rangle$ is reproduced and an excellent fit
obtained for $|{v}_{12}^\downarrow| = 73\pm4~\mu$eV, $g^\ave_1 =
-0.03\pm0.04$, $g^\ave_2 = -0.10\pm0.06$, $\tilde g^\eff_1 =
1.90\pm0.04$, and $\tilde g^\eff_2 = 1.74\pm0.04$.  This yields a
difference of $\tilde g^\eff_j - g^\eff_j = 0.06$ for $j=1,2$, which
is consistent with \Eq{eq:pert-g}, since $4|{v}_{12}^\downarrow|^2 /
(\varepsilon_{2,0} - \varepsilon_{1,0})^2 = 0.06$.  This amount is the
contribution to $g^\eff_j$ of spin-orbit coupling between states $|1
\sigma \rangle$ and $|2 \bar \sigma \rangle$, and can be used as
estimate for $ 4 \overline{|v_{ij}^\sigma|^2}/d^2 = (2/\pi) \lambda_\so$;
the result, $\lambda_\so \simeq 0.09$, confirms that we are in the
regime of weak spin-orbit scattering.  Since the amount of 0.06
accounts for only 40\% (or 20\%) of the total reduction $g^\pure -
g^\eff_j$ of 0.16 (or 0.32) for orbital state $1$ (or 2), the
remainder must come from spin-orbit coupling to other $(i \neq 1,2)$
states, whose contribution is therefore very significant.

The eigenstates of $\hat H_\tot^\tildezp$ have the general form
\begin{eqnarray}
  \label{eq:true-eigenstates}
  | \oneup \rangle = \gamma  | \oneup \rangle_\tildezp
 + \eta  | \twodown \rangle_\tildezp \; , \qquad 
 | \twodown \rangle = - \eta^\ast  | \oneup \rangle_\tildezp
 + \gamma^\ast | \twodown \rangle_\tildezp \; , 
\end{eqnarray}
with $ |\gamma|^2 + |\eta|^2 = 1 $ (and similarly for $|\onedown
\rangle$, $|\twoup \rangle$).  The coefficients $\gamma(H)$ and
$\eta(H)$ are fully determined by the parameters obtained from the
above-mentioned fit, and are plotted in \Fig{fig:spin-orbit}(d).
Evidently, their relative magnitudes are ``interchanged'' as $H$ is tuned
through the avoided crossing, and the character of, \eg, $| \oneup
\rangle$ changes from being predominantly $| \oneup \rangle_\tildezp$
for small $H$ to predominantly $| \twodown \rangle_\tildezp$ for large
$H$.

This feature is also the reason for property (v).  Salinas \etalia\ 
were able to confirm this quantitatively by analyzing the
$H$-dependence of, say, the  rates $\Gamma^\ssL_{\oneup}
\equiv \Gamma^{\ssL -}_{G, \oneup}$ and $\Gamma^{\ssL}_{\twodown}
\equiv \Gamma^{\ssL -}_{G,\twodown}$ for a tunneling-off transition via
barrier L from one of the odd eigenstates $|\oneup \rangle_{N+1}$ or
$|\twodown\rangle_{N+1}$ into the even ground state $|G\rangle_N$.
According to \Eq{eq:Gamma-short}, these rates have the general
form\footnote{ The rightmost expressions in \Eqs{eq:somrates-1} and
  (\ref{eq:somrates-2}) were obtained by neglecting terms containing
  factors of the form ${}_\tildezp \langle j \sigma| c^\dag_{l \bar
    \sigma} | \G \rangle$.  These are not completely zero (because, by
  \Eq{eq:spin-orbit-eigenstates-pert}, $|j\sigma\rangle_\tildezp$ is
  not a pure spin-$\sigma$ state and contains some spin $\bar \sigma $
  components), but are smaller than those kept by a factor of order
  ${v}_{ij}^\sigma / (\varepsilon_{j \sigma, 0} - \varepsilon_{i
    \bar \sigma, 0})$. }
\begin{eqnarray}
  \label{eq:somrates-1}
  \Gamma^{\ssL}_{\oneup} & \, = \, &
  \sum_{l \sigma} \Gamma^\ssL_{l \sigma} 
  | \langle \G | c_{l \sigma} | \oneup \rangle|^2
 \, \simeq \, |\gamma|^2 \tilde \Gamma^{\ssL}_{\oneup} + 
 |\eta|^2 \tilde \Gamma^{\ssL}_{\twodown} \; , 
\\
\label{eq:somrates-2}
  \Gamma^{\ssL}_{\twodown} & \, = \, &
  \sum_{l \sigma} \Gamma^\ssL_{l \sigma} 
  | \langle \G  | c_{l \sigma} | \twodown \rangle|^2
 \simeq |\eta|^2 \tilde \Gamma^{\ssL}_{\oneup} + 
 |\gamma|^2 \tilde \Gamma^{\ssL}_{\twodown} \; , 
\\
\label{eq:somrates-3}
\tilde \Gamma^{\ssL}_{\oneup} &\, \equiv \, &
  \sum_{l} \Gamma^\ssL_{l \uparrow}
  | \langle \G  | c_{l \uparrow} |  \oneup \rangle_\tildezp |^2 
  \; , \qquad
\tilde  \Gamma^{\ssL}_{\twodown} \, \equiv \,
  \sum_{l} \Gamma^\ssL_{l\downarrow } 
  |  \langle \G  | 
c_{l \downarrow} | \twodown \rangle_\tildezp |^2  \; .
\end{eqnarray}
The $H$-dependence of $\Gamma^{\ssL}_{\oneup}$ and $
\Gamma^{\ssL}_{\twodown} $ is therefore contained completely in the
factors $|\gamma|^2$ and $|\eta|^2$, since $\tilde
\Gamma^{\ssL}_{\oneup}$ and $ \tilde \Gamma^{\ssL}_{\twodown} $ are
$H$-independent (and fully determined by $H=0$ data). Salinas \etalia\ 
compared the predictions of the \som\ for $\Gamma^{\ssL}_{\oneup}$ and
$ \Gamma^\ssL_{\twodown} $ [solid lines in \Figs{fig:spin-orbit}(e,f)]
with estimates [markers in \Figs{fig:spin-orbit}(e,f)] that were
extracted from the measured conductance peak areas mentioned in
property (v).  These estimates involved inverting
\Eq{eq:currentexpectation} for the current in order to express the
rates $\Gamma^{\ssL}_{\oneup}$ and $ \Gamma^{\ssL}_{\twodown} $
occurring therein in terms of the (measured) current and the
probabilities $P_\alpha$ that the grain is in eigenstate
$|\alpha\rangle$, and determining the $P_\alpha$ by solving a
corresponding master equation [\Eq{generalmasterequation}].  The
analysis is complicated by the fact \emph{all} accessible states
$|\alpha\rangle$ must be considered, not only $|\oneup\rangle$ and
$|\twodown \rangle$, including nonequilibrium excitations\footnote{
  The shifting of single-particle excitation energies due to
  nonequilibrium excitations discussed in
  \Sec{sec:beyound-orthodox-model} could not be resolved for the
  present grain, however.} that can be created when the bias voltage
$|eV|$ is sufficiently large. To keep the model tractable
nevertheless, some simplifying assumptions were made about unknown
model parameters, \eg\ that the ratios
$\Gamma^{\ssL}_{\alpha}/\Gamma^{\ssR}_{\alpha}$ are the same for all
$\alpha$ (see \Ref{salinas99} for details). The results depend, in
particular, on whether the relaxation rates $\Gamma^\inel$ for
nonequilibrium excitations are assumed smaller or larger than the
average tunneling rate $\Gamma_\tun$.  The ``slow-relaxation''
assumption [$\Gamma^\inel \ll \Gamma_\tun$, \Fig{fig:spin-orbit}(e)]
yields good agreement between the prediction (solid lines) of the
\som\ and the experimental estimates (markers).  For the
``fast-relaxation'' assumption [$\Gamma^\inel \gg \Gamma_\tun$,
\Fig{fig:spin-orbit}(f)] the agreement is not so good, though
qualitatively still reasonable.  The fact that the slow-relaxation
assumption works better is in accord with estimates \cite{agam97a}
(see \Sec{sec:estimating-relaxation}) that the energy relaxation rate
due to phonons is $\Gamma^\inel \approx 10^8 \mbox{s}^{-1}$, an order
of magnitude less than the estimates of $\Gamma_\tun \simeq 10^9
\mbox{s}^{-1} $ for the tunneling rates extracted from the detailed
analysis.

\subsection{Distributions for the effective $g$ factors}

The statistics of the fluctuations in $g_j^\eff$ have recently been
investigated theoretically by Matveev, Glazman and Larkin
\cite{matveev00}, and independently by Brouwer, Waintal and Halperin
\cite{brouwer00}, who studied the dependence of $g_j^\eff$ on the
direction of the applied field. In this section, we summarize those
aspects of these results that are relevant for ultrasmall metallic
grains.

\subsubsection{Distribution of $g_j^\eff$ for a random field direction}
\label{sec:estimate-g-eff}

\begin{figure}[t]
\centerline{\epsfig{figure=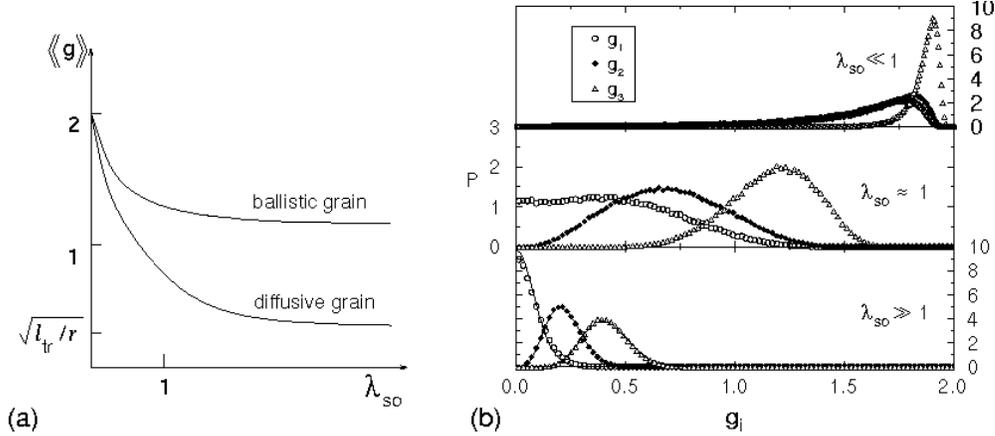,%
width=0.94\linewidth}}
\caption[Distribution of effective $g$ factors]
{Properties of ensemble-averaged $g$ factors of ultrasmall
metallic grains, as function of the strength
  $\lambda_\so \equiv \hbar \Gamma_\so / d$ of the spin-orbit
  interaction.  (a) Sketch \cite{matveev00} of $\llangle g \rrangle$
  as function of $\lambda_\so$, for a magnetic field applied in a
  random direction; the limits $\lambda_\so \ll 1$ and $\gg 1$
  illustrate \Eqs{eq:g-fluctuations-so-weak} and
  (\ref{eq:average-g^2}), respectively. (b) Distributions of the
  principal $g$ factors $g_1$, $g_2$, $g_3$ for weak, intermediate and
  strong spin-orbit coupling \cite{brouwer00}.  The data points were
  obtained from numerical simulations of a random-matrix version of
  the \som\ of \Sec{subsec:spin-orbit-theory} (see \cite{brouwer00} for
  details; the parameter $\lambda_\so$ corresponds to $\lambda^2$
  there, up to numerical coefficients). }
\label{fig:gfactor}
\end{figure}

Matveev, Glazman and Larkin \cite{matveev00} studied the statistics of
$g^\eff_j$, as defined by \Eq{g-effective}, for a magnetic field
applied in a random direction, in the two limits that the spin-orbit
interaction is weak or strong.  Their results are as follows:

(i) For \emph{weak spin-orbit scattering} ($ \lambda_\so \ll 1$), the
average $\llangle g \rrangle $ of the effective $g^\eff_j$ factors,
averaged either over an ensemble of grains or over different levels of
a single grain, can be calculated from the perturbative expression
(\ref{g-effective-pert}). MGL's calculation corresponds to replacing
$|{v}_{ij}^\sigma|^2$ by $\lambda_\so d^2/2 \pi $ and performing the
$\sum_i$ sum by assuming the bare energy levels $\varepsilon_{j,0}$ to
obey Wigner-Dyson statistics in the orthogonal ensemble. This leads to
the following result, in which the logarithm arises due to level
repulsion:\footnote{Level repulsion suppresses the probability $p_0$
  of two levels being very close by a factor $p_0 (\varepsilon_{i,0} -
  \varepsilon_{j,0}) = \pi^2 | \varepsilon_i - \varepsilon_j| / 6
  d^2$; the logarithmic divergence in the $\sum_j$ sum is cut off
  above by $d$, and below at the energy scale $|\varepsilon_{i,0} -
  \varepsilon_{j,0}| \sim |{v}_{ij}^\sigma | \propto d \sqrt
  \lambda_\so$, because of the additional level repulsion caused by
  the weak spin-orbit coupling.}
\begin{eqnarray}
  \label{eq:g-fluctuations-so-weak}
  \llangle g \rrangle = g^\pure \left[1 - {\pi\over 12}
  \lambda_\so \ln {1 \over \lambda_\so} \right] \; ,
\end{eqnarray}

(ii) For \emph{strong spin-orbit scattering} ($\lambda_\so \gg 1$),
\Eq{g-effective-pert} is not an adequate starting point.  Instead, MGL
calculated the linear term in \Eq{g-effective} perturbatively in the
magnetic field $H$, exploiting general properties of the exact $H=0$
eigenstates $|j \sigma \rangle$, which were treated using random
matrix theory for an ensemble of symplectic matrices.  They found the
distribution $P(g)$ of $g^\eff_i$ factors to be given by
\begin{eqnarray}
  \label{eq:P(g)-distribution}
  P (g) & = & 3 \sqrt{6 \over \pi} 
  {g^2 \over \llangle g^2 \rrangle^{3/2}}  \exp\left( - 
      {3 g^2 \over 2 \llangle g^2 \rrangle } \right),
\quad
 \llangle g \rrangle = (8/3 \pi)^{1/2} \llangle g^2 \rrangle^{1/2} , 
\\
\label{eq:average-g^2}
  \llangle g^2 \rrangle &=&  {6 \over \pi}{1\over \lambda_\so} 
+ b \, {\ltr \over r} \;   .
\end{eqnarray}
Here $b$ is a geometrical constant of order unity and $r$ is the system
size (\eg\ radius for a spherical or hemispherical grain); for a
diffusive grain, $\ltr = 3 \Ddiff / \vF$ is the transport mean free
path, and for a ballistic grain, $\ltr$ should, roughly speaking, be
replaced by $r$.  \Fig{fig:gfactor}(a) schematically illustrates the
behavior of $\llangle g \rrangle$ as function of $\lambda_\so$.  The
parametric dependencies of the two contributions to $\llangle g^2
\rrangle$ in \Eq{eq:average-g^2} can be interpreted intuitively as
follows \cite{matveev00}: the linear-in-$H$ contribution to the
energies $\varepsilon_{j \pm}$ of \Eq{g-effective} of the
time-reversed states $|j \pm \rangle$ is given by $ \langle j \pm |
\hat M_z | j \pm \rangle \muB H \equiv \pm \langle \hat M_z \rangle_j
\muB H $, where $\hat M_z = \hat l_z + g^\pure \hat s_z$ is the total
magnetic moment, and $\hat l_z$ and $\hat s_z$ are the orbital angular
momentum and spin, respectively (all in units of $\hbar$).  Therefore,
by \Eq{g-effective}, $g^\eff_j = 2 \langle \hat M_z \rangle_j $.  Now,
the angular momentum of an electron traversing the ``closed
trajectory'' corresponding to a discrete quantum level can be
estimated as the typical (directed) area $A_\typ$ covered by its
trajectory divided by the period $\hbar/ d$ of its motion, $\langle
\hat l_z \rangle \approx m A_\typ d/\hbar^2$, and its spin as $\langle
\hat s_z \rangle \approx 1/\sqrt N$, where $N \approx (\hbar / d)
\Gamma_\so$ is the number of spin-flips due to spin-orbit scattering
which it undergoes during one period. It follows that $\langle \hat
s_z \rangle^2 \approx 1/ \lambda_\so$, which corresponds to the first
term of \Eq{eq:average-g^2}.  Moreover, using $A_\typ \approx r^2
\sqrt \gdc $ [cf.\ \Sec{sec:orbitalmagnetism}] and
\Eqs{eq:g-dimensionlessconductance} and (\ref{eq:define-Thouless}) for
the dimensionless conductance $\gdc$, one finds that $\langle \hat l_z
\rangle^2 \approx {\ltr /r}$ [or ${\cal O} (1)$] for the diffusive [or
ballistic] case, which corresponds to the second term of
\Eq{eq:average-g^2}.

Note that, in contrast to the case of weak spin-orbit scattering considered in
most of this section, the orbital diamagnetic contribution, though small,
cannot be neglected for a grain with $\lambda_\so \gg 1$, since the spin
contribution to $\llangle g \rrangle^2$ is small as $1/ \lambda_\so$. MGL
suggested that for such a grain the parametrically small factor ${ \ltr / r}$
distinguishing the diffusive from the ballistic case [as illustrated in
\Fig{fig:gfactor}(a)] can be used to judge which of the two cases applies,
depending on whether its tunneling resonances are measured to have effective
$g$ factors of $g^\eff_j \ll 1$ or $1 \lesssim g^\eff_j \le 2$, respectively.

Note also that if one defines the correlation field ($H_\orb$) at
which orbital diamagnetism becomes dominant by equating the orbital
level splitting to the mean level spacing, $2 \langle \hat l_z
\rangle_j \muB H_\orb \approx d$, and substitutes $\langle \hat l_z
\rangle \approx m A_\typ d/\hbar^2$, one arrives at $H_\orb \approx
\Phi_0 / (\pi A_\typ)$, which is  (up to a factor of $\pi$) 
the criterion used in \Sec{sec:orbitalmagnetism} to derive
\Eq{eq:Horbital}. If, more formally,  $H_\orb$ is associated
with the field at which the crossover between
the symplectic and unitary ensembles,
driven by the orbital effects of the magnetic field,
is complete  \cite{Kravtsov92}, the result is 
again \Eq{eq:Horbital}.

\subsubsection{Distribution for anisotropic tensor of $g$ factors}
\label{sec:g-anisotropic}

Very recently, Brouwer, Waintal and Halperin (BWH) \cite{brouwer00}
pointed out that since spin-orbit coupling provides a mechanism for
the spin to ``notice'' ani\-so\-tro\-pies in the orbital wave
functions, the splitting of a Kramers doublet $\varepsilon_{j\sigma}$,
and hence the effective $g^\eff_j$ factor of a given conductance
resonance, should measurably depend on the \emph{direction} of the
applied magnetic field $\vec H$.  In general,
the splitting
of the Kramers doublet has the form  \cite{Slichter}
\begin{eqnarray}
(\varepsilon_{j+} - \varepsilon_{j-})^2 = 
 \muB^2 \vec H \cdot {\cal G}_j 
  \cdot \vec H \, = \,
 \muB^2 ( g_1^2 H_1^2 + g_2^2 H_2^2 + g_3^2 H_3^2) \; .
  \label{eq:deltaE}
\end{eqnarray}
Here $({\cal G}_{j})$ is a $3 \times 3$ tensor, the $g_a^2$
($a=1,2,3$) are its eigenvalues, called ``principal $g$ factors'' of
level $j$, and the $H_a$ are the components of $\vec H$ along the
level's ``principal axes'' (defined as the coordinate axes which
diagonalize ${\cal G}_j$).  ${\cal G}_j$ is isotropic (all three $g_a
= 2$) only in the absence of spin-orbit scattering. In its presence,
it becomes anisotropic, is subject to mesoscopic fluctuations, and for
large $\lambda_\so$ decreases as $1/ \lambda_\so$.  BWH calculated the
distribution $P(g_1,g_2,g_3)$ with respect to an ensemble of small
metallic grains of roughly equal size, neglecting orbital effects
(\ie\ assuming $\langle \hat l_z \rangle = 0$, as applicable for a
diffusive grain with $\ltr/r \ll 1$).  For sufficiently strong
spin-orbit scattering ($\lambda_\so \gg 1$), they found it to be of
the form
\begin{equation}
\label{eq:PgGSE0}
  P(g_1,g_2,g_3) \propto \prod_{a<b} |g_a^2 - g_b^2| \prod_{a} 
  e^{-3 g_a^2/2 \llangle \bar g^2 \rrangle} \; . 
\end{equation}  
Here $\bar g^2 = {1\over 3}(g_1^2 + g_2^2 + g_3^2)$ is the average of
$( \varepsilon_{j +} - \varepsilon_{j-})^2 / (\mu_B |H|)^2$ over all
directions of $\vec H$, and its average over the ensemble of grains,
$\llangle \bar g^2 \rrangle$, is $\propto 1 / \lambda_\so$ [in
agreement with \Eq{eq:average-g^2}].  Importantly, the factor $|g_a^2
- g_b^2|$ in \Eq{eq:PgGSE0} suppresses the probability for the ${\cal
  G}_j$ tensor to be isotropic (\ie\ $g_a$'s all equal), and hence
\emph{favors an anisotropic response} of a given level
$\varepsilon_{j\sigma}$ to an applied magnetic field.
\Fig{fig:gfactor}(b), which shows the distributions for the $g_a$'s
for weak, intermediate and strong spin-orbit coupling
\cite{brouwer00}, illustrates both how increasing $\lambda_\so$ leads
to smaller $g_a$'s, and the tendency for $g_1$, $g_2$ and $g_3$ to be
unequal.  By changing variables from $g_1$, $g_2$ and $g_3$ 
(with the convention that $g_1 < g_2 < g_3$) to $\bar
g$, $r_{12} = |g_1/g_2|$ and $r_{23} = |g_2/g_3|$,
 in terms of which
 \Eq{eq:PgGSE0} becomes
\begin{eqnarray}
  P (\bar g, r_{12}, r_{23})  
  &\propto&  
  {r_{23}^3 (1-r_{23}^2) (1-r_{23}^2 r_{12}^2)
  (1-r_{12}^2) \over (1 + r_{23}^2 + r_{23}^2 r_{12}^2)^{9/2}}\, 
  \bar g^8 e^{-9 \bar g^2/2 \llangle \bar g^2 \rrangle} \; , 
\label{eq:Pr12}
\end{eqnarray}
the anisotropy is seen to be rather strong: 
the typical value for $r_{12}$ (or $r_{23}$) 
is of order 1/3 (or 1/2), so if the direction of $\vec H$ is
varied arbitrarily, changes in $g^\eff_j$ by
a factor of  8 to 10 can be expected (independently
of the actual value of $\lambda_\so$, as long as it is $\gg 1$). 

\Eq{eq:PgGSE0} can be used to recover the results of MGL
\cite{matveev00} by associating their $g$ with $[({\cal
  G}_j)_{zz}]^{1/2}$, the $g^\eff_j$-factor for a magnetic field in
the $z$-direction (which is a random direction with respect to the
grain's principal axes). For
$g$ thus defined, the resulting distribution $P(g)$  is 
found to be given by \Eq{eq:P(g)-distribution}.

\subsection{Experimental results for strong spin-orbit interaction}

Experimentally, three reports of 
grains in the regime $\lambda_\so \gg 1$ 
have been published  to date, all involving Au,
whose large atomic number leads to a
much stronger spin-orbit interaction than for Al:
\begin{itemize}
\item[(1)] Salinas \etalia\ \cite{salinas99} observed values of
  $g_j^\eff$ in the range 0.5--0.8, as well as the occurrence of
  avoided crossings, in large Al grains doped with 4\% of Au
  impurities.
\item[(2)] Davidovi\'c and Tinkham \cite{davidovich99} observed
  $g_j^\eff = 0.28$ and 0.44 in two Au grains whose estimated radii
  (assuming hemispherical shapes) were 4.5~nm and 3~nm, respectively
  [these grains correspond to samples~1 and 3 in \Fig{fig:drago1}].
\item[(3)] Davidovi\'c and Tinkham \cite{davidovich00} observed
  $g^\eff_j$ values between 0.2 and 0.3 in an Au grain whose estimated
  radius (assuming hemispherical shape) was 1.5~nm. \ 
\end{itemize}
The grain of case (3) had the further interesting feature that the
measured level spacing was very much larger than the free-electron
estimate, and that the spacing between three subsequent levels
\emph{increased} with applied field in the large-$H$ regime, in a way
reminiscent of a spin multiplet.  Davidovi\'c and Tinkham have
suggested the possibility that this might perhaps reflect a ground
state having a \emph{total spin $s$ larger than} 1/2, favored by the
Coulomb interaction: if, \eg, the spacing between two orbital energy
levels at $\eF$ is smaller than the Coulomb interaction, say $u$,
between two electrons in the same orbital state, then, in analogy to
Hund's first rule in atomic physics, the ground state would be a $s=1$
spin triplet in which both orbital states are singly occupied by
electrons with parallel spins, because this allows their coordinate
wave functions to be antisymmetrized, reducing their Coulomb
interaction energy.  This is well-known to occur in
cylindrically-shaped semiconductor quantum dots \cite{tarucha96}, and
was predicted to be possible in metallic grains too by Brouwer, Oreg
and Halperin \cite{brouwer99}: they calculated the statistical
distribution $P(s)$ of the ground-state spin $s$ for an ensemble of
small normal metallic grains without spin-orbit coupling, and found an
appreciable probability for values other than 0 or 1/2 already for
interaction strengths well below the Stoner criterion, \ie\ $u$ well
below $d$ (see also \cite{Baranger} for similar work for quantum
dots).  For example, already at the quite modest interaction strength
of $u/d \simeq 0.4$, a ground-state spin of $s=1$ was found to be more
likely than $s=0$.  The probability to find non-minimal $s$ is reduced
if spin-orbit coupling is present, but may still be appreciable if the
Coulomb interaction is strong \cite{brouwer99b}.  More work is
required to demonstrate definitively that the clustering observed in
\cite{davidovich00} is due to excitations within spin multiplet and
not a non-equilibrium effect (of the sort described in
\Sec{sec:nonequilibrium}).

\subsection{Spin-orbit interaction in superconducting grains}

\begin{figure}[t]
\centerline{\epsfig{figure=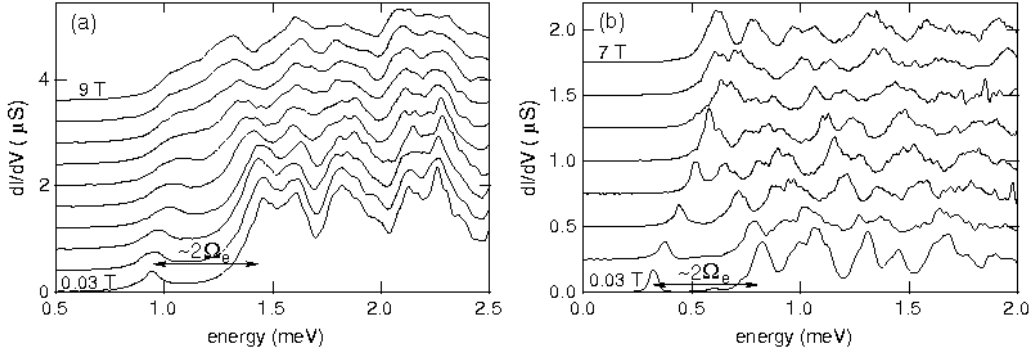,%
width=0.98\linewidth}}
\caption[The effect of spin-orbit scattering on
a superconducting grain]{(a) Odd-to-even tunneling spectrum
  \cite{salinas99} for an Al grain containing 4\% Au impurities, for a
  sequence of magnetic fields from 0.03 to 9~Tesla in 1~Tesla
  increments, at $T\!=\!15$ mK.  (b) Comparison data for a pure Al
  grain \cite{rbt97}, the same as in \Figs{fig:generic-IV},
  \ref{fig:Vg-V},
  \ref{fig:sc-spectra(h=0)}, \ref{fig:sc-magneticfield} and
  \ref{fig:sugap-scclusters}.  The curves are artificially offset for
  visibility.  }
\label{fig:sc-spin-orbit}
\end{figure}

The Au-doped Al grain mentioned in case (1) of the previous section
was sufficiently large that its odd-to-even tunneling spectrum, shown
in \Fig{fig:sc-spin-orbit}(a), exhibited the sizeable spectroscopic
gap typical of superconducting pairing correlations [cf.\ 
\Figs{fig:sc-spectra(h=0)}, \ref{fig:sc-magneticfield}].
Interestingly, this allowed Salinas \etalia\ \cite{salinas99} to study
how pairing correlations are affected by spin-orbit scattering, by
comparing the magnetic-field dependence of this spectrum to that of a
pure Al grain of similar size, shown in \Fig{fig:sc-spin-orbit}(b)
[the same grain as that discussed in \Secs{sec:gap-in-spectrum} and
\ref{sec:sub-gap-structures}].  We note the following features:

(i) \emph{Nonequilibrium broadening:}
For the Au-doped grain (a), which had no gate,
the resonance peaks  are somewhat broader than in the undoped grain (b).
This is believed {\em not} to be related to the Au impurities, 
but is instead a nonequilibrium effect,
since the spectrum of grain (b), which had a gate, showed a
similar broadening when purposefully 
tuned into nonequilibrium [cf.\ \Fig{fig:sugap-scclusters}(a)].

(ii) \emph{Reduced $g^\eff$ factors:}
In contrast to grain (b), for which  $g^\eff = 2 \pm 0.05$,
for grain (a) the first two peaks
move at low $H$ with slopes 
$g^\eff_1 /2+ g^\ave_1 = 0.41 \pm 0.03$ and $-g^\eff_2/2+ 
g^\ave_2 = -0.27 \pm 0.03$, suggesting values for $g^\eff$ in 
the range 0.5-0.8. This  behavior is typical of
strong spin-orbit scattering.

(iii) \emph{Spectroscopic gap unchanged:} At $H =0$, the
  spectroscopic gaps between the first two states are very similar for
  grains (a) and (b), namely $2 \Omega_e \simeq 0.26$ and 0.25~meV,
  respectively.  This nicely illustrates that spin-orbit coupling does
  not disrupt the superconducting pairing of time-reversed states, since
  it does   not break time-reversal symmetry.

(iv) \emph{Increased critical field:}
Whereas for grain (b) the spectroscopic gap between the
  lowest two states has disappeared already at 4~T, for grain (a) it
   decreases much slower with $H$, due to the reduced $g^\eff$ factors,
  and is non-zero even at 9~T.  This implies that the critical field
  for the paramagnetic breakdown of superconductivity [cf.\ 
  \Sec{sec:paramagnetic-breakdown}] is much increased by spin-orbit
  scattering, an effect familiar from thin films in a parallel magnetic
  field \cite{Meservey-94}.

(v) \emph{Avoided crossing:} At fields above 6~Tesla, the
  slope of the energy vs.~$H$ curve of the first peak of grain
  (a) changes sign (with the energy decreasing with increasing $H$ at
  high fields), suggesting an avoided crossing (with a minimum gap of
  130 $\mu$eV) with the higher-lying levels [similar to
  \Fig{fig:spin-orbit}(a)]. Interestingly, this implies that spin-orbit
  scattering modifies the details of the paramagnetic breakdown of
  pairing correlations: in contrast to a pure grain, where this
  breakdown is expected to occur rather abruptly when the Zeeman
  energy of a spin-1 state $|1\rangle$ (with one broken pair) crosses
  below that of the spin-0 state (with no broken pairs), here the
  first two levels never really cross, and are not pure spin
  states to begin with. Therefore the disruption of pairing
  correlations must occur gradually, as the spin content of the
  particle's ground state changes continuously in the avoided crossing
  region.

\newpage \section{Ferromagnetic grains}
\label{sec:ferromagnetism}

If the leads and/or island of a single-electron transistor are made
from an itinerant ferromagnetic material such as Co, Fe or Ni, which
have different densities of states for spin-up (majority-band) or
spin-down (minority-band) electrons, transport becomes
\emph{spin-dependent}. This leads to a number of interesting new
phenomena \cite{Ono}, such as a tunneling magnetoconductance (the
conductance of the SET depends on the relative orientation of the
magnetic moments of leads and island) or Coulomb oscillations as a
function of the applied magnetic field (which can shift the chemical
potential of the island, analogous to the effect of $\Vg$, if spin-up
and spin-down electrons have different density of states near $\eF$).
Such effects had previously been studied in micron-size ferromagnetic
islands
\cite{Ono,Fert1,Takahashi98,Hershfield,Shimada1,Korotkov1,%
  Barnas,Brataas,Imamura,Takahashi99}, and also in nm-scale cobalt
grains \cite{Schelp,Desmicht} at or above helium temperatures. 

This work has recently been extended to smaller size and/or lower
temperatures by Gu\'eron, Deshmukh, Myers and Ralph 
\cite{gueron99}, who studied individual nm scale Co grains using
tunneling spectroscopy. They found that these differed in several
interesting ways from non-magnetic grains, showing, in particular,
hysteretic behavior and a larger-than-expected number of low-energy
excitations.  Since a complete understanding of the observed phenomena
and a reliable theoretical framework for their analysis is still
lacking, we shall here just briefly mention the most important
experimental features, referring the reader to \Ref{gueron99} for details.

\subsection{Experimental results}
\label{sec:Co-experiments}

\begin{figure}[t]
\centerline{\epsfig{figure=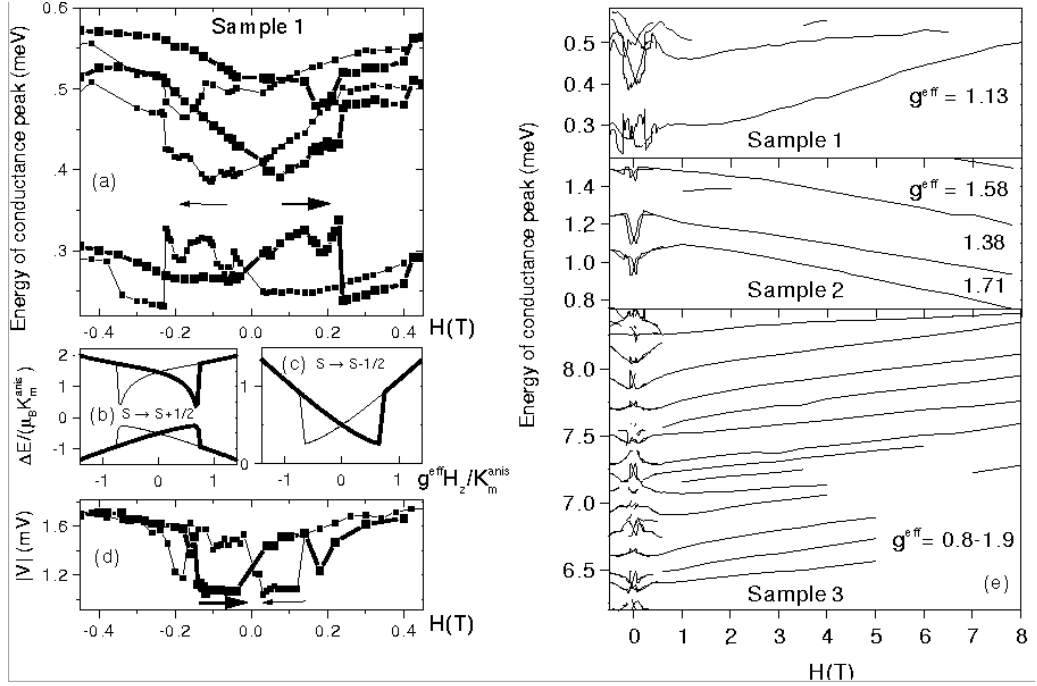,%
width=0.98\linewidth}}
  \caption[Magnetic-field dependence of the tunneling
  spectra of ultrasmall Co grains]{ Magnetic-field dependence of the
    tunneling spectra of several Co grains \cite{gueron99}.
  (a) Hysteresis curves
    showing the dependence of tunneling energies on $H$ for sample~1,
    at $T=20$~mK.  (b,c) Lowest-energy transitions calculated using the
    Hamiltonian $\Hspin$ of \Eq{eq:spin-Hamiltonian}, for $S=50$ and $H$
    oriented 45$^{\circ}$ from the easy axis, for the case where $S$
    increases during tunneling (b) and decreases (c).  The qualitative
    features are independent of the value of $S$.  (d) Voltage
    threshold for tunneling in one sample which exhibits
    antihysteretic behavior.  (e) Tunneling energies over a larger
    range of $H$, for three different samples.  }
   \label{fig:gueron} 
\end{figure}

Gu\'eron \etalia\ studied Co grains connected to Al leads via
Al$_2$O$_3$ tunnel junctions, without a gate.  Grain radii were
estimated to vary in the range 0.5-2~nm, which, assuming hemispherical
grains and a magnetic moment of 1.7$\muB$/atom (the value for bulk
Co), would imply magnetic moments of 90-6000~$\muB$.  The tunneling
spectra showed well-resolved peaks due to tunneling via discrete
electronic levels within each particle, qualitatively similar to those
of Al and Au grains, but with a very different magnetic-field
dependence.  The latter is summarized in \Fig{fig:gueron}, which
exhibits the following representative features:

(i) \emph{Hysteresis:} \Fig{fig:gueron}(a) shows the small-$H$
behavior of the transition energies for the lowest three conductance
resonances of a Co grain (sample 1).  Each line shows hysteresis as
$H$ is swept from $-0.45$~T to 0.45~T (thick lines) and back (thin
lines); it changes smoothly with $H$, except at two
``switching-fields'' $\pm \Hsw$, at which all three transition
energies simultaneously show sudden jumps. The qualitative explanation
for these features is that the direction of the magnetic moment vector
$\vec m$ of the grain changes as $H$ is increased from $-0.45$~T:
initially $\vec m$ is parallel to $\vec H$, then rotates toward and
past the easy direction (say $\hat z$) as the field approaches zero
and changes sign, and at $\Hsw$ abruptly changes direction to align
itself with $\vec H$ again.  We shall call this abrupt directional
change ``\magrev''. 

(ii) \emph{Antihysteresis:} \Fig{fig:gueron}(d) shows data for a
different grain which exhibits antihysteresis, \ie\ magnetization
reversal occurs before $H$ changes sign.  This can be explained by the
influence of a dipolar magnetic field oriented opposite to the applied
$H$, produced by a second magnetic grain adjacent to the one through
which electron tunneling occurs.  The reversed field from the second
particle can shift the hysteresis curve of the first so that its value
of $H_{\rm{sw}}$ can be negative, while the non-monotonic shifts at
large positive $H$ ($\simeq 0.2$ T) are understood as the magnetization
reversal of the second particle.  

(iii) \emph{Strong spin-up-down asymmetry:} \Fig{fig:gueron}(e) shows
the large-$H$ behavior of the Co tunneling spectra. Once $H$ is
significantly larger than $\Hsw$ (say $H >2$~T), all measurable
transition energies within a given sample have the \emph{same sign of
  slope} for $\delta E (H)$ vs.\ $H$. This contrasts with Al and Au
grains, which have lines of both slopes, with comparable conductance
amplitudes, and with a degeneracy at $H\!=\!0$. Thus, for Co there is
a strong asymmetry between the tunneling probability for spin-up or
spin-down electrons.

(iv) \emph{$g^\eff$ factors:} The effective $g^\eff_j$ factors
for individual transitions vary between 0.8 and 1.9
and fluctuate quite strongly from level to level.

(v) \emph{Many low-energy excitations:} The energy spacing between
tunneling peaks is somewhat less than 0.2~meV, much smaller than
expected from the mean level spacing $d$ for non-interacting
electrons: For grains of radii 0.5-2~nm, the calculated density of
states (including both sp and d bands) in Co, namely 0.88~eV$^{-1}{\rm
  atom}^{-1}$ \cite{papa}, implies $d$-values between 0.75 and 40~meV.
The level spacing for individual sp or d bands would be even larger.
Note also that the peak spacings of the three samples in
\Fig{fig:gueron} are surprisingly  similar to each other. 

\subsection{In search of a model}
\label{sec:Co-model?}

At present, a theoretical framework within
which all these features can be consistently understood is still
lacking. The
hysteretic behavior of point (i) can be understood qualitatively in
terms of a simple model Hamiltonian \cite{gueron99}:
\begin{equation}
\label{eq:spin-Hamiltonian}
\Hspin =- g^\eff \muB \vec{H}\cdot\vec{S}
-K^\anis_{m}\mu_BS_z^2/\sqrt{S(S+1)} \; , 
\end{equation}
It describes a quantum-mechanical spin $\vec S$ (representing $\vec
m$), Zeeman-coupled to a magnetic field, with an easy-axis anisotropy
in the $\hat z$ direction, where $K^\anis_m$ is an anisotropy energy
per unit $|\vec{m}|$. The magnitude of $S$ of the spin is assumed not
to change with $H$, since the exchange energy $U$ in Co is huge
(several eV), and $U$ determines the energy cost for changing $S$
(analogous to the role of the charging energy $\Ec$ for changing $N$).
By diagonalizing $\Hspin$ for a grain with $N$ or $N\pm 1$ electrons
(and spin $S$ or $S \pm 1/2$), the tunneling spectrum $\delta \E (H)
\equiv \E_\alpha^{N\pm 1} - \E_{\alpha'}^{N}$ can be calculated for
spin-increasing and -decreasing transitions (see \cite{gueron99} for
details).  The results, shown in \Fig{fig:gueron}(b,c), do exhibit
hysteresis and jumps reminiscent of the measured ones. In particular,
they show that the scale of the anisotropy constant $K^\anis_m$ can be
estimated as $\muB K^\anis_m \approx \E^{\rm jump}$, where $\E^{\rm
  jump}$ is the size of the jump in the tunneling energy at
$H_{\rm{sw}}$. This yields $\muB K^\anis_m \simeq 0.05$~meV for the
jumps of \Fig{fig:gueron}(a). Thus, for ultrasmall Co grains the
``characteristic magnetic energy scale'', which governs the low-lying
excitation spectrum in the meV range (and in that sense plays a role
analogous to the pairing parameter $\tilde \Delta$ for superconducting
grains), is the \emph{anisotropy} energy (and not the exchange energy
$U$).

This model fails to capture an important detail, however: it predicts
that the jumps occurring for $\delta \E (H)$ as $|H|$ increases through
$\Hsw$ will always have the same sign $(\pm)$ as the ensuing
large-$|H|$ slope of $ \delta \E (H)$, but the lowest line of
\Fig{fig:gueron}(a) is a counterexample (jump goes downward, then line
moves upward).  Gu\'eron \etalia\ proposed that the latter type of
behavior could be explained by assuming either that $K^\anis_m$
changes significantly, due to mesoscopic fluctuations, if the grain's
electron number changes, or that the magnitude of $\vec S$
(classically $\vec m$) does in fact change with $H$ (despite the
largeness of $U$), but no theoretical estimates for the likelihood of
these possibilities exist yet.  Making such estimates would have to
involve a somewhat more microscopic approach \cite{Kleff}, perhaps in
the spirit of a Stoner model, in which the degrees of freedom are
individual electrons in discrete energy levels, not just their total
spin $\vec S$.

Next, note that the spin asymmetry of point (iii) is not surprising
for Co, since the exchange field breaks the symmetry between spin-up
and down.  However, since Co is not fully spin-polarized ($P \approx
30\%$, \cite{Meservey-94}), \emph{a priori} lines of both slopes would
be expected to occur, even if in unequal numbers and with different
amplitudes.  Recent data (unpublished) do occasionally show a line
with a slope quite different in magnitude from that of the majority of
lines, although the sign of the slope is still the same.  The reason
why all observed transitions shift as a function of magnetic field
with the same sign of the slope is still a mystery.  Its resolution
may involve a better understanding of whether the states involved in
tunneling are primarily sp or primarily d states (whose density of
states near $\eF$ differ vastly), whether there is a difference in the
nature of the matrix elements for spin-up or spin-down electrons to
enter the many-body eigenstates of the grain, or whether spin-orbit
coupling within the magnetic nanoparticle is sufficiently strong that
the states involved in tunneling are coherent mixtures of spin states
(rather than being purely spin-up or spin-down).

A possible candidate for explaining the small level spacing of point
(iv), apart from nonequilibrium effects (which cannot be ruled out in
ungated devices), are ``spatially uniform\footnote{The contribution of
  exchange energy to the lowest-energy {\em non-uniform} spin-wave
  modes can be estimated by quantizing the spin-wave dispersion curve
  of Co within the size of a nanoparticle.  This gives an energy
  $(300$~meV$) (a/2r)^2$ where $a$ is a lattice spacing and $r$ is the
  grain radius \cite{ashcroft-mermin}, or $\approx$ 1~meV for a 2-nm
  grain.}  ($\vec k=0$) spin-wave modes'': classically speaking, these
involve fluctuations in the direction of $\vec m$ about its ground
state direction (say $\hat m_0$), quantum-mechanically speaking, they
involve different $s_m$ eigenvalues of the operator $\hat m_0 \cdot
\vec S$. The spin-wave excitation energy can be estimated as $\approx
2 \muB K^\anis_m \approx 0.1$~meV, which is comparable to the observed
inter-peak spacing. Note, though, that spin-waves of different quantum
numbers $s_m$ could be expected to have $s_m$-dependent slopes for
$\delta \E (H)$ vs.\ $H$, whereas no systematic tendencies for the
slopes are discernable in \Fig{fig:gueron}(e).  This might be held
against the spin-wave interpretation, but might also be a consequence
of mesoscopic fluctuations in $g^\eff_j$ factors [cf.\ point (iv)].

\subsection{Dynamics of magnetization reversal}

As the above discussion indicates, a number of features of the
tunneling spectra of Co grains still await detailed clarification.
This reflects the fact that the electronic structure of Co is much
more involved than that of Al or Au. However, independently of whether
all open questions can be resolved or not, the fact that magnetic
anisotropy causes each resonance energy in the tunnel spectrum to
shift reproducibly by on the order of 0.1~meV as $H$ is swept about
the hysteresis loop has a very interesting potential application: it
could be used as a tool to perform detailed studies of the
\emph{dynamics of magnetization reversal} in individual nm-scale
grains, since the jumps occuring in the tunneling spectra upon
reversal of the magnetization allow one to monitor precisely when, as
function of ramped applied field or of waiting time, this
magnetization reversal occurs. Thus this method would be complementary
to magnetic force microscopy \cite{Lederman}, Hall magnetometry
\cite{Lok}, and SQUID techniques \cite{Wernsdorfer97,Wernsdorfer99}
for studying magnetization reversal.

Overcoming the energy barrier between two different directions for the
magnetization occurs by thermal activation for large temperatures, or quantum
tunneling for sufficiently small $T$.  For Co grains of the present size ($m
\approx 90$-$ 6000 \muB$), the latter case would be an example of
\emph{macroscopic quantum tunneling} (MQT) of the magnetization, which has
been studied in great detail theoretically \cite{MQT}.  It should be very
interesting to use the relevant information in the literature in order to
predict the feasibility for observing MQT in an individual Co grain in a SET
geometry, and to estimate what grain sizes would be optimal for this purpose.
Wernsdorfer \etalia\ \cite{Wernsdorfer97} found deviations from purely
thermally activated behavior when studying an individual Co grain with a
diameter of 20~nm, using highly sensitive SQUID techniques. Since the grains
of Gu\'eron \etalia\ are much smaller, it is quite likely that they would show
such deviations too. It will be important, however, to clarify to what extent
damping, due to the coupling to the leads, reduces the chances for seeing MQT.
Conversely, if it turns out that MQT can indeed be observed, it might be
possible to investigate the effects of damping on MQT in a controlled way by
purposefully tuning the grain out of equilibrium.

\newpage \section{The Kondo box: a magnetic impurity in
an ultrasmall grain}
\label{sec:kondo-box}

In this section we briefly discuss what is expected to happen if a
normal ultrasmall grain contains a single magnetic impurity.  This
system, to be called a ``Kondo box'', was studied theoretically by
Thimm, Kroha and von Delft (TKvD) \cite{Thimm}. The Kondo box is
another example of the general rule of thumb that we repeatedly
encountered in previous sections: when the mean level spacing $d$
becomes larger than the energy scale characterizing the system's
correlations, in this case the Kondo temperature $\Tk$, interesting
new effects occur. TKvD showed that for $d \gtrsim \Tk$, level
discreteness strongly affects the Kondo resonance, in a
parity-dependent way, and predicted that this should lead to
measurable anomalies in the conductance through the grain.

For the impurity concentrations of 0.01\% to 0.001\% that yield a
detectable Kondo effect in bulk alloys, an ultrasmall grain of
typically $10^4-10^5$ atoms will contain only a single impurity, so
that inter-impurity interactions need not be considered. TKvD thus
considered a single impurity in an ultrasmall metallic grain,
described by the impurity Anderson model \cite{Andersonmodel,Hewson}
with a discrete conduction band:
\begin{eqnarray}
H= H_0  + \varepsilon_d \sum _{\sigma } c^{\dag}_{d \sigma }
c^{\phantom{\dag}}_{d\sigma } + v \sum _{j,\sigma} 
(c^{\dag}_{j\sigma}  c^\ds_{d \sigma} + c^\dag_{d\sigma}  c^{\ds}_{j\sigma})  
+ U c^\dagger_{d+} c^\ds_{d+} c^\dagger_{d-} c^\ds_{d-} \; 
\label{H-kondo}
\end{eqnarray}
where $H_0 = \sum _{j,\sigma } \varepsilon_j
c_{j\sigma}^{\dag}c^{\phantom{\dag}}_{j\sigma}$.  Here $\sigma$
denotes spin and the $c^\dagger_{j \sigma}$ create conduction
electrons in the discrete, delocalized eigenstates $|j \sigma \rangle$
of the ``free'' system (i.e.\ without impurity).  Their energies,
measured relative to the chemical potential $\mu$, are taken uniformly
spaced for simplicity: $\varepsilon_j = j \KDelta + \bar \varepsilon_0
- \mu$.  $c^\dag_{d \sigma}$ creates a spin-$\sigma$ electron in the
localized level of the magnetic impurity, which has bare energy
$\varepsilon_d$ far below $\eF$, and a Coulomb energy cost $U (\to
\infty)$ for being doubly occupied.  The hybridization matrix element
$v$ between the local impurity level and the conduction band is an
overlap integral between a localized and a delocalized wave-function,
and, due to the normalization of the latter, scales as $v \propto
\Vol^{- 1/2}$.  Thus the effective width of the impurity level,
$\Gamma = \pi v^2 /\KDelta$, is volume-independent, as is the bulk
Kondo temperature, $\Tk = \sqrt{2\Gamma D/ \pi }\ {\rm
  exp}(-\pi\varepsilon _d/2\Gamma)$, where $D$ is a high energy band
cutoff.  To distinguish, within the grand canonical formalism, grains
for which the \emph{total} number of ($c_{j\sigma}$ \emph{and}
$c_{d\sigma}$) electrons is even or odd, $\mu$ is
chosen either on ($\mu = \bar \varepsilon_0$) or half-way between two
($\mu = \bar \varepsilon_0 + \KDelta/2$) single-particle levels,
respectively \cite{vondelft96,braun97}.

\begin{figure}[t]
\centerline{\epsfig{figure=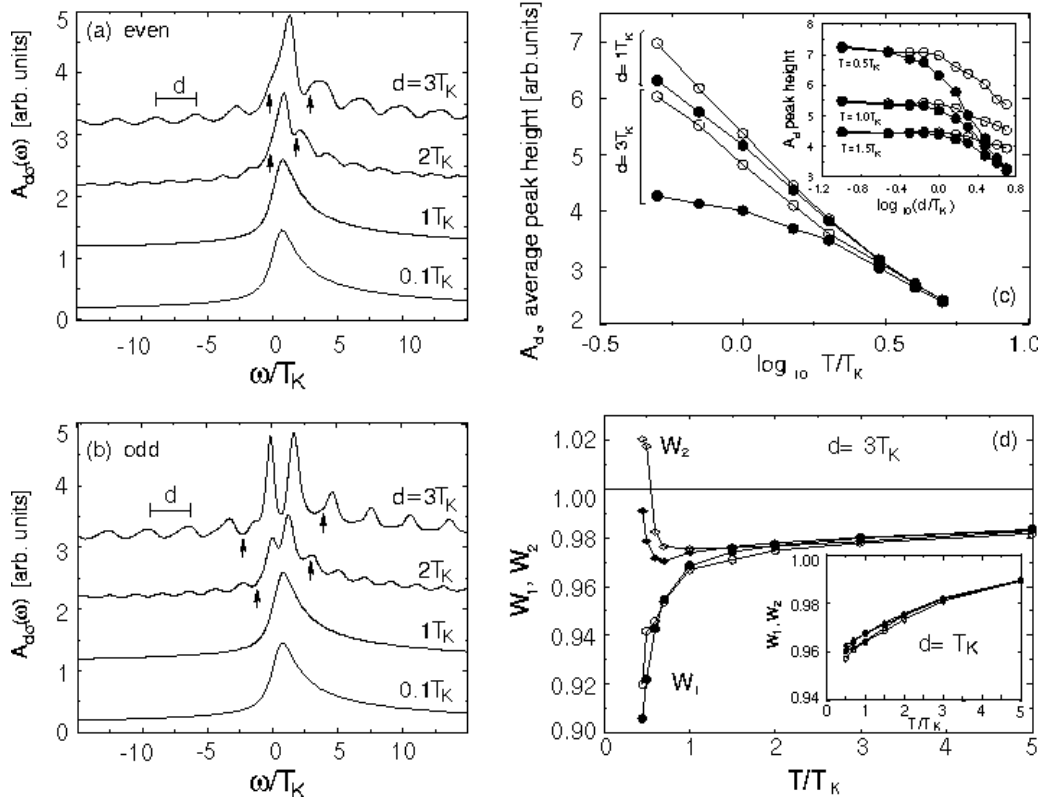,%
width=0.98\linewidth}}
  \caption[Results for a magnetic impurity 
  in an ultrasmall metallic grain.]{Results for a magnetic impurity in
    an ultrasmall metallic grain. (a,b) Impurity spectral function
    $A_{d\sigma}(\omega )$ for various values of $\KDelta $ at
    $T=0.5\Tk$; (a) even, (b) odd total number of electrons. The
    individual curves are vertically offset by one unit each.  (c)
    Even/odd dependence of the average peak height of the Kondo
    resonance, as function of $T/\Tk$.  For an even box ($\circ$),
    $A_{d\sigma}$ was averaged over a range $\KDelta$ centered on its
    central sub-peak, for an odd box ($\bullet$) over a range $2
    \KDelta$ centered on its central two sub--peaks [as indicated by
    arrows in (a) and (b)].  The inset shows the same quantity as function
    of $\KDelta/\Tk$.  Numerical uncertainties are smaller than the
    symbol sizes.  (d) Anomalous temperature dependence of the weights
    $W_1$ (circles) and $W_2$ (diamonds) of the first two conductance
    peaks of an even grain ($\circ$,$\lozenge$) and an odd grain
    ($\bullet$,$\blacklozenge$), for $\KDelta = 3 \Tk$ and $\KDelta =
    \Tk$ (inset).  }
   \label{fig:kondo}
\end{figure}

The system's correlations can be characterized in terms of 
the spectral density $A_{d\sigma}(\omega)$ of 
the impurity Green's function $G_{d\sigma} (t) = -i \theta(t)
 \langle \{ c_{d \sigma} (t), c_{d \sigma}^\dagger (0)\} \rangle$,
which TKvD calculated using the 
standard noncrossing approximation (NCA) \cite{NCA.ref,kuramoto}. Their
results are summarized in \Fig{fig:kondo}.

\begin{itemize}
\item[(i)] \emph{Bulk limit:}
For $\KDelta \ll T$, the impurity's spectral density is
indistinguishable from the bulk case ($\KDelta \to 0$) [lowest lines
in \Figs{fig:kondo}(a,b)]: it shows the familiar Kondo resonance near
$\omega =0$, with width of order $\Tk$ when $T \ll \Tk$, which arises
due to coherent virtual transitions between the impurity level and the
conduction band.  
\item[(ii)] \emph{Subresonances:}
When $\KDelta$ is increased well beyond $T$,
however, the Kondo resonance
 splits up into a set of individual sub-peaks, reflecting
the discreteness of the conduction band [higher lines of
\Figs{fig:kondo}(a,b)].
Nevertheless, the Kondo resonance retains its main feature, namely
significant spectral weight within a few $\Tk$ around the Fermi
energy, up to the largest ratios of $\KDelta/\mbox{max}(T,\Tk)$
($\simeq 5$) that were considered.  This implies that the Kondo
correlations induced by the spin-flip transitions between the
impurity level and the lowest-lying unoccupied $j$-levels persist up to
remarkably large values\footnote{Actually, remnants 
of Kondo correlations have been demonstrated to exist even in 
molecular systems  \cite{Fulde1,Fulde2}.}
 of $\KDelta/\mbox{max}(T,\Tk)$.
\item[(iii)] The \emph{Kondo correlations} do \emph{weaken}
systematically with increasing $\KDelta$, however, as
can be seen in the inset of \Fig{fig:kondo}(c), which shows the
average peak height of the Kondo resonance (which quantifies the
``strength'' of the Kondo correlations) as function of $\KDelta$ at
fixed $T$: the peak height drops logarithmically with increasing
$\KDelta$ once $\KDelta$ becomes larger than about $T$.  Conversely,
at fixed $\KDelta$, it drops logarithmically with increasing $T$ once
$T$ becomes larger than about $0.5 \KDelta$ [main part of
\Fig{fig:kondo}(c)], thus reproducing the familiar bulk beha\-vior.
Qualitatively, these features are readily understood in perturbation
theory, where the logarithmic divergence of the spin flip amplitude,
$t(\omega ) \propto \sum_{ j \neq \omega} {f(\varepsilon_j) \over
  \omega - \varepsilon_{j}}$, is cut off by either $T$ or $\KDelta$,
whichever is largest.
\item[(iv)] {\em Parity Effects:}\/ For $\KDelta \gg T$, the even and
  odd spectral functions $A_{d \sigma}$ in \Fig{fig:kondo}(a) and (b)
  differ strikingly: the former has a single central main peak,
  whereas the latter has two main peaks of roughly equal weight.  This
  can be understood as follows: For an even grain, spin-flip
  transitions lower the energy by roughly $\Tk$ by binding the $d$
  electron and the conduction electrons into a Kondo singlet, in which
  the topmost, {\em singly}\/-occupied $j$ level of the free Fermi sea
  carries the dominant weight, hence the single dominant peak in
  $A_{d\sigma}$.  For an odd grain, in contrast, the free Fermi sea's
  topmost $j$ level is {\em doubly}\/ occupied, blocking
  energy-lowering spin-flip transitions.  To allow the latter to
  occur, the two electrons in this topmost level are redistributed
  with roughly equal weights between this and the next-higher-lying
  level, causing two main peaks in $A_{d\sigma}$ and reducing the net
  energy gain from $\Tk$ by an amount of order $\KDelta$.  This energy
  penalty intrinsically weakens Kondo correlations in odd relative to
  even grains; indeed, the average $A_{d \sigma}$ peak heights in
  \Fig{fig:kondo}(c) are systematically lower in odd than in even
  grains, and more so the larger $\KDelta$ and the smaller $T$.
\item[(v)] \emph{Anomalous conductance:} The Kondo correlations for
  $A_{d\sigma}(\omega)$ affect the properties of the conduction
  electron density of states on the grain, in a way which should be
  detectable by using single-electron-tunneling spectroscopy and
  measuring the differential conductance $G(V)$ through the grain.
  TKvD showed that contributions to $G(V)$ due to the bare conduction
  electron density of states interfere with contributions coming from
  the Kondo resonance in $A_{d\sigma}(\omega)$, in a way reminiscent
  of a Fano resonance \cite{fano.61}.  This interference causes the
  conductance to acquire a distinct, anomalous $T$ dependence: the
  \emph{weights} $W_j$ under the individual peaks (labeled by $j$) of
  $G(V)$ become $T$ dependent. (In contrast, the weight $W_0$ under an
  individual peak of the bare conductance, say $G_0(V)$, is $T$
  independent, since the $T$ dependence of the peak shapes of $G_0$ is
  determined solely by the derivative of the Fermi function.)  This
  anomalous behavior is illustrated in \Fig{fig:kondo}(d), which shows
  the $T$ dependence of the weights $W_1$ and $W_2$ of the 1st and 2nd
  conductance peaks that occur when $V$ is increased from 0: When $T$
  decreases at fixed $\KDelta = \Tk$, both $W_1$ and $W_2$ decrease,
  while at fixed $\KDelta= 3 \Tk$, $W_1$ decreases whereas $W_2$
  increases.  The fact that the weights can either increase or
  decrease with decreasing $T$ results from the constructive or
  destructive Fano-like interference effects mentioned above.
  Moreover, at the larger value for $\KDelta$, both $W_1$ and $W_2$
  develop a {\em parity effect in the strength of their $T$
    dependence.}\/
\item[(vi)] \emph{Strength of anomalies:} TKvD estimated
  that the Kondo-induced $T$
  dependence in peak weights in \Fig{fig:kondo} should be strong
  enough to be experimentally detectable.  For e.g.\ 
  an Fe impurity in an even Cu grain of size $(3\mbox{nm})^3$
  ($\KDelta = 30 \mbox{K}$, $\Tk\simeq 10\mbox{K}$), cooling from $T=
  2\Tk$ to $0.5\Tk$ should change $W_1$ by $\simeq 7\%$.  (TKvD
  expect   $W_1$ to change some more as $T$ is lowered further, but
  their   numerics became unreliable in this regime.)
\end{itemize}

To conclude this section, we note that since the effects discussed
above result, above all, from the discrete density of states near
$\eF$, they should be generic for ultrasmall grains.  In other words,
they should be robust against including randomness in the model, like
$j$-dependent level spacings $\KDelta_j$ and hybridization matrix
elements $v_j$.

\newpage \section{Summary and outlook}
\label{sec:summary}

The technique of single-electron-tunneling spectroscopy on ultrasmall metallic
grains has proved to be a very fruitful way of probing electron correlations
in metals, and the way in which these are modified by mesoscopic fluctuations
and level discreteness.  Let us briefly summarize the main conclusions reached
in each of the preceding sections:

\Sec{sec:normal-grain-magfield}: For normal grains, the effect of an
applied magnetic field is simply to lift the Kramers degeneracy of
time-reversed states. This can be used to determine the grain's
\emph{number parity}.  In ultrasmall grains, the effect of a magnetic
field on orbital motion is negligible.

\Sec{sec:superconductivity}: For largish Al grains, the observation of
a distinct spectral gap in even grains and its absence in odd grains
is clear evidence for the presence of \emph{superconducting pairing
  correlations}. 
 These can be satisfactorily described using the
simple \dbcsm\ introduced in \Sec{sec:model}.  The blocking of some
levels by unpaired electrons leads to various measurable parity
effects; among these, a pairbreaking-energy parity effect should be
observable in experiments of the present kind, provided the grain size
can be better controlled.  The dominant mechanism by which a magnetic
field destroys pairing correlations in ultrasmall grains is Pauli
paramagnetism.  Decreasing the grain size softens the first-order
transition observed for thin films in a parallel field, by reducing
the number of spins flipped from being ma\-cros\-copically large for
$d \ll \tilde \Delta$ to being of order one for $d \simeq \tilde
\Delta$.  The grand-canonical variational BCS approach fails for $d
\gtrsim \tilde \Delta$; nevertheless, it yields a useful framework for
a qualitative analysis of the experiments, which had $d \lesssim
\tilde \Delta$.

\Sec{sec:crossover}: The \emph{crossover} of the behavior of
superconducting pairing correlations from the bulk limit $(d \ll
\tilde \Delta)$ to the fluctuation-dominated regime $(d \gg \tilde
\Delta)$ is parity dependent and completely smooth. This remains true
for systems with non-uniform rather than uniform level spacings. --
Very remarkably, the \dbcsm\ has an exact solution, due to Richardson,
with which $T=0$ properties can be calculated rather easily.
Finite-temperature properties for finite-sized systems can be
calculated quite reliably with the correlated static-path
approximation (provided $T> T_{\rm CSPA}$).  Thus,
the only remaining open problem is the development of
canonical finite-$T$ methods that remain quantitatively reliable for
$d \gtrsim \tilde \Delta$. 
 -- The spin susceptibility $\chi (T)$ of an odd grain shows
an interesting re-entrant behavior even for $d \gg \tilde \Delta$,
which might be a way to detect remnants of pairing correlations in the
fluctuation-dominated regime.

\Sec{sec:nonequilibrium}: The excitation spectra of
\emph{nonequilibrium} grains show \emph{clusters of resonances}. The
spacing between clusters is comparable to the mean level spacing, but the
spacings between subresonances of the same cluster are much smaller than
$d$.  Each cluster of resonances can be identified with one excited
single-electron state of the metal grain.  Subresonances reflect
energy shifts (of order $d/\gdc$) due to different nonequilibrium
occupancy configurations of other single-electron states. These energy
shifts are manifestations of mesoscopic fluctuations of the matrix
elements of the electron-electron interaction. -- Superconducting
grains that are tuned out of equilibrium by a gate voltage show subgap
structures in odd-to-even but not in even-to-odd tunneling spectra;
these features can be attributed to nonequilibrium excitations
generated by \emph{cotunneling} processes. -- As a function of increasing
excitation energy, the nature of the excitation spectrum has been observed
to change from consisting of discrete peaks to being continuous,
the crossover scale being $\varepsilon \sim \ETh$;
 the observed details of this crossover are in
accord with theoretical expectations for the behavior of quasiparticle
lifetimes in disordered interacting  systems.

\Sec{sec:spin-orbit}: The \emph{spin-orbit interaction} leads to
reduced $g_j^\eff$ factors for the splitting of Kramers doublets in a
magnetic field, and to avoided crossings. -- For grains with
superconducting pairing correlations, the critical field for which
these are expected to break down is increased significantly relative
to pure samples; the $H=0$ spectroscopic gap remains unchanged,
however, as expected due to time-reversal symmetry.  -- The $g^\eff_j$
factors exhibit mesoscopic fluctations, which can be studied
theoretically using random matrix theory.  Interestingly, the
Zeeman-response to a magnetic field has been  predicted to be anisotropic in
the presence of spin-orbit scattering, implying that 
upon changing the direction of the applied field, the measured
$g^\eff_j$ factors should vary strongly (by a factor of up to 8-10). 

\Sec{sec:ferromagnetism}: \emph{Ferromagnetic Co grains} have
hysteretic tunneling spectra. These can be used as a tool to measure
the switching field at which the grain's magnetization undergoes a
sudden change in direction as an applied magnetic field is ramped.
There is a strong asymmetry for the tunneling probability of spin-up
versus spin-down electrons. The spacing between low-lying excitations
is much smaller than expected from the estimated free-electron mean
level spacing $d$ (and remarkably uniform from
one grain to the next). However, it is comparable to the grains'
anisotropy energy (per unit magnetization), which for ultrasmall Co
grains is the ``characteristic magnetic energy scale'' governing the
low-lying excitation spectrum in the meV range.  Many features of
these measurements are not yet understood in detail, and further work,
both experimental and theoretical, will be needed to clarify them.

\Sec{sec:kondo-box}: For an ultrasmall grain containing a \emph{magnetic
Kondo impurity},  the Kondo resonance is strongly affected,
in a parity-dependent way, when the mean level spacing in the grain
becomes larger than the Kondo temperature.  The tunneling spectrum is
predicted to show Kondo-induced Fano-type resonances of measurable
size, with an anomalous dependence on temperature and level spacing.

\emph{Prospects for future work:} 

\emph{Experiment:}
In the current generation of experiments, the grain's actual size and
shape cannot be determined very accurately.  It would be a great
advance if fabrication techniques could be developed to the point that
grains can be used which have been custom-made, by chemical
techniques, to have well-defined sizes and shapes (\eg\ spherical).
This would significantly reduce the uncertainties which one presently
encounters when estimating characteristic parameters of the grain,
such as the single-particle mean level spacing $d$, the dimensionless
conductance $\gdc$ or the Thouless energy. Moreover, it would allow
systematic studies of the dependence of various quantities on grain
size or mean level spacing [for example, it would be interesting to
try to do this for the pairbreaking energies $\Omega_e, \Omega_o$ of
\Fig{fig:spectral-gap}(b)].  Encouragingly, the feasibility of using
chemically-prepared grains in SETs has already been demonstrated
several times \cite{Klein-97,sunmurray,{Schmid97}}, though the
resulting devices have not yet been used for single-electron-tunneling
spectroscopy.

Moreover, most of the devices that have been studied so far had no
gate, because it is technically very challenging to add one. It would,
however, be very useful if ways could be found for routinely
fabricating gates, because that allows both the grain's number parity
and the amount of nonequilibrium to be adjusted controllably.

\emph{Theory:} The behavior of superconducting pairing correlations in
an individual ultrasmall grain can now be regarded as a subject that
is well understood.  It would be interesting to try to use the
insights that have been gained for a single grain in order to now
study systems of several coupled grains: what, for example, is the
fate of the Josephson effect between two coupled grains as their sizes
are reduced to the point that $d \sim \tilde \Delta$?

The experiments on nonequilibrium effects and spin-orbit interactions
revealed the importance of mesoscopic fluctuations in remarkably
direct ways, and are well understood in terms of theories that treat
these fluctuations  via random matrix theory.  The success of the
latter approach stems from the fact that for chaotic systems such as
irregularly-shaped grains, ``details do not matter'' much, so that a
description in terms of random matrices is appropriate.

At present, the most intruiging open theoretical problem seems to be
that of finding a suitable framework within which to analyze and
interpret the experiments on ferromagnetic Co grains. The observed
phenomena are rather rich and complex, and it is currently not at all
clear which details ``matter'' and which do not. More experimental
studies on this system will be needed as guidelines for theoretical
attempts to construct a useful model for it.

A further stimulus for theoretical work on ferromagnetic grains might
come from the prospect of exploiting the hysteretic properties of
their tunneling spectra to study the \emph{dynamics of magnetization
  reversal}: it would be interesting to estimate the feasibility for
seeing macroscopic quantum tunneling in this system.  In particular,
one should try to determine whether (or to what extent) this
phenomenon can survive when a current flows through the grain.

\newpage
\appendix

\section{Superconducting leads}
\label{sec:superconductingleads}

For the case of superconducting leads  in the absence of an external
magnetic field, the analysis of Sections~\ref{sec:generalhamiltonian}
and \ref{sec:tunnelingcurrent} has to be modified as follows (for
generality, we continue to allow for the presence of a gate; if none
is present, set $C_\ssg = 0$ below): Firstly, the lead Hamiltonian
(\ref{eq:generalhamiltonian-leads}) must be replaced by the BCS
Hamiltonian for Bogoljubov quasiparticles,
\begin{eqnarray}
  \label{eq:schamiltonian-leads}
  H_r &=& \sum_{k \sigma} (E_{kr} + e V_r)
  \gamma^\dagger_{k \sigma r} \gamma_{k \sigma r}  \; , \qquad
   E_{kr} = \sqrt{ \varepsilon_{kr}^2 + \Delta^2_r} \,  ,
\\
  \label{eq:Bogoljubov-app}
   \gamma_{k \uparrow r} &=&
     u_{kr} c_{k \uparrow r} -  v_{kr} c^\dagger_{- k \downarrow r} 
      \; , \qquad 
   \gamma_{k \downarrow r} =
    u_{kr} c_{k \downarrow r } +  
v_{kr} c^\dagger_{ -k \uparrow r} \; , 
\\ u_{kr} &=& \mbox{[} 
\half \left( 1 + \varepsilon_{kr} / E_{kr} \right) 
\mbox{]}^{1/2} \,  \qquad 
v_{kr}=  \mbox{[} \half \left( 1 - \varepsilon_{kr} / E_{kr} \right)
     \mbox{]}^{1/2} \, ,
\end{eqnarray}
and $H_\tun$ of (\ref{eq:generalhamiltonian-tunnel-ss}) should be
rewritten in terms of the $\gamma$'s.  Secondly, the golden rule
expressions for the tunneling rates now are [instead of
(\ref{eq:sigmar+}) to (\ref{eq:sigmar-short})]:
\begin{eqnarray}
  \label{eq:sigmar+.sc}
  \Sigma_{\alpha \alpha'}^{r+} &=& {2 \pi \over \hbar} \sum_{k \sigma} 
|T_{k \sigma, \alpha \alpha'}^{r \ast} |^2   \left[ f(E_{kr}) |u_{kr}|^2 
  \delta(\E_\alpha - \E_{\alpha'} - E_{kr} - e \bar V_r) \right. 
\\ & & + \left.  
\bigl(1 - f(E_{-kr})\bigr) |v_{kr}|^2 
  \delta(\E_\alpha - \E_{\alpha'} + E_{-kr} - e \bar V_r) \right]
\\
\label{eq:sigmar+short.sc}
 & \simeq &  f (\E_\alpha - \E_{\alpha'} - e \bar V_r) \, 
\N_{r,\qp} (\E_\alpha - \E_{\alpha'} - e \bar V_r) \, 
  \Gamma_{\alpha \alpha'}^{r+} \; , 
\\
  \label{eq:sigmar-.sc}
  \Sigma_{\alpha \alpha'}^{r-} &=&    {2 \pi \over \hbar} \sum_{k\sigma}
|T_{k \sigma, \alpha \alpha'}^{r} |^2 
\left[\bigl(1 - f(E_{kr})\bigr) |u_{kr}|^2 
  \delta(\E_\alpha - \E_{\alpha'} + E_{kr} - e \bar V_r) \right. 
\\ & & + \left.   f(E_{-kr}) |v_{kr}|^2 
  \delta(\E_\alpha - \E_{\alpha'} - E_{-kr} - e \bar V_r) \right]
\\
\label{eq:sigmar-short.sc}
 &= & f (\E_\alpha - \E_{\alpha'} + e \bar V_r)] \, 
\N_{r,\qp} (\E_\alpha - \E_{\alpha'} + e \bar V_r) \, 
  \Gamma_{\alpha \alpha'}^{r-} \; , 
\end{eqnarray}
where $\N_{r,\qp} (\omega)$ is the BCS quasiparticle density of
states in lead $r$ [see \Eq{eq:sc-densityofstates}].  To arrive at
(\ref{eq:sigmar+short.sc}) and (\ref{eq:sigmar-short.sc}), we took the
limit $T \ll \Delta_r$ [so $f(E_{kr}) \simeq 0$] and used [instead of
\Eq{eq:spectralGamma}]
\begin{eqnarray}
  \label{eq:spectralGamma-sc}
{2 \pi \over \hbar} \sum_{k}
   T^{r \ast}_{kl \sigma} T^r_{kl' \sigma} \, 
{\textstyle 
\left\{ \begin{array}{c} |u_{kr}|^2 \\ |v_{kr}|^2 \end{array} \right\}}
\, \delta( \omega - E_{kr} )  =  \delta_{l l'} \, 
 \Gamma_{l\sigma }^r  \,  \N_{r,\qp} (\omega) 
\end{eqnarray}
(the $\varepsilon$-antisymmetric part of $|u_{kr}|^2$ and $|u_{kr}|^2$
does not contribute).

\Eqs{eq:sigmar+short.sc} and (\ref{eq:sigmar-short.sc})
are the results used in \Eq{eq:sigmarplusminusshort.sc}
of \Sec{sec:sc-leads}.

\newpage \section{Richardson's exact solution of discrete BCS model}
\label{app:richardson}

This appendix summarizes some of the main features of
Richardson's exact solution of the \dbcsm.
We begin in \App{sec:derive-rich}
by deriving in pedagogical detail, following 
\cite{vondelft-ankara99}, some of the key results of Richardson's exact
solution of the \dbcsm\ of \Sec{sec:model}. 
In \App{app:electrostatic-analogy} we transform 
the eigenvalue equation derived below
[\Eq{eq:richardson-eigenvalues}] 
into a less singular form more convenient 
for numerical solution, and in 
\App{app:correlators} we present analytic 
formulas needed to calculate the correlators 
$\langle b^\dag_j b_j^\ds\rangle $ and $\langle b^\dag_i
b_j^\ds\rangle $ exactly.

\subsection{Derivation of eigenstates and eigenvalues}
\label{sec:derive-rich}

\subsubsection{The hard-core boson problem}
For presented purposes it suffices to 
consider the pair Hamiltonian\footnote{
In this appendix the shorthand $g = \lambda d$ is useful,
since Richardson's solution depends only on this combination.}
 [cf.\ \Eq{1}]
\begin{eqnarray}
 &&  H_\U = \sum_{ij}^\U \left( 2 \varepsilon_j \delta_{ij} -
 \; g \right)  b_i^\dagger b^\ds_j \; ,
\qquad (\mbox{where}
\; g = \lambda d) , 
 \end{eqnarray}
 where $j$ runs over a set $U$ of $N_U$ of non-singly-occupied single-particle
 levels $\U$ (degenerate levels with $\varepsilon_i = \varepsilon_{j}$
 for $i \neq j$ are allowed), on which the pair-creation operators
 $b^\dagger_j = c^\dagger_{j+} c^\dagger_{j-}$ satisfy ``hard-core
 boson'' relations,
\begin{eqnarray}
\label{hard-core-boson-1-app}
  b^{\dagger 2}_j = 0, \qquad
\label{hard-core-boson-2-app}
\mbox{[} b^\ds_j, b^\dagger_{j^{\prime}}\mbox{]} =
\delta_{j j^{\prime}} (1 - 2 b^\dagger_j b^\ds_j), \qquad 
\label{hard-core-boson-3-app}
\mbox{[} b^\dagger_{j} b^\ds_{j},  
b^\dagger_{j'} \mbox{]} &=& \delta_{j j'} b^\dagger_j \; .
\end{eqnarray}
We seek eigenstates that contain
$n$ pairs,  satisfy $(H_\U - \E_n) |\Psi_n \rangle$
and are of the general form 
\begin{equation}
  \label{eq:general-eigenstates-app}
 |\Psi_n \rangle  =  \sum_{j_1, \dots, j_n}^\U
\psi (j_1, \dots , j_n) \prod_{\nu=1}^n b_{j_\nu}^\dagger  
|\Vac \rangle \; .
\end{equation}
In his original publications \cite{richardson63a,richardson63b,richardson64},
Richardson derived a Schr\"odinger equation for $\psi (j_1, \dots , j_n)$ and
showed that its exact solution was simply a generalization of the form that
$\psi (j_1, \dots , j_n)$ would have had if the $b$'s had been true (not
hard-core) bosons.  With the benefit of hindsight, we shall here follow an
alternative, somewhat shorter root, also due to Richardson
\cite{richardson-private-99}: we first consider the related but much simpler
case of {\em true\/} bosons and write down the generic form of its
eigenstates; we then clarify why this form fails to produce eigenstates of the
{\em hard-core\/} boson Hamiltonian; and
having identified the reason for the failure, we show that (remarkably)
only a slight generalization is needed to repair it and to obtain the
sought-after hard-core-boson eigenstates.

\subsubsection{True bosons}
Let $\tilde b_j$ denote a set of true bosons (i.e.\
$[ \tilde b_j, \tilde b_{j'}^\dagger] = \delta_{j j'}$), 
governed by a Hamiltonian $\tilde H_\U$ of precisely the form
(\ref{1}), with  $b_j \to \tilde b_j$. 
 This problem, being quadratic, can be solved straightforwardly by
any  number of methods. The solution is as follows: 
$\tilde H_\U$ can be written as
\begin{eqnarray}
  \label{eq:trueosons}
  \tilde H_\U = \sum_J \tilde E_J \tilde B_J^\dagger \tilde B_J  + \mbox{const.}
\end{eqnarray}
where the new bosons $\tilde B_J^\dagger$ 
(with normalization constants $C_J$) are given by 
\begin{eqnarray}
  \label{eq:newbosons}
 \tilde B^\dagger_J 
= g C_J \sum_j^\U {\tilde b^\dagger_j \over 2 \varepsilon_j -  \tilde E_J}
 \; , \qquad
  \label{eq:normalization}
  {1 \over (g C_J)^2} = 
\sum_j^\U {1 \over (2 \varepsilon_j -  \tilde E_J)^2} \; ,
\end{eqnarray} 
and the boson eigenenergies $ \tilde E_J$ are the roots of the eigenvalue
equation
\begin{eqnarray}
  \label{eq:eigenvalue}
  1 - \sum_j^\U {g \over 2 \varepsilon_j -  \tilde E_J} \; = \; 0 \; .
\end{eqnarray}
This is an equation of order $N_\U$ in $ \tilde E_J$. It thus has $N_\U$
roots, so that the label $J$ runs from 1 to $N_\U$.  As the coupling $g$ is
turned to 0, each $E_J$ smoothly evolves to one of the bare eigenenergies $2
\varepsilon_j$.  A general $n$-boson eigenstate of $\tilde H_\U$ and its
eigenenergy $\tilde {\cal E}_n$ thus have the form
\begin{eqnarray}
  \label{eq:truebosoneigenstates-1}
  |\tilde \Psi_n \rangle = \prod_{\nu=1}^n \tilde 
B_{J_\nu}^\dagger |0\rangle \, ,
\qquad
\label{Etotal} 
\tilde {\cal E}_n = \sum_{\nu=1}^n  \tilde E_{J_\nu} \; ,
\end{eqnarray} 
where the $n$ indices $J_1, \dots, J_n$ that characterize this state need
not all be distinct, since the $B_J^\dagger$ are true bosons.

\subsubsection{Complications arising for hard-core bosons}
Let us now return to the hard-core boson Hamiltonian $H_\U$.  Its
eigenstates will obviously {\em not\/} be identical to the true-boson
eigenstates just discussed, since matters are changed considerably by
the hard-core properties of $b_j$.  To find out exactly {\em what\/}
changes they produce, it is very instructive to take an Ansatz for $|
\Psi_n \rangle$ similar to (\ref{eq:truebosoneigenstates-1}) (but
suppressing the normalization constants and taking all 
boson indices to be
distinct), namely
\begin{eqnarray}
  \label{eq:truebosoneigenstates}
  | \Psi_n \rangle = \prod_{\nu=1}^n 
B_{\nu}^\dagger |0\rangle \, ,
\qquad \mbox{with} \qquad 
  B_\nu^\dagger = \sum_j^\U {b_j^\dagger \over 2 \varepsilon_j - E_\nu} \; ,
\end{eqnarray} 
and to check explicitly whether or not it could be an eigenstate of
$H_\U$, i.e.\ to check under what conditions $(H_\U - {\cal E}_n)
|\Psi_n \rangle $ would equal zero, where ${\cal E}_n = \sum_{\nu}^n
E_{\nu}$.  To this end, we commute $H_\U$ to the right past all the
$B^\dagger_{\nu}$ operators in $|\Psi_n \rangle$, using
\begin{eqnarray}
  \label{eq:commuteHtofront-1}
 \left[ H_\U,  \prod_{\nu = 1}^{n} B_{\nu}^\dagger  \right] =
\sum_{\nu = 1}^n \left\{ 
\left( \prod_{\eta = 1}^{\nu -1} B_{\eta}^\dagger \right)
[H_\U,  B_{\nu}^\dagger ] 
\left( \prod_{\mu = \nu+1}^n B_{\mu}^\dagger \right) \right\} .
\end{eqnarray}
To evaluate the commutators appearing here, we write
$H_\U$ as 
\begin{eqnarray}
  \label{eq:checkAnsatztrueboson}
  H_\U = \sum_{j}^\U 2 \varepsilon_j  b_j^\dagger b_j \; - \;
   g B^\dagger_0  B_0 \; , \qquad
 \mbox{where} \quad B^\dagger_0 = \sum_j^\U  b^\dagger_j \; , 
\end{eqnarray}
and use the following relations:
\begin{eqnarray}
  \label{eq:bjBJ}
  [b_j^\dagger b_j, B^\dagger_\nu] = 
     {b_j^\dagger \over    2 \varepsilon_j - E_\nu} \; , \qquad
  [B_0, B^\dagger_\nu] = \sum_j^\U
     {1- 2 b_j^\dagger b_j \over    2 \varepsilon_j - E_\nu} \; , 
\end{eqnarray}
\begin{eqnarray}
\label{HUBJ}
  [H_\U, B^\dagger_\nu] =  E_\nu B_\nu^\dagger \; + \;
     B_0^\dagger \left[ 1 - g \sum_j^\U {1 - 2 b_j^\dagger b_j 
  \over 2 \varepsilon_j - E_\nu}    \right] \; .
\end{eqnarray}
Inserting these into (\ref{eq:commuteHtofront-1}) and using $H_\U
|0\rangle =0$ and ${\cal E}_n = \sum_{\nu}^n E_{\nu}$, we find
\begin{eqnarray}
\nonumber
\lefteqn{
  H_\U |\Psi_n\rangle  =   {\cal E}_n |\Psi_n\rangle 
+ \sum_{\nu=1}^n \left[ 1 - \sum_j^\U 
{g  \over 2 \varepsilon_j - E_{\nu}}    \right]
\! B_0^\dagger \! 
\left( \prod_{\eta =1 (\neq \nu)}^{n} B_{\eta}^\dagger \right) |0\rangle} 
\qquad \qquad \qquad \phantom{.} \\
  \label{eq:commuteHtofront-2}
& & \phantom{.} \hspace{-1.5cm} 
+ \sum_{\nu=1}^n
\left\{ \left( \prod_{\eta = 1}^{\nu -1} B_{\eta}^\dagger \right)
\left[   \sum_j^\U { 2 g B_0^\dagger  \, b_j^\dagger b_j 
  \over 2 \varepsilon_j - E_{\nu} }    \right] 
\left( \prod_{\mu = \nu+1}^n B_{\mu}^\dagger \right) \right\} |0\rangle \; .
\end{eqnarray}
Now, suppose we do the same calculation for true instead of hard-core
bosons (i.e.\ run through the same steps, but place a $\tilde
{\phantom{.}}$ on $H_\U$, $b_j$,  $E_\nu$ and ${\cal E}_n$). 
Then the second line of
(\ref{eq:commuteHtofront-2}) would be absent (because the
$b_j^\dagger b_j$ terms in the second of
Eqs.~(\ref{hard-core-boson-2-app}) and (\ref{eq:bjBJ})
and in (\ref{HUBJ}) would be absent); and 
the first line of (\ref{eq:commuteHtofront-2}) would imply that 
$(\tilde H_\U - \tilde {\cal   E}_n) |\tilde \Psi_n\rangle = 0$
provided that the term in square brackets vanishes, 
which is nothing but the condition that  the $\tilde E_\nu$
satisfy the 
the true-boson eigenvalue equation of
(\ref{eq:eigenvalue})! In other words, we have just verified
explicitly that all true-boson states of the form
(\ref{eq:truebosoneigenstates-1}) are indeed eigenstates of $\tilde
H_\U$, provided that the $\tilde E_\nu$ satisfy (\ref{eq:eigenvalue}).
Moreover, we have identified the term in second line of
(\ref{eq:commuteHtofront-2}) as the extra complication that arises for
hard-core bosons.

\subsubsection{The cure: a generalized eigenvalue equation}

Fortunately, this extra complication is tractable:
first, we note that 
\begin{eqnarray}
  \label{eq:newtermBJ}
  \left[   \sum_j^\U { 2 g B_0^\dagger \, b_j^\dagger b_j 
  \over 2 \varepsilon_j - E_{\nu}} , B^\dagger_{\mu}   \right]
& = & \sum_j^\U {  2 g B_0^\dagger 
  \over 2 \varepsilon_j - E_{\nu}} 
 {  b_j^\dagger \over 2 \varepsilon_j - E_{\mu}}
\\
 \label{eq:newtermBJ-2}
&=& 2  g B_0^\dagger \, {B_{\nu}^\dagger - B_{\mu}^\dagger
  \over  E_{\nu} -  E_{\mu}}  .
\end{eqnarray}
The second line follows via a partial fraction expansion,
and remarkably, contains only $B^\dagger$ operators
and no more $b^\dagger_j b_j$s. This enables
us to eliminate the  $b^\dagger_j b_j$s from 
the second line of (\ref{eq:commuteHtofront-2}), by
rewriting it  as follows (we commute its
term in square brackets to the right, using a
relation similar to (\ref{eq:commuteHtofront-1}),
but with the commutator (\ref{eq:newtermBJ}) instead
of $[H_\U,  B_{\mu}^\dagger]$):
\begin{eqnarray}
&& 
\sum_{\nu=1}^n \!
\left\{ \! \! \left( \prod_{\eta = 1}^{\nu-1} B_{\eta}^\dagger \right)\!\!
\sum_{\mu = \nu +1}^n \! \!
\left\{ \! \! 
\left( \prod_{\eta' = \nu + 1}^{\mu -1} \!\! B_{{\eta'}}^\dagger \right) \!\!
\left[   2  g B_0^\dagger \, {B_{\nu}^\dagger - B_{\mu}^\dagger
  \over  E_{\nu} -  E_{\mu}} \right] \!\!
\left( \prod_{\mu' = \mu +1}^n \!\! B_{{\mu'}}^\dagger
\right) \! \! \right\} \! \! \right\}  \!
|0\rangle \nonumber \\
&& = 
\sum_{\mu = 1}^n \! \!
\left[ \sum_{\nu = 1}^{\mu - 1} {2g \over E_{\nu} - E_{{\mu}}} \right]
\! B_0^\dagger \! \left( \prod_{\eta =1 (\neq  \mu)}^{n} 
\!\! B_{\eta}^\dagger \right) |0 \rangle \nonumber
\\
&& \qquad  - \sum_{\nu = 1}^n \! \!
\left[ \sum_{\mu = \nu+1}^n {2g \over E_{\nu} - E_{{\mu}}} \right]
\! B_0^\dagger \! \left( \prod_{\eta =1 (\neq  \nu)}^{n} 
\!\! B_{\eta}^\dagger \right)  |0 \rangle   \nonumber \\
&&= 
   \sum_{\nu = 1}^n \! \!
\left[ \sum_{\mu =1 (\neq \nu)}^n {2g \over E_{\mu} - E_{{\nu}}} \right]
B_0^\dagger \left( \prod_{\eta =1 (\neq  \nu)}^{n} 
\!\! B_{\mu}^\dagger \right) |0 \rangle  \; . 
  \label{eq:commuteHtofront-3}
\end{eqnarray}
(The last line follows by renaming the dummy indices
$\nu \leftrightarrow \mu$ in the second line.)
Substituting (\ref{eq:commuteHtofront-3}) for the second line of
(\ref{eq:commuteHtofront-2}),  we conclude that $ ( H_\U - {\cal E}_n)
|\Psi_n\rangle$ will equal zero provided that
\Eq{eq:richardson-eigenvalues-cc} holds, namely: 
\begin{eqnarray}
  \label{eq:richardson-eigenvalues}
  1 - \sum_j^\U 
{g  \over 2 \varepsilon_j - E_{\nu}}    
+ \sum_{\mu =1 (\neq \nu)}^n {2g \over E_{{\mu}} - E_{{\nu}}} = 0 \; ,
\qquad \mbox{for}\quad \nu = 1, \dots, n \; .
\end{eqnarray}
This consitutes a set of $n$ coupled equations for the $n$ parameters
$E_{1}, \dots,$ $E_{n}$, which may be thought of as
self-consistently-determined pair energies.
Eq.~(\ref{eq:richardson-eigenvalues}) can be regarded as a
generalization of the true-boson eigenvalue equation
(\ref{eq:eigenvalue}). It is truly remarkable that the exact
eigenstates of a complicated many-body problem can be constructed by
such a simple generalization of the solution of a quadratic (i.e.\ 
non-interacting) true-boson Hamiltonian!

\subsection{Solving the eigenvalue equation numerically}
\label{app:electrostatic-analogy}

In this section we discuss some algebraic transformations invented by
Richardson to render \Eqs{eq:richardson-eigenvalues-cc} or
(\ref{eq:richardson-eigenvalues}) less singular and thereby simplify
their numerical solution.

First note that the solutions of \Eqs{eq:richardson-eigenvalues-cc} can
be given an electrostatic interpretation \cite{gaudin,richardson77},
since they are the extrema w.r.t. $E_\nu$ of the function
\begin{equation}
  \label{eq:electrostatic-analogy}
  W [ \{ E_\nu \} ] \equiv  \sum_\nu  {E_\nu \over 2 g}
+ {1 \over 2} \sum_\nu \sum_{j}^{U} \log | 2 \varepsilon_j - E_\nu|
 - \sum_{\nu < \mu} \log |E_\mu - E_\nu |  ,
\end{equation}
where $E_\nu$ is regarded as a complex number, say $E_{\nu x} + i
E_{\nu y}$, and $|E_\nu | = \sqrt{ E_{\nu x}^2 + E_{\nu y}^2}$, etc.
$W[ \{ E_\nu \} ] $ can thus be interpreted as the potential energy of
$n$ free unit charges at the locations $(E_{\nu x} , E_{\nu y})$ in
the $xy$-plane (actually lines of charge perpendicular to the plane),
under the combined influence of a uniform external field $- (1/2g)
\hat x$ parallel to the $x$-axis, and the field of a set of fixed
charges of strength $-1/2$, located at the points $2 \varepsilon_j$ on
the $x$-axis. The task at hand is to find the (unstable) equilibrium
positions of $n$ free unit charges.

We shall consider only the case that all the $\varepsilon_j$ are
non-degenerate (degeneracies are considered in \cite{gaudin}).  Then every
solution of \Eqs{eq:richardson-eigenvalues-cc}, \ie\ every set $\{ E_\nu \}$,
coincides at $g=0$ with a set of $n$ bare energies, say $\{ 2
\varepsilon_{j_\nu} \}$ (by inspection; or in the electrostatic analogy, the
field $-1/2g$ is so strong that the only stable configurations are those where
each positive unit charges sits infinitely close to some fixed $-1/2$ ``host''
charge).  Richardson showed \cite{richardson66} that as $g$ is turned on, all
$E_\nu$'s initially smoothly evolve toward lower values (as the field weakens,
the free charges leave their hosts, pulled in the negative $x$-direction by
the field); but as $g$ increases further, singular points are reached at which
two roots approach both each other (and a bare energy), say $E_\mu = E_\nu (=
2 \varepsilon_{\bar \jmath})$, and then turn complex, becoming a complex
conjugate pair, so that ${\cal E}_n$ remains real (as the field weakens some
more, the repulsion between the positive charges begins to become important;
it tends to push some of them off the $x$-axis into the $xy$ plane, but does
so without breaking reflection symmetry about the $x$-axis).

Due to this complication, it is convenient to parametrize
the roots in terms of purely real variables, proceeding as follows:
Denote roots destined to become conjugates by 
$(E_{2a-1},E_{2a})$ [with $g=0$ values
$(\varepsilon_{j_{2a-1}},\varepsilon_{j_{2a}})$, say], with $a = 1,
\dots n/2$ for even $n$, with one further purely real root, say
$E_{0}$, for odd $n$.  Let us write $E_{2a-1} = \xi_a - i \eta_a$,
$E_{2a} = \xi_a + i \eta_a$, where $\xi_a$ and $\eta_a^2$ are purely
real, with $\eta^2_a \llogg 0$ for $g \llogg g_a$, the critical value
where this pair of roots becomes complex. Rewriting  Richardson's
eigenvalue equation (\ref{eq:richardson-eigenvalues-cc})
in terms of these purely real variables by adding and subtracting
the equations for $\nu = 2a -1$ and $\nu = 2a$, one readily finds,
for each $a$, 
\begin{eqnarray}
 \nonumber
& & {1 \over g} - \sum_{j \neq 2a, 2a-1}^U 
{ 2 \varepsilon_j - \xi_a \over (2 \varepsilon_j - \xi_a)^2
+ y_a(x^2_a - \rho^2_a)}  -
{ 2 x_a (1 + y_a) \over \rho_a^2 (1 - y_a)^2 - x_a^2 (1+y_a)^2} 
\\  \label{eq:non-singular-richardson-1}
& & \qquad + {2 (E_0 - \xi_a) \over (E_0 - \xi_a)^2 + \eta_a^2}
+ \sum_{b \neq a}^{n/2} { 4 \xi_{ba} (\xi^2_{ba} + \eta^2_b +
  \eta^2_a) \over 
(\xi^2_{ba} + \eta^2_b + \eta^2_a)^2 - 4 \eta^2_b \eta^2_a} = 0 \; ,
\\ \nonumber
& & {(1 - y_a^2) \over \rho_a^2(1-y_a)^2 - x_a^2(1 + y_a)^2}
+ \sum_{j \neq 2a, 2a-1} {y_a \over
(2 \varepsilon_j - \xi_a)^2 + \eta_a^2} 
\\  \label{eq:non-singular-richardson-2}
& & \qquad - {y_a \over (E_0 - \xi_a)^2 + \eta_a^2 }
- \sum_{b \neq a} { (\xi^2_{ba} - \eta_b^2 + \eta_a^2) 4 y_a \over
(\xi^2_{ba} + \eta_b^2 + \eta_a^2)^2 - 4 \eta_b^2 \eta_a^2 } = 0 \; , 
\end{eqnarray}
where we introduced the further real variables
\begin{eqnarray}
  \label{eq:further-real-variables}
x_a & = &  \xi_a - \varepsilon_{2a-1} 
- \varepsilon_{2a} \; , 
\qquad y_a =  \eta_a^2 /(x_a^2 - \rho_a^2) \; , \\
\rho_a & = & \varepsilon_{2a} - \varepsilon_{2a-1} \; ,
\quad \hspace{1.5cm} \xi_{ba} \equiv \xi_b - \xi_a \; ,
\end{eqnarray}
and factored out an overall factor of $(x_a^2 - \rho_a^2)$ from
\Eq{eq:non-singular-richardson-2}. Moreover, the terms involving
$E_0$ occur only if $n$ is odd and should be omitted if $n$ is even.

Compared to \Eqs{eq:richardson-eigenvalues-cc},
\Eqs{eq:non-singular-richardson-1} and
(\ref{eq:non-singular-richardson-2}) have the advantage that their
roots are always real and that no singularities occur any more. They
can thus be solved straightforwardly for the $x_a$'s and $y_a$'s
by standard numerical techniques
(\eg\ Broyden's algorithm \cite{broyden} in the form given in
\Ref{numerical-recipes}), using the set ${\cal R} = \{
(\varepsilon_{j_{2a-1}},\varepsilon_{j_{2a}}),\varepsilon_{j_0} \}$ as
``initial solution''. Note, however, that the choice of initial
pairings in ${\cal R}$ is crucial: an incorrect choice results in the
failure of solutions with real $x_a$, $y_a$ to exist beyond a certain
$g$-value. Since making this choice requires knowledge of which
roots end up as conjugate pairs, some trial and error may be involved
in finding the correct ${\cal R}$ (see, \eg, Ref.~15 of \cite{sierra99}).

\subsection{Correlation functions}
\label{app:correlators}

In \Ref{richardson65b}, Richardson derived 
the following explicit results
for the normalization constant ${\cal N}$ of
(\ref{eq:truebosoneigenstates-cc}) and the occupation probabilities
$\bar v_j^2$ and correlators $C_{ij}$ of \Eq{eq:C_ij}:
\begin{eqnarray}
\label{eq:normalization-result}
  {\cal N} & = &  \mbox{[} \mbox{det}(M)\mbox{]}^{-1/2} \; , 
\\
  \label{eq:occupation-probability}
  \bar v^2_j & \equiv & \langle \Psi_n | b^\dag_j b^\ds_j | \Psi_n \rangle  = 
  \sum_\mu \alpha_\mu E_{j \mu}^2 \; , 
\qquad \alpha_\mu \equiv - g^2 ({\d} E_\mu / {\d} g) \; , 
\\
  \label{eq:Cij-correlators}
  C_{ij} & = & {\cal N}^2 \sum_{\mu \nu} 
 E_{i \mu} E_{j \nu} I_{\mu \nu}(ij) \; , 
\end{eqnarray}
where we used the following abbreviations:
\begin{eqnarray}
  \label{eq:normalization-matrix}
   M_{\mu \nu} & = & \left\{ \begin{array}{ccl}
                 C_{\mu^2} - 2 S_{\mu^2} \qquad & \mbox{for} & \mu = \nu
\\
                 2 E_{\mu \nu}^2 \qquad & \mbox{for} & \mu \neq \nu \; , 
\end{array} \right. 
\\
  \label{eq:abbreviations}
  E_{j \mu} & \equiv & (2 \varepsilon_j  - E_\mu)^{-1} \; , 
\qquad
  E_{\mu \nu}  \equiv  (E_\mu - E_\nu)^{-1} \; , 
\\
  C_{\mu^2} &  \equiv & \sum_j E_{j \mu}^2 \; , 
\qquad 
  S_{\mu} \equiv \sum_{\eta }^{n \backslash \mu} E_{ \eta \mu} \; ,
\qquad
  S_{\mu^2} \equiv \sum_{\eta }^{n \backslash \mu} E^2_{ \eta \mu} \; ,
\\ \label{eq:I_{11'}} 
I_{1 1'}(ij) &  \equiv & S'_{n-1}
 \left| \begin{array}{cccccc}
1 \; 
& \; 2 E_{12} E_{1' 2'} \; & \; \dots \; & \;
 2 E_{1 n} E_{1' n'} \\
1 \; & \; \; \tilde D_{22'}(ij)  - 2 S_2 S_{2'} \; \;
& 
 \; \dots & 2 E_{2 n} E_{2' n'} \\
\vdots & & & 
\\
1 \;  & \;  2 E_{ n2 } E_{ n' 2'} \;  & \; \dots \; & \; \;
 \tilde D_{nn'} (ij) - 2 S_n S_{n'} 
\end{array} \right|
\\
\tilde D_{\mu \nu} (ij) & \equiv & D_{\mu \nu} +  ( E_{i \mu} E_{j\nu} 
- E_{i \mu} E_{i \nu} - E_{j \mu} E_{j \nu}) \; ,
\\
D_{\mu \nu}  & \equiv &  \sum_j E_{j \mu} E_{j \nu} = - 4 E_{\mu \nu}^2  +
\sum_{\eta}^{n \backslash \mu,\nu} 2 E_{\eta \mu} E_{\eta\nu} \; , 
\end{eqnarray}
Here $M$ is a matrix with matrix indices $\mu, \nu$.  The notation
$\sum_{\mu}^{n\backslash \nu}$ means a sum over all $\mu = 1, \dots, n$,
excluding $\nu$.  The coefficients $\alpha_\mu $ in
\Eq{eq:occupation-probability} can most conveniently be found by solving the
set of $n$ algebraic equations
\begin{eqnarray}
  \label{eq:alpha-equations}
  (C_{\mu^2} - 2 S_{\mu^2} ) \alpha_\mu +  
  \sum_{\eta }^{n \backslash \mu} E_{\eta \mu}^2 \alpha_\eta 
  = 1 \; , \qquad (\mu = 1, \dots, n) \; ,
\end{eqnarray}
which follow from differentiating \Eq{eq:richardson-eigenvalues-cc}
with respect to $g$.  In \Eq{eq:I_{11'}}, $S'_{n-1}$ is an operator
which symmetrizes the $(n-1)$ primed indices $ \eta' = 2', \dots , n'$
occuring in the determinant for $I_{1 1'} (ij)$, and then sets $\eta'
= \eta$; in other words, $S'$ stands for the sum over the $(n-1)!$
permutations of these indices, and for $\eta' = 2', \dots n'$, the
prime on $\eta'$ is only a mark to distinguish it from $\eta$ for the
purposes of symmetrization.  (To obtain $I_{\mu \mu'} (ij)$, simply
make the replacements $1 \leftrightarrow \mu$ and $1' \leftrightarrow
\mu'$ in the determinant written in \Eq{eq:I_{11'}} for $I_{1 1'}
(ij)$, then symmetrize over the indices $\eta' = 1', \dots n'$,
excluding $\mu'$.) The need to evaluate $(n-1)!$ permutations
makes $C_{ij}$  practically impossible to calculate numerically for
$n$  much larger than, say, 10. However,  Richardson showed
that \Eq{eq:Cij-correlators} simplifies to the (tractable) 
\begin{equation}
  \label{eq:Cij-shortcut}
  C_{ij}  \simeq   \sum_{\mu \nu} \alpha_\mu 
 E_{i \mu} E_{j \mu}  \; ,
\end{equation}
if $\tilde D_{\mu \nu} (ij)$ is approximated by $D_{\mu \nu}$ in
\Eq{eq:I_{11'}}.  The accuracy of this approximation can be jugded by
using the fact \cite{richardson66} that when used to evaluate ${\cal
  E}_n - \langle \Psi_n | H_U | \Psi_n \rangle$, it produces, instead
of zero, the result $ (1/g) \sum_\nu^n \alpha_\nu [g^2 ( \sum_{j}^U
E_{j \nu} )^2 - 1 ] $, which can easily be evaluated numerically.

\newpage

\begin{ack}

We have benefitted from discussions and collaborations
with a great many physicists. We would like to thank
F. Braun,
P. Brouwer,
D. Davidovi\'c,
L. Glazman,
M. Greiter, 
R. Rossignoli,
M. Schechter,
G. Sch\"on and
X. Waintal
for reading and commenting on various parts of the draft. 
We acknowledge fruitful collaborations with
O. Agam,
B. Altshuler,
E. Bonet Orozco,
R. Buhrman,
C. Black,
F. Braun, 
P. Brouwer,
M. Deshmukh, 
C. Dobler,
J. Dukelsky, 
G. Dussel, 
D. Golubev, 
S. Gu\'eron,
S. Kleff,
J. Kroha,
E. Myers,
A. Pasupathy,
J. Petta,
M. Pirmann, 
D. Salinas,
M. Schechter, 
G. Sch\"on,
G. Sierra, 
W. Thimm, 
W. Tichy, 
M. Tinkham, 
N. Wingreen,
A. Zaikin.
Furthermore, we acknowledge  helpful and stimulating discussions with
I. Aleiner,
V. Ambegaokar,
S. Bahcall,
Y. Blanter,
S. B\"ocker,
C. Bruder,
T. Costi,
F. Evers,
G. Falci, 
R. Fazio,
P. Fulde,
A. Garg,
L. Glazman,
B. Halperin,
C. Henley,
J. Hergenrother,
M. Itzler,
B. Janko, 
J. K\"onig,
A. Larkin,
K. Likharev,
A. MacDonald,
A. Mastellone,
K. Matveev,
A. Mirlin,
Y. Oreg,
T. Pohjola, 
A. Rosch,
J. Siewert,
R. Smith,
F. Wilhelm,
P. W\"olfle,
G. Zar\'and,
F. Zawadowski
and W. Zwerger. 

J.v.D. was supported in part by the Deutsche Forschungsgesellschaft
through Sonderforschungsbereich 195, and also acknowledges support
from the DFG-Pro\-gram, ``Semiconductor and Metallic Clusters'',
and from the DAAD-NSF.  
D.R. was supported in part by the grants ONR (N00014-97-1-0745), NSF
(DMR-9705059) and by the MRSEC program of the NSF (DMR-0632275).

\end{ack}

\end{document}